\documentclass[12pt]{gsasthesis} 

\usepackage{etex} 

\usepackage[margin={1.2in}]{geometry}

\usepackage[titletoc]{appendix}

\usepackage{rotating}

\usepackage{longtable}

\usepackage{natbib}
\usepackage{bibentry}
\nobibliography*

\usepackage{pdfpages}

\usepackage[utf8x]{inputenx}

\usepackage[polutonikogreek,english]{babel}

\usepackage{amsmath}

\usepackage{deluxetable}

\usepackage{epigraph}

\makeatletter
\newcommand\footnoteref[1]{\protected@xdef\@thefnmark{\ref{#1}}\@footnotemark}
\makeatother

\usepackage{mathabx}

\usepackage{mathrsfs}

\usepackage{microtype}

\usepackage[stable,multiple]{footmisc}

\RequirePackage[font=small,format=plain,labelfont=bf,textfont=it]{caption}
\usepackage{aas_macros}

\newlength{\halffigurewidth}
\newlength{\threequartersfigurewidth}
\newlength{\figurewidth}
\newcommand{\original}[1]{This thesis chapter has been originally published as
\bibentry{#1}.}
\newcommand{\publishedarxiv}[1]{This thesis chapter has been originally
published as an arXiv e-print, and has also been published after the submission
of this thesis as \bibentry{#1}.}
\newcommand{\publishedafter}[1]{This thesis chapter has been published after
the submission of this thesis as \bibentry{#1}.}
\newcommand{\chapterabstract}{\section*{Abstract}}
\newcommand{\acknowledgements}{\paragraph{Acknowledgements}}

\newcommand{\ads}{ADS 16402}
\newcommand{\apriori}{\textit{a priori}}
\newcommand{\arcsec}{\ensuremath{''}}
\newcommand{\circleangle}{\texttt{circleangle}}
\newcommand{\ellipseangle}{\texttt{ellipseangle}}
\newcommand{\hatpelevenb}{HAT-P-11b}
\newcommand{\hatpeleven}{HAT-P-11}
\newcommand{\hatponeb}{HAT-P-1b}
\newcommand{\hatpone}{HAT-P-1}
\newcommand{\hatptwentysevenb}{HAT-P-27b}

\newcommand{\hdegyb}{HD 189733b}
\newcommand{\hdegy}{HD 189733}
\newcommand{\hdkettob}{HD 209458b}
\newcommand{\hdketto}{HD 209458}
\newcommand{\hipparcos}{\textit{Hipparcos}}
\newcommand{\integratetransit}{\texttt{integratetransit}}
\newcommand{\keplerseventeenb}{Kepler-17b}
\newcommand{\keplerseventeen}{Kepler-17}
\newcommand{\kepler}{\textit{Kepler}}
\newcommand{\keplerthirteenb}{Kepler-13b}
\newcommand{\keplerthirteen}{Kepler-13}
\newcommand{\macula}{\texttt{macula}}
\newcommand{\prism}{\textsc{prism}}
\newcommand{\python}{\texttt{Python}}
\newcommand{\rmeffect}{Rossiter--Mc\-Laugh\-lin effect}
\newcommand{\spotrod}{\texttt{spotrod}}
\newcommand{\strobo}{\textit{stroboscopic effect}}
\newcommand{\sun}{\ensuremath\odot}
\newcommand{\tauboo}{\ensuremath{\tau} Boo}
\newcommand{\Teq}{\ensuremath{T_\mathrm{eq}}}
\newcommand{\totalspots}{203}
\newcommand{\xs}{\ensuremath{x_\mathrm s}}
\newcommand{\ys}{\ensuremath{y_\mathrm s}}
\newcommand{\zcrit}{\ensuremath{z_\mathrm{crit}}}

\title{Development and Application of Tools \\ to Characterize Transiting Astrophysical Systems}
\author{Bence Béky}
\degreename{Doctor of Philosophy}
\degreefield{Astronomy and Astrophysics}
\department{The Department of Astronomy} 
\degreemonth{April} 
\degreeyear{2014} 
\principaladvisor{Dr.~Matthew Holman}

\begin{document}


\pagenumbering{roman} 

\includepdf{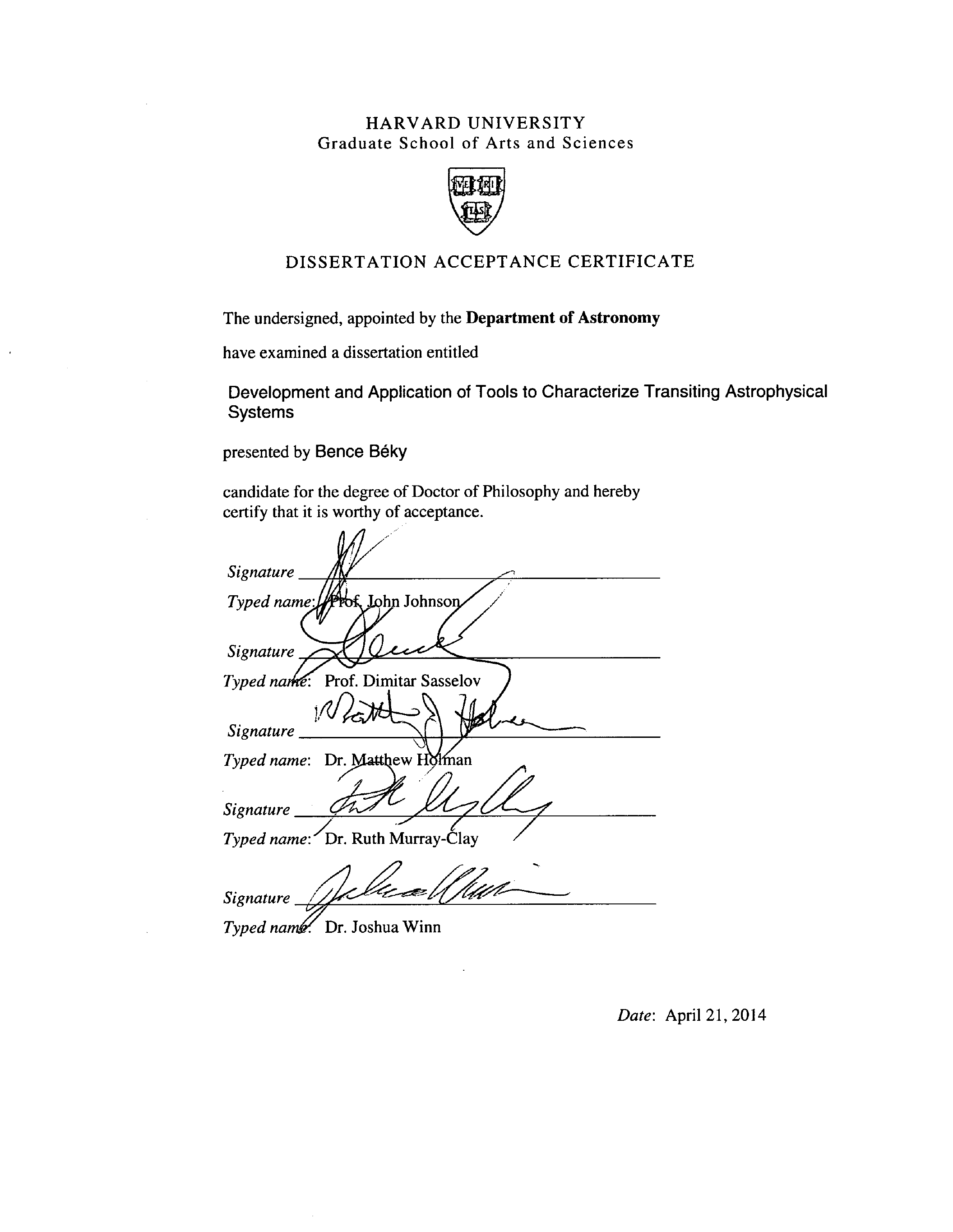}
\intentionallyblank
\thesistitlepage
\copyrightpage
\begin{abstract}
Since the discovery of the first exoplanets (planets outside our Solar System) more than 20 years ago, there has been an increasing need for photometric and spectroscopic models to characterize these systems.  While imaging has been used extensively for Solar System bodies and extended objects like galaxies, the small angular extent of typical planetary systems makes it difficult or impossible to resolve them.  Spatially integrated observations like measuring the total brightness or spectrum, however, can be conducted at a resonable cost.  This thesis focuses on photometric models in the context of transiting systems, which exhibit a number of phenomena that can be exploited for characterization.

First, we showcase the popular methods of transiting exoplanet discovery and characterization by ground based observations on the hot Jupiter \hatptwentysevenb{}.  We demonstrate how transits allow us to constrain planetary mass, radius, and orbital inclination, which would not be possible based only on, for example, radial velocity measurements.

Next, we perform reflection spectroscopy on \hatponeb{}, another hot Jupiter, using the binary companion of the host star as a reference to remove systematic errors from the signal.  Here the transiting nature of the system allows us to look for the very faint light reflected by the planet.

We also apply the idea of planetary transits to investigate the feasibility of transit observations in astrophysical systems of very different scale: stars in galactic nuclei potentially transiting the accretion disk of the supermassive black hole in the galactic center.

Finally, we focus on mapping spots on the stellar surface using transits.  This method has been used for a decade, and helped constrain stellar rotation or orbital geometry in a number systems.  We study starspots on \hatpeleven{} not only to learn more about stellar rotation, but also to investigate the size and contrast of the spots themselves.
\end{abstract}

\renewcommand{\contentsname}{\protect\centering\protect\Large Contents}
\renewcommand{\listtablename}{\protect\centering\protect\Large List of Tables}
\renewcommand{\listfigurename}{\protect\centering\protect\Large List of Figures}

\tableofcontents 

\listoftables
\listoffigures
\begin{acknowledgments}

I shall start by giving due credit to people who helped me on the journey
before I started my career as an astronomer.  I thank my parents, my sisters,
and other members of my extended but close family; fellow students and
educators during my formal and informal education, especially high school; and
friends.  These people play a key role in my life, and are responsible for
that when I enrolled to Harvard four years ago, I felt I was ready.

I would like to express gratitude to a large number of astronomers who
accepted my younger self with no formal training in their discipline as one of
them, and helped me along my apprenticeship.  First of all, I thank Gáspár
Bakos, who recruited me as a research assistant and introduced me to
exoplanets and people in this field.  I thank Matt Holman for taking over the
role of my advisor after Gáspár left for another university and was not
successful in convincing me to leave the CfA behind. 

I am also grateful to a large number of people who volunteered their time and
advice without being labeled advisors: Dave Kipping and Bob Noyes, with whom I
had regular weekly meetings throughout the last two years, Bence Kocsis, who
gave me the opportunity to lead the research paper based on his idea, and
many, many other people, whom I had the opportunity to consult with
regularily or occasionally.  This includes collaborators and coauthors,
mentors, lecturers, and people I met at SSP coffee, exoplanet pizza lunch, and
other scientific gatherings.  I am especially indebted to the Mount Hopkins
Observatory staff for their infinite patience with me as I made my debut as an
observer.

When one thinks about whom they learned the most from, it is easy to think
only about senior people and forget about peers.  However, my confusion was
often so great that I would have been ashamed to seek help of anyone other
than the folks who still remember vividly their own struggles with astronomy:
fellow graduate students.  They did not leave a single question of mine
unanswered, and every interaction left me humbled by how much more there
was for me to learn.

I would like to acknowledge the hard work of Peg Herlihy, Robb Scholten, and
Donna Adams, without whom I would not have had an office, an advisor, stipend,
healthcare, I could not have taken a single class, I could not have taught, I
could not have loaned a single book from the library, I could not have entered
the building after 17:30, and I would have been deported from the US, just to
name a few things.

Being an astronomer is about more than just astronomy, and being a graduate
student is about more than just being in graduate school.  I am grateful for
all the people who kept reminding me of this in the past years, including
friends that I met through CouchSurfing, through the local Hungarian
community, and through other stochastic processes; fellow graduate students
who showed me that they also have a non-astronomer side; but most of all,
Katherine Beaty, who has earned a special place in my heart and on my ring
finger.  And I thank Owen Gingerich and Charles Alcock for making it possible
that she put that ring on my finger in the dome of the Great Refractor.

\medskip
\begin{center}
$\heartsuit$
\end{center}
\medskip

\epigraph{So long, and thanks for all the fish}{Douglas Adams (1984)}

\end{acknowledgments}

\begin{dedication}
To my family
\end{dedication}

\pagenumbering{arabic} 


\chapter{Introduction}
\label{ch:intro}

\section{History of Solar System planets}

\epigraph{History always has a few tricks up its frayed sleeve.  It's been
around a long time.}{Terry Pratchett (1987)} 

The first planets known to humanity are those visible to the naked eye:
Mercury, Venus, Mars, Jupiter, and Saturn.  The earliest written observations
of these objects come from the Babylonian civilization, dating back to
second millenium BC \citep{1974RSPTA.276...43S}.  The word ``planet''
originates from ancient Greek {\selectlanguage{polutonikogreek}πλανήτης}
(planētēs), meaning wanderer, as they move across the sky with respect to
stars.  Mercury, Venus, Mars, Jupiter, and Saturn were the five planets in
Greek astronomy after Pythagoras or Parmenides correctly identified the
evening star and morning star as the same object, Venus (Aphrodite in Greek)
\citep{burnet2007greek}.  Paradoxially, Earth, the
planet closest to humankind, with the largest apparent size, has only been
proposed to be a planet in the early 1500s, by the Polish astronomer Mikołaj
Kopernik (Latinized as Nicolaus Copernicus).  His framework, known as the
heliocentric system, only became widely accepted two centuries later, with the
advent of telescopic observations.

The rest of the Solar System planets were discovered using what were
considered large telescopes at
the time: Uranus in 1781 by William Herschel, and Neptune by Urbain Le Verrier
and Johann Gottfried Galle in 1846.

Pluto was discovered in 1930 by Clyde Tombaugh, although it had been captured
on photographic plates fifteen times before since as early as 1909.  It was
reclassified as a dwarf planet in 2006, leaving us with eight Solar System
planets.  However, it has not been the first object demoted from being a
planet: the Sun and the Moon were thought of as planets until the acceptance
of the heliocentric system, although the concept of planet had necessarily
been less refined at the time.  After the discovery of Uranus, a number of other
celestial bodies were discovered and labelled as planets, most notably Ceres,
Pallas, Juno, and Vesta, between 1801 and 1807.  At the time of the discovery
of Neptune, it was considered the 13th planet, until around 1850, when the
others were reclassified as asteroids.  Ceres was reclassified again in 2006,
as a dwarf planet, together with Pluto \citep{1971JHA.....2..195F,
2007JAHH...10...21H}.

The quest for discovering exoplanets (or extrasolar planets, that is, planets
outside our Solar System) was held back until very recently by their distance:
$\alpha$ Centauri Bb, the closest known potential exoplanet, is ten thousand
times as far from the Earth as Neptune, the furthest planet within the Solar
System.  It is not possible to look up the night sky and spot such a remote
planet with the naked eye, and direct observations even with very powerful
telescopes are rare.

However, not all Solar System planets were discovered by scouring around the
sky: the presence of Neptune was deducted by the irregular motion of Uranus.
In 1843, John Couch Adams started to examine observations of Uranus to predict
the position of Neptune, which was discovered a few years later based on the
independent calculations of Urbain Le Verrier.  This teaches us an important
lesson: looking into your telescope pointed at a random field on the sky is
not the only way to look for planets.  Sometimes, observing another object
like a planet or a star will provide indirect evidence of a new planet, even
without being able to directly see it.

Another important lesson from history is that sometimes what was thought to be
a planet turns out to be something else.  In case of the Solar System, only
the meaning of the word ``planet'' evolved, without actually changing the
physical nature of the Sun, the Moon, Ceres, Pallas, Juno, Vesta, Pluto, and
many others.  For indirectly discovered exoplanets, however, reanalysis of
data or gathering more observations might reveal that what was thought to be a
planet is too massive to be one, does not exist at all, or is indeed a planet
but with a different orbital period.

\section{Exoplanet detection methods}

\epigraph{L'essentiel est invisible pour les yeux.}
{Antoine de Saint Exupéry (1943)} 

Exoplanets can be detected with a number of methods.  We briefly review the most
important ones in this section.  The takeaway message is that no method is
superior: they have different biases, limitations, and costs.

Biases are important, because we need to account for them when inferring the
ensemble distribution of planet parameters from the distribution of known
planets.  They are also important when choosing a method, designing an
instrument, and setting up an observing strategy for the specific goal of
finding planets of certain properties, for example, ones that might
potentially harbor life.

\subsection{Direct imaging}

Just like Uranus was discovered as a small bright object in the field of the
telescope, it is possible to resolve an exoplanet, that is, separate the light
from two objects: the planet and its host star.  For a telescope with an
optical element of diameter $d$, at wavelength $\lambda$, diffraction of light
limits the angular resolution to
\begin{align*}
\theta &\approx \frac {1.22\lambda} d.
\end{align*}
A quick order-of-magnitude calculation tells us that for a nearby star at
$10\;\textrm{pc}$, observed by a 10 meter telescope at $1\;\mu\textrm m$, this
translates to a
projected star--planet distance of $0.25\;\textrm{AU}$.  At this distance, the
starlight reflected by the planet is too faint compared to the star to be
detected with current technology, therefore direct imaging is limited to
the thermal radiation originating from the planets themselves, typically
observed in the infrared.  This is essentially different from the
situation in our Solar System, where every planet on the sky was first
detected using light that originated from the Sun and reflected off the
planet.  Direct imaging observations favor young, massive planets, because
they have more residual heat, and
a larger surface to radiate.  These biases are reflected by the fact that until
recently, all directly imaged exoplanets were younger than $50\;\textrm{Ma}$
\citep{2013ApJ...774...11K}, and with the exception of Fomalhaut b, all have a
inferred planetary mass at least four times that of
Jupiter.  So far, eight planets have been discovered by direct
imaging.\footnoteref{footnote:exoplanets}

\subsection{Timing variations}

While direct imaging is the natural extension of observations of Solar System
planets, the first exoplanets were discovered using different methods.  One of
the first discoveries is based on timing variation of radio signals.  The
millisecond pulsar PSR 1527 + 12 is a rapidly rotating object
that emits radio waves synchronously to its rotation at a highly regular rate,
every $6.2\;\mathrm{ms}$.  As a planet orbits the pulsar, the pulsar does not
remain stationary: in fact, both of them orbit their common barycenter.
During this orbit, both objects periodically approach and recede as seen by the
observer.  Since the
radio emission rate of the pulsar is very stable, this radial displacement can
be detected in the timing of the radio signals, because it takes light a
little longer to reach the observer when the pulsar is furher away.  Two
planets were detected around PSR 1527 + 12 via this effect, even though they
change the pulsation period by only $\pm15\;\mathrm{ps}$
\citep{1992Natur.355..145W}.

Timing effects are also used to discover non-transiting planets via the
transit timing variations (TTV) of transiting planets in the same system.  For
example, the transit times of Kepler-19b deviate from strictly periodic by up
to approximately five minutes, revealing the presence of Kepler-19c
\citep{2011ApJ...743..200B}.  Dynamical arguments give a lower limit of 0.1
$M_\Earth$ for the mass of Kepler-19c.  Another example of analyzing TTVs to
constrain planetary masses is the Kepler-88 system
\citep{2013ApJ...777....3N}.  This method can sometimes also be applied to
targets that are too faint for radial velocity measurements.

Timing variations are limited to systems exhibiting very regularly periodic
behaviour.  In the above cases, we are limited to planets transiting a
millisecond pulsar, and planets that strongly interact with already discovered
transiting ones.  To date, less than two dozen planets have been discovered
via some kind of timing phenomenon.\footnoteref{footnote:exoplanets}

\subsection{Radial velocity}

Just like with millisecond pulsar timing, planets orbiting stars can be
detected by their gravitational pull on the star, if we can measure the radial
velocity of the host star precisely enough.  This can be done via
spectroscopic observations: the observed wavelength of absorption lines in the
stellar spectrum are influenced by the radial velocity of the star via
the Doppler effect \citep{1992PASP..104..270M, 2011exop.book.....S}.
Targeted surveys of main sequence stars are fruitful
methods of discovering new planets, albeit very expensive in terms of
observations, because typically dozens of observations with medium to large
telescopes are required for discovering such planets.  So far, hundreds of
planets have been discovered via radial velocity
observations.\footnoteref{footnote:exoplanets}

Radial velocity measurements are important not only for discovering planets,
but also for the mass determination of planets discovered by other methods,
like transits.  However, when other methods are not available for
characterizing the planet, radial velocity observations by themselves are not
able to resolve the degeneracy between planetary mass $m$ and orbital
inclination $i$, as they only constrain the product $m\sin i$.

The biases associated with radial velocity measurements are the following:
first, since the primary measured quantity is radial velocity, massive
planets and those with short periods are favored as they result in larger
stellar radial velocity amplitude.  Second, as rapidly rotating stars have
broad spectral features that hinder precise radial velocity determination,
rapidly rotating stars like massive main sequence A stars and very young stars
are usually excluded from observing campaigns.  Third, spectroscopic reference
lines from ionine cells and ThAr lamps are densest at optical wavelengths,
therefore RV surveys were historically biased against cooler stars like M
dwarfs, because their emission mostly falls in the infrared.

\subsection{Transits}

The orbital orientation of a planet might be such that it gets in between the
star and the observer once every orbit.  In this case, the planet blocks part
of the star's light, which we observe as a periodical dip in the stellar
lightcurve (brightness as a function of time).  This method is fairly unique
in that the depth of the transit reveals the planetary radius relative to the
stellar radius.  In addition, the transit lightcurve by itself allows for
measuring the density of the host star \citep{2003ApJ...585.1038S,
2011exop.book.....S}.

Compared to other methods, transits are relatively inexpensive to observe: a
transiting survey can monitor tens of thousands of stars in a single field
simultaneously for brightness variations.  It is therefore not suprising that
most known transiting exoplanets have been discovered by transits,
more than a thousand by now.\footnoteref{footnote:exoplanets}

However, special configurations of background stars or hierachical triple
systems might mimic the behaviour a transiting system, thus usually other
observations are required to confirm the planetary nature of the system.
Therefore it is not uncommon for transiting exoplanet surveys to focus their
efforts on targets that are known to be feasible targets for radial velocity
measurements, that is, relatively bright stars that are not fast rotators.
This is also beneficial because transits tell us about the planetary size, and
radial velocity measurements confine the mass, providing us with the unique
opportunity to determine planetary density and thus potentially learn about
the composition of the planet.

Ground-based transiting exoplanet surveys will be discussed in detail in
Section \ref{sec:factories}.

\subsection{Gravitational microlensing}

Gravitational microlensing happens when two stars are almost perfectly aligned
as seen by the observer.  In this case, the gravitational field of the
foreground star acts a magnifying lens, enhancing the light from the
background star.  Such events usually last for days of weeks, because of the
motion of the background star, the foreground star, and the Solar System with
the observer in it, in space.  If the lensing star has a planet, it may be
revealed as a second spike in the microlensing light curve if the planet is in
a favorable location during the event.  To
date, 18 planets have been discovered\footnoteref{footnote:exoplanets} via
this phenomenon: 10 by the MOA project \citep{2013ApJ...763...67S}, and 8 by
OGLE \citep{2013A&A...552A..70K}.

The unique charactestic of microlensing is that these are rare events, usually
only happening once to a system.  Therefore this method cannot constrain
orbital period directly, only planetary mass and projected separation.

\subsection{Orbital brightness modulation}

A planet might influence the total brightness of the system even if it does
not transit the star.  For example, as the planet goes through phases during the
orbit, the amount of reflected starlight changes periodically.  This has been
measured for a number of known planets, but the first planets discovered
through this effect are Kepler-70b and Kepler-70c \citep{2011Natur.480..496C}.
This method is biased toward large, close-in planets with high albedos,
because they reflect more starlight.

In addition, radial motion of stars result in tiny brightness variations due
to relativistic beaming, allowing for detection of planets.  The first such
discovery was Kepler-76b by \citet{2013ApJ...771...26F}.  In the lightcurve
analysis, the authors also account for starlight reflected by the planet, and
variations due to the ellipsoidal shape of the star caused by tidal forces.

\subsection{Polarimetry}

Starlight is unpolarized, whereas reflection from the planetary atmosphere is
partially polarized.  This effect has been observed, for example, for the
exoplanet HD 189733b \citep{2008ApJ...673L..83B}, but no previously unknown
planets have been discovered yet with this method.

\subsection{Astrometry}

Just like a planet induces periodic motion of its host star in the radial
direction (unless the orbit is in the sky plane), it does so in the sky plane
too.  Therefore astrometry, that is, careful observation of the position of
the star, could potentially reveal exoplanets.  This effect has not been
observed for any planetary host yet.

However, astrometry sometimes plays an important role in ruling out transiting
binary systems mimicking transit behaviour.  For the transit depth to be
reasonable for a planet, the light of a binary system must be blended together
with light from a field star.  This system might not be resolved by the
instrument, but astrometric measurements could reveal that the light is coming
more from the field star during transit than at other times.  Note, however,
that it is not physical motion of a star that is detected in this case, but an
artifact due to the structure of the blended source.  Not detecting such
``astrometric'' variations decreases the likelihood of a false positive
\citep{2011ApJ...727...24T}.

\section{Planet factories}
\label{sec:factories}

The most successful transiting exoplanet discovery project by the number of
planets is the \kepler{} Space Satellite, with hundreds of confirmed and
validated detections to date \citep{2010Sci...327..977B, 2011ApJ...736...19B,
2013ApJS..204...24B, 2014ApJS..210...19B, 2014ApJ...784...45R}.  
While there are other satellites looking
for exoplanets, the next most fruitful efforts are ground-based surveys: WASP
\citep{2006PASP..118.1407P} with 64 planets, and HATNet
\citep{2004PASP..116..266B} and HATSouth \citep{2013PASP..125..154B} with a
total of 43 planets discovered to date.\footnote{\label{footnote:exoplanets}\url{http://exoplanets.org}, retrieved 2014-03-25} In
this section, we highlight a few aspects of these ground-based surveys.

\subsection{Instrument design}

The HATNet, HATSouth, and WASP surveys use commercially available telephoto
lenses or telescopes measuring 11 cm, 18 cm, and 20 cm in diameter,
respectively.  While part of the reason for the relatively small size
compared to other professional telescopes is to keep the design simple and
costs low, the planetary discovery rate does not seem to suffer from the
relatively small apertures.  The reason is that transiting system candidates
are typically confirmed using radial velocity (RV) follow-up measurements,
which would become costly for stars fainter than $\sim14$ visual magnitude,
and for brighter stars, these apertures already collect enough light for
successful identification.  Brighter stars require shorter exposures not only
for RV measurements, but also for spectroscopic observations of transits,
which would allow us to learn about the composition of the planetary
atmosphere.  With such observations, longer exposures can be overly
complicated: with the Hubble Space Telescope, for example, each exposure is
limited to approximately 40 minutes, before the spacecraft hides behind the
Earth on its orbit.  Also, if the required exposure time exceeds the length of
the transit, observations taken at multiple transits need to be combined.

Given that the scientific goal of HATNet, HATSouth, and WASP is to find a
large number of transiting exoplanets that are suitable for various follow-up
observations, and that the entire sky has not been surveyed yet for transiting
exoplanets, they were designed to be wide angle, shallow surveys.  They employ
fast optics, with focal ratios between $f/1.8$ and $f/2.8$, resulting in a
field between 4$^\circ$ and 8$^\circ$ on the side, with tens of thousands of
stars down to fourteenth magnitude for each telescope.

In a ground-based survey, the observed brightness of targets is influenced by
systematic effects due to the atmosphere.  The amount of light taken away by
the atmosphere (extinction) depends on the amount of Earth's atmosphere the
light from the star has to traverse to reach the telescope (airmass).
Extinction is function of humidity, but also zenith angle, so it changes
throughout the night as the target star rises and sets, and as weather
conditions change.  To separate this effect from the astrophysical brightness
variations, a number of stable stars are
selected in the same field, and their brightness is compared.  To complicate
things, the wavelength-dependent scattering in air results in different
extinction for stars of different color, a phenomenon known as differential
extinction.  This necessitates a large number of stars in the field among
which appropriate ones can be chosen as photometric reference, also justifying
wide fields for photometric surveys.  The HATNet and HATSouth surveys use the
Trend Filtering Algorithm \citep{2005MNRAS.356..557K} to remove systematics
using lightcurves of reference stars, and the External Parameter Decorrelation
\citep{2007ApJ...670..826B} to account for effects based on where the
telescope was pointing and where the star fell on the detector.

\subsection{Observing strategy}

After deciding on the hardware: shallow, wide field versus deep, narrow field,
the next major decision is in the observing strategy: observe one field for
many months, or a new field every week.  At the two extremes, a very short
campaign could not cover two consecutive transits that are indispensible for
determining the orbital period, and a very long campaign would have a small
marginal discovery rate, because all short period planets are already
discovered, while long period transiting planets are a lot less frequent due
to the lower geometric probability of a transit.

It is important to note that typical transits last for a few hours, therefore
a single-longitude survey has a good chance of missing any particular one.
In light of this, it is not surprising that the continuous sky coverage
(weather permitted) provided by the three HATSouth sites spread out in
longitude results in an expected planet recovery fraction exceeding three times
that of a single site \citep{2013PASP..125..154B}.

\subsection{Follow-up observations}

Once a planet host candidate is identified from the small aperture photometric
observations, typically three kinds of follow-up observations are performed:
reconnaissance spectroscopy to determine the spectral type of the star,
photometry to refine transit parameters, and RV measurements to constrain the mass and
orbital eccentricity of the planet.

Reconnaissance spectroscopy is typically done on a one meter class telescope.
A small number of exposures are usually enough to tell a main sequence star
from a giant.  This is very important, because the object that causes a 1\%
deep transit in the lightcurve of a giant star is too large to be a planet.
Many eclipsing binary systems can also be ruled out based on their
double-lined spectra.

At this step, the spectral type of the star is also identified.  Usually, A
stars are not followed up, because they rotate fast, therefore their spectral
lines broaden because of the Doppler shift, making them unsuitable to detect
RV variations due to low-mass planets.  However, evolved A stars eventually
spin down, making them suitable targets.  These ``retired'' A stars are among
the most massive known planetary hosts \citep{2007ApJ...665..785J,
2008ApJ...675..784J, 2010ApJ...709..396B, 2010PASP..122..701J,
2010ApJ...721L.153J, 2011AJ....141...16J, 2011ApJS..197...26J}.
Recently, however, it has been suggested that the evolved stars targeted in
these surveys are not as massive as originally thought 
\citep{2011ApJ...739L..49L, 2013ApJ...772..143S}.

Follow-up photometry can also be performed with a one meter class telescope.
Surveying stars looking for transits would be too expensive with such
telescopes, but once there is a transit prediction, high precision photometry
is valuable in measuring the shape and depth of the transit, sometimes in
multiple bands, to constrain orbital parameters.  In addition, the shape of
the transit might provide further confirmation to the planetary nature of the
system, as those exhibit transits with mostly flat bottoms, as opposed to the
V-shaped transits of eclipsing binaries.

Once a planetary candidate passes these two tests with modest size telescopes,
the last step is to perform radial velocity measurements.  Depending on the
brightness of the target and the instruments available to the research group,
this might be done with telescopes with an aperture diameter starting from a
few meters.  The amplitude of the RV variations is determined by the planet to
star mass ratio, which together with the planet size inferred from transit
depth allows for constraining the planetary density.  In addition, RV data
points help determine the orbital eccentricity and the argument of periastron.
However, measurements are typically taken at quadrature, that is, at the
predicted peak radial velocity values, because such data points will yield the
best constaints on the mass ratio.  Unfortunately, such observations are the
least useful in determining the orbital eccentricity.

\section{Planetary system characterization}

\epigraph{Nature shows us only the tail of the lion. But there is no doubt in
my mind that the lion belongs with it even if he cannot reveal himself to the
eye all at once because of his huge dimension.}{Albert Einstein (1914)}

In this section, we highlight a few very specific aspects of characterizing
transiting planetary systems.

\subsection{Lightcurve models}

A central theme in exoplanetary explorations is the inability to resolve the
planetary system (except for planets at large orbital separation).  Usually,
all information we have about the planet and the host star has to be inferred
from observations without spatial resolution.
Measuring the lightcurve, that is, brightness as a function of time, plays an
important role in the field of exoplanets, partially because brightness
measurements are less costly than spectroscopic ones.

It is important to note that even though the observer cannot resolve the
transited object, a good understanding of its spatial structure is crutial for
generating an accurate lightcurve model.

Since the first observation of planetary transits \citep[HD
209458,][]{2000ApJ...529L..41H,2000ApJ...529L..45C}, there has been a need for
fast and accurate modeling of transit lightcurves.  Speed is an important
issue, because parameters of the system (radius ratio, impact parameter, etc.)
can be constrained by generating a large number of model lightcurves and
comparing them to the observations.  A popular and robust method for exploring
parameter uncertainties and correlations is Monte Carlo Markov Chain
(MCMC) calculations, which require an even larger number of model evaluation.

To model a transit lightcurve, one has to calculate not only what fraction of the
planet is in front of the star at any given moment, but also account for the
fact that the stellar radiation is not uniform across the stellar disk.  Most
photons we see come from a certain distance under the stellar surface, and at
the center of the star, this distance corresponds to a larger depth than
closer to the limb, where we see the surface at an angle.  Since the star is
hotter the deeper we go under the surface, and hotter material emits more,
the center of the disk will be seen to be brighter, and the limb darker.  This
phenomenon is called limb darkening.

A pioneering paper for lightcurve models is written by
\citet{2002ApJ...580L.171M}.
They deal with the nonlinear limb darkening in the form
\begin{align*}
I(r) = 1 - c_0 - c_1\left(1-\mu^{\frac12}\right) - c_2\left(1-\mu\right) -
c_3\left(1-\mu^{\frac32}\right) - c_4\left(1-\mu^2\right),
\end{align*}
and its special case the quadratic limb darkening, when $c_1=c_3=0$
\citep{2000A&A...363.1081C}.  Here $I(r)$ is the intensity of the stellar disk
as seen by the observer as a function of $r$, the distance from the center, in
stellar radius units, so that $0 \leqslant r \leqslant 1$.

The main idea in their paper is essentially a change of variables: the stellar
disk can be thought of a continuous family of concentric disks with radii
ranging from zero to one, each with uniform intensity distribution,
superimposed.  For the smaller disks, the planet-star separation and the
planetary radius are proportionally blown up when expressed in units of disk
radii.  This treatment allows the authors to evaluate the integral
analytically, creating the fast lightcurve model that has became known as the
Mandel--Agol model.

Building on this model, lightcurve calculations have been developed for other
situations or applications.  \citet{2008MNRAS.390..281P} presents analytical
formulae for the partial derivatives of flux, which speeds up certain fitting
algorithms by almost an order of magnitude.  \citet{2012MNRAS.420.1630P}
investigates elaborate configurations of multiple transiting bodies.
\citet{2013MNRAS.432.2216A} derive efficient analytic and numerical formula
for a large range of limb darkening laws.

Models that require numerical integration in one dimension are more
computationally expensive than analytic ones, and models that require
numerical integration in two dimensions are much, much more computationally
expensive.  In Chapter \ref{ch:spotrod}, we present a one-dimensional
numerical integral for transits of spotted stars, a problem for which only
two-dimensional numerical integral models existed before.

Chapter \ref{ch:agn} applies some methods and understanding from the domain of
planetary transits to a different scenario: stars transiting accretion disks
around supermassive black holes in galactic centers.  An important difference
from the usual treatment of planetary transits is that in this case, the
source has a very complicated structure, so most computation time is spent on
generating the image as the observer would see the accretion disk unocculted.

\subsection{Obliquity}

In our Solar System, planetary orbits are aligned with the equator of the Sun
within 7$^\circ$.  The angle between the orbit of an exoplanet and the equator
of its host star is commonly called obliquity, and from the Solar System, one
might guess that it is typically low.

The most popular method for measuring obliquity is the \rmeffect{}
\citep{1924ApJ....60...15R, 1924ApJ....60...22M}.  It exploits the redshift of
the receding half of the star and the blueshift of the approaching half, to
infer obliquity from time-resolved spectroscopic measurements during transit.
It follows from the nature of this phenomenon that it requires a large
telescope for taking exposures shorter than the transit, that
its sensitivity depends on the rotation rate and the inclination
of the host star and the transit impact parameter, and that it in fact
measures projected obliquity (that is, projected onto the sky plane).

The first observation of the \rmeffect{} on a transiting exoplanet is
described by \citet{2000A&A...359L..13Q}, who find that \hdketto{} has a low
projected obliquity with respect to the orbit of \hdkettob{}, as expected by
extrapolation of the Solar System observations.  However, later observations
unveiled a number of planets on polar and retrograde orbits, which has
important implications on planetary formation and migration theories.
\citet{2010ApJ...718L.145W} review projected obliquity measurements, find a
correlation with host star effective temperature, and explore possible
explanations.

\subsection{Eccentricity}

Measurement results of most physical parameters, like planetary radius or semi-major
axis, are often given as a Gaussian posterior distribution.  Even if the underlying
distribution is log-normal, a normal distribution usually provides an adequate
approximation as long as the relative uncertainty is small.  In such cases,
the result can be parametrized with a single error bar, like $10.2\pm0.3$.  In
some cases, the distribution is skewed, and characterization like
$10.2^{+0.4}_{-0.3}$ is customary.

However, orbital eccentricity is usually more complicated: for bound orbits,
$0.0 \leqslant e < 1.0$, so the distribution is truncated.  Also, highly
eccentric orbits are rare, so the distribution is usually skewed.  Last, but
not at least, close systems often circularize due to tidal interaction,
yielding in a bimodal eccentricity distribution: $e=0.0$ for a positive
fraction of objects, and $e>0.0$ for the rest of them.

Since it is an important question to decide whether a planetary orbit is
circular, sometimes two models are investigated: one where the orbit is
restricted to be circular, and another with two extra parameters: eccentricity
$e$ and argument of periastron (observer-star-periastron angle) $\omega$.
The two models can be compared by
statistical methods, for example, using the Bayesian Information Criterion
\citep[BIC, see e.g.,][]{Schwarz1978}, or calculating the Bayesian evidence
\citep[see e.g.,][]{2009MNRAS.398.1601F}.

However, occasionally only the eccentric case is analyzed, and the resulting
eccentricity is quoted with a single error bar.  \citet{1971AJ.....76..544L}
investigate this case and provide a method to estimate the posterior
likelihood of a circular orbit.

When evaluating the eccentric model, there is still a need to quantify the
uncertainty of eccentricity.  Instead of citing the maximum likelihood value
and standard deviation for $e$ itself, a better proxy is to give the results in
terms of the orbital parameters $k=e\cos\omega$ and $h=e\sin\omega$
\citep{2005AJ....129.1706F}.  These describe the case $e=0$ seamlessly, without
either parameters being truncated at this limit.  Such characterization has
been used for planetary orbits, for example, by \citet{2009ApJ...696.1950B} and
\citet{2010ApJ...710.1724B}.

Finally, it is important to note that when performing Monte Carlo Markov Chain
(MCMC) calculations, this method in fact provides a biased prior for the
eccentricity: in $k-h$ space, each eccentricity value has a measure
proportional to $e$.  Using $\sqrt e\cos\omega$ and $\sqrt e\sin\omega$ instead
of $k$ and $h$ to describe $e$ and $\omega$ result in a flat prior in $e$,
which is preferable to one proportional to $e$.  In the context of exoplanet
orbits, this characterization has been used, for example, by
\citet{2011ApJ...726L..19A, 2011A&A...533A..88G}.

\section{Atmospheres}

\epigraph{With insufficient data it is easy to go wrong.}{Carl Sagan (1980)}

The first order approximation of transits assumes that the planet is a
completely dark sphere.  This treatment is sufficient for measuring planetary
radius, impact parameter, and other characteristics based on the lightcurve of
the star.

In reality, however, planets are not completely dark.  They emit thermal
radiation, and extinguish part of the starlight traveling through their
atmosphere.  We already encountered the first phenomenon when discussing
direct imaging of young planets in the infrared.  Just like those planets
exhibit thermal radiation due to residual heat since their formation, planets
on close orbits exhibit thermal radiation due to their elevated equilibrium
temperature caused by the large stellar flux.

When discussing equilibrium temperature, it is usually assumed that heating
from incoming stellar flux get efficiently redistributed throughout the entire
planetary atmosphere, unless measurements indicate otherwise.  In this case,
the effective temperature of the planet $T_\mathrm{planet}$ can be calculated
from that of the star $T_\mathrm{star}$ according to the energy balance:
\begin{align*}
\sigma T_\mathrm{star}^4 4\pi R_\mathrm{star}^2
\frac{\pi R_\mathrm{planet}^2}{4\pi a^2} (1-A_\mathrm B) &=
\sigma T_\mathrm{planet}^4 4\pi R_\mathrm{planet}^2 \\
T_\mathrm{star}^4 \frac{R_\mathrm{star}^2}{4 a^2} (1-A_\mathrm B) &=
T_\mathrm{planet}^4 
\end{align*}
Here $\sigma$ is the Stefan--Boltzmann constant, $a$ is the orbital radius (we
assume a circular orbit in this derivation), and $A_\mathrm B$ is the Bond
albedo that tells us what fraction of the incoming electromagnetic radiation
power is reflected by the planet.  The left hand side of the equation is the
total electromagnetic power emitted by the star times a geometric factor that
tells us what fraction of that radiation falls on the planet times a factor
that tells us what fraction is absorbed.  The right hand side is the total
power emitted by the planet.

However, many close planets are presumably tidally locked, and in this case,
it is
possible that not all incoming power get redistributed over the entire
planetary surface.  Therefore the dayside temperature, that is, of the side
facing the star, is more elevated than the nightside temperature.  See, for
example, \citet{2013ApJ...776..134P} for a detailed theoretical model of heat
redistribution by the atmosphere, and \citet{2011ApJ...729...54C} for heat
redistribution coefficient measurements based on photometric observations.

If the planet has a constant effective temperature on its entire surface, the
only way to distinguish its thermal radiation from that of the star is
observing the occultation of the planet by the star, also called a secondary
eclipse.  On the other hand, if there is a difference between the dayside and
nightside temperature, that shows up as an orbital modulation of the lightcurve.

Depending on the wavelength of the observations, the thermal emission of the
planet can be negligible, and the secondary eclipse depth can be dominated by
the reflected light.  Such observations allow direct measurement of the
geometric albedo, which in turns reveals information about the atmosphere.
For example, a cloud-free atmosphere is expected to have low albedo due to
pressure-broadened absorption lines of neutral sodium and potassium
\citep{2000ApJ...538..885S}.  On the other hand, high altitude clouds reflect
off most incoming starlight, resulting in a high albedo value.

When performing multiband or spectroscopic observations of secondary eclipses,
emission lines can also be detected \citep[e.g.,][]{2008ApJ...673..526K}.

Another exciting method for probing planetary atmospheres is transmission
spectroscopy: detecting the extinction of starlight as it travels through the
atmosphere, as first theoretically described by \citet{2000ApJ...537..916S}.
Since most scattering phenomena depend smoothly on wavelength,
the extinction spectrum will in fact reveal absorption features, allowing us to
learn about atmospheric composition.  One way to think about this phenomenon
is to consider that observed at a strong absorption feature, the atmosphere
blocks light, therefore the planet effectively seems to be larger, resulting
in a deeper transit.  On the other hand, a flat transmission spectrum, like
the one observed for GJ1214b \citep{2014Natur.505...69K}, might indicate a
grey opacity source, suggesting the presence of clouds.

\section{Stellar activity}

\epigraph{The sun comes up just about as often as it goes down, in the long
run, but this doesn't make its motion random.}{Donald E.~Knuth (1969)}

An idealized transit lightcurve model might assume a spherical, homogeneously
radiating star.  An idealized radial velocity model might assume narrow lines
emitted uniformly from the stellar surface.  In reality, limb darkening
results in an axisymmetric intensity profile of the stellar disk, and spectral
lines are broadened by temperature and by rotation.  Stellar activity, which
includes flares, granulation, and starspots, further complicates the
lightcurve or spectrum of the star.  In the rest of this section, we focus on
the effect of starspots.

Stellar activity is linked to magnetic processes.  In fully convective stars
or stars with a convective envelope, the magnetic field interacts with material
flow close to the surface.  Stronger magnetic fields can effectively block
convection in areas, resulting in dark features on the surface known as spots.
The best studied starspots are those on the Sun, but observations show that
spots can differ greatly in their size, lifetime, latitude, and other
characteristics, even on a Sun-like star like EK Draconis \citep[see,
e.g.,][]{1998A&A...330..685S}.  These observations are based on separating
spots on the approaching (blueshifted) side of the star from those on the
receding (redshifted) side, a technique called Doppler imaging
\citep{1958IAUS....6..209D,1987ApJ...321..496V,1989A&A...208..179R}.

Starspots rotate with the stellar surface, causing variability on the
timescale of stellar rotation.  While this is a nuisance, for example, when
modeling planetary transits, one can actually use this variation to model
spots \citep[see, e.g.,][]{2012MNRAS.427.2487K,2012A&A...547A..37B}, or to
measure the stellar rotational period \citep[see, e.g.,][]{2013MNRAS.432.1203M}.

Spots on transiting planet hosts can actually be eclipsed by the planet,
resulting in an anomalous rebrightening of the transit lightcurve.  This
effect has been first observed by \citet{2003ApJ...585L.147S}, and has been
extensively studied since.  This phenomenon is the central point of Chapters
\ref{ch:commensurability}--\ref{ch:spotrod}.



\chapter{Discovery of the transiting exoplanet \hatptwentysevenb{}}
\label{ch:hatp27}

\original{2011ApJ...734..109B}

\newcommand{\lc}{light curve}
\newcommand{\lcs}{light curves}
\newcommand{\Lc}{Light curve}
\newcommand{\Lcs}{Light curves}
\newcommand{\avg}[1]{\ensuremath{\langle #1\rangle}}
\newcommand{\med}[1]{\ensuremath{\langle #1\rangle_{med}}}
\newcommand{\dpt}{data-point}
\newcommand{\dpts}{data-points}
\newcommand{\tel}{telescope}
\newcommand{\magn}{magnitude}
\newcommand{\stan}{standard}
\newcommand{\aper}{aperture}
\newcommand{\oot}{out-of-transit}
\newcommand{\OOT}{Out-of-Transit}
\newcommand{\cfa}{Harvard--Smithsonian Center for Astrophysics (CfA)}
\newcommand{\cfadigi}{CfA Speedometers}
\newcommand{\cmd}{color--magnitude diagram}
\newcommand{\diam}{\ensuremath{\oslash}}
\newcommand{\ccdsize}[1]{\ensuremath{\rm #1\times\rm#1}}
\newcommand{\fovsize}[2]{\ensuremath{\rm #1 #2\times\rm#1 #2}}
\newcommand{\tsize}[1]{\mbox{\rm #1 m}}
\newcommand{\band}[1]{\ensuremath{#1}-band}
\newcommand{\ordo}{\ensuremath{\mathcal{O}}}
\newcommand{\chisq}{\ensuremath{\chi^2}}
\newcommand{\RA}[3]{\ensuremath{#1^{\mathrm h}#2^{\mathrm m}#3^{\mathrm s}}}
\newcommand{\DEC}[3]{\ensuremath{#1^{\mathrm d}#2^{\mathrm m}#3^{\mathrm s}}}
\newcommand{\ghr}{\ensuremath{^h}}
\newcommand{\gmin}{\ensuremath{^m}}
\newcommand{\Ks}{\ensuremath{K_s}}
\newcommand{\masy}{\ensuremath{\rm mas\,yr^{-1}}}
\newcommand{\kms}{\ensuremath{\rm km\,s^{-1}}}
\newcommand{\ms}{\ensuremath{\rm m\,s^{-1}}}
\newcommand{\mss}{\ensuremath{\rm m\,s^{-2}}}
\newcommand{\gcmc}{\ensuremath{\rm g\,cm^{-3}}}
\newcommand{\ergscmsq}{\ensuremath{\rm erg\,s^{-1}\,cm^{-2}}}
\newcommand{\C}{\ensuremath{^{\circ}C\;}}
\newcommand{\el}{\ensuremath{e^-}}
\newcommand{\sqarcsec}{\ensuremath{\Box^{\prime\prime}}}
\newcommand{\sqarcdeg}{\ensuremath{\Box^{\circ}}}
\newcommand{\pxs}{\ensuremath{\rm \arcsec pixel^{-1}}}
\newcommand{\aduel}{\ensuremath{\lbrack ADU/\el \rbrack}}
\newcommand{\eladu}{\ensuremath{\lbrack \el/ADU \rbrack}}
\newcommand{\adupixs}{\ensuremath{\rm ADU/(pix\,s)}}
\newcommand{\elpixs}{\ensuremath{\rm \el/(pix\,s)}}
\newcommand{\masyr}{\ensuremath{\rm mas\,yr^{-1}}}
\newcommand{\msini}{\ensuremath{m \sin i}}
\newcommand{\mplsini}{\ensuremath{\mpl\sin i}}
\newcommand{\teff}{\ensuremath{T_{\rm eff}}}
\newcommand{\logg}{\ensuremath{\log{g}}}
\newcommand{\vsini}{\ensuremath{v \sin{i}}}
\newcommand{\feh}{\ensuremath{\rm [Fe/H]}}
\newcommand{\logl}{\ensuremath{\log{L}}}
\newcommand{\vmac}{\ensuremath{v_{\rm mac}}}
\newcommand{\vmic}{\ensuremath{v_{\rm mic}}}
\newcommand{\rhk}{\ensuremath{R^{\prime}_{HK}}}
\newcommand{\logrhk}{\ensuremath{\log\rhk}}
\newcommand{\Savg}{\ensuremath{\langle S\rangle}}
\newcommand{\rsun}{\ensuremath{R_\sun}}
\newcommand{\msun}{\ensuremath{M_\sun}}
\newcommand{\lsun}{\ensuremath{L_\sun}}
\newcommand{\loglsun}{\ensuremath{\log{L_\sun}}}
\newcommand{\teffsun}{\ensuremath{T_{eff,\sun}}}
\newcommand{\rhosun}{\ensuremath{\rho_\sun}}
\newcommand{\loggsun}{\ensuremath{\log{g_{\sun}}}}
\newcommand{\rstar}{\ensuremath{R_\star}}
\newcommand{\mstar}{\ensuremath{M_\star}}
\newcommand{\fehstar}{\ensuremath{\feh_\star}}
\newcommand{\lstar}{\ensuremath{L_\star}}
\newcommand{\astar}{\ensuremath{a_\star}}
\newcommand{\loglstar}{\ensuremath{\log{L_\star}}}
\newcommand{\teffstar}{\ensuremath{T_{\rm eff\star}}}
\newcommand{\rhostar}{\ensuremath{\rho_\star}}
\newcommand{\loggstar}{\ensuremath{\log{g_{\star}}}}
\newcommand{\rearth}{\ensuremath{R_\Earth}}
\newcommand{\mearth}{\ensuremath{M_\Earth}}
\newcommand{\learth}{\ensuremath{L_\Earth}}
\newcommand{\teffearth}{\ensuremath{T_{\rm eff,\earth}}}
\newcommand{\rhoearth}{\ensuremath{\rho_\earth}}
\newcommand{\rpl}{\ensuremath{R_{\mathrm p}}}
\newcommand{\mpl}{\ensuremath{M_{\mathrm p}}}
\newcommand{\lpl}{\ensuremath{L_{\mathrm p}}}
\newcommand{\teffpl}{\ensuremath{T_{\rm eff,{\mathrm p}}}}
\newcommand{\rhopl}{\ensuremath{\rho_{\mathrm p}}}
\newcommand{\ipl}{\ensuremath{i_{\mathrm p}}}
\newcommand{\epl}{\ensuremath{e_{\mathrm p}}}
\newcommand{\gpl}{\ensuremath{g_{\mathrm p}}}
\newcommand{\loggpl}{\ensuremath{\log g_{\mathrm p}}}
\newcommand{\arstar}{\ensuremath{a/\rstar}}
\newcommand{\zrstar}{\ensuremath{\zeta/\rstar}}
\newcommand{\rjup}{\ensuremath{R_{\rm J}}}
\newcommand{\mjup}{\ensuremath{M_{\rm J}}}
\newcommand{\ljup}{\ensuremath{L_{\rm J}}}
\newcommand{\teffjup}{\ensuremath{T_{eff,{\rm J}}}}
\newcommand{\rhojup}{\ensuremath{\rho_{\rm J}}}
\newcommand{\gjup}{\ensuremath{\g_{\rm J}}}
\newcommand{\rjuplong}{\ensuremath{R_{\rm Jup}}}
\newcommand{\mjuplong}{\ensuremath{M_{\rm Jup}}}
\newcommand{\ljuplong}{\ensuremath{L_{\rm Jup}}}
\newcommand{\teffjuplong}{\ensuremath{T_{eff,{\rm Jup}}}}
\newcommand{\rhojuplong}{\ensuremath{\rho_{\rm Jup}}}
\newcommand{\gjuplong}{\ensuremath{\g_{\rm Jup}}}
\newcommand{\pack}[1]{\textsc{\lowercase{#1}}}
\newcommand{\prog}[1]{\texttt{\lowercase{#1}}}
\newcommand{\iraf}{\pack{iraf}}
\newcommand{\todcor}{\prog{todcor}}
\newcommand{\xcsao}{\prog{xcsao}}
\newcommand{\daophot}{\pack{daophot}}
\newcommand{\fihat}{\pack{fihat}}
\newcommand{\fistar}{\prog{fistar}}
\newcommand{\fiphot}{\prog{fiphot}}
\newcommand{\grmatch}{\prog{grmatch}}
\newcommand{\grtrans}{\prog{grtrans}}
\newcommand{\refp}[1]{p.~\pageref{#1}}
\newcommand{\reffig}[1]{Fig.~\ref{fig:#1}}
\newcommand{\refsec}[1]{\mbox{\S\ \ref{sec:hatp27:#1}}}
\newcommand{\refeq}[1]{Eq.~\ref{eq:#1}}
\newcommand{\reftab}[1]{Tab.~\ref{tab:#1}}
\newcommand{\reffigl}[1]{Figure~\ref{fig:#1}}
\newcommand{\refsecl}[1]{\mbox{Section \ref{sec:hatp27:#1}}}
\newcommand{\refeql}[1]{Equation~\ref{eq:#1}}
\newcommand{\reftabl}[1]{Table~\ref{tab:#1}}
\newcommand{\reffigp}[1]{\reffig{#1} on \pref{fig:#1}}
\newcommand{\refsecp}[1]{\refsec{#1} on \pref{sec:hatp27:#1}}
\newcommand{\refeqp}[1]{\refeq{#1} on \pref{eq:#1}}
\newcommand{\reftabp}[1]{\reftab{#1} on \pref{tab:#1}}
\newcommand{\flwof}{\mbox{FLWO 1.2 m}}
\newcommand{\flwos}{\mbox{FLWO 1.5 m}}
\newcommand{\flwot}{\mbox{TopHAT 0.25 m}}
\newcommand{\mmt}{\mbox{MMT 6.5 m}}
\newcommand{\ssts}{{\em Spitzer}}
\newcommand{\sstL}{{\em Spitzer Space Telescope}}
\newcommand{\hst}{{\em HST}}
\newcommand{\wom}{\mbox{Wise 1 m}}
\newcommand{\piszkessch}{Konkoly 0.6 m Schmidt}
\newcommand{\piszkesrcc}{Konkoly 1 m RCC}
\newcommand{\dscu}{\mbox{$\delta$ Scuti}}
\newcommand{\gdor}{\mbox{$\gamma$ Dor}}
\newcommand{\hj}{hot Jupiter}
\newcommand{\vhj}{very hot Jupiter}
\newcommand{\hd}[1]{\mbox{HD #1}}
\newcommand{\BD}[1]{\mbox{BD #1}}
\newcommand{\hip}[1]{\mbox{HIP #1}}
\newcommand{\gj}[1]{\mbox{GJ #1}}
\newcommand{\tbn}[1]{\tablenotemark{#1}}

\newcommand{\hatcurhtr}{HTR378-001}                                    
\newcommand{\hatcurfieldA}{377}                                        
\newcommand{\hatcurfieldB}{378}                                        
\newcommand{\hatcurCCra}{\ensuremath{14^{\mathrm h}51^{\mathrm m}04.32^{\mathrm s}}}   
\newcommand{\hatcurCCdec}{\ensuremath{+05^\circ56'50.5''}}          
\newcommand{\hatcurCCmag}{12.214}                                      
\newcommand{\hatcurCCtwomass}{2MASS~14510418+0556505}                  
\newcommand{\hatcurCCgsc}{GSC~0333-00351}                              
\newcommand{\hatcurCCtassmv}{12.214}                                   
\newcommand{\hatcurCCtwomassJmag}{\ensuremath{10.626\pm0.026}}         
\newcommand{\hatcurCCtwomassHmag}{\ensuremath{10.249\pm0.023}}         
\newcommand{\hatcurCCtwomassKmag}{\ensuremath{10.109\pm0.021}}         
\newcommand{\hatcurCCcitJmag}{\ensuremath{10.635\pm0.026}}             
\newcommand{\hatcurCCcitHmag}{\ensuremath{10.242\pm0.024}}             
\newcommand{\hatcurCCcitKmag}{\ensuremath{10.133\pm0.021}}             
\newcommand{\hatcurCCbbJmag}{\ensuremath{10.696\pm0.028}}              
\newcommand{\hatcurCCbbHmag}{\ensuremath{10.265\pm0.024}}              
\newcommand{\hatcurCCbbKmag}{\ensuremath{10.153\pm0.021}}              
\newcommand{\hatcurCCesoJmag}{\ensuremath{10.700\pm0.030}}             
\newcommand{\hatcurCCesoHmag}{\ensuremath{10.262\pm0.031}}             
\newcommand{\hatcurCCesoKmag}{\ensuremath{10.151\pm0.022}}             
\newcommand{\hatcurCCesoJHmag}{\ensuremath{0.437\pm0.041}}             
\newcommand{\hatcurCCesoJKmag}{\ensuremath{0.550\pm0.037}}             
\newcommand{\hatcurCCesoHKmag}{\ensuremath{0.112\pm0.038}}             
\newcommand{\hatcurLCdip}{\ensuremath{10.6}}                           
\newcommand{\hatcurLCrprstar}{\ensuremath{0.1186\pm0.0031}}            
\newcommand{\hatcurLCbsq}{\ensuremath{0.690_{-0.032}^{+0.029}}}        
\newcommand{\hatcurLCimp}{\ensuremath{0.831_{-0.020}^{+0.017}}}        
\newcommand{\hatcurLCzeta}{\ensuremath{38.20\pm0.80}}                  
\newcommand{\hatcurLCdur}{\ensuremath{0.0705\pm0.0019}}                
\newcommand{\hatcurLCdurshort}{\ensuremath{0.0705}}                    
\newcommand{\hatcurLCdurhr}{\ensuremath{1.693\pm0.046}}                
\newcommand{\hatcurLCdurhrshort}{\ensuremath{1.693}}                   
\newcommand{\hatcurLCq}{\ensuremath{0.0232\pm0.0006}}                  
\newcommand{\hatcurLCqshort}{\ensuremath{0.023}}                       
\newcommand{\hatcurLCingdur}{\ensuremath{0.0214\pm0.0053}}             
\newcommand{\hatcurLCP}{\ensuremath{3.039586\pm0.000012}}              
\newcommand{\hatcurLCPprec}{\ensuremath{3.039586}}                     
\newcommand{\hatcurLCPshort}{\ensuremath{3.0396}}                      
\newcommand{\hatcurLCT}{\ensuremath{2455186.01879\pm0.00054}}          
\newcommand{\hatcurLCTA}{\ensuremath{2454882.06016\pm0.00125}}         
\newcommand{\hatcurLCTB}{\ensuremath{2455258.96886\pm0.00061}}         
\newcommand{\hatcurLChatnetmA}{\ensuremath{11.9625\pm0.0001}}          
\newcommand{\hatcurLCiblendA}{\ensuremath{0.78\pm0.07}}                
\newcommand{\hatcurLChatnetmB}{\ensuremath{11.9627\pm0.0001}}          
\newcommand{\hatcurLCiblendB}{\ensuremath{0.80\pm0.07}}                
\newcommand{\hatcurSMEiteff}{\ensuremath{5350\pm90}}                   
\newcommand{\hatcurSMEizfeh}{\ensuremath{+0.31\pm0.08}}                
\newcommand{\hatcurSMEizfehshort}{\ensuremath{+0.31}}                  
\newcommand{\hatcurSMEilogg}{\ensuremath{4.61\pm0.06}}                 
\newcommand{\hatcurSMEivsin}{\ensuremath{2.7\pm0.5}}                   
\newcommand{\hatcurSMEivmac}{\ensuremath{3.37}}                        
\newcommand{\hatcurSMEivmic}{\ensuremath{0.85}}                        
\newcommand{\hatcurSMEiiteff}{\ensuremath{5300\pm90}}                  
\newcommand{\hatcurSMEiizfeh}{\ensuremath{+0.29\pm0.10}}               
\newcommand{\hatcurSMEiizfehshort}{\ensuremath{+0.29}}                 
\newcommand{\hatcurSMEiilogg}{\ensuremath{4.50\pm0.06}}                
\newcommand{\hatcurSMEiivsin}{\ensuremath{0.4\pm0.4}}                  
\newcommand{\hatcurSMEiivmac}{\ensuremath{3.29}}                       
\newcommand{\hatcurSMEiivmic}{\ensuremath{0.85}}                       
\newcommand{\hatcurDSteff}{\ensuremath{NULL\pmNULL}}                   
\newcommand{\hatcurDSzfeh}{\ensuremath{NULL\pmNULL}}                   
\newcommand{\hatcurDSlogg}{\ensuremath{NULL\pmNULL}}                   
\newcommand{\hatcurDSvsini}{\ensuremath{NULL\pmNULL}}                  
\newcommand{\hatcurDSgamma}{\ensuremath{NULL\pmNULL}}                  
\newcommand{\hatcurDSnumspec}{\ensuremath{0}}                          
\newcommand{\hatcurDSspan}{\ensuremath{0}}                             
\newcommand{\hatcurDSrvrms}{\ensuremath{0.00}}                         
\newcommand{\hatcurTRESteff}{\ensuremath{5250\pm100}}                  
\newcommand{\hatcurTRESzfeh}{\ensuremath{0.0\pm0.0}}                   
\newcommand{\hatcurTRESlogg}{\ensuremath{4.75\pm0.25}}                 
\newcommand{\hatcurTRESvsini}{\ensuremath{1.5\pm1.0}}                  
\newcommand{\hatcurTRESgamma}{\ensuremath{-15.765\pm0.51}}             
\newcommand{\hatcurTRESnumspec}{\ensuremath{2}}                        
\newcommand{\hatcurTRESspan}{\ensuremath{223}}                         
\newcommand{\hatcurTRESrvrms}{\ensuremath{0.05}}                       
\newcommand{\hatcurFIESteff}{\ensuremath{NULL\pmNULL}}                 
\newcommand{\hatcurFIESzfeh}{\ensuremath{NULL\pmNULL}}                 
\newcommand{\hatcurFIESlogg}{\ensuremath{NULL\pmNULL}}                 
\newcommand{\hatcurFIESvsini}{\ensuremath{NULL\pmNULL}}                
\newcommand{\hatcurFIESgamma}{\ensuremath{NULL\pmNULL}}                
\newcommand{\hatcurFIESnumspec}{\ensuremath{0}}                        
\newcommand{\hatcurFIESspan}{\ensuremath{0}}                           
\newcommand{\hatcurFIESrvrms}{\ensuremath{0.00}}                       
\newcommand{\hatcurLBiz}{\ensuremath{0.2799}}                          
\newcommand{\hatcurLBiiz}{\ensuremath{0.2987}}                         
\newcommand{\hatcurLBii}{\ensuremath{0.3627}}                          
\newcommand{\hatcurLBiii}{\ensuremath{0.2816}}                         
\newcommand{\hatcurLBiI}{\ensuremath{0.3355}}                          
\newcommand{\hatcurLBiiI}{\ensuremath{0.2871}}                         
\newcommand{\hatcurLBig}{\ensuremath{0.7139}}                          
\newcommand{\hatcurLBiig}{\ensuremath{0.1117}}                         
\newcommand{\hatcurLBikep}{\ensuremath{0.1000}}                        
\newcommand{\hatcurLBiikep}{\ensuremath{0.1000}}                       
\newcommand{\hatcurISOm}{\ensuremath{0.94\pm0.04}}                     
\newcommand{\hatcurISOmshort}{\ensuremath{0.94}}                       
\newcommand{\hatcurISOmlong}{\ensuremath{0.945\pm0.035}}               
\newcommand{\hatcurISOr}{\ensuremath{0.90_{-0.04}^{+0.05}}}            
\newcommand{\hatcurISOrshort}{\ensuremath{0.90}}                       
\newcommand{\hatcurISOrlong}{\ensuremath{0.898_{-0.039}^{+0.054}}}     
\newcommand{\hatcurISOrho}{\ensuremath{1.84\pm0.26}}                   
\newcommand{\hatcurISOlogg}{\ensuremath{4.51\pm0.04}}                  
\newcommand{\hatcurISOlum}{\ensuremath{0.57_{-0.07}^{+0.09}}}          
\newcommand{\hatcurISOlumshort}{\ensuremath{0.57}}                     
\newcommand{\hatcurISOmv}{\ensuremath{5.55\pm0.17}}                    
\newcommand{\hatcurISOvi}{\ensuremath{0.846\pm0.023}}                  
\newcommand{\hatcurISOage}{\ensuremath{4.4_{-2.6}^{+3.8}}}             
\newcommand{\hatcurISOsigma}{\ensuremath{0.00120\pm0.00019}}           
\newcommand{\hatcurISOMJ}{\ensuremath{4.13\pm0.14}}                    
\newcommand{\hatcurISOMH}{\ensuremath{3.70\pm0.13}}                    
\newcommand{\hatcurISOMK}{\ensuremath{3.62\pm0.12}}                    
\newcommand{\hatcurISOJK}{\ensuremath{0.50\pm0.04}}                    
\newcommand{\hatcurISOJKred}{\ensuremath{0.514\pm0.04}}                
\newcommand{\hatcurISOspec}{G8}                                        
\newcommand{\hatcurRVK}{\ensuremath{96.1\pm4.5}}                       
\newcommand{\hatcurRVk}{\ensuremath{0.036\pm0.031}}                    
\newcommand{\hatcurRVh}{\ensuremath{0.066\pm0.048}}                    
\newcommand{\hatcurRVgamma}{\ensuremath{-1.1\pm3.0}}                   
\newcommand{\hatcurRVjitter}{\ensuremath{6.3}}                         
\newcommand{\hatcurRVeccen}{\ensuremath{0.078\pm0.047}}                
\newcommand{\hatcurRVomega}{\ensuremath{63\pm64}}                      
\newcommand{\hatcurRVfitrms}{\ensuremath{6.6}}                         
\newcommand{\hatcurPPi}{\ensuremath{84.7_{-0.7}^{+0.4}}}               
\newcommand{\hatcurPPg}{\ensuremath{15.1\pm1.8}}                       
\newcommand{\hatcurPPlogg}{\ensuremath{3.18\pm0.05}}                   
\newcommand{\hatcurPPar}{\ensuremath{9.65_{-0.54}^{+0.40}}}            
\newcommand{\hatcurPParel}{\ensuremath{0.0403\pm0.0005}}               
\newcommand{\hatcurPPrho}{\ensuremath{0.73\pm0.13}}                    
\newcommand{\hatcurPPm}{\ensuremath{0.66\pm0.03}}                      
\newcommand{\hatcurPPmshort}{\ensuremath{0.66}}                        
\newcommand{\hatcurPPmlong}{\ensuremath{0.660\pm0.033}}                
\newcommand{\hatcurPPme}{\ensuremath{209.7\pm10.4}}                    
\newcommand{\hatcurPPmeshort}{\ensuremath{209.7}}                      
\newcommand{\hatcurPPmelong}{\ensuremath{209.65\pm10.39}}              
\newcommand{\hatcurPPr}{\ensuremath{1.04_{-0.06}^{+0.08}}}             
\newcommand{\hatcurPPrshort}{\ensuremath{1.04}}                        
\newcommand{\hatcurPPrlong}{\ensuremath{1.038_{-0.058}^{+0.077}}}      
\newcommand{\hatcurPPre}{\ensuremath{11.6_{-0.6}^{+0.9}}}              
\newcommand{\hatcurPPreshort}{\ensuremath{11.6}}                       
\newcommand{\hatcurPPrelong}{\ensuremath{11.63_{-0.65}^{+0.86}}}       
\newcommand{\hatcurPPmrcorr}{\ensuremath{0.310}}                        
\newcommand{\hatcurPPteff}{\ensuremath{1207\pm41}}                     
\newcommand{\hatcurPPtheta}{\ensuremath{0.054\pm0.004}}                
\newcommand{\hatcurPPfluxperi}{\ensuremath{5.61_{-0.87}^{+1.64}}}      
\newcommand{\hatcurPPfluxperidim}{\ensuremath{8}}                      
\newcommand{\hatcurPPfluxap}{\ensuremath{4.09\pm0.48}}                 
\newcommand{\hatcurPPfluxapdim}{\ensuremath{8}}                        
\newcommand{\hatcurPPfluxavg}{\ensuremath{4.79_{-0.56}^{+0.78}}}       
\newcommand{\hatcurPPfluxavgdim}{\ensuremath{8}}                       
\newcommand{\hatcurXsecphase}{\ensuremath{0.5229\pm0.0197}}            
\newcommand{\hatcurXsecondary}{\ensuremath{2455187.608\pm0.060}}       
\newcommand{\hatcurXsecdur}{\ensuremath{0.0739\pm0.0061}}              
\newcommand{\hatcurXsecingdur}{\ensuremath{0.0368\pm0.0066}}           
\newcommand{\hatcurPPphiconj}{\ensuremath{0.0632_{-0.1394}^{+0.1004}}} 
\newcommand{\hatcurPPperi}{\ensuremath{2455185.83\pm0.37}}             
\newcommand{\hatcurPPaequiv}{\ensuremath{0.0535\pm0.0035}}             
\newcommand{\hatcurPPtcirc}{\ensuremath{188.8\pm60.6}}                 
\newcommand{\hatcurPPtinfall}{\ensuremath{3489.3\pm878.5}}             
\newcommand{\hatcurXdist}{\ensuremath{204\pm14}}                       
\newcommand{\hatcurCCpmra}{\ensuremath{-24.7\pm5.8}}                   
\newcommand{\hatcurCCpmdec}{\ensuremath{4.0\pm5.8}}                    
\newcommand{\hatcurCCpm}{\ensuremath{25.0218\pm8.20244}}               

\newcommand{\hatcur}{HAT-P-27}
\newcommand{\hatcurb}{\hatcur\lowercase{b}}
\newcommand{\hatcurRVgammaabs}{\hatcurDSgamma}                           
\newcommand{\hatcurRVgammarel}{\hatcurRVgamma}                           
\newcommand{\hatcurCCtassvi}{\ensuremath{0.527\pm0.12}}                  
\newcommand{\hatcurSMEversion}{ii}                                       
\newcommand{\hatcurSMEteff}{\ifthenelse{\equal{\hatcurSMEversion}{i}}{\hatcurSMEiteff}{\hatcurSMEiiteff}}
\newcommand{\hatcurSMEzfeh}{\ifthenelse{\equal{\hatcurSMEversion}{i}}{\hatcurSMEizfeh}{\hatcurSMEiizfeh}}
\newcommand{\hatcurSMEzfehshort}{\ifthenelse{\equal{\hatcurSMEversion}{i}}{\hatcurSMEizfehshort}{\hatcurSMEiizfehshort}}
\newcommand{\hatcurSMElogg}{\ifthenelse{\equal{\hatcurSMEversion}{i}}{\hatcurSMEilogg}{\hatcurSMEiilogg}}
\newcommand{\hatcurSMEvsin}{\ifthenelse{\equal{\hatcurSMEversion}{i}}{\hatcurSMEivsin}{\hatcurSMEiivsin}}
\newcommand{\hatcurSMEvmac}{\ifthenelse{\equal{\hatcurSMEversion}{i}}{\hatcurSMEivmac}{\hatcurSMEiivmac}}
\newcommand{\hatcurSMEvmic}{\ifthenelse{\equal{\hatcurSMEversion}{i}}{\hatcurSMEivmic}{\hatcurSMEiivmic}}

\newcommand{\hatcurisoshort}{YY}
\newcommand{\hatcurisofull}{Yonsei--Yale (YY)}
\newcommand{\hatcurisocite}{2001ApJS..136..417Y}
\newcommand{\hatcurlumind}{\arstar}
\newcommand{\hatcurjhkfilset}{ESO}

\chapterabstract

We report the discovery of \hatcurb{}, an exoplanet transiting
the moderately bright \hatcurISOspec\ dwarf star \hatcurCCgsc{}
($V=\hatcurCCtassmv$). The orbital period is \hatcurLCP{} d,
the reference epoch of transit is \hatcurLCT{} (BJD),
and the transit duration is \hatcurLCdur{} d.
The host star with its effective temperature \hatcurSMEteff{} K is
somewhat cooler than the Sun,
and is more metal-rich with a metallicity of \hatcurSMEzfeh{}.
Its mass is \hatcurISOm{} \msun{}
and radius is \hatcurISOr{} \rsun{}.
For the planetary companion we determine
a mass of \hatcurPPmlong{} \mjup{}
and radius of \hatcurPPrlong{} \rjup.
For the 30 known transiting exoplanets between 0.3 \mjup{} and 0.8 \mjup,
a negative correlation between host star metallicity and planetary radius,
and an additional dependence of planetary radius on equilibrium temperature
are confirmed at a high level of statistical significance.

\section{Introduction}
\label{sec:hatp27:introduction}

Studying exoplanets is vital for understanding our own Solar
System, particularly its formation. The sample of more than 500
confirmed exoplanets\footnote{According to
\url{http://www.exoplanet.eu/catalog-all.php}.}
so far enables us, for example, to
test accretion and migration theories \citep{2008ApJ...673..487I},
study tidal interactions \citep{2007MNRAS.382.1768M},
examine atmospheric structures \citep{2010DPS....42.4701F},
and investigate correlations between the existence of
planetary companions and the host star's metallicity \citep{2004DDA....35.0105I},
and between the mass of close-in planets
and the spectral type of their host star \citep{2005ApJ...626.1045I}.

Among these planets, transiting ones are
of special significance, because
the transit parameters yield planetary mass and radius estimates.
They also provide a means to determine some of the
stellar parameters more accurately than is possible with spectroscopy
alone, such as the stellar surface gravity. The more than 100 transiting
exoplanets confirmed to date provide a sample large enough
to draw meaningful conclusions about the planetary parameters
that could not be determined by radial velocity (RV)
data alone; for example,
the correlation between stellar chromospheric activity and
planetary surface gravity \citep{2010ApJ...717L.138H}, or
the correlation of planetary parameters with host star metallicity
and planetary equilibrium temperature,
as described in \refsecl{discussion}.

The Hungarian-made Automated Telescope Network
\citep[HATNet;][]{2011EPJWC..1101002B} is a system of fully automated
wide-field small telescopes designed to detect the small
photometric dips when exoplanets transit their host stars.
Since 2006, HATNet has announced and 
published 26 planetary systems with 28 planets in total, 26 of which
transit their host stars.

Here we report the detection of our
twenty-seventh transiting exoplanet, named \hatcurb{}, around
the relatively bright \hatcurISOspec{} dwarf known as
\hatcurCCgsc{}.
This planet is a textbook example of a transiting exoplanet with its
radius \hatcurPPrshort{} \rjup{} and orbital period \hatcurLCPshort{} d
being close to the median values for currently known transiting exoplanets,
and with its mass of \hatcurPPmshort{} \mjup{}
being not much less than the median mass of transiting exoplanets.

In \refsecl{obs} we present the photometric detection of the transit,
along with photometric and spectroscopic follow-up observations of
the host star \hatcur{}.
In \refsecl{analysis}, we describe the analysis of the data,
first ruling out false positive scenarios,
then determining parameters of the host star,
and finally performing a global fit for all observational data.
We conclude the paper by discussing \hatcurb{} in the context of other
known transiting exoplanets and investigating correlations of planetary
parameters with host star metallicity and equilibrium temperature
in \refsecl{discussion}.

\section{Observations}
\label{sec:hatp27:obs}

\subsection{Photometric detection}
\label{sec:hatp27:detection}

Transits of \hatcurb{} were detected in two HATNet fields
containing its host star \hatcurCCgsc{}, also known as
\hatcurCCtwomass{}; $\alpha = \hatcurCCra$, $\delta = \hatcurCCdec$,
J2000, V=\hatcurCCtassmv{} \citep{2006PASP..118.1666D}; hereafter \hatcur{}.
These fields were observed in Sloan \band {r} on a nightly basis,
weather conditions permitting, from 2009 January to August,
with the HAT-6 and HAT-10 instruments on Mount Hopkins,
and with the HAT-9 instrument on Mauna Kea.
In total, we took 10\,600 science frames
with 5 minute exposure time and 5.5 minute cadence.
For approximately 1200 of the images,
photometric measurements of individual stars had significant error,
therefore these frames were rejected.
Each image contains approximately 20\,000 stars down to $r \approx 14.5$.
For the brightest stars, the per-point photometric precision
is approximately 4.5 mmag.

\begin{figure}
\includegraphics*[width=\figurewidth]{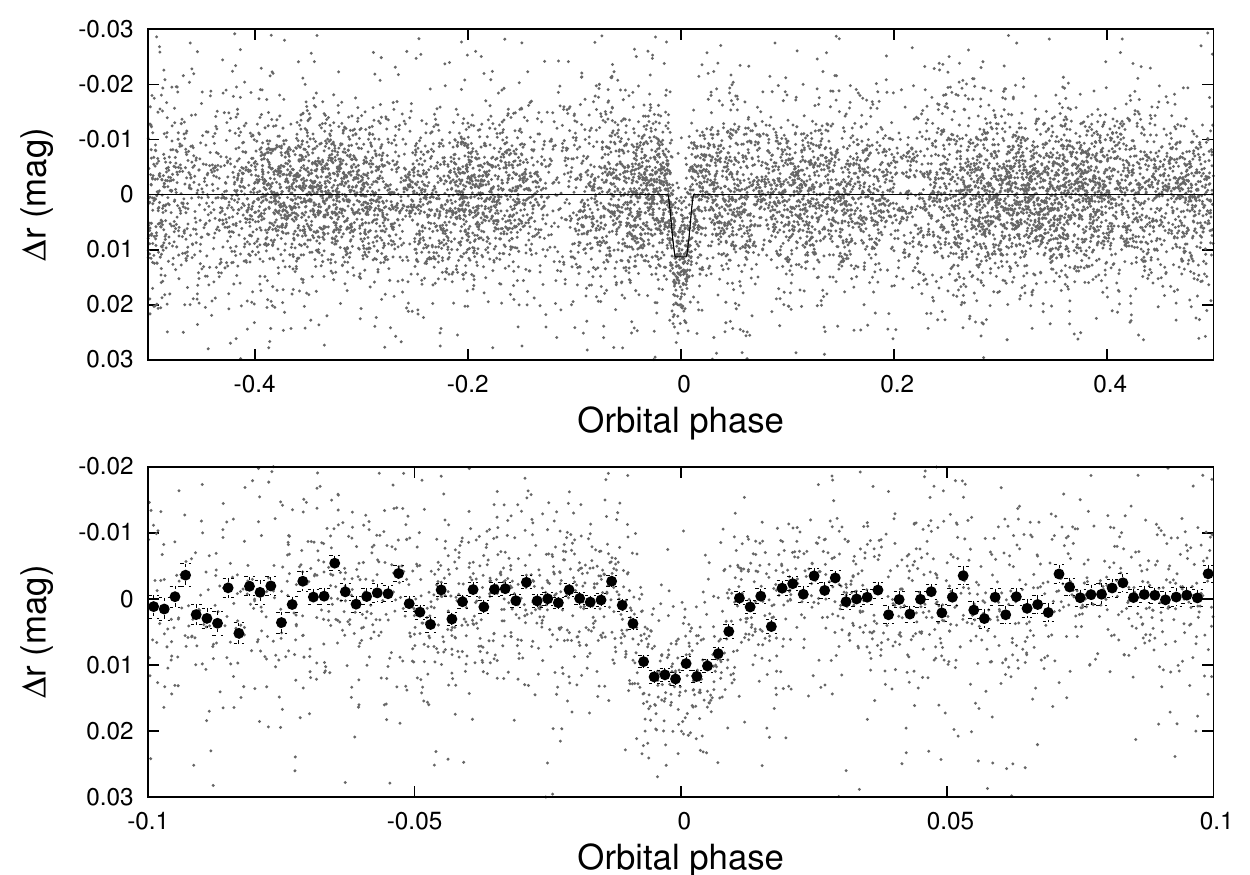}
\caption{
    {\em Top panel:} Unbinned photometric data of \hatcur{} consisting
        of 9\,400 Sloan \band{r} 5.5 minute cadence measurements
        obtained with HATNet telescopes, folded with period $P =
        \hatcurLCPprec$ d.  A simple transit curve fit to the data
        points is displayed with a solid line.  See \refsecl{globmod} for
        details.
    {\em Bottom panel:} A close-up view of the transit. Small gray
        dots are the same folded data as above; large black dots
        show the light curve binned in phase using a
        bin size of 0.002.
\label{fig:hatnet}}
\end{figure}

Calibration, astrometry, aperture photometry, External Parameter
Decorrelation (EPD), the Trend Filtering Algorithm (TFA) and the Box-fitting
Least Squares method were applied to the data as described in
\citet{2010ApJ...710.1724B}.
We detected a transit signature
in the \lc{} of \hatcur, with a 
signal depth of $\hatcurLCdip$ mmag and period of
$P=\hatcurLCPshort$ days.  This presumed transit
has relative first-to-last-contact duration $q =
\hatcurLCq$, corresponding to total duration
\hatcurLCdur~days (\hatcurLCdurhr~hours).
The folded \lc{} is presented in \reffigl{hatnet}.

\subsection{Reconnaissance Spectroscopy}
\label{sec:hatp27:recspec}

Reconnaissance spectra \citep{2009ApJ...704.1107L} of \hatcur{} were taken using three
facilities: the Tillinghast Reflector Echelle Spectrograph
\citep[TRES;][]{2008FureszPHD} on the 1.5 m Tillinghast Reflector at
FLWO, the echelle spectrograph on the 2.5 m du Pont telescope at Las
Campanas Observatory (LCO) in Chile, and the echelle spectrograph on
the Australian National University (ANU) 2.3 m telescope
at Siding Spring Observatory (SSO) in Australia. 
We gathered two spectra of \hatcur{} with TRES in
2009 July and 2010 February,
two spectra with the du Pont telescope
in 2009 July, and fourteen spectra with the ANU
2.3 m telescope in 2009 July. The exact dates and the results of these
observations are summarized in \reftabl{reconspecobs}.

\begin{deluxetable}{llccccr}
\tablewidth{0pc}
\tabletypesize{\scriptsize}
\tablecaption{
    Summary of reconnaissance spectroscopy observations of \hatcur{}
    \label{tab:reconspecobs}
}
\tablehead{
    \multicolumn{1}{c}{Instrument}          &
    \multicolumn{1}{c}{Date}                &
    \multicolumn{1}{c}{Number of}           &
    \multicolumn{1}{c}{$\teffstar$}         &
    \multicolumn{1}{c}{$\loggstar$}         &
    \multicolumn{1}{c}{$\vsini$}            &
    \multicolumn{1}{c}{$\gamma_{\rm RV}$\tablenotemark{a}} \\
    &
    &
    \multicolumn{1}{c}{Spectra}             &
    \multicolumn{1}{c}{(K)}                 &
    \multicolumn{1}{c}{(cgs)}               &
    \multicolumn{1}{c}{(\kms)}              &
    \multicolumn{1}{c}{(\kms)}
}
\startdata
TRES      & 2009 Jul 05 & 1 & $5250$  & $4.5$   & $2$     & $-15.75$ \\
du Pont   & 2009 Jul 10 & 1 & $5250$  & $4.0$   & $0$     & $-16.03$ \\
du Pont   & 2009 Jul 11 & 1 & $5000$  & $3.5$   & $0$     & $-18.03$ \\
ANU 2.3 m & 2009 Jul 16 & 5 & \nodata & \nodata & \nodata & $-20.58$ \\
ANU 2.3 m & 2009 Jul 17 & 6 & \nodata & \nodata & \nodata & $-21.32$ \\
ANU 2.3 m & 2009 Jul 18 & 3 & \nodata & \nodata & \nodata & $-20.61$ \\
TRES      & 2010 Feb\,13& 1 & $5250$  & $5.0$   & $1$     & $-15.78$
\enddata 
\tablenotetext{a}{
    The mean heliocentric RV of the target. Systematic differences
    between the velocities from different instruments are consistent
    with the velocity zero-point uncertainties.
}
\end{deluxetable}

Following \citet{2012ApJ...745...80Q} and \citet{2010ApJ...720.1118B},
we calculated the mean radial velocity, effective temperature,
surface gravity, and projected rotational velocity of the host star,
based on spectra taken by TRES.
The inferred radial velocity RMS residual of \hatcurTRESrvrms{} \kms{}
is consistent with no detectable RV
variation within the precision of the measurements.
We established the following atmospheric parameters for \hatcur{}:
effective temperature $\teffstar =
\hatcurTRESteff$ K, surface gravity $\loggstar = \hatcurTRESlogg$
(cgs), and projected rotational velocity $\vsini =
\hatcurTRESvsini$ \kms, indicating a
\hatcurISOspec\ dwarf. The mean heliocentric RV of \hatcur{} after
subtracting the gravitational redshift of the Sun is
$\gamma_{\rm RV} = \hatcurTRESgamma$ \kms.

Because this is the first time we used the du Pont 2.5 m and ANU 2.3 m
telescopes for reconnaissance spectroscopy of HATNet targets, we
briefly describe the instruments and our data reduction procedure.

The spectrograph on the du Pont telescope was used with a $4\arcsec$
long and $1.5\arcsec$ wide slit. The obtained spectra have
wavelength coverage $\approx$ 3700--7000 \AA{} at a resolution of
$\lambda/\Delta\lambda \approx 26\,000$. During the first observation
the seeing ranged between 2--3\arcsec\ and we used an exposure time of
1200 s, which provided $\sim 3000$ electrons per resolution element at 
the wavelength of 5187 \AA\@. The seeing during the second observation
was $\approx
1.8\arcsec$ and we used an exposure time of 150 s to obtain a lower
S/N spectrum sufficient to detect a velocity variation of several
\kms. We obtained a ThAr lamp spectrum before and after each
observation to use in determining the wavelength solution. We used the
CCDPROC package of IRAF\footnote{IRAF is distributed by the National
  Optical Astronomical Observatory, which is operated by the
  Association of Universities for Research in Astronomy (AURA) under
  cooperative agreement with the National Science Foundation.} to
perform overscan correction and flat-fielding of the images,
and the ECHELLE package to extract the spectra
and to determine and apply the dispersion corrections.

The extracted du Pont spectra were then cross-correlated against a
library of synthetic stellar spectra to estimate the effective
temperature, surface gravity, projected rotational velocity, and
radial velocity of the star. We followed a procedure similar to that
described by \citet{2002AJ....123.1701T}, using the same synthetic templates,
but broadened to the resolution of the du Pont echelle. These
templates, which were generated for the CfA Digital Speedometer
\citep{1992ASPC...32..110L}, only cover a wavelength range of
5150--5360 \AA{}, so we restricted our analysis to a single order of
the spectrum covering a similar range.

Spectra were also taken using the echelle spectrograph on the ANU 2.3 m
telescope.  The echelle was used in a standard configuration with a
$1.8\arcsec$ wide slit and $300\;\mathrm{mm}^{-1}$ cross-disperser setting
of $5^{\circ}50'$, which delivered 27 full orders between 
3900--6720 \AA{} with a nominal spectral resolution of
$\lambda/\Delta\lambda \approx 23\,000$.  The CCD is a 2K$\times$2K
format with $13.5\;\mu\mathrm m \times 13.5\;\mu\mathrm m$ pixels.
The gain was two electrons per ADU resulting in a read noise
of approximately 2.3 ADU for each pixel.  The spectra were binned by
two along the spatial direction.
A total of fourteen 1200 s exposures were taken.  
The seeing ranged from $2\arcsec$ to $3\arcsec$.  
The signal-to-noise on a single pixel was between 5 and 10 for each
individual exposure.  ThAr lamp calibration exposures were taken every
hour for wavelength calibration.  A high signal-to-noise exposure was
also taken of the bright radial velocity standard star \hd{223311}.

Spectra were reduced using tasks in the IRAF packages CCDPROC and
ECHELLE.  The spectra were cross-correlated against the radial velocity
standard star \hd{223311} using the IRAF task FXCOR in the RV package.
We used at least 20, typically 25 of the 27 orders for the
cross-correlation, rejecting the bluest orders for many exposures where
the signal-to-noise was too low.  Each spectral order was treated
separately and the mean of the velocities from the individual orders
was calculated. Their standard deviations were less than $0.65\;\kms$
for each exposure.

Each night, the exposures were taken within a two hour interval, much
shorter than the orbital period of the presumed companion.
For detecting large radial velocity variations to rule out the
possibility of an eclipsing binary, we consider the mean radial
velocities per night. The standard deviations between exposures were
less than $0.75\;\kms$ for each night.
Stellar parameters could have been estimated only with large
uncertainty based on data with such low signal-to-noise. Therefore
these parameters are not calculated from the
ANU 2.3 m observations. 

The results of the observations taken with these three telescopes
are listed in \reftabl{reconspecobs}. Note that for each telescope,
the radial velocity measurement uncertainty is much higher than the
radial velocity variations of the Sun due to Solar System bodies.
Therefore we only calculated heliocentric radial velocities of the
target. For the more accurate measurements described in the next section,
we will use barycentric radial velocities instead.
This accuracy, however, is enough to rule out the case of an eclipsing
binary star with great certainty. The largest RV variation {\em within an
instrument} was only 2 \kms, much less than the orbital speed in a 
typical binary system. Note that the zero-point shift between
instruments is as large as $\sim5$ \kms{}, due to the different
methods used for data reduction.
Also, all eighteen spectra were single-lined
and spectral lines were symmetric, providing no evidence for
additional stars in the system up to the precision of the measurements.

\subsection{High resolution, high S/N spectroscopy}
\label{sec:hatp27:hispec}

We acquired high-resolution, high-S/N spectra of \hatcur{} using the
HIRES instrument \citep{1994SPIE.2198..362V} on the Keck~I telescope located on
Mauna Kea, Hawaii, between 2009 December and 2010 June.
The spectrometer was configured with the $0.86\arcsec$ wide slit,
yielding a resolving power of 
$\lambda/\Delta\lambda \approx 55\,000$ over the wavelength range
of $\approx$ 3800--8000 \AA\@.

\begin{figure}
\begin{center}
\includegraphics*[width=\halffigurewidth]{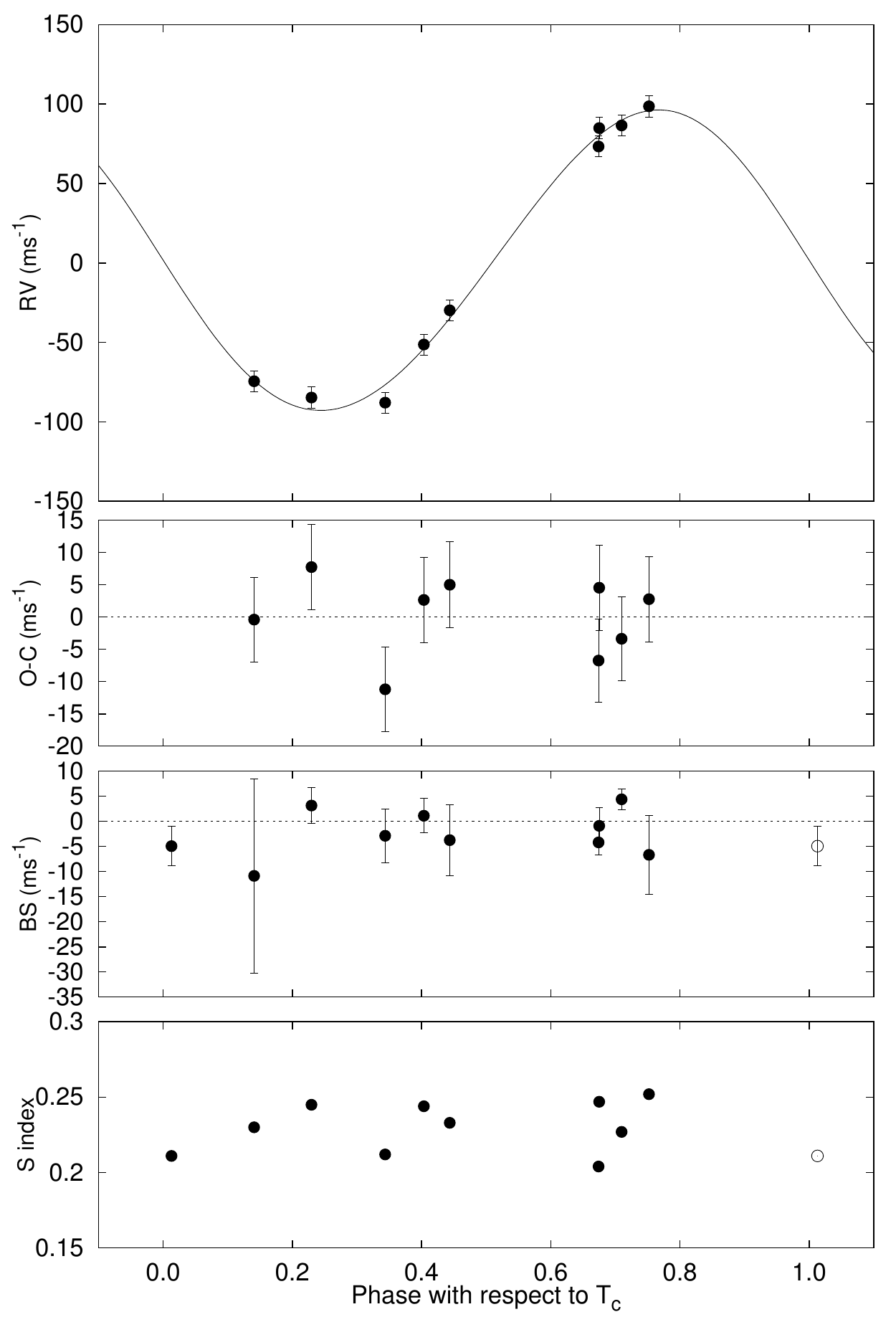}
\end{center}
\caption{
	{\em Top panel:} Keck/HIRES RV measurements for \hbox{\hatcur{}}
    shown as a function of orbital phase, along with our best-fit model
    (see \reftabl{planetparam}).  Zero phase corresponds to the time of
    mid-transit.  The center-of-mass velocity has been subtracted.
	{\em Second panel:} Velocity $O\!-\!C$ residuals from the best fit. 
    The error bars include a component from astrophysical/instrumental
    jitter ($\hatcurRVjitter$ \ms) added in quadrature to the formal
    errors (see \refsecl{globmod}).
	{\em Third panel:} Bisector spans (BS), with the mean value
        subtracted, and corrected for sky contamination.  The
        measurement from the template spectrum is included (see
        \refsecl{blend}).
	{\em Bottom panel:} Chromospheric activity index $S$.  Again, the
        measurement from the template spectrum is included.
	{\em Note:} Panels have different vertical scales. The data point
    replotted in the second period is represented by an open symbol.
\label{fig:rvbis}}
\end{figure}

Nine exposures were taken using an $\mathrm{I}_2$ gas cell
\citep[see][]{1992PASP..104..270M},
and a single template exposure was obtained without the absorption cell.
We followed \cite{1996PASP..108..500B} to establish RVs in the Solar System
barycentric frame.
We also calculated the $S$ index for each spectrum
\citep[following][]{2010ApJ...725..875I}.
The resulting values and their uncertainties are listed in \reftabl{rvs}.
They are plotted period-folded in \reffigl{rvbis}, together with the
fit established in \refsecl{analysis}.

The effective temperature established later in \refsecl{stelparam}
compared to Figure 4~of \citet{2005ApJS..159..141V} implies $B-V=0.800$ for the star.
This can in turn be used in the formula given in \citet{1984ApJ...279..763N}
together with the median $S_\mathrm{HK}$ of 0.231 to conclude
$\log R'_\mathrm{HK}=-4.785$.
This activity value is consistent with the RV jitter value
\hatcurRVjitter{} \ms{} established in \refsecl{globmod},
according to the observations given by \citet{2005PASP..117..657W}.
Based on Figure 9 in \citet{2010ApJ...725..875I}, this value qualifies
\hatcur{} as chromospherically active relative to California
Planet Search targets of the same spectral class.
The
activity index does not correlate significantly
with orbital phase.

\begin{deluxetable}{lrrrrrr}
\tablewidth{0pc}
\tablecaption{
	Relative radial velocities, bisector spans, and activity index
	measurements of \hatcur{}
	\label{tab:rvs}
}
\tablehead{
	\colhead{BJD (UTC)} & 
	\colhead{RV\tablenotemark{a}} & 
	\colhead{\ensuremath{\sigma_{\rm RV}}\tablenotemark{b}} & 
	\colhead{BS\tablenotemark{c}} & 
	\colhead{\ensuremath{\sigma_{\rm BS}}} & 
	\colhead{$S$\tablenotemark{d}} & 
	\colhead{Phase}\\
	\colhead{\hbox{($2\,400\,000+$)}} & 
	\colhead{(\ms)} & 
	\colhead{(\ms)} &
	\colhead{(\ms)} &
    \colhead{(\ms)} &
	\colhead{} &
	\colhead{}
}
\startdata
$ 55192.13748 $ & \nodata {\ } & \nodata      & $    -4.96 $ & $     3.92 $ & $    0.211 $ & $   0.013 $\\
$ 55193.14273 $ & $   -87.86 $ & $     1.73 $ & $    -2.89 $ & $     5.30 $ & $    0.212 $ & $   0.344 $\\
$ 55194.14692 $ & $    73.31 $ & $     1.51 $ & $    -4.20 $ & $     2.47 $ & $    0.204 $ & $   0.674 $\\
$ 55252.00750 $ & $    86.53 $ & $     1.75 $ & $     4.37 $ & $     2.03 $ & $    0.227 $ & $   0.710 $\\
$ 55257.15592 $ & $   -51.27 $ & $     1.84 $ & $     1.11 $ & $     3.44 $ & $    0.244 $ & $   0.404 $\\
$ 55261.02101 $ & $    84.94 $ & $     2.02 $ & $    -0.91 $ & $     3.63 $ & $    0.247 $ & $   0.675 $\\
$ 55290.06295 $ & $   -84.71 $ & $     1.89 $ & $     3.14 $ & $     3.57 $ & $    0.245 $ & $   0.230 $\\
$ 55312.92780 $ & $    98.63 $ & $     2.08 $ & $    -6.69 $ & $     7.87 $ & $    0.252 $ & $   0.752 $\\
$ 55374.90154 $ & $   -74.42 $ & $     1.85 $ & $   -10.88 $ & $    19.39 $ & $    0.230 $ & $   0.141 $\\
$ 55375.82176 $ & $   -29.80 $ & $     2.10 $ & $    -3.76 $ & $     7.04 $ & $    0.233 $ & $   0.444 $
\enddata
\tablenotetext{a}{
	The zero-point of these velocities is arbitrary. An overall offset
    $\gamma_{\rm rel}$ fitted to these velocities in \refsecl{globmod}
    has {\em not} been subtracted.
}
\tablenotetext{b}{
    Internal errors excluding the component of
    astrophysical/instrumental jitter considered in \refsecl{globmod}.
}
\tablenotetext{c}{
    The bisector spans have been corrected for sky contamination
    following \citet{2011ApJ...726...52H}.
}
\tablenotetext{d}{
	Relative chromospheric activity index, calibrated to the
	scale of \citet{1978PASP...90..267V}.
}
\tablecomments{
    For the iodine-free template exposures there is no RV
    measurement, but the BS and $S$ index can still be determined.
}
\end{deluxetable}

\subsection{Photometric follow-up observations}
\label{sec:hatp27:phot}

\begin{figure}
\includegraphics*[width=\figurewidth]{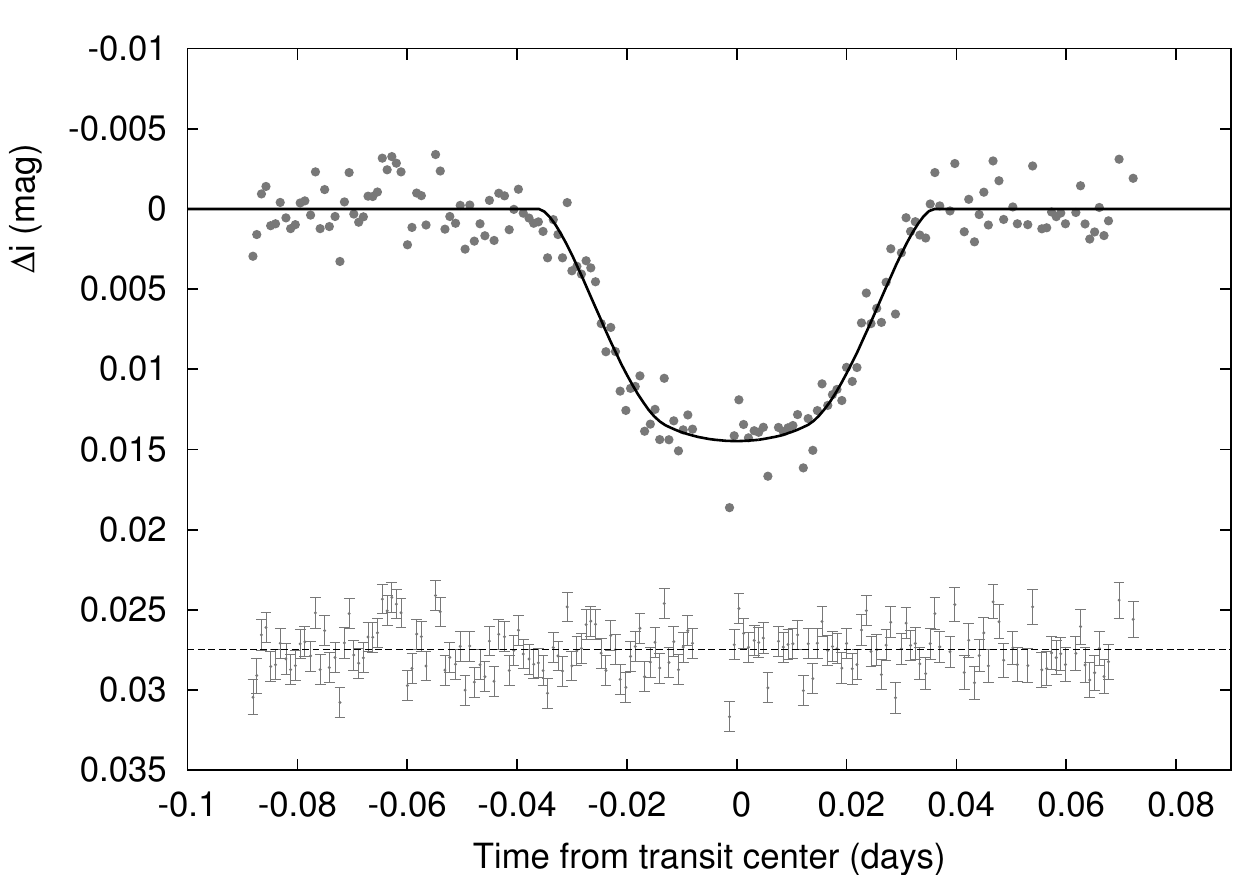}
\caption{
	Unbinned Sloan \band{i} transit light curve, acquired
	with KeplerCam on the \flwof{} telescope on 2010 March 2.
	The light curve has been EPD and TFA-processed.
    Our best fit from the global modeling is shown by the solid line.
    See \refsecl{globmod} for details.  Residuals from the
    fit are displayed at the bottom.  The error bars represent the
    photon and background shot noise, plus the readout noise.
\label{fig:lc}}
\end{figure}

A high-precision photometric follow-up of a complete transit was
carried out, permitting refined estimates of the light curve parameters
and thus orbital and planetary properties: The transit of
\hatcur{} was observed on the night of 2010 March 2 with the
KeplerCam CCD camera on the \flwof{} telescope.
We acquired 165 science frames in Sloan \band {i}
with 60 second exposure time, 73 second cadence.

Following the procedure described by \citet{2010ApJ...710.1724B},
these images were first calibrated, then astrometry and aperture photometry
was performed to arrive at \lcs, which were finally cleaned of trends
using EPD and TFA, carried out simultaneously with the global modeling
described in \refsecl{globmod}.
The result is shown in \reffigl{lc}, along with the best-fit
transit \lc{}; the individual measurements are
reported in \reftabl{phfu}.

\begin{deluxetable}{lrrrr}
\tablewidth{0pc}
\tablecaption{High-precision differential photometry of \hatcur\label{tab:phfu}}
\tablehead{
	\colhead{BJD (UTC)} & 
	\colhead{Mag\tablenotemark{a}} & 
	\colhead{\ensuremath{\sigma_{\rm Mag}}} &
	\colhead{Mag(orig)\tablenotemark{b}} & 
	\colhead{Filter} \\
	\colhead{\hbox{($2\,400\,000+$)}} & 
	\colhead{} & 
	\colhead{} &
	\colhead{} & 
	\colhead{}
}
\startdata
$ 55258.88089 $ & $   0.00295 $ & $   0.00109 $ & $  10.86800 $ & $ i$\\
$ 55258.88157 $ & $   0.00160 $ & $   0.00109 $ & $  10.86810 $ & $ i$\\
$ 55258.88242 $ & $  -0.00094 $ & $   0.00095 $ & $  10.86430 $ & $ i$\\
$ 55258.88327 $ & $  -0.00140 $ & $   0.00094 $ & $  10.86320 $ & $ i$\\
$ 55258.88412 $ & $   0.00104 $ & $   0.00095 $ & $  10.86640 $ & $ i$\\
$ 55258.88499 $ & $   0.00093 $ & $   0.00094 $ & $  10.86500 $ & $ i$\\
$ 55258.88584 $ & $  -0.00041 $ & $   0.00095 $ & $  10.86470 $ & $ i$\\
$ 55258.88689 $ & $   0.00056 $ & $   0.00095 $ & $  10.86350 $ & $ i$\\
$ 55258.88773 $ & $   0.00123 $ & $   0.00094 $ & $  10.86580 $ & $ i$\\
$ 55258.88859 $ & $   0.00098 $ & $   0.00094 $ & $  10.86540 $ & $ i$
\enddata
\tablenotetext{a}{
	The out-of-transit level has been subtracted. These magnitudes have
	been obtained by the EPD and TFA procedures, carried out
	simultaneously with the transit fit.
}
\tablenotetext{b}{
	Raw magnitude values without application of the EPD and TFA
	procedures.
}
\tablecomments{
    This table is available in a machine-readable form in the online
    journal.  A portion is shown here for guidance regarding its form
    and content.
}
\end{deluxetable}

\section{Analysis}
\label{sec:hatp27:analysis}

\subsection{Excluding blend scenarios}
\label{sec:hatp27:blend}

To further exclude possible blends, we perform bisector analysis
the same way as in \S 5 of \cite{2007ApJ...670..826B}.  A significant scatter
is found, strongly correlated to the presence of moonlight, which we
account for using the method described by \cite{2011ApJ...726...52H}.  The
bisector spans, corrected for the effect of the moonlight, are shown 
in the third panel of \reffigl{rvbis}.  They do not exhibit
significant correlation with the RV values, and the RMS scatter of the
bisector spans (4.6 \ms{}) is significantly smaller than the RV
amplitude.  These findings rule out a blend scenario with high
certainty, implying that the measured photometric and spectroscopic
features are due to a planet orbiting \hatcur{}.

\subsection{Properties of the parent star}
\label{sec:hatp27:stelparam}

We first determine spectroscopic parameters of \hatcur{},
which will allow us to calculate stellar mass and radius.
The Spectroscopy Made Easy analysis package \citep[SME;][]{1996A&AS..118..595V}
is used to establish the effective temperature, metallicity and
stellar surface gravity
based on the Keck/HIRES template spectrum of \hatcur{},
using atomic line data from the database of \cite{2005ApJS..159..141V}.
After an initial estimate for these parameters,
we perform a Monte Carlo calculation,
relying also on the normalized semi-major axis \arstar{}
inferred from transit \lcs{},
for the reasons described by \citet{2011ApJ...742..116B}.
The final values adopted after two iterations are
$\teffstar = \hatcurSMEiiteff$ K, 
$\feh = \hatcurSMEiizfeh$, and 
$\vsini = \hatcurSMEiivsin$ \kms, also listed in \reftabl{stellar}.

\begin{figure}
\includegraphics*[width=\figurewidth]{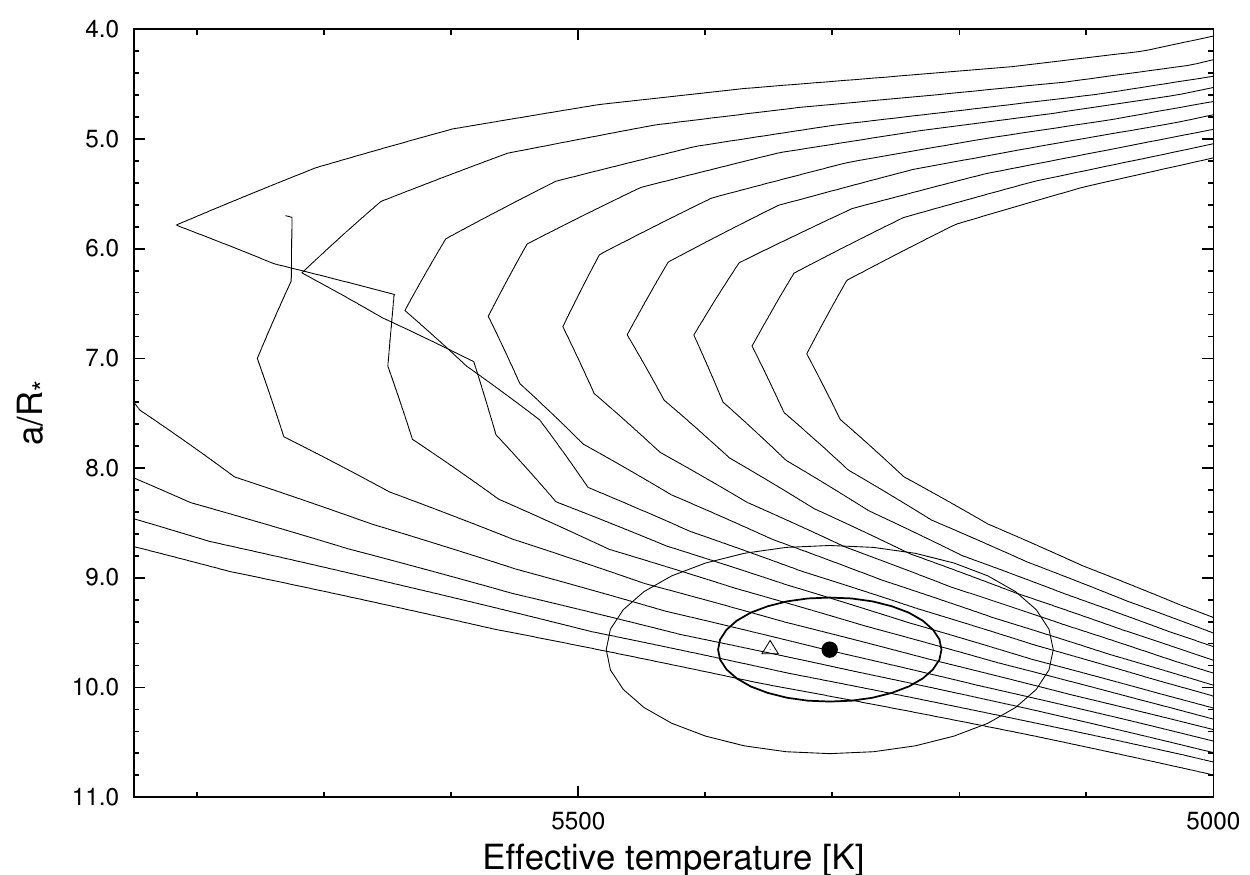}
\caption{
    Model isochrones from \cite{\hatcurisocite} for the measured
    metallicity of \hatcur, \feh = \hatcurSMEiizfehshort, and ages
    running from 1{} Gyr to 14{} Gyr in 1{} Gyr increments,
    left to right.  The adopted values of $\teffstar$ and \arstar{}
    are shown as the solid dot, surrounded by $1\sigma$ and 
    $2\sigma$ confidence ellipsoids.  The
    initial values of \teffstar\ and \arstar\ from the first SME and
    \lc\ analyses are represented with a triangle.
\label{fig:iso}}
\end{figure}

Based on the final spectroscopic parameters the model isochrones yield
a stellar mass \mstar\ = \hatcurISOmlong{} \msun{} and radius
\rstar\ = \hatcurISOrlong{} \rsun{} for \hatcur{},
along with other properties listed at the bottom of \reftabl{stellar}.
These values classify the star as a \hatcurISOspec{} dwarf,
and suggest an age of \hatcurISOage{} Gyr.
The model isochrones of \cite{\hatcurisocite} for a metallicity of
\hatcurSMEiizfehshort{} are plotted in \reffigl{iso}, along with
the best estimate of \arstar{} and \teffstar{} of \hatcur{},
and their $1\sigma$ and $2\sigma$ confidence ellipsoids. 
For comparison, the initial SME result, corresponding to a somewhat
younger state, is also indicated.

The intrinsic absolute magnitude predictions of this model
(given in the ESO photometric system)
can be compared to observations to calculate the distance of \hatcur{}.
For this we use the near-infrared brightness measurements from
the 2MASS Catalogue \citep{2006AJ....131.1163S}:
$J_{\rm 2MASS}=\hatcurCCtwomassJmag$, 
$H_{\rm 2MASS}=\hatcurCCtwomassHmag$ and 
$K_{\rm 2MASS}=\hatcurCCtwomassKmag$.
These values are converted to ESO
\citep{2001AJ....121.2851C}, then
compared to the absolute magnitude estimates to calculate the
distance. We account for interstellar dust extinction in the line of
sight using $E(B\!-\!V)=0.036 \pm 0.010$ from the dust map by
\citet{1998ApJ...500..525S}\footnote{
\url{http://irsa.ipac.caltech.edu/applications/DUST}}
with a conservative uncertainty estimate. This has to be multiplied
by a factor depending on the distance of the star and its Galactic latitude
\citep[see][]{2000AJ....120.2065B}. Assuming diffuse interstellar medium
and no dense clouds along the line of sight, we use the value
$R_V=3.1$, along with the coefficients given in Table 3 in
\citet{1989ApJ...345..245C}. These let us calculate extinction parameters
for each band based on the reddening. Finally, comparing extinctions,
absolute magnitude predictions and 2MASS apparent magnitudes for $J$,
$H$ and $K$ bands, we arrive at separate distance estimates. These 
are in good agreement, yielding an average distance of
$\hatcurXdist$ pc. The uncertainty does not account for possible systematics
of the stellar evolution model. Note that this value
is only 1 pc less than the more simple estimate ignoring extinction.
The model predicts an unreddened color index of
$(J-K)_\mathrm{model} = \hatcurISOJK$.
Reddening would change it to an estimated observed value of
$(J-K)_\mathrm{red} = \hatcurISOJKred$ using the above parameters.
This matches the actual observed color index
$(J-K)_\mathrm{ESO} = \hatcurCCesoJKmag$ within $1\sigma$.

\begin{deluxetable}{lrl}
\tablewidth{0pc}
\tabletypesize{\scriptsize}
\tablecaption{
	Stellar parameters for \hatcur{}
	\label{tab:stellar}
}
\tablehead{
	\colhead{~~~~~~~~Parameter~~~~~~~~}	&
	\colhead{Value} &
	\colhead{Source}
}
\startdata
\noalign{\vskip -3pt}
\sidehead{Spectroscopic properties}
~~~~$\teffstar$ (K)\dotfill         &  \hatcurSMEteff       & SME\tablenotemark{a}\\
~~~~$\feh$\dotfill                  &  \hatcurSMEzfeh       & SME                 \\
~~~~$\vsini$ (\kms)\dotfill         &  \hatcurSMEvsin       & SME                 \\
~~~~$\vmac$ (\kms)\dotfill          &  \hatcurSMEvmac       & SME                 \\
~~~~$\vmic$ (\kms)\dotfill          &  \hatcurSMEvmic       & SME                 \\
~~~~$\gamma_{\rm RV}$ (\kms)\dotfill&  \hatcurTRESgamma     & TRES                \\
\sidehead{Photometric properties}
~~~~$V$ (mag)\dotfill               &  \hatcurCCtassmv      & TASS                \\
~~~~$V\!-\!I_C$ (mag)\dotfill       &  \hatcurCCtassvi      & TASS                \\
~~~~$J$ (mag)\dotfill               &  \hatcurCCtwomassJmag & 2MASS           \\
~~~~$H$ (mag)\dotfill               &  \hatcurCCtwomassHmag & 2MASS           \\
~~~~$K_s$ (mag)\dotfill             &  \hatcurCCtwomassKmag & 2MASS           \\
\sidehead{Derived properties}
~~~~$\mstar$ ($\msun$)\dotfill      &  \hatcurISOmlong      & \hatcurisoshort+\hatcurlumind+SME \tablenotemark{b}\\
~~~~$\rstar$ ($\rsun$)\dotfill      &  \hatcurISOrlong      & \hatcurisoshort+\hatcurlumind+SME         \\
~~~~$\loggstar$ (cgs)\dotfill       &  \hatcurISOlogg       & \hatcurisoshort+\hatcurlumind+SME         \\
~~~~$\lstar$ ($\lsun$)\dotfill      &  \hatcurISOlum        & \hatcurisoshort+\hatcurlumind+SME         \\
~~~~$M_V$ (mag)\dotfill             &  \hatcurISOmv         & \hatcurisoshort+\hatcurlumind+SME         \\
~~~~$M_K$ (mag, \hatcurjhkfilset)\dotfill & \hatcurISOMK    & \hatcurisoshort+\hatcurlumind+SME         \\
~~~~Age (Gyr)\dotfill               &  \hatcurISOage        & \hatcurisoshort+\hatcurlumind+SME         \\
~~~~Distance (pc)\dotfill           &  \hatcurXdist         & \hatcurisoshort+\hatcurlumind+SME
\enddata
\tablenotetext{a}{
	SME = Spectroscopy Made Easy package for the analysis of
	high-resolution spectra \citep{1996A&AS..118..595V}.  These parameters
	rely primarily on SME, but have a small dependence also on the
	iterative analysis incorporating the isochrone search and global
	modeling of the data, as described in the text.
}
\tablenotetext{b}{
	\hatcurisoshort+\hatcurlumind+SME = Based on the \hatcurisoshort\
    isochrones \citep{\hatcurisocite}, \hatcurlumind\ as a luminosity
    indicator, and the SME results.
}
\end{deluxetable}

\subsection{Global modeling of the data}
\label{sec:hatp27:globmod}

We fit the combined model described by \citet{2010ApJ...710.1724B} 
to the HATNet photometry, follow-up photometry,
and radial velocity measurements simultaneously.
The eight main parameters describing the model are
the time of the first observed transit center, 
the time of the follow-up transit center, 
the normalized planetary radius $p \equiv \rpl/\rstar$,
the square of the impact parameter $b^2$,
the reciprocal of the half duration of the transit $\zrstar$,
the RV semi-amplitude $K$,
and the Lagrangian elements $k \equiv e\cos\omega$ and $h \equiv e\sin\omega$
(where $\omega$ is the longitude of periastron).

Instrumental parameters of the model include
the HATNet \oot{} magnitude
and the relative zero-point
of the Keck RVs.  The joint fit provides the full {\em a posteriori}
probability distributions of all variables (including \loggstar), which
are used to update the limb-darkening coefficients for another
iteration of the joint fit.  This leads to estimated distributions for
the stellar, \lc, and RV parameters, which are combined to calculate
values for planetary parameters.  These final values are summarized in
\reftabl{planetparam}.

The orbital eccentricity is consistent with zero (using the method of
\citealt{1971AJ.....76..544L}, we find that there is a 25\% conditional
probability of detecting an eccentricity of at least 0.078 given a
circular orbit and an uncertainty of 0.047).
Nevertheless, we stress that throughout 
the global modeling, the orbit was allowed to be eccentric, and all
system parameters and their respective errors 
inherently contain the uncertainty arising from the
floating $k$ and $h$ values.

\begin{deluxetable}{lr}
\tablewidth{0pc}
\tabletypesize{\scriptsize}
\tablecaption{Orbital and planetary parameters\label{tab:planetparam}}
\tablehead{
	\colhead{~~~~~~~~~~~~~~~Parameter~~~~~~~~~~~~~~~} &
	\colhead{Value}
}
\startdata
\noalign{\vskip -3pt}
\sidehead{\Lc{} parameters}
~~~$P$ (days)             \dotfill    & $\hatcurLCP$              \\
~~~$T_c$\tablenotemark{a} (BJD, UTC)    
                          \dotfill    & $\hatcurLCT$              \\
~~~$T_{14}$ (days)
      \tablenotemark{b}   \dotfill    & $\hatcurLCdur$            \\
~~~$T_{12} = T_{34}$ (days)
      \tablenotemark{c}   \dotfill    & $\hatcurLCingdur$         \\
~~~$\arstar$              \dotfill    & $\hatcurPPar$             \\
~~~$\zrstar$              \dotfill    & $\hatcurLCzeta$           \\
~~~$\rpl/\rstar$          \dotfill    & $\hatcurLCrprstar$        \\
~~~$b^2$                  \dotfill    & $\hatcurLCbsq$            \\
~~~$b \equiv a \cos i/\rstar$
                          \dotfill    & $\hatcurLCimp$            \\
~~~$i$ (degree)           \dotfill    & $\hatcurPPi$              \\

\sidehead{Limb-darkening coefficients \tablenotemark{d}}
~~~$a_i$ (linear term)    \dotfill    & $\hatcurLBii$             \\
~~~$b_i$ (quadratic term) \dotfill    & $\hatcurLBiii$            \\

\sidehead{RV parameters}
~~~$K$ (\ms)              \dotfill    & $\hatcurRVK$              \\
~~~$k_{\rm RV}$\tablenotemark{e} 
                          \dotfill    & $\hatcurRVk$              \\
~~~$h_{\rm RV}$\tablenotemark{e}
                          \dotfill    & $\hatcurRVh$              \\
~~~$e$                    \dotfill    & $\hatcurRVeccen$          \\
~~~$\omega$ (degree)      \dotfill    & $\hatcurRVomega$          \\
~~~RV jitter (\ms)
    \tablenotemark{f}     \dotfill    & \hatcurRVjitter           \\

\sidehead{Secondary eclipse parameters}
~~~$T_s$ (BJD)            \dotfill    & $\hatcurXsecondary$       \\
~~~$T_{s,14}$             \dotfill    & $\hatcurXsecdur$          \\
~~~$T_{s,12}$             \dotfill    & $\hatcurXsecingdur$       \\

\sidehead{Planetary parameters}
~~~$\mpl$ ($\mjup$)       \dotfill    & $\hatcurPPmlong$          \\
~~~$\rpl$ ($\rjup$)       \dotfill    & $\hatcurPPrlong$          \\
~~~$C(\mpl,\rpl)$
    \tablenotemark{g}     \dotfill    & $\hatcurPPmrcorr$         \\
~~~$\rhopl$ (\gcmc)       \dotfill    & $\hatcurPPrho$            \\
~~~$\log g_p$ (cgs)       \dotfill    & $\hatcurPPlogg$           \\
~~~$a$ (AU)               \dotfill    & $\hatcurPParel$           \\
~~~$T_{\rm eq}$ (K)       \dotfill    & $\hatcurPPteff$           \\
~~~$\Theta$\tablenotemark{h}\dotfill  & $\hatcurPPtheta$          \\
~~~$F_{per}$ ($10^{\hatcurPPfluxperidim}$ \ergscmsq)
    \tablenotemark{i}     \dotfill    & $\hatcurPPfluxperi$       \\
~~~$F_{ap}$  ($10^{\hatcurPPfluxapdim}$ \ergscmsq)
    \tablenotemark{i}     \dotfill    & $\hatcurPPfluxap$         \\
~~~$\langle F \rangle$ ($10^{\hatcurPPfluxavgdim}$ \ergscmsq) 
    \tablenotemark{i}     \dotfill    & $\hatcurPPfluxavg$
\enddata
\tablenotetext{a}
    {Reference epoch of mid transit that minimizes the
    correlation with the orbital period.}
\tablenotetext{b}
    {Total transit duration, time between first to last contact.}
\tablenotetext{c}
    {Ingress/egress time, time between first 
	and second, or third and fourth contact.
	}
\tablenotetext{d}
    {Adopted from the tabulations by
    \cite{2004A&A...428.1001C} according to the spectroscopic (SME) parameters
    listed in \reftabl{stellar}.}
\tablenotetext{e}
    {Lagrangian orbital parameters derived from the global modeling, and
    primarily determined by the RV data.}
\tablenotetext{f}
    {The contribution of the intrinsic stellar jitter
    and possible instrumental errors
    that needs to be added in quadrature
    to the calculated RV uncertainties
    so that $\chi^{2}/{\rm dof} = 1$ in the joint fit.}
\tablenotetext{g} 
	{Correlation coefficient between 
	the planetary mass and radius.}
\tablenotetext{h}
   {The Safronov number is given by \citet{2007ApJ...671..861H} as $\Theta =
   \frac{1}{2}(V_{\rm esc}/V_{\rm orb})^2 = (a/\rpl)(\mpl/\mstar)$.}
\tablenotetext{i}
	{Stellar irradiation flux per unit surface area at periastron,
	apastron and time-averaged over the orbit, respectively.}
\end{deluxetable}

\section{Discussion}
\label{sec:hatp27:discussion}

\subsection{Properties of \hatcurb{}}

\begin{figure}
\includegraphics*[width=\figurewidth]{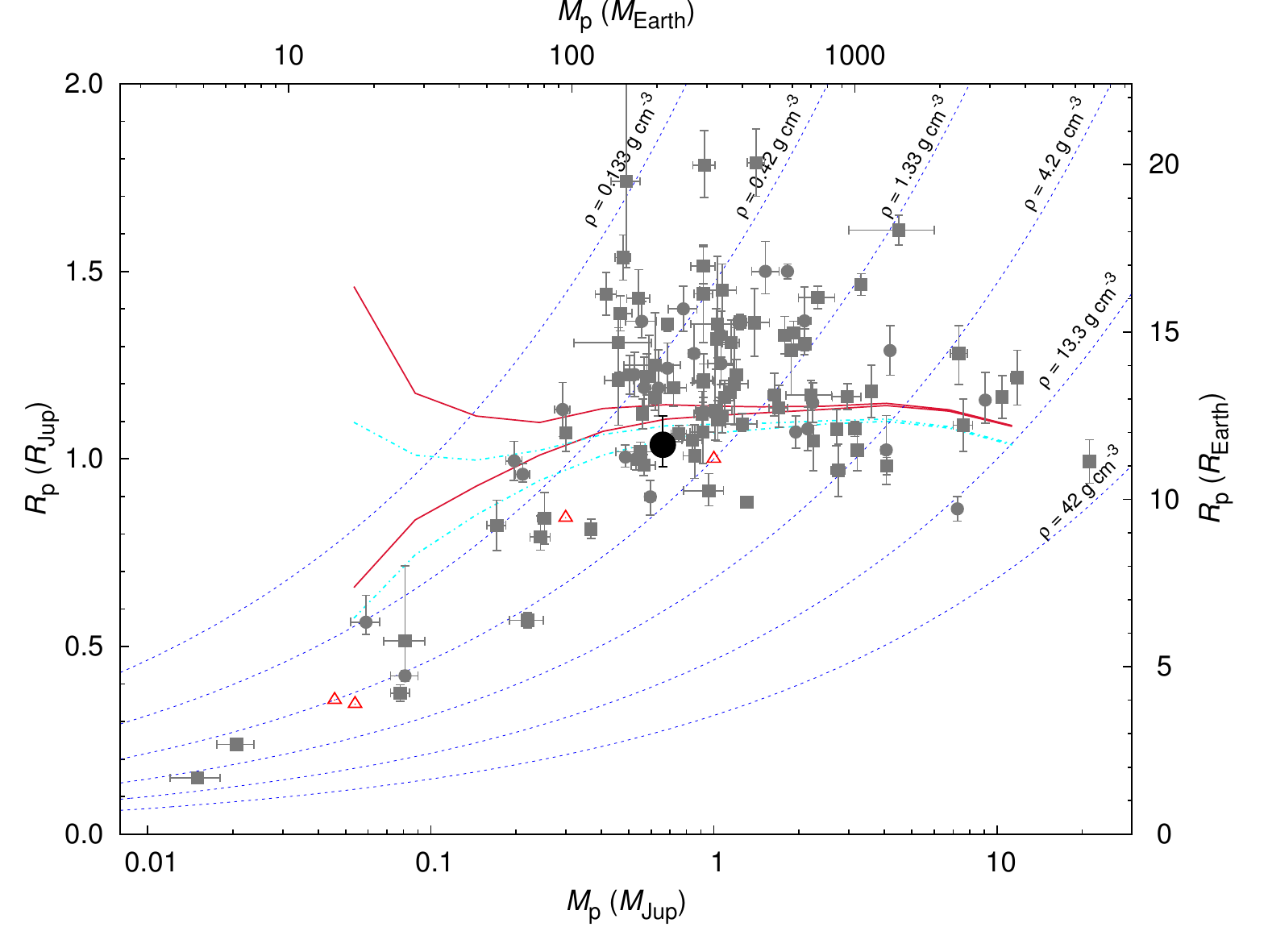}
\caption{
    Mass--radius diagram showing \hatcurb{} (solid black circle),
    other HATNet planets (solid gray circles),
    other known transiting exoplanets (solid gray squares),
    and Solar System gas giants (empty red triangles).
    Overlaid are \citet{2007ApJ...659.1661F} planetary isochrones interpolated
    to the solar equivalent semi-major axis of \hatcurb{} for ages of
    1 Gyr (solid crimson lines) and 4 Gyr (dashed-dotted cyan lines)
    and core masses of 0 and 10 \mearth{} (upper and lower pairs of lines
    respectively). Isodensity curves are shown for 0.133, 0.42,
    1.33 (Jupiter density), 4.2, 13.3, and 42 \gcmc (dashed lines).
\label{fig:exomr}}
\end{figure}

\reffigl{exomr} presents the currently known transiting exoplanets and
Solar System gas planets on a mass---radius diagram, with \hatcurb{}
highlighted. Also shown are the planetary isochrones of
\citet{2007ApJ...659.1661F} interpolated to the insolation of \hatcur{}
at the orbit of \hatcurb{}. Taking into consideration the age
established in \refsecl{stelparam}, the planetary parameters are
consistent with a hot Jupiter with a 10 \mearth{} core in a 4 Gyr
old system.

\begin{figure}
\begin{center}
\includegraphics*[width=\halffigurewidth]{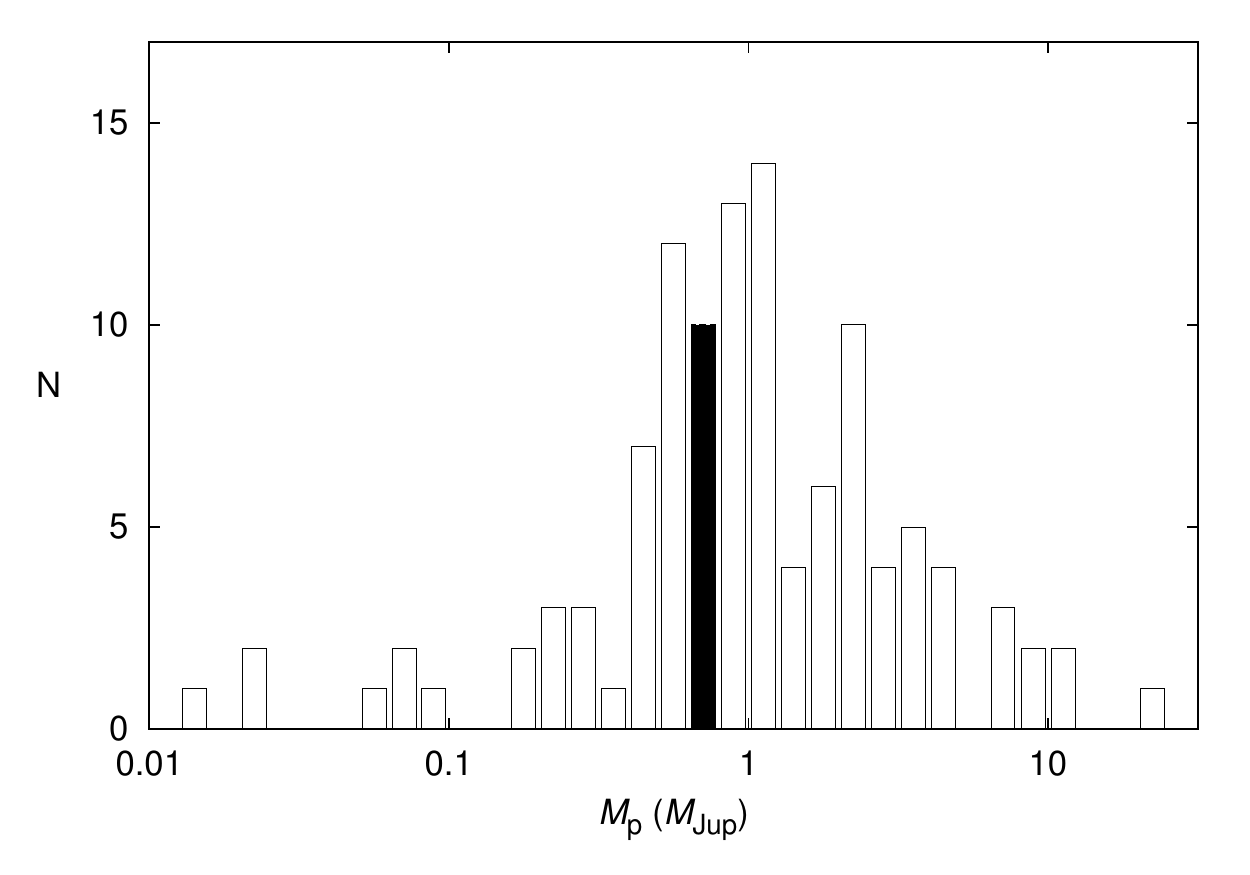}
\includegraphics*[width=\halffigurewidth]{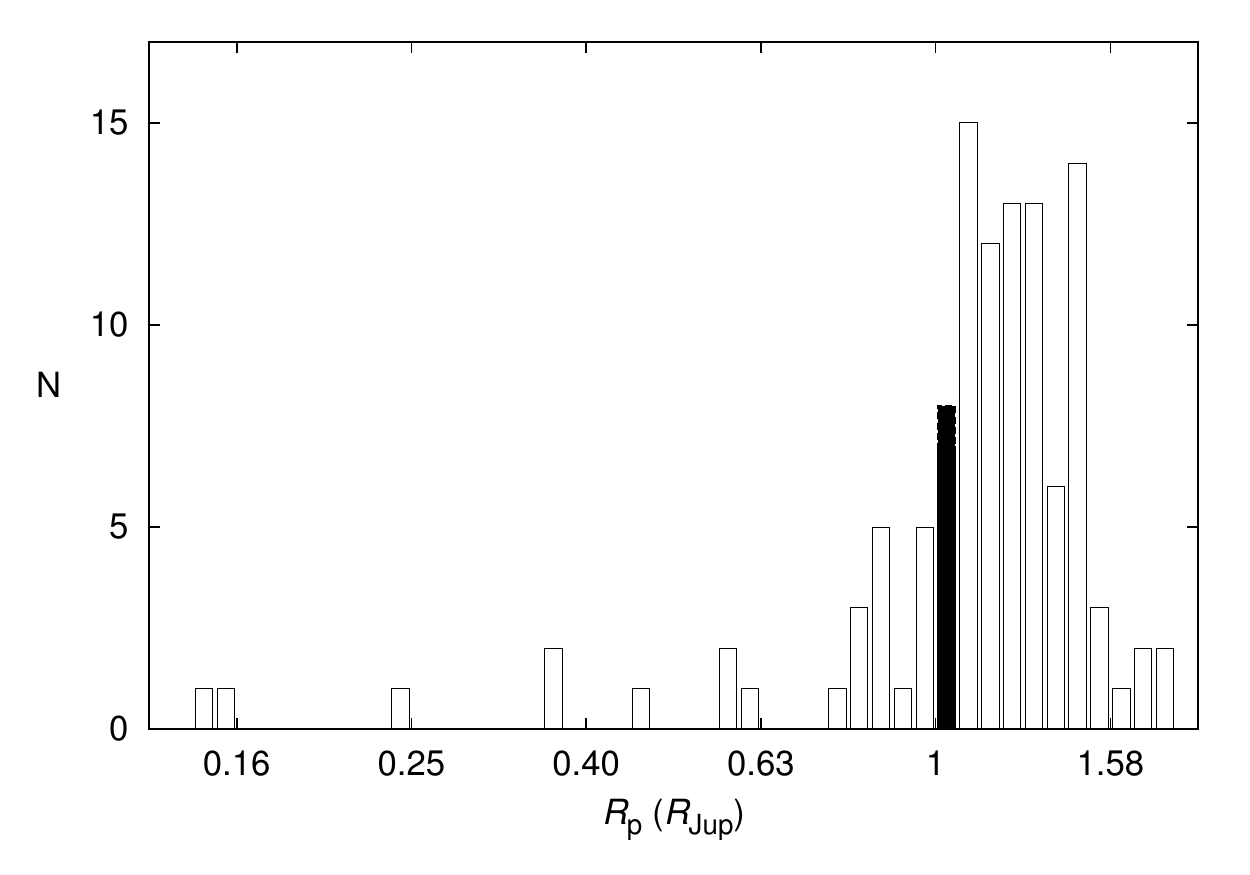} \\
\includegraphics*[width=\halffigurewidth]{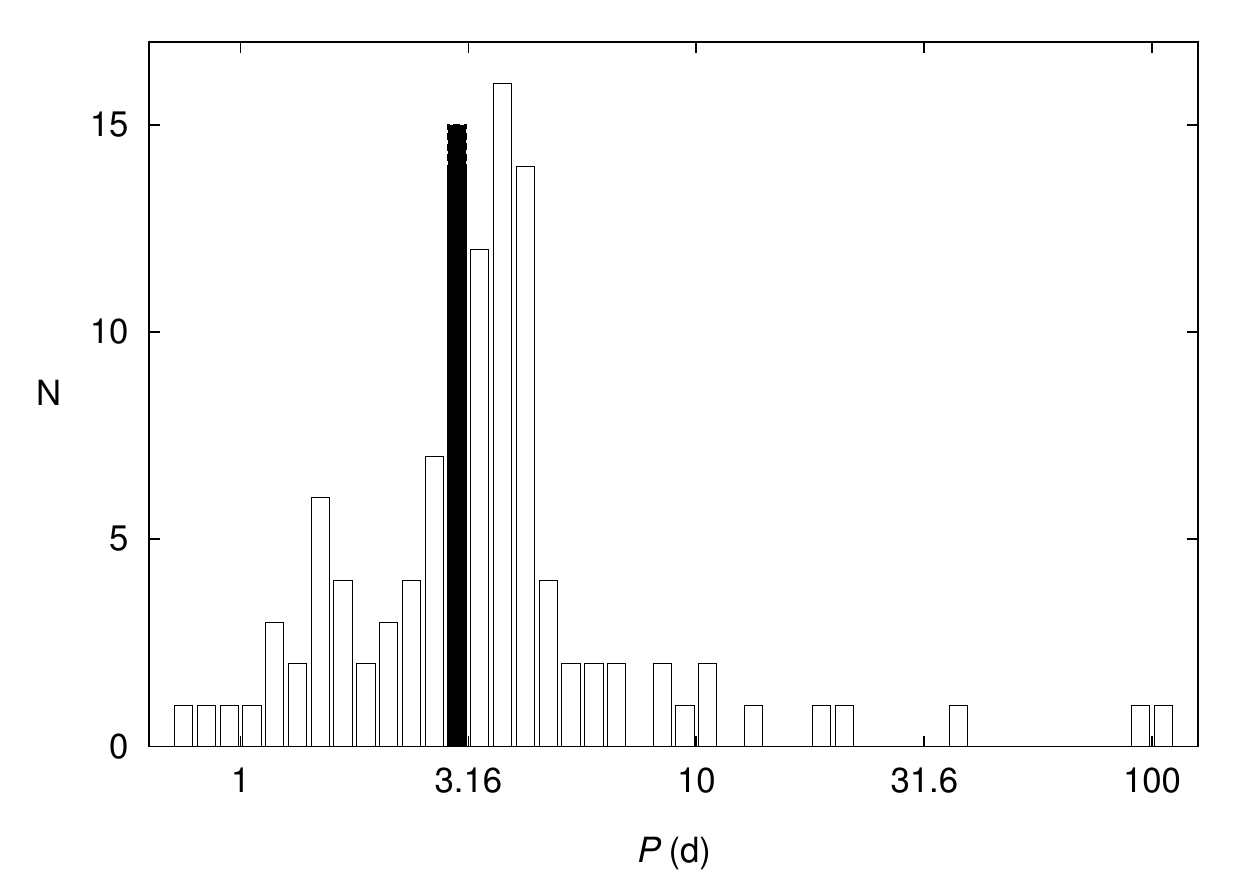}
\includegraphics*[width=\halffigurewidth]{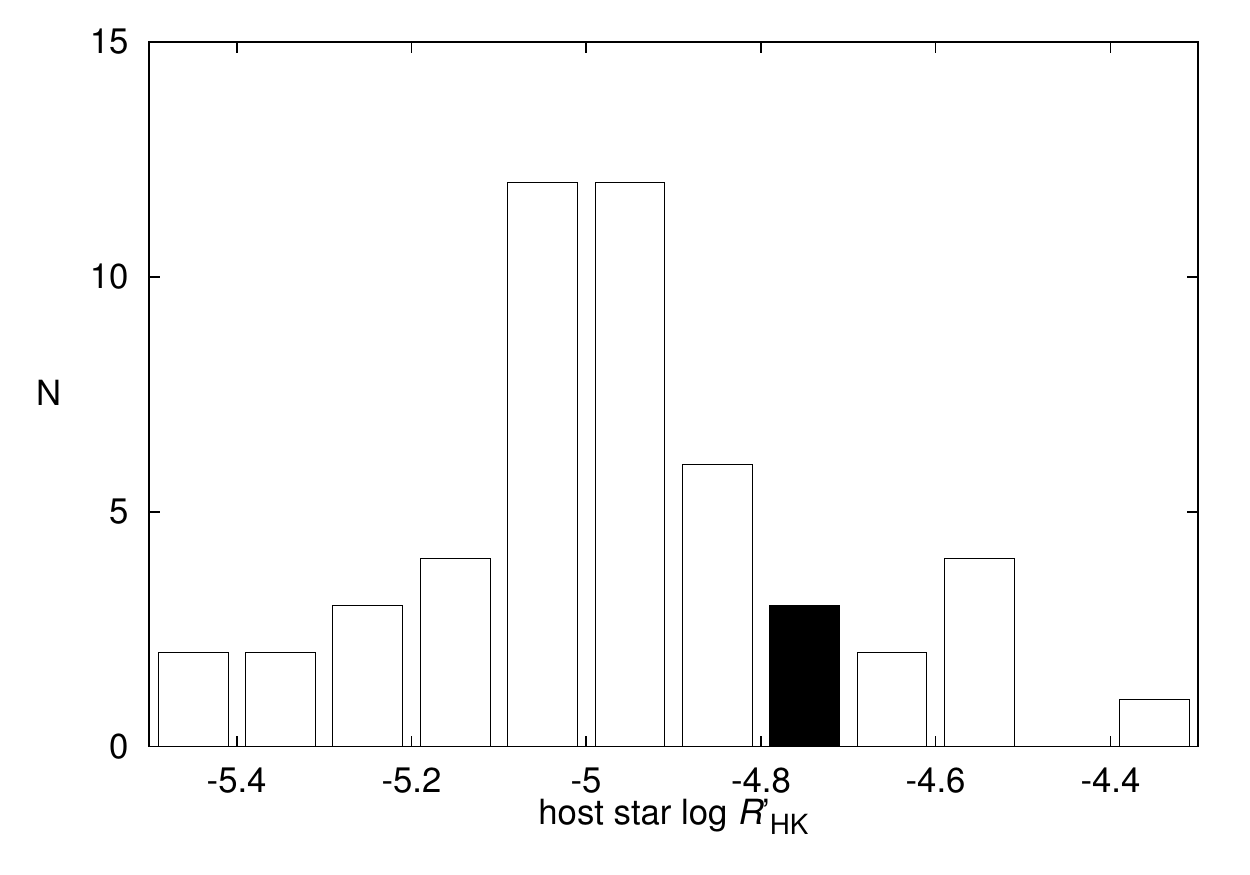}
\end{center}
\caption{
    Distribution of planetary and stellar parameters for 
    transiting exoplanets known to date.
    Horizontal axes and bins are logarithmic except for $\log R'_\mathrm{HK}$.
    Vertical axes show the number of planets in each bin,
    and the bin containing \hatcurb{} has solid filling.
	{\em Top left panel:} histogram of planetary mass
	in Jupiter masses, with logarithmic bin size 0.2.
	{\em Top right panel:} histogram of planetary radius
	in Jupiter radii, with logarithmic bin size 0.025.
	{\em Bottom left panel:} histogram of period in days,
	with logarithmic bin size 0.05.
	{\em Bottom right panel:} histogram of $\log R'_\mathrm{HK}$,
	with bin size 0.1. This index has only been reported for 52 host stars.
\label{fig:exohist}}
\end{figure}

\hatcurb{} can be seen to lie inside the large
accumulation of planets with similar masses and radii.
To further compare it to other Hot Jupiters, in \reffigl{exohist} we
present histograms of mass, radius and period
for the 112
transiting exoplanets confirmed to date.

When comparing these parameters, we note that
there is only one transiting exoplanet known that is more massive
and has a smaller radius and a smaller period than \hatcurb{}:
this is HAT-P-20b with 7.246 \mjup{}, 0.867 \rjup{} on a 2.875 d orbit
\citep{2011ApJ...742..116B}.
This means
that \hatcurb{} is less inflated than other planets of similar mass
and orbital period, possibly due to a larger than average core.

Regarding eccentricity, there are 31 transiting exoplanets known under
8 \mjup{} with an orbital period within 0.5 d of that of \hatcurb{},
out of which 8 -- more than a quarter of them -- are thought to be
eccentric.
This hints that there is a possibility
for the orbit of \hatcurb{} to be eccentric as well,
justifying our choice not to fix eccentricity to zero in \refsecl{analysis}.
Future observations of radial velocity or occultation
timing would be required to determine whether the orbit is indeed
eccentric.

The impact parameter of \hatcurb{} is unusually large. 
As \citet{2008ApJ...677L..59R} pointed out, such a near grazing transit
has the advantage of its depth and duration being more sensitive
to the presense of further planetary companions on inclined orbits.
This makes \hatcur{} a promising target for transit timing variation
and transit duration variation studies.

\citet{2010ApJ...720.1569K} found a strong negative correlation between
chromospheric activity of the host star and temperature inversion in
the planetary atmosphere.  However, since early type stars dominate
magnitude limited surveys, cool, that is, active planetary hosts are
rare.  The bottom right panel of \reffigl{exohist} shows that \hatcur{}
is relatively active compared to known planetary hosts for which $\log
R'_\mathrm{HK}$ has been reported, making it an exciting target for
{\em Spitzer Space Telescope} to test this correlation.

\subsection{Correlation of planetary parameters with host star metallicity}

The relation between host star metallicity (denoted as $\fehstar$ for
clarity, not to be confused with the assumed metal content of the planet) 
and planetary composition was studied by \citet{2006A&A...453L..21G}.  
A positive correlation, with
Pearson correlation coefficient $r=0.78$, was found between the
inferred mass of the planetary core and stellar metallicity for the
seven transiting exoplanets known at that time with 
positive
inferred
core mass.  The idea is that planets have formed from the same cloud as
their host stars, 
and
therefore their metal content should correlate. 
However, it is not clear how stellar metallicity is connected to
planetary metallicity, especially because a larger rocky core is likely
to accrete more gas during the planet's formation.

\citet{2007ApJ...661..502B} also investigated this relation, based on
12 transiting exoplanets known at the time. They
used an atmospheric opacity dependent core mass model to
explain radius anomalies. 
They also found a strong
correlation between host star metallicity and inferred core mass, but the
correlation coefficient was not reported.

\citet{2011MNRAS.410.1631E} found that there is a
strong negative correlation with $r=-0.53$ between \fehstar\ and \rpl\ for
the 18 known transiting exoplanets below 0.6 \mjup{},
whereas this correlation is negligible for more massive planets.
This can be explained by noticing that
the theoretical planet models of \citet{2007ApJ...659.1661F}, \citet{2003ApJ...592..555B},
and \citet{2008A&A...482..315B} all suggest that the radius of a planet is more sensitive
to its composition for low mass planets than it is for more massive ones.

\begin{deluxetable}{llllll}
\tablewidth{0pc}
\tabletypesize{\scriptsize}
\tablecaption{
    Parameters of 30 transiting exoplanets between 0.3 \mjup{} and 0.8 \mjup{} in increasing order of mass
    \label{tab:planets}
}
\tablehead{
  \multicolumn{1}{c}{name}                 &
  \multicolumn{1}{c}{$ \mpl (\mjup) $}     &
  \multicolumn{1}{c}{$ \rpl (\rjup) $}     &
  \multicolumn{1}{c}{$ \teff (\mathrm K)$} &
  \multicolumn{1}{c}{$ \fehstar $}         &
  \multicolumn{1}{c}{reference}
}
\startdata
WASP-21b          & $0.3\pm0.01$              & $1.07\pm0.05$             & $1262\pm31$          & $-0.4\pm0.1$    & \citet{2010AA...519A..98B} \\
HD 149026b        & $0.368^{+0.013}_{-0.014}$ & $0.813^{+0.027}_{-0.025}$ & $1792^{+44}_{-32}$   & $+0.36\pm0.05$  & \citet{2009AA...507..523A},\\&&&&&\citet{2009ApJ...696..241C} \\
Kepler-7b         & $0.416^{+0.036}_{-0.035}$ & $1.439^{+0.058}_{-0.056}$ & $1565^{+31}_{-30}$   & $+0.11\pm0.03$  & \citet{2010ApJ...713L.140L},\\&&&&&\citet{2011ApJ...730...50K} \\
WASP-13b          & $0.46^{+0.056}_{-0.05}$   & $1.21^{+0.14}_{-0.12}$    & $1417^{+62}_{-58}$   & $+0.0\pm0.2$    & \citet{2009AA...502..391S} \\
Kepler-8b         & $0.46\pm0.14$             & $1.31^{+0.076}_{-0.08}$   & $1628^{+52}_{-53}$   & $-0.055\pm0.03$ & \citet{2010ApJ...724.1108J},\\&&&&&\citet{2011ApJ...730...50K} \\
CoRoT-5b          & $0.467^{+0.047}_{-0.024}$ & $1.388^{+0.046}_{-0.047}$ & $1438\pm39$          & $-0.25\pm0.06$  & \citet{2009AA...506..281R} \\
WASP-31b          & $0.478\pm0.03$            & $1.54\pm0.06$             & $1568\pm33$          & $-0.19\pm0.09$  & \citet{2011AA...531A..60A} \\
WASP-11/HAT-P-10b & $0.487\pm0.018$           & $1.005^{+0.032}_{-0.027}$ & $1020\pm17$          & $+0.13\pm0.08$  & \citet{2009ApJ...696.1950B} \\
WASP-17b          & $0.49^{+0.059}_{-0.056}$  & $1.74^{+0.26}_{-0.23}$    & $1662^{+113}_{-110}$ & $-0.25\pm0.09$  & \citet{2010ApJ...709..159A} \\
WASP-6b           & $0.503^{+0.019}_{-0.038}$ & $1.224^{+0.051}_{-0.052}$ & $1194^{+58}_{-57}$   & $-0.20\pm0.09$  & \citet{2009AA...501..785G} \\
HAT-P-1b          & $0.524\pm0.031$           & $1.225\pm0.059$           & $1306\pm30$          & $+0.21\pm0.03$  & \citet{2008ApJ...677.1324T},\\&&&&&\citet{2009AA...507..523A} \\
HAT-P-17b         & $0.53\pm0.019$            & $1 293\pm0.03$            & $787\pm15$           & $+0.0\pm0.08$   & \citet{2012ApJ...749..134H} \\
WASP-15b          & $0.542\pm0.05$            & $1.428^{+0.077}_{-0.077}$ & $1652\pm28$          & $-0.17\pm0.11$  & \citet{2009AJ....137.4834W} \\
OGLE-TR-111b      & $0.55\pm0.1$              & $1.019^{+0.026}_{-0.026}$ & $1025^{+26}_{-25}$   & $+0.19\pm0.07$  & \citet{2006AA...450..825S},\\&&&&&\citet{2008ApJ...677.1324T},\\&&&&&\citet{2010ApJ...714...13A} \\
HAT-P-4b          & $0.556\pm0.068$           & $1.367^{+0.052}_{-0.044}$ & $1686^{+30}_{-26}$   & $+0.24\pm0.08$  & \citet{2007ApJ...670L..41K},\\&&&&&\citet{2008ApJ...677.1324T},\\&&&&&\citet{2011AJ....141...63W} \\
WASP-22b          & $0.56\pm0.02$             & $1.12\pm0.04$             & $1430\pm30$          & $-0.05\pm0.08$  & \citet{2010AJ....140.2007M}  \\
XO-2b             & $0.566^{+0.055}_{-0.055}$ & $0.983^{+0.029}_{-0.028}$ & $1319^{+24}_{-23}$   & $ 0.44\pm0.04$  & \citet{2008ApJ...677.1324T},\\&&&&&\citet{2009AA...507..523A} \\
HAT-P-25b         & $0.567\pm0.022$           & $1.19^{+0.081}_{-0.056}$  & $1202\pm36$          & $+0.31\pm0.08$  & \citet{2012ApJ...745...80Q} \\
WASP-25b          & $0.58\pm0.04$             & $1.22^{+0.06}_{-0.05}$    & $1212\pm35$          & $-0.07\pm0.1$   & \citet{2011MNRAS.410.1631E} \\
WASP-34b          & $0.59\pm0.01$             & $1.22^{+0.11}_{-0.08}$    & $1250\pm30$          & $-0.02\pm0.1$   & \citet{2011AA...526A.130S} \\
HAT-P-3b          & $0.596^{+0.024}_{-0.026}$ & $0.899^{+0.043}_{-0.049}$ & $1127^{+49}_{-39}$   & $+0.27\pm0.04$  & \citet{2007ApJ...666L.121T},\\&&&&&\citet{2008ApJ...677.1324T} \\
HAT-P-28b         & $0.636\pm0.037$           & $1.189^{+0.102}_{-0.075}$ & $1371\pm50$          & $+0.12\pm0.08$  & \citet{2011ApJ...733..116B} \\
HAT-P-27b         & $0.660\pm0.033$           & $1.038^{+0.077}_{-0.058}$ & $1207\pm41$          & $+0.29\pm0.1$   & this paper \\
HAT-P-24b         & $0.685\pm0.033$           & $1.242\pm0.067$           & $1637\pm42$          & $-0.16\pm0.08$  & \citet{2010ApJ...725.2017K} \\
HD 209458b        & $0.685^{+0.015}_{-0.014}$ & $1.359^{+0.016}_{-0.019}$ & $1449\pm12$          & $+0.01\pm0.03$  & \citet{2005ApJ...629L.121L},\\&&&&&\citet{2008ApJ...677.1324T} \\
Kepler-6b         & $0.62^{+0.025}_{-0.028}$  & $1.164^{+0.025}_{-0.017}$ & $1459^{+25}_{-24}$   & $+0.34\pm0.04$  & \citet{2010ApJ...713L.136D},\\&&&&&\citet{2011ApJ...730...50K} \\
OGLE-TR-10b       & $0.62\pm0.14$             & $1.25^{+0.14}_{-0.12}$    & $1481^{+71}_{-55}$   & $+0.15\pm0.15$  & \citet{2008ApJ...677.1324T} \\
CoRoT-4b          & $0.72\pm0.08$             & $1.19^{+0.06}_{-0.05}$    & $1074\pm19$          & $+0.05\pm0.07$  & \citet{2008AA...488L..47M} \\
TrES-1b           & $0.752^{+0.047}_{-0.046}$ & $1.067^{+0.022}_{-0.021}$ & $1140^{+13}_{-12}$   & $+0.02\pm0.05$  & \citet{2008ApJ...677.1324T} \\
HAT-P-9b          & $0.780\pm0.090$           & $1.40\pm0.06$             & $1530\pm40$          & $+0.12\pm0.2$   & \citet{2009ApJ...690.1393S}
\enddata
\end{deluxetable}

In this subsection, we examine further the
correlation between host star metallicity and planetary mass or planetary
radius. We use a substantially expanded sample of 
30 known transiting exoplanets with masses
between 0.3 \mjup{} and 0.8 \mjup{}, see \reftabl{planets}.
The upper limit is selected to exclude more massive planets whose radius
is expected to depend less on the composition, see above.
We explain the role of the lower limit and the effect of the two lowest mass
planets in \reftabl{planets} later.

The null hypothesis is that the host star metallicity
and the selected planetar parameter are independent.  The alternative
hypothesis is that they are related by some underlying phenomenon.  A
false positive, also known as an error of the first kind, 
is rejection
of the null hypothesis in spite of it being true.  We implement three
independent statistical methods to estimate the false positive
probability: $t$-test, bootstrap technique and $F$-test.  We denote the
probability estimates by $p_1$, $p_2$ and $p_3$, respectively.  This is
the statistical significance of the correlation: the lower this
probability is, the more confidently the null hypothesis (i.e., no
correlation) can be rejected.

For the $t$-test, we assume that the investigated parameters have
normal distribution.  For each set of data pairs,
we calculate the $t$
value from the correlation coefficient $r$ and sample size $n$ using
the formula
$$ t = r \sqrt {\frac {n-2}{1-r^2}}\;.$$
The conditional distribution of this variable given the null hypothesis
is Student's $t$ distribution with $n-2$ degrees of freedom
\citep[p.~640]{1992nrfa.book.....P}. Then the estimate $p_1$ for
false positive probability is determined
by looking up the two-tailed probability of this distribution
yielding larger absolute value that the one measured.
For comparison, we also performed the $t$-test for the samples
and parameters studied by \citet{2006A&A...453L..21G} and \citet{2011MNRAS.410.1631E}.
The resulting values are 
listed in \reftabl{corr}.

For the sample set of 
\reftabl{planets},
we also implement the bootstrap resampling technique
\citep{efron1994introduction}.  This has the advantage that no assumption about
the {\em a priori} distribution of the data is necessary.  
To perform
bootstrap resampling, consider the data $(x_1, y_1), (x_2, y_2),
\ldots, (x_n, y_n)$, where $x_i$ is the host star metallicity, and
$y_i$ is the corresponding planetary mass or radius.  Again, we would
like to calculate an estimate $p_2$ of the probability that a sample of
similar distribution, but independent parameters for each pair, has a
correlation coefficient that exceeds that of our measurements in
absolute value.  For this, we build 10\,000\,000 sample sets of $n$
pairs by drawing $x$ and $y$ values independently with replacements
from the set of measured $x$ and $y$ values, respectively.  The
percentile rank of the absolute value of the correlation coefficient
for the measured data among the random samples gives our estimates
$p_2$, listed in \reftabl{corr}.

Finally, we test these correlations with an additional method, the $F$-test
\citep[see e.g.][p.~100]{1993stp..book.....L}.  This requires that the null
hypothesis (no correlation) be nested in the tested hypothesis (linear
correlation), which indeed is the case.  Let RSS$_1$ denote the
residual sum of squares for the best fit of the null hypothesis, that
is, the variance of $y_i$ about its mean, and RSS$_2$ denote the
residual sum of squares of the linear fit.  The no correlation model
has $l_1=1$ free parameters: the mean, whereas the linear fit has
$l_2=2$.  The conditional distribution of
$$ F = \frac {\frac {\mathrm{RSS}_1 - \mathrm{RSS}_2} {l_2-l_1}} 
{\frac {\mathrm{RSS}_2}{n-l_2}} $$
given the null hypothesis is the $F$-distribution with $(l_2-l_1,n-l_2)$
degrees of freedom.  This enables us to calculate $p_3$, the third
estimate for the false positive probability.

The estimates $p_1$, $p_2$, $p_3$ given by the three methods
are listed in \reftabl{corr}. 
For each correlation, they coincide up to the uncertainty of the
methods. 
The values reflect the significant
$\fehstar\textrm{---}M_{core}$ and $\fehstar\textrm{---}\rpl$
correlations found by \citet{2006A&A...453L..21G} and \citet{2011MNRAS.410.1631E}.

As for the 30 planets listed in \reftabl{planets},
it is important to note that the correlations depend strongly on the
choice of the lower mass limit. The two least massive planets in the table
are WASP-21b with a mass of $0.3\;\mjup$, very low host star metallicity of $-0.4$,
and average radius of $1.07\;\rjup$;
and the dense HD 149026b with a mass of $0.368\;\mjup$, 
high host star metallicity of $+0.36$,
and low radius of $0.813\;\rjup$.
Increasing the lower mass limit for our sample first excludes WASP-21b,
which would much support the positive $\fehstar$---$\mpl$ correlation
with its low mass and low host star metallicity.
Further increasing the limit then excludes HD 149026b, which would
much weaken it with its low mass and high host star metallicity.
To have an unbiased result, outliers cannot be excluded
without a justified reason, therefore we need to compare the false positive
probabilities of the three nested samples. They scatter between 15\% and 58\%,
neither supporting, nor rejecting a $\fehstar$---$\mpl$ correlation.

Similarily, both WASP-21b and HD 149026b have a strong effect
on the negative $\fehstar$---$\rpl$ correlation, because of the extreme
value of their host star metallicities. In this case, we see that the
maximum of the false positive probabilities is $0.44\%$, therefore
this correlation is statistically significant for all our choices
of lower mass limits. This is at least a fivefold improvement over the sample
investigated by \citet{2011MNRAS.410.1631E}, due to the larger sample size.

\begin{deluxetable}{r@{}c@{}lccr@.lr@.lr@.lr@.l}
\tablewidth{0pc}
\tabletypesize{\scriptsize}
\tablecaption{
    Correlation between the host star metallicity and planetary
    parameters for known transiting exoplanets
    \label{tab:corr}
}
\tablehead{
    \multicolumn{3}{c}{Restriction}          &
    \multicolumn{1}{c}{$n$\tablenotemark{a}} &
    \multicolumn{1}{c}{Planetary}            &
    \multicolumn{2}{c}{$r$\tablenotemark{b}} &
    \multicolumn{2}{c}{$p_1$\tablenotemark{c}}&
    \multicolumn{2}{c}{$p_2$\tablenotemark{c}}&
    \multicolumn{2}{c}{$p_3$\tablenotemark{c}}\\
    \multicolumn{3}{c}{on planets}
&&   \multicolumn{1}{c}{parameter}
}
\startdata
       $0<$  & $M_\mathrm{core}$ &  &  7 & $M_\mathrm{core}$ &  0&78\tablenotemark{d} &  3&9\%      \\
             &  \mpl & $<0.6$ \mjup & 18 & \rpl              & -0&53\tablenotemark{e} &  2&4\%       \\
0.3  $\mjup\leqslant$ & \mpl & $\leqslant0.8$ \mjup & 30 & \mpl &  0&270  & 15&0\% & 15&0\%  & 15&0\% \\
0.35 $\mjup\leqslant$ & \mpl & $\leqslant0.8$ \mjup & 29 & \mpl &  0&106  & 58&4\% & 58&2\%  & 58&4\%  \\
0.4  $\mjup\leqslant$ & \mpl & $\leqslant0.8$ \mjup & 28 & \mpl &  0&247  & 20&4\% & 20&4\%  & 20&4\%   \\
0.3  $\mjup\leqslant$ & \mpl & $\leqslant0.8$ \mjup & 30 & \rpl & -0&505  &  0&44\% & 0&43\% &  0&44\%   \\
0.35 $\mjup\leqslant$ & \mpl & $\leqslant0.8$ \mjup & 29 & \rpl & -0&620  &  0&03\% & 0&03\% &  0&034\%   \\
0.4  $\mjup\leqslant$ & \mpl & $\leqslant0.8$ \mjup & 28 & \rpl & -0&575  &  0&14\% & 0&14\% &  0&14\%
\enddata
\tablenotetext{a}{sample size}
\tablenotetext{b}{correlation coefficient}
\tablenotetext{c}{estimates for false positive probability given by $t$-test, bootstrap method and $F$-test, respectively}
\tablenotetext{d}{reported by \citet{2006A&A...453L..21G}}
\tablenotetext{e}{reported by \citet{2011MNRAS.410.1631E}}
\end{deluxetable}

\subsection{Dependence on planetary equilibrium temperature}

Other factors like insolation are likely to influence planetary radius
as well, see e.g.~\cite{2007ApJ...659.1661F}, \cite{2010ApJ...724..866K}, \cite{2011MNRAS.410.1631E},
and \cite{2011A&A...531A..40F}. To
further investigate this relation, we compare two models: for null hypothesis,
we accept the linear planetary radius---host star metallicity relation
of the previous section:
\begin {equation}
\tilde \rpl^\mathrm{I} = R_0^\mathrm{I} + \alpha^\mathrm{I} \cdot\fehstar.
\end {equation}

The second model -- alternative hypothesis -- is similar to that
of \citet {2011MNRAS.410.1631E}, accounting for the equilibrium temperature
$T_\mathrm{eq}$ in addition to the host star metallicity:
\begin {equation}
\label{eq:secondmodel}
\tilde \rpl^\mathrm{II} = R_0^\mathrm{II} + \alpha^\mathrm{II} \cdot\fehstar + \beta^\mathrm{II} \cdot T_\mathrm{eq}.
\end {equation}
The equilibrium temperature of the planet is calculated from
the time-averaged stellar flux on its orbit,
assuming gray body spectrum for the planets,
and neglecting tidal and other heating mechanisms.
For simplicity, we now include all 30 planets of \reftabl {planets} in our models.
With the best fit parameters, the two models are
\begin{eqnarray*}
\tilde \rpl^\mathrm{I}  & = & 1.235\,\rjup{}-0.478\,\rjup{} \cdot \fehstar,\\
\tilde \rpl^\mathrm{II} & = & 0.690\,\rjup{}-0.431\,\rjup{} \cdot \fehstar + 0.398 \,\rjup \cdot \frac {T_\mathrm{eq}}{1000\,\mathrm K}.
\end{eqnarray*}

To quantify the statistical significance,
we apply the $F$-test again.
The resulting false positive probability is 0.18\%.
This means that once we accept the dependence of planetary radius
on host star metallicity, then the probability of such a correlation
with planetary equilibrium temperature if they were not physically
related is only 0.18\%. This strongly supports the three-parameter
linear fit model.
For reference, the residual sums of
squares for the one, two and three-parameter fits of the
\fehstar---\rpl\ data are $1.13\;\rjup^2$, $0.84\;\rjup^2$ and $0.58\;\rjup^2$, respectively.

\begin{figure}
\includegraphics*[width=\figurewidth]{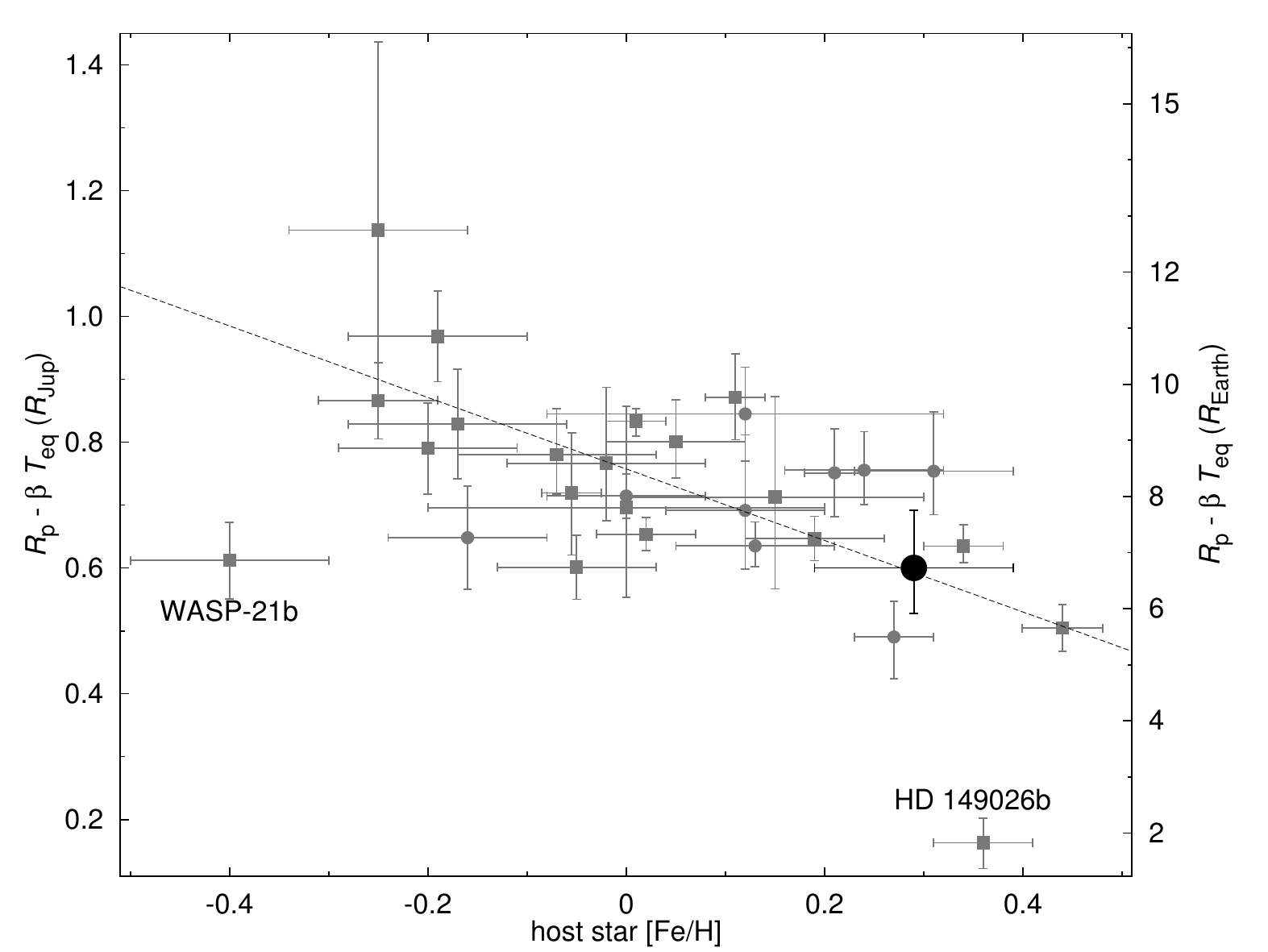}
\caption{
	Planetary radius corrected for linear equilibrium temperature dependence
	in Jupiter radii versus host star metallicity for
	the 30 known transiting exoplanets with masses between 0.3 \mjup{} and
	0.8 \mjup{}, including
    \hatcurb{} (solid black circle),
    other HATNet planets (solid gray circles),
    and other known transiting exoplanets (solid gray squares,
    WASP-21b and HD 149026b labeled);
    with best linear fit overlaid (dashed line).
\label{fig:exocorr}}
\end{figure}

\reffigl{exocorr} presents $\rpl-\beta^\mathrm{II}\cdot T_\mathrm{eq}$
versus metallicity for the 30 planets.  Equation (\ref{eq:secondmodel})
predicts this quantity to be
$R_0^\mathrm{II}+\alpha^\mathrm{II}\cdot\fehstar$, which is also plotted. 
\hatcurb{} apparently follows the model's prediction.  For reference,
the correlation coefficient between the displayed transformed variables
is now $r^\mathrm{II}=-0.536$, which has a larger absolute value than
$r^\mathrm{I}=-0.505$ between $\rpl$ and metallicity, as expected.

This analysis supports the statement that planetary radius depends on
equilibrium temperature in addition to host star metallicity, as found
by \citet {2011MNRAS.410.1631E} and \citet{2011A&A...531A..40F}.  However, this
correlation does not imply that insolation itself would inflate
planets: the underlying phenomenon could be related to anything
correlated to equilibrium temperature, or equivalently, orbital radius. 
For instance, \citet {2010ApJ...714L.238B} suggest that it is Ohmic
dissipation in the interior of the planet that inflates hot Jupiters. 
This theory is further supported by \citet{2011ApJ...729L...7L}.

Altogether, \hatcurb{} is an important addition to the growing sample
of low-mass Jupiters. It orbits a metal rich star, and supports the
suggested correlations between host star metallicity,
planetary equilibrium temperature, and planetary radius. Also,
\hatcur{} is chromospherically active, providing an excellent case
for refining the confidence level of the hypothesized correlation
between stellar activity and planetary temperature inversion.

\acknowledgements
HATNet operations have been funded by NASA grants NNG04GN74G,
NNX08AF23G and SAO IR\&D grants.  Work of G.\'A.B.~was
supported by the Postdoctoral Fellowship of the NSF Astronomy and
Astrophysics Program AST-0702843.
G.T.~acknowledges partial support from NASA grant NNX09AF59G.
A.J.~acknowledges support from Fondecyt project 1095213, BASAL CATA
PFB-06, FONDAP CFA 15010003, MIDEPLAN ICM Nucleus P07-021-F and Anillo
ACT-086.
G.K.~thanks the Hungarian Scientific Research Foundation (OTKA)
for support through grant K-81373.
L.L.K.~is supported by the ``Lend\"ulet'' Young Researchers Program
of the Hungarian Academy of Sciences
and the Hungarian OTKA grants K76816, K83790 and MB08C 81013.
Tam\'as Szalai (Univ.~of Szeged) is acknowledged for his
assistance during the ANU 2.3 m observations.
We acknowledge partial support also from the Kepler Mission under NASA
Cooperative Agreement NCC2-1390 (D.W.L., PI).
This research has made use of Keck telescope time
granted through NOAO (A201Hr) and NASA (N018Hr, N167Hr).
We also thank Mount Stromlo Observatory and Siding Spring Observatory
for the ANU 2.3 m telescope time.

\chapter{High precision relative photometry with HST}
\label{ch:hst}

\original{2013AJ....145..166B}

\chapterabstract

We present HST STIS observations of two occultations of the transiting exoplanet \hatponeb{}. By measuring the planet to star flux ratio near opposition, we constrain the geometric albedo of the planet, which is strongly linked to its atmospheric temperature gradient. 
An advantage of \hatpone{} as a target is its binary companion \ads{} A, which provides an excellent photometric reference, simplifying the usual steps in removing instrumental artifacts from HST time-series photometry. We find that without this reference star, we would need to detrend the lightcurve with the time of the exposures as well as the first three powers of HST orbital phase, and this would introduce a strong bias in the results for the albedo. However, with this reference star, we only need to detrend the data with the time of the exposures to achieve the same per-point scatter, therefore we can avoid most of the bias associated with detrending. Our final result is a $2\sigma$ upper limit of 0.64 for the geometric albedo of \hatponeb{} between 577 and 947 nm. 


\section {Introduction}

The effective temperature of Jovian extrasolar planets in close orbits is strongly influenced by irradiation from their host stars. It is the Bond albedo, defined as the reflected fraction of incident electromagnetic power, that determines how much of this irradiation contributes to the thermal balance of the planet. Atmospheric temperature, in turn, determines -- for a given composition -- the presence and position of absorbers, clouds and other structures determining the albedo. Once the albedo is inferred from observations, the goal is to find a self-consistent atmospheric model and temperature.

Even though it is the Bond albedo that can be used directly in the calculation of effective temperature, the geometric albedo $A_\mathrm g$ is more accessible observationally. It is defined as the ratio of reflected flux at opposition (zero phase angle) to the reflected flux by a hypothetical flat, fully reflecting, diffusely scattering (Lambertian) surface of the same cross-section at the same position. The geometric albedo of Lambertian surfaces is at most one: for example, a fully reflecting Lambertian sphere has a geometric albedo of $\frac23$ \citep {1992essi.book.....H}. However, some surfaces reflect light preferentially in the direction where it came from (a phenomenon known as opposition surge), and thus can exhibit geometric albedos exceeding one \citep[e.g.][]{1990Icar...88..407H}.

A simple way to measure albedo is to observe an occultation (secondary eclipse) of a transiting exoplanet and directly compare the brightness of the star only (while the planet is occulted) to the total brightness of the star and the planet near opposition (shortly before or after the occultation). Examples of other phenomena that can be used to constrain albedo but are not discussed in this paper are phase variations \citep[e.g.][]{2006Sci...314..623H}, Doppler-shift of reflected starlight \citep[e.g.][]{2007arXiv0711.2304L}, and polarization of reflected starlight \citep[e.g.][]{2006PASP..118.1302H}.

To mention specific examples, the exoplanet \hdkettob{} has mass $M=0.69\;M_\mathrm J$, radius $R=1.36\;R_\mathrm J$, orbital period $P=3.52$ days, and zero-albedo equilibrium temperature $\Teq=1450$ K \citep{2008ApJ...677.1324T}. When calculating \Teq, perfect heat redistribution on the planetary surface is assumed. \citet {2008ApJ...689.1345R,2009IAUS..253..121R} observe the occultation of \hdkettob{} in the 400--700 nm bandpass with the Microvariability and Oscillations of Stars (MOST) satellite. They find a geometric albedo of $A_\mathrm g=0.038\pm0.045$, and infer a $1\sigma$ upper limit of 0.12 on the Bond albedo, indicating the absence of reflective clouds. Based on atmospheric models, this constrains the atmospheric temperature to between 1400 K and 1650 K. Normally, a cloud-free atmosphere exhibits low albedo due to the strong pressure-broadened absorption lines of neutral sodium and potassium \citep {2000ApJ...538..885S}. However, Spitzer Space Telescope observations of \hdkettob{} at 3.6, 4.5, 5.8, and 8.0 $\mu$m indicate water emission features, suggesting temperature inversion in the higher atmosphere, which hints that an unknown absorber is present at low pressure \citep {2008ApJ...673..526K,2007ApJ...668L.171B}.

Another well-studied example is \hdegy{} \citep[$M=1.14\;M_\mathrm J$, $R=1.14\;R_\mathrm J$, $P=2.22$ days, $T_\mathrm{eq}=1200$ K,][]{2008ApJ...677.1324T}. \citet {2008Natur.456..767G} and \citet{2008ApJ...686.1341C} observe its occultations with the Spitzer Space Telescope. They find strong water absorption features, indicating the lack of temperature inversion in the atmosphere. \citet {2011MNRAS.416.1443S} perform transmission spectroscopy on this planet, and interpret the results as indicating high altitude haze. This would cause the planet to exhibit high geometric albedo in the visible. \citet {2009IAUS..253..121R} report on MOST observations of this planet's occultation, but cannot constrain the albedo due to the high activity of the host star. Note that \hdkettob{} and \hdegyb{} only differ by a few hundred kelvins in terms of \Teq, yet seem to exhibit very different atmospheres.

The subject of this work is \hatponeb{}, a transiting exoplanet with mass $M=0.524\;M_\mathrm J$, radius $R=1.225\;R_\mathrm J$, orbital period $P=4.47$ days \citep{2008ApJ...686..649J}, and zero-albedo equilibrium temperature $\Teq=1300$ K \citep {2008ApJ...677.1324T}. This last value is between those of \hdkettob{} and \hdegyb{}. Indeed, \citet {2010ApJ...708..498T} observe two occultations of \hatponeb{} with Spitzer and infer a modest temperature inversion in the atmosphere from the occultation depths at 3.6, 4.5, 5.8, and 8.0 $\mu$m. In the light of these observations, constraining the geometric albedo of this planet in the visible and near infrared would be useful for better understanding its atmospheric structure, and for refining atmospheric models.

In this paper, we report on observations of two occultations of \hatponeb{}. Section \ref {sec:hst:observations} presents the details of the observations, flux extraction, and detrending. We describe how we calculate the geometric albedo in Section \ref {sec:hst:albedo}, with special attention to handling each uncertainty source, and comparing relative photometry results to those without using the reference star. We discuss our findings in Section \ref {sec:hst:discussion}.


\section {Observations and data analysis}
\label {sec:hst:observations}

The exoplanetary host star \hatpone{} is a member of the wide binary system \ads, with $11.2"$ projected separation at a distance of 139 pc. \hatpone, also known as \ads{} B, with $V=10.4$ is only 0.4 magnitude fainter than its binary companion \ads{} A of the same G0V spectral type. This allows for relative (differential) photometry to mitigate the effect of systematic errors.

\subsection {Observations and data preprocessing}
A proposal was accepted as GO 11069 to observe two occultations (secondary eclipses) of \hatponeb{} with the Hubble Space Telescope (HST) Advanced Camera for Surveys (ACS) High Resolution Channel (HRC). However, ACS failed in 2007 January, before the observations would have been carried out, and HRC remains inoperational to date. Instead, another HST instrument, the Space Telescope Imaging Spectrograph (STIS) carried out the observations, as program GO 11617. Two occultations of \hatponeb{} were observed during two visits, including two orbits before and one after each occultation. Because of the brightness of the targets, spectroscopy was required to allow for reasonably long exposures. To capture both stars without very tight constraints on spacecraft orientation, we did not use a slit. Table \ref {tab:observations} summerizes the details of the observations. With these settings, the largest electron count in each exposure was about half the well size, well below saturation. However, it is interesting to note that longer exposures would not have posed a problem in terms of linearity either: when STIS pixels get saturated, the excees charge bleeds into surrounding pixels with virtually no loss, and summing these pixel counts still results in a linear response \citep {1999PASP..111.1009G}.

\begin{deluxetable}{lcc}
\tablewidth{0pc}
\tabletypesize {\scriptsize}
\tablecaption {HST/STIS program GO 11617 observation parameters\label {tab:observations}}
\tablehead {& visit 1 & visit 2}
\startdata
HJD at beginning of first exposure & 2\,455\,544.8269 & 2\,455\,888.6602 \\
HJD at end of last exposure & 2\,455\,545.1181 & 2\,455\,888.9497 \\
number of orbits & 5 & 5 \\
number of spectra for each orbit & 19+23+23+23+23 & 19+23+23+23+23 \\
number of spectra in total & 111 & 111 \\
grating & \multicolumn2c{\texttt{G750L}} \\
slit & \multicolumn2c{slitless} \\
exposure time & \multicolumn2c{100 s} \\
cadence & \multicolumn2c{128 s} \\
subarray size & \multicolumn2c{380x1024 pixels} \\
gain & \multicolumn2c{4}
\enddata
\end{deluxetable}

\begin {figure}
\begin {center}
\includegraphics*[width=\figurewidth]{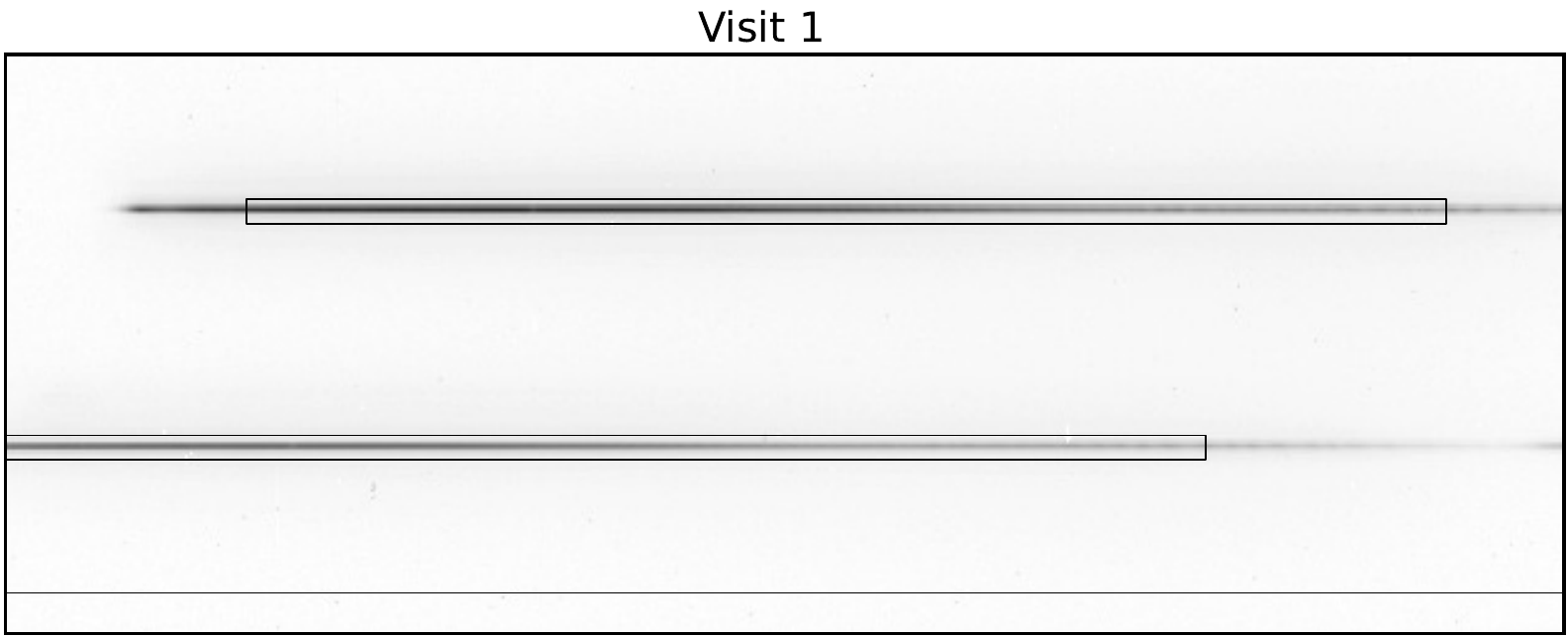}
\includegraphics*[width=\figurewidth]{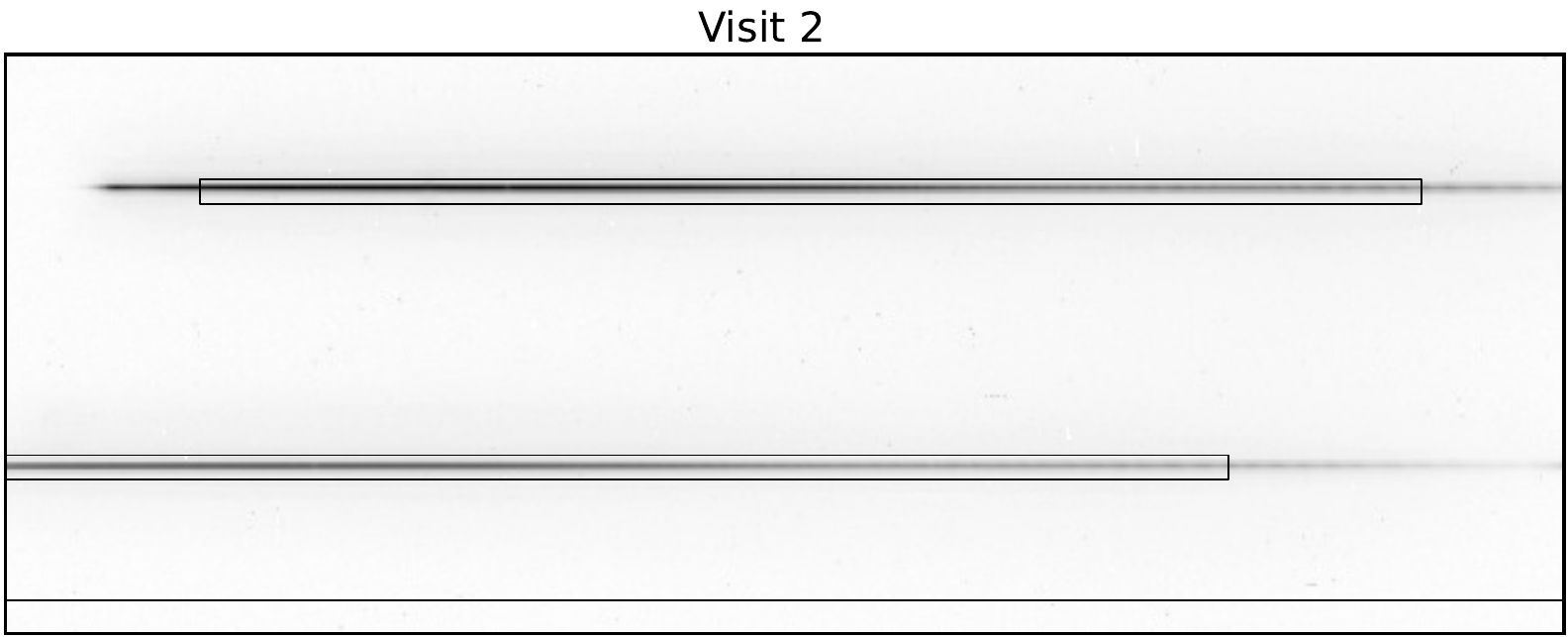}
\end {center}
\caption {The first spectrum used in the final analysis (the second exposure of the second orbit) of the first and second visits (left and right panels, respectively). The upper star is \ads{} A, the lower one is the planetary host \ads{} B. The rectangular apertures around the spectra are also shown. The bottom rectangle is the background aperture. Note the different cross-dispersion distance and dispersion direction shift between the two stars' spectra for the two visits due to spacecraft orientation. The blue end of the stellar apertures is determined by where the detector edge cuts the specrum of \ads{} B.}
\label {fig:spectra}
\end {figure}

We identify hot pixels and exclude them from our apertures, and identify cosmic rays and substitute them by the average of the previous and next frame values. Then we perform rectangular aperture photometry on the two stars, and define an aperture in the entire length of the detector in the dispersion direction for sky background estimation. For each exposure, we subtract the background aperture photon count from the stellar aperture photon counts, scaled by the number of pixels. Figure \ref {fig:spectra} shows a typical exposure from each visit, with the apertures used. The blue end of the stellar apertures is 564 nm for the first visit and 557 nm for the second. The difference is caused by different orientations of the telescope around its optical axis, resulting in the STIS detector edge cutting the \hatpone{} spectrum at different positions for the two visits. These wavelength values are calculated after identifying the H$\alpha$ and Na I D lines in the stellar spectra. We extract the same spectral range for the two stars within a visit to fight wavelength-dependent systematics. However, we allow for different blue end cuts between visits, otherwise we would lose too many photons in the second visit due to the more restrictive wavelength limit of the first one. We do not expect the geometric albedo to vary significantly due to this small change in blue end wavelength cut.

\subsection {Detrending}

The next step is to detrend the data, that is, to mitigate instrumental effects by subtracting multiples of vectors describing circumstances of the observations. We try detrending with time (to remove the overall linear trend within a visit), HST orbital phase and its powers (to remove orbitwise periodic variations), CCD housing temperature (CCD chip temperature is not available with the current Side-2 electronics), a focus model provided by HST Observatory Support, fine pointing data available from telemetry, and fine pointing data based on a fit for the position of the spectra on the CCD. To detrend, we simultaneously fit for free parameters of the lightcurve model (reference flux and planet-to-star flux ratio) as well as detrending vector coefficients using a linear algebraic least square method. We actually use the magnitude of \hatpone{} or the magnitude difference of the two stars, that is, we assume that instrumental effects are multiplicative. Since both the planet-to-star flux ratio and detrending corrections are very small, this is equivalent to assuming additive effects and fitting for the flux or flux ratio. Each detrending vector is mean subtracted so that they do not change the average stellar magnitude. To quantify the effect of detrending and avoid overfitting, we minimize the Bayesian Information Criterion (BIC), which is the sum of $\chi^2$ and a term penalizing extra model parameters \citep {Schwarz1978}.

When analyzing STIS data to perform photometry on exoplanetary host stars, \citet {2007ApJ...655..564K} find it justified to detrend with a cubic polynomial of HST orbital phase, whereas \citet {2001ApJ...552..699B} and \citet {2011MNRAS.416.1443S} use fourth order polynomials. Indeed, if we only consider the lightcurve of \hatpone, we find the lowest BIC when detrending with time at mid-exposure and the first three powers of HST orbital phase. However, if we divide the lightcurve of the planetary host by that of the reference star ADS 16402 A, we find the lowest BIC when detrending only with mid-exposure time. This shows that relative photometry is less sensitive to systematics, and can mitigate HST orbital effects enough that detrending with orbital phase is not justified. The Bayesian Information Criterion does not justify detrending with the temperature, focus, or jitter vectors in either case.

Figure \ref {fig:lightcurves} presents the raw lightcurves (panels a--d), without detrending, which indeed show strong orbitwise periodic variations. Panels (e, f) show the lightcurve of \hatpone{} detrended with time at mid-exposure and the first three powers of HST phase, demonstrating that this procedure corrects for the overall linear trend and most of the orbitwise periodic variation. On the other hand, the raw relative lightcurves shown on panels (g, h) do not exhibit such large variations, and we only need to remove the linear trend (panels i, j). These panels all have logaritmic vertical axes with the same scaling, so that relative scatter is directly comparable. Note that since we perform a simultaneous fit of the occultation lightcurve and detrending vectors, the resulting $\chi^2$ tells us how close the data are to a model accounting for both the occultation and systematics, without the danger of misinterpreting the occultation as scatter.

The strength of these observations is the presence of the reference star, which already proves to be advantageous. In order to quantify how much it improves the albedo limits, we perform a full analysis both without and with this reference data, independently tuning all extraction parameters.

\subsection {Aperture parameters and data omission}

If we only extract the flux of \hatpone, we find that we obtain the least scatter after detrending (with mid-exposure time and first three powers of HST orbital phase) if the stellar apertures are 14 and 16 pixels wide in cross-dispersion direction for the two visits, respectively, and the stellar spectra are cut off at 788 nm. Redward of this wavelength there is extra scatter due to fringing, which is difficult to combat in slitless mode. If we use the flux ratio of the two stars, we get the least scatter after detrending (this time with mid-exposure time only) if the stellar apertures are 16 and 18 pixels wide in cross-dispersion direction for the two visits respectively, and the stellar spectra are cut off further in the near infrared at 947 nm. Using the reference star thus allows us to extract photons from a larger aperture. The optimal background aperture is 27 pixels wide for the first visit and 22 for the second in both cases.

We estimate sky background using an aperture placed as far from the stellar spectra as possible, to avoid contamination by starlight. Varying the background aperture width by a few tens of pixels introduces scatter of 0.02 in the best fit albedo as long as the aperture is not too narrow and not too close to the stars either. Given the other uncertainty sources discussed in Section \ref {sec:hst:uncertainty}, this means that the results are practically insensitive to the exact choice of the background aperture. However, if we place the background aperture between the two spectra on the detector, or expand it on the side to get within 50 pixels of \hatpone, we find a larger average error, of 0.05, in the geometric albedo due to stray starlight.

\begin {figure}
\includegraphics*[width=\halffigurewidth]{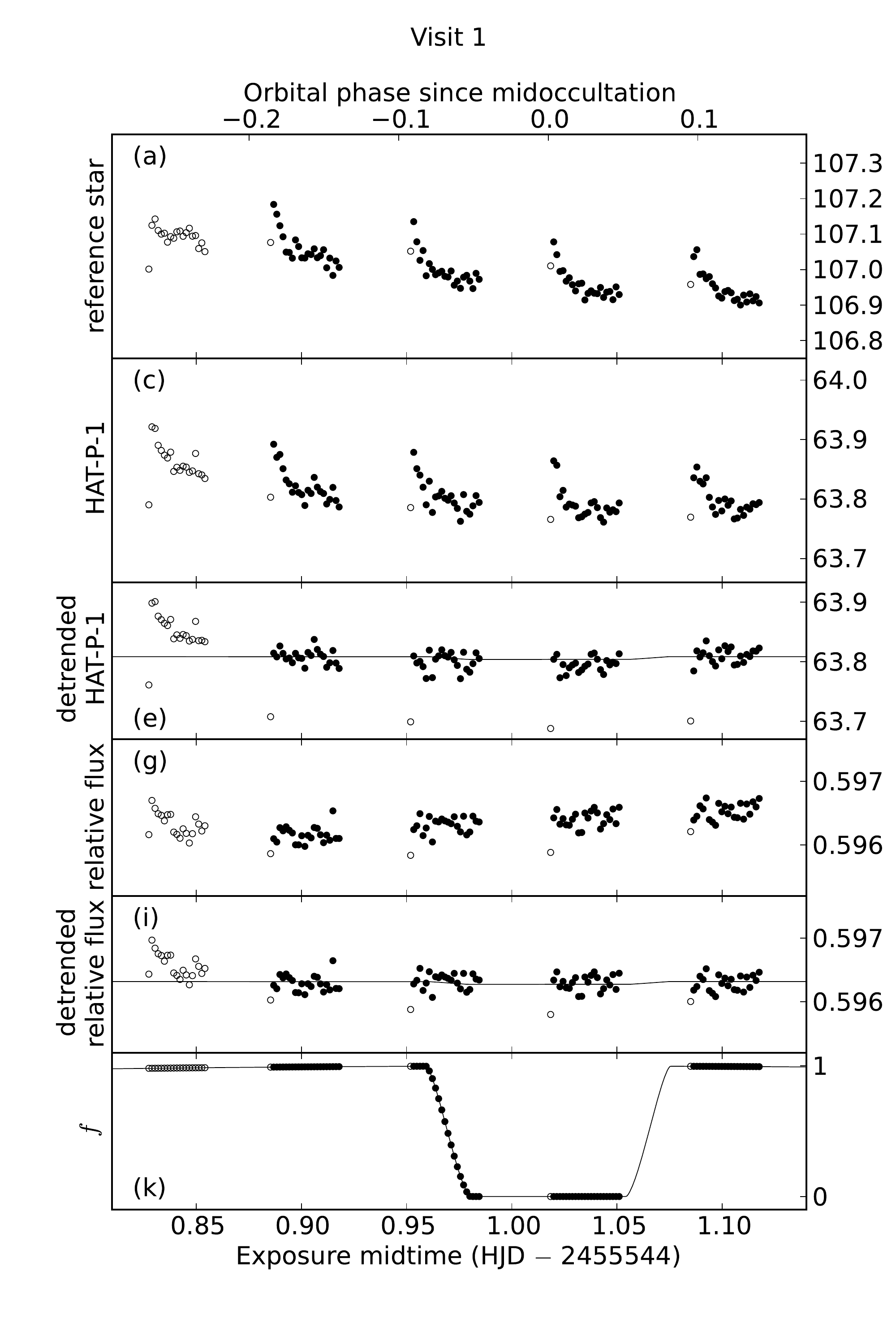}
\includegraphics*[width=\halffigurewidth]{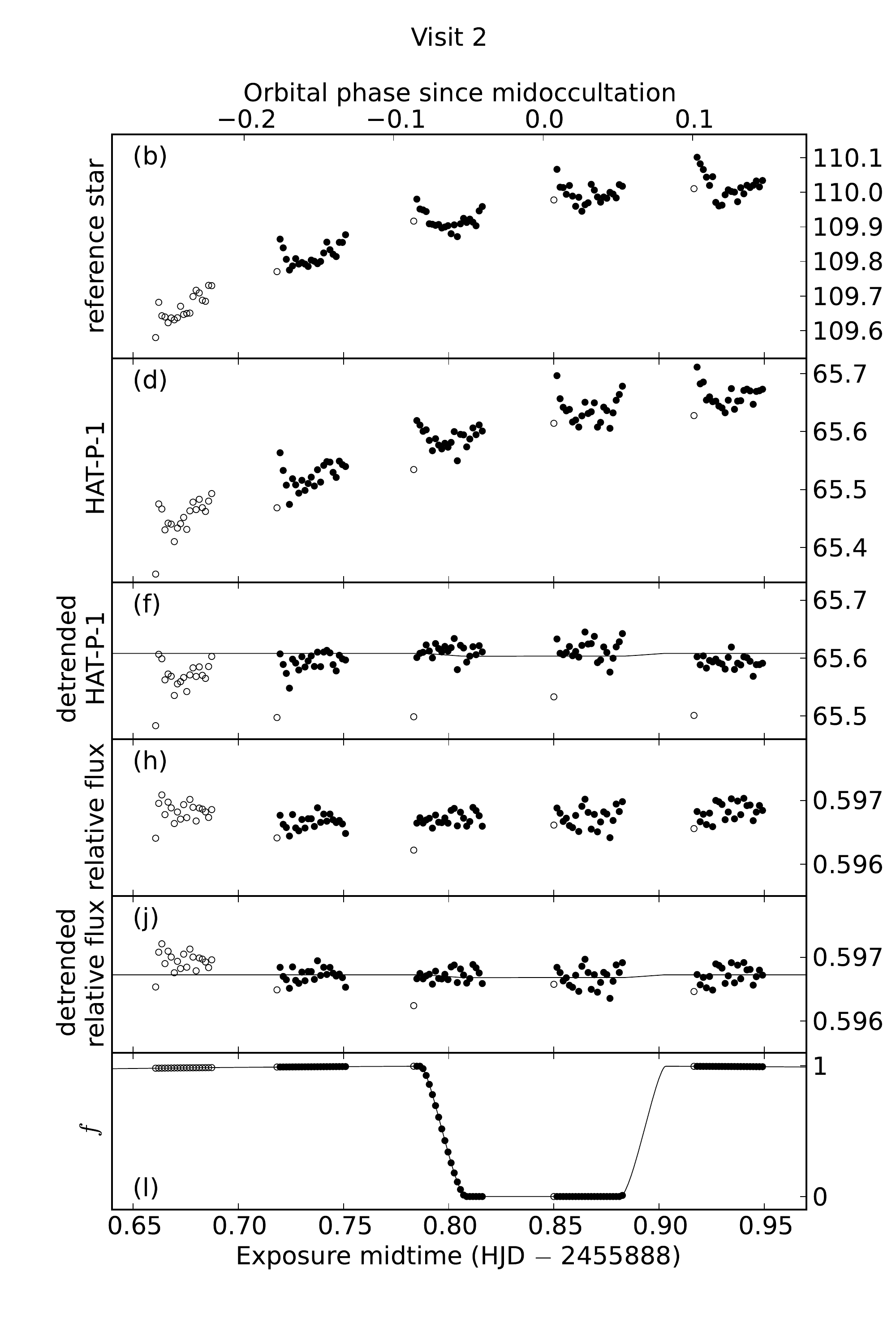}
\caption {Panels (a, b): background-subtracted photon count per exposure for reference star \ads{} A, in million photons. Panels (c, d): same for planetary host star \hatpone. Panels (e, f): photon count of \hatpone{} detrended with time at mid-exposure and first three powers of HST orbital phase, in million photons. The model lightcurve of a fully reflecting planet is overplotted with a solid line. Panels (g, h): relative flux, that is, background-subtracted photon count of \hatpone{} divided by background-subtracted photon count of reference star. Panels (i, j): relative flux detrended with time at mid-exposure. The model lightcurve of a fully reflecting planet is overplotted with a solid line. Panels (k, l): the fraction $f$ of the planetary surface that is illuminated and unobscured, with the values for each exposure overlaid on the continuous curve. The bottom horizontal axes show mid-exposure time in HJD, the top horizontal axes show $\phi-\phi_0$, the planetary orbital phase since midoccultation. Panels (a--j) have logarithmic vertical axes with the same scaling, so that relative scatter is directly comparable. Panels (a), (c), (e), (g), (i) and (k) display the first visit, (b), (d), (f), (h), (j), and (l) the second. Filled circles represent data points included in the analysis, empty circles the omitted ones.}
\label {fig:lightcurves}
\end {figure}

Finally, we investigate whether it is justified to omit data points. For example, \citet {2011MNRAS.416.1443S} and \citet {2007ApJ...655..564K}
both omit the first orbit of each five orbit visit, also the first exposure of each subsequent orbit, because these data points exhibit larger scatter. The scatter of the first orbit might be attributed to the thermal settling of the spacecraft after its new pointing. We therefore calculate the scatter per data point for all twelve possible combinations of omitting the first orbit or not, omitting the first or first two exposures of each orbit or not, and omitting the last exposure of each orbit or not. Comparing these results, we find that both for the analysis of \hatponeb{} only and for relative photometry using the reference star, it justified to omit the first orbit of each visit and the first exposure of each subsequent orbit, but not more, consistently with \citet {2011MNRAS.416.1443S} and \citet {2007ApJ...655..564K}. We therefore keep 88 data points per visit. These data points are represented with filled circles on Figure \ref {fig:lightcurves}, whereas omitted data points are represented with empty ones.

Note that the data analysis parameter space has many dimensions: stellar and background aperture geometry, data omission, and choice of detrending parameters. The choices presented above were found after multiple iterations, and they yield the minimum residual sum of squares in each dimension, but strictly speaking, they might not represent the global minimum.

\section {Upper limit on geometric albedo}
\label {sec:hst:albedo}

\subsection {Lightcurve model}

To model the occultation lightcurve, we adopt the transit ephemeris, the planetary radius and orbital semi-major axis in stellar radius units, and transit impact parameter values and uncertainties reported by \citet {2008ApJ...686..649J}. We assume that the orbit is circular, consistently with theoretical expectations, radial velocity measurements \citep {2007ApJ...656..552B}, and occultation timing \citep {2010ApJ...708..498T}. We account for the light travel time of 55 s across the planetary orbit, and neglect the thermal radiation of the planet. We calculate the projected area of the unobscured part of the dayside of the planet by assuming that both the stellar and planetary disks are circular, and the terminator line on the planet is an arc of an ellipse. Let $f$ denote this area relative to the total planetary disk. Then the observed flux at any given time is $F_\star + fF_\mathrm p$, where $F_\star$ is the stellar flux, and $F_\mathrm p$ is the flux of the planet in opposition. Note, however, that we can never observe the theoretical maximum flux $F_\mathrm p$ from the planet, because opposition happens during occultation (therefore $f$ is always smaller than one). We derive the following model for $f$:
\begin {align}
\nonumber
d &= \sqrt {R_\star^2 b^2 + a^2 \sin^2 (\phi-\phi_0)} \\
\nonumber
u &= \arccos\frac {d^2 + R_\star^2 - R_\mathrm p^2}{2dR_\star}, \quad\quad
v = \arccos\frac {d^2 + R_\mathrm p^2 - R_\star^2}{2d R_\mathrm p} \\
f &=  \begin {cases}
0 & \genfrac{}{}{0pt}{1}{\textrm {if } d \leqslant R_\star - R_\mathrm p\quad}{\textrm{(occultation),}} \\
\max \left( \frac {1+\sqrt{1-\frac{d^2}{a^2}}} 2 - \left(\frac u\pi - \frac{\sin2u}{2\pi}\right)\frac{R_\star^2}{R_\mathrm p^2} - \left(\frac v\pi - \frac{\sin2v}{2\pi}\right), 0 \right) & \genfrac{}{}{0pt}{1}{\textrm {if } R_\star - R_\mathrm p < d \leqslant R_\star + R_\mathrm p}{\textrm{(ingress and egress),}} \\
\frac {1+\sqrt{1-\frac{d^2}{a^2}}} 2 & \genfrac{}{}{0pt}{1}{\textrm {if } R_\star + R_\mathrm p < d\quad\quad\quad}{\textrm{(out of occultation).}}
\end {cases}
\end {align}
Here 
$R_\star$ is the stellar radius,
$R_\mathrm p$ is the planetary radius,
$a$ is the planetary orbital radius, 
$\phi$ is the planetary orbital phase,
$\phi_0$ is the orbital phase at midoccultation, 
$b$ is the impact parameter, 
$d$ is the projected distance of the center of the planetary and stellar disks, 
and $u, v \in [0,\pi]$ are auxiliary functions for the case of ingress and egress.
The bottom panels of Figure \ref {fig:lightcurves} show $f$ as a continuous function of time, with the value in the middle of each exposure overlaid.

This model gives the exact result as long as the boundary of the stellar disk and the terminator line on the planet do not intersect in projection, and is extended in a monotonic and continuous manner at the end of ingress and beginning of egress when they do. This introduces a small error which only affects a few data points. Note, however, that there are other sources of error: when determining the terminator line, we assume that the planet is irradiated by parallel rays from the direction of the center of the star, instead of properly calculating irradiation in the belt from where the star is partially seen on the horizon of the planet. Also, we do not account for that the substellar point is closer to the star, therefore it is exposed to more irradiation. These errors are in the order of $\frac{R_\star^2}{a^2}\approx0.01$, that is, they bias our geometric albedo estimate by the order of 1\%.

We assume Lambertian reflectance of the incident light. Note that the maximum angle of stellar irradiance on the planet to the line of sight during our science exposures used in the final analysis is $\approx11^\circ$. In case of reflection from materials like regolith, this would be very different from reflection in opposition, but we expect the reflected flux for a gas giant like \hatponeb{} to depend only weakly on the incident angle.

Using this model lightcurve, we then fit simultaneously for the planet to star flux ratio at opposition and the coefficients of the detrending vectors. Then the geometric albedo can be calculated using the expression \citep[e.g.][]{2008ApJ...689.1345R}
\begin {equation}
A_\mathrm g = \frac {F_\mathrm p}{F_\star} \frac{a^2}{R_\mathrm p^2}. 
\end {equation}
To illustrate the magnitude of the effect with respect to the scatter of the data, Figure \ref {fig:lightcurves} panels (e, f) and (i, j) feature a plot of a model lightcurve, assuming that the planet is a fully reflecting Lambertian sphere with a geometric albedo of $\frac23$.

\subsection {Uncertainty sources}
\label {sec:hst:uncertainty}

There are two sources of errors in the inferred geometric albedo: observational errors and contribution of parametric uncertainties of the system. We estimate the total observational uncertainty $u_\mathrm{obs}$ by bootstrapping. Since instrument parameters like temperature might not have a normal distrubtion or might not influence the data in a linear fashion, we cannot assume that the error distribution is normal. Therefore we apply the bias corrected accelerated bootstrap method \citep {1987JASA..82..171}, a generalized bootstrap algorithm that partially corrects for effects due to non-normal error distributions.

The total observational uncertainty is due to photon noise and other uncertainties due to instrument thermal instability, pointing jitter, and stellar activity in both stars. (Quantization noise due to the gain set to 4 can be neglected.) We estimate the uncertainty $u_\mathrm{photon}$ due to photon noise by independently redrawing every data point (\hatpone, reference star, and background for each exposure) from a Poisson distribution with a parameter given by the original photon count, and recalculating the geometric albedo. Finally, we get the estimate of the uncertainty $u_\mathrm{other}$ due to other sources by subtracting the photon noise estimate from the total observational uncertainty estimate in quadrature.

When quantifying the uncertainty contributions of the planetary system parameters, we divide them into two groups: ephemeris (mid-transit time and period), and geometry (planetary radius and orbital semi-major axis relative to the stellar radius, and impact parameter). This division is important for two reasons: first, the two visits took place 1181 and 1525 days after the reference mid-transit ephemeris of \citet {2008ApJ...686..649J}, respectively, therefore we expect that the second visit will suffer more from the uncertainty in mid-occultation time. Second, it is easier to refine the ephemeris by photometric follow-up transit observations than to refine the geometric parameters, so we want to assess how much this would improve the results. Note that geometric parameters not only influence the occultation lightcurve shape that we use for fitting, but also factor in when converting the planet-to-star flux ratio to geometric albedo. We estimate the uncertainties $u_\mathrm{ephem}$ in the geometric albedo due to uncertainties in ephemeris, $u_\mathrm{geom}$ due to those in geometric parameters, and their total contribution $u_\mathrm{param}$ due to all parametric uncertainties by redrawing the respective parameters from independent normal distributions defined by their best fit values and uncertainties, and recalculating the geometric albedo in each case. We confirm that $u_\mathrm{ephem}$ and $u_\mathrm{geom}$ add up in quadrature to $u_\mathrm{param}$. We use 1\,000\,000 bootstrap iterations to estimate the observational uncertainty $u_\mathrm{obs}$, and 1\,000\,000 random drawings to estimate each of $u_\mathrm{photon}$, $u_\mathrm{ephem}$, $u_\mathrm{geom}$, and $u_\mathrm{param}$.

Finally, we add the observational and parametric uncertainty estimates in quadrature to estimate the total uncertainty $u_\mathrm{total}$, and add this to the best fit geometric albedo to calculate the upper limits. The breakdown of uncertainty sources is represented in a tree structure in Tables \ref{tab:only} and \ref{tab:relative}: the upper limit is the sum of its children nodes best fit and total uncertainty, and uncertainties are quadrature sums of their children nodes.

We find that most lower albedo limits of the final analysis are negative, therefore we only present upper limits for the geometric albedo. The $1\sigma$ and $2\sigma$ upper limits are determined so that the probability of the geometric albedo being smaller than this is $0.6827$ and $0.9545$, respectively. As a comparison, the one-sided $1\sigma$ and $2\sigma$ upper limits of a normal distribution occur at $0.475\sigma$ and $1.690\sigma$ above the mean, respectively, and we expect our corresponding uncertainties to have a similar ratio. 

Table \ref{tab:only} presents the best fit values, uncertainties due to various phenomena, and upper limits for the geometric albedo, based on the lightcurve of \hatpone{} only. We performed the calculations for the two visits separately, and also for a joint model, where a single geometric albedo value was fit for the two visits, but we allowed for different coefficients of the detrending vectors for the two visits. The upper limits highlight the weakness of detrending: the $2\sigma$ upper limit based on the first visit data is meaningless (greater than one), and even worse, the limit based on the second visit data only is unphysical (negative). The reason for this is that the detrending vectors are not orthogonal to the occultation signal, that is, systematic effects have a component that mimics the occultation lightcurve. Therefore even though detrending is justified by the Bayesian Information Criterion, it introduces a bias in the geometric albedo values.

On the other hand, if we feed the flux ratio of the planetary host star \hatpone{} and the reference star \ads{} A into the occultation lightcurve model, much less detrending is justified, thus we expect less bias in the result. Indeed, Table \ref{tab:relative} shows that the results from the two visits are much closer to each other, and all upper limits are positive. Even though the uncertainties in the two cases are very similar, as we can tell by comparing values in Tables \ref{tab:only} and \ref{tab:relative}, in the second case, this is achieved by using only one detrending vector instead of four. This shows the enormous advantage of the reference star: to diminish the need for detrending, therefore arrive at similar uncertainties with much less bias. We adopt the upper limits of the joint fit as our final result.

\begin {deluxetable} {llllcccccc}
\tablewidth{0pc}
\tabletypesize {\scriptsize}
\tablecaption {Best fit values, uncertainties due to different sources, and upper limits for the geometric albedo, based on \hatpone{} lightcurve only, for comparison\label{tab:only}}
\tablehead {\; & \; & \; & \; & \multicolumn2c {visit 1} & \multicolumn2c {visit 2} & \multicolumn2c {joint fit} \\
&&&& $1\sigma$ & $2\sigma$ & $1\sigma$ & $2\sigma$ & $1\sigma$ & $2\sigma$}
\startdata
\multicolumn4l {$A_g$ upper limit} & 1.44 & 1.94 & $-1.99$ & $-1.46$ & $-0.38$ & 0.03 \\
& \multicolumn3{|@{}l} {-- best fit $A_g$} & \multicolumn2c {$1.25$} & \multicolumn2c {$-2.19$} & \multicolumn2c {$-0.54$} \\
& \multicolumn3{|@{}l} {-- $u_\mathrm{total}$} & 0.19 & 0.69 & $\phantom{-}0.21$ & $\phantom{-}0.74$ & $\phantom{-}0.16$ & 0.57 \\
&& \multicolumn2{|@{}l} {-- $u_\mathrm{obs}$} & 0.19 & 0.68 & $\phantom{-}0.20$ & $\phantom{-}0.72$ & $\phantom{-}0.16$ & 0.57 \\
&& \multicolumn1{|l}{\;} & \multicolumn1{|@{}l} {-- $u_\mathrm{photon}$} & 0.15 & 0.52 & $\phantom{-}0.14$ & $\phantom{-}0.51$ & $\phantom{-}0.10$ & 0.36 \\
&& \multicolumn1{|l}{\;} & \multicolumn1{|@{}l} {-- $u_\mathrm{other}$} & 0.12 & 0.44 & $\phantom{-}0.14$ & $\phantom{-}0.51$ & $\phantom{-}0.12$ & 0.44 \\
&& \multicolumn2{|@{}l} {-- $u_\mathrm{param}$} & 0.03 & 0.11 & $\phantom{-}0.05$ & $\phantom{-}0.17$ & $\phantom{-}0.01$ & 0.05 \\
&&& \multicolumn1{|@{}l} {-- $u_\mathrm{ephem}$} & 0.00 & 0.00 & $\phantom{-}0.03$ & $\phantom{-}0.08$ & $\phantom{-}0.01$ & 0.04 \\
&&& \multicolumn1{|@{}l} {-- $u_\mathrm{geom}$} & 0.03 & 0.11 & $\phantom{-}0.04$ & $\phantom{-}0.15$ & $\phantom{-}0.01$ & 0.03
\enddata
\end {deluxetable}

\begin {deluxetable} {llllcccccc}
\tablewidth{0pc}
\tabletypesize {\scriptsize}
\tablecaption {Best fit values, uncertainties due to different sources, and upper limits for the geometric albedo, based on the lightcurves of \hatpone{} and reference star \ads{} A, our final results\label{tab:relative}}
\tablehead {\; & \; & \; & \; & \multicolumn2c {visit 1} & \multicolumn2c {visit 2} & \multicolumn2c {\textbf{joint fit}} \\
&&&& $1\sigma$ & $2\sigma$ & $1\sigma$ & $2\sigma$ & $\mathbf1\pmb\sigma$ & $\mathbf2\pmb\sigma$}
\startdata
\multicolumn4l {$A_g$ upper limit} & 0.08 & 0.59 & 0.55 & 1.14 & $\mathbf{0.24}$ & $\mathbf{0.64}$ \\
& \multicolumn3{|@{}l} {-- best fit $A_g$} & \multicolumn2c {$-0.13$} & \multicolumn2c {$0.31$} & \multicolumn2c {$\mathbf{0.09}$} \\
& \multicolumn3{|@{}l} {-- $u_\mathrm{total}$} & 0.20 & 0.71 & 0.23 & 0.83 & 0.15 & 0.54 \\
&& \multicolumn2{|@{}l} {-- $u_\mathrm{obs}$} & 0.19 & 0.71 & 0.23 & 0.83 & 0.15 & 0.53 \\
&& \multicolumn1{|l}{\;} & \multicolumn1{|@{}l} {-- $u_\mathrm{photon}$} & 0.15 & 0.57 & 0.16 & 0.56 & 0.11 & 0.39 \\
&& \multicolumn1{|l}{\;} & \multicolumn1{|@{}l} {-- $u_\mathrm{other}$} & 0.12 & 0.42 & 0.17 & 0.61 & 0.10 & 0.36 \\
&& \multicolumn2{|@{}l} {-- $u_\mathrm{param}$} & 0.03 & 0.10 & 0.01 & 0.07 & 0.01 & 0.07 \\
&&& \multicolumn1{|@{}l} {-- $u_\mathrm{ephem}$} & 0.02 & 0.08 & 0.00 & 0.05 & 0.01 & 0.05 \\
&&& \multicolumn1{|@{}l} {-- $u_\mathrm{geom}$} & 0.01 & 0.05 & 0.01 & 0.06 & 0.01 & 0.05
\enddata
\end {deluxetable}

By comparing the different contributions to the uncertainty of the geometric albedo, we see that the observational uncertainties are much larger than the parametric ones. This means that performing further photometric observations of occultations could significantly improve the albedo upper limit, whereas using additional transit observations to refine the ephemeris and geometric parameters and use them to reanalyze these data would not. As for the observational uncertainties, residual systematic uncertainty $u_\mathrm{other}$ and photon noise contribution $u_\mathrm{photon}$ are within a factor of two, which tells us that further improving our data analysis could push the upper albedo limits down only by a small amount. Surprisingly, we do not always find $u_\mathrm{ephem}$ to be larger for the second visit than for the first, but we confirm that $1\sigma$ and $2\sigma$ uncertainties of each kind have a ratio close to what is expected for a normal distribution.

\section {Discussion and summary}
\label {sec:hst:discussion}

Based on HST STIS observations, we established $0.24$ as the $1\sigma$, and $0.64$ as the $2\sigma$ upper limit for the geometric albedo of \hatponeb{} in the 557--947 nm band. Unfortunately, this limit is not tight enough to determine whether there is temperature inversion in the atmosphere. This question is relevant because \hatponeb{} has an equilibrium temperature between that of \hdkettob{} (thought to exhibit temperature inversion) and \hdegyb{} (thought not to). In addition, a better constrained albedo would provide information about the actual atmospheric temperature of the planet, as well as indicate the presence or absence of reflective clouds or high-altitude haze.

Our data analysis demonstrates that the reference star \ads{} A helps us greatly to reduce systematic effects. Even though detrending with powers of HST orbital phase would equally reduce scatter in the signal \citep[just like demonstrated by][]{2001ApJ...552..699B,2007ApJ...655..564K,2011MNRAS.416.1443S}, we find that it introduces a bias in the geometric albedo estimate. We attribute this effect to the fact that these detrending vectors are not orthogonal to the occultation signal. Most of this bias can be avoided by performing relative photometry, so that much less detrending is necessary to mitigate systematic effects. This is possible because \ads{} A has a similar brightness and same spectral type as \hatpone{}, and their angular distance is small enough so that they fit in the STIS field of view, but large enough so that their PSFs do not overlap.

We found that the uncertainties of the system parameters have a negligible effect on the geometric albedo uncertainty. The dominant uncertainty sources are photon noise and other noise effects (thermal instability, pointing jitter, other systematics, and astrophysical noise of the two stars), the contributions of which scale inversely with the square root of the number of observations. This also means that additional observations would improve the upper limit, without being too limited by how precisely we know the geometry and ephemeris of the planetary system. For example, if the geometric albedo of \hatponeb{} was 0.1, then approximately seven times more observations would be required to arrive at 0.4 as a $3\sigma$ upper limit, enough to infer the absence of an omnipresent reflective cloud layer.

It is interesting to note that these observations were originally proposed for the grism instrument ACS/HRC, which has a total throughput of 0.15--0.25 in this wavelength range, as opposed to 0.04--0.08 for STIS with the \texttt{G750L} grating. 
Thus ACS/HRC observations would have resulted in roughly three times more photons. Assuming the same $u_\mathrm{other}$ and $u_\mathrm{param}$ values, this approximately translates to a $u_\mathrm{total}$ of $0.12$ instead of $0.15$ for the $1\sigma$ limit, and $0.43$ instead of $0.54$ for $2\sigma$.

The binary companion star can help data analysis of further \hatpone{} observations (e.g., Wakeford et al., submitted). Also, similar methods could be used for other planetary hosts in binary systems. A suitable example is XO-2, of magnitude $V=11.2$, with the companion XO-2 S of $V=11.1$ at 31" separation. This companion star has been used as a reference for transmission spectroscopy both from the ground with GTC \citep {2012MNRAS.426.1663S}, and with HST NICMOS \citep {2012ApJ...761....7C}. XO-2 is a potential target for relative photometry during occultation with HST STIS in slitless mode: the $P=2.6$ day orbital period of XO-2b \citep {2007ApJ...671.2115B} would mean a larger planet-to-flux ratio than in case of \hatponeb{} for the same geometric albedo, and its zero-albedo equilibrium temperature $\Teq=1300$ K \citep {2008ApJ...677.1324T}, similar to that of \hatponeb{}, would make such a measurement interesting in terms of atmospheric models.

\acknowledgements
M.J.H. and J.N.W. gratefully acknowledge support from NASA Origins grant NNX09AB33G.
G.Á.B. acknowledges support from NSF grant AST-1108686 and NASA grant NNX12AH91H.

\chapter{Modeling prospective AGN transits}
\label{ch:agn}

\original{2013ApJ...762...35B}

\chapterabstract

Supermassive black holes (SMBH) are typically surrounded by a dense stellar population in galactic nuclei.
Stars crossing the line of site in active galactic nuclei (AGN)
produce a characteristic transit lightcurve, just like extrasolar planets do when they transit their host star.
We examine the possibility of finding such AGN transits in deep optical, UV, and X-ray surveys.
We calculate transit lightcurves using the Novikov--Thorne thin accretion disk model,
including general relativistic effects.
Based on the expected properties of stellar cusps,
we find that around $10^6$ solar mass SMBHs,
transits of red giants are most common for stars on close orbits
with transit durations of a few weeks and orbital periods of a few years.
We find that detecting AGN transits requires repeated observations of thousands of low mass AGNs
to 1\% photometric accuracy in optical, or $\sim10\%$ in UV bands or soft X-ray.
It may be possible to identify stellar transits in the Pan-STARRS and LSST optical and the eROSITA X-ray surveys.
Such observations could be used to constrain black hole mass, spin, inclination and accretion rate.
Transit rates and durations could give valuable information on the circumnuclear stellar clusters as well.
Transit lightcurves could be used to image accretion disks with unprecedented resolution,
allowing to resolve the SMBH silhouette in distant AGNs.


\section {Introduction}
\label {sec:agn:introduction}

Photometric transits have been successfully used to identify and characterize transiting extrasolar planets since the discovery of the first one, HD 209458b \citep {2000ApJ...529L..45C, 2000ApJ...529L..41H}. Transit shape is a telltale of planetary, orbital, and stellar parameters.
Moreover, the apparent time-dependent redshift of the star due to the planet covering its receding or approaching side during transit can reveal the projected angle between the planetary orbital axis and the stellar rotational axis \citep[Rossiter--McLaughlin effect,][] {1924ApJ....60...15R,1924ApJ....60...22M}. These methods show that transits are very powerful in probing planetary systems.

Similarly, active galactic nuclei (AGN) accretion disks can be probed by observing occultations in X-ray due to broad line region clouds: optically thick clouds orbiting the supermassive black hole (SMBH) in the region from where broad emission lines of the AGN are thought to originate. These occultations have a large covering factor of $\sim0.1$ to $1$ \citep [see e.g.] [] {1998ApJ...501L..29M, 2008A&A...483..161T, 2009ApJ...695..781B, 2010A&A...517A..47M, 2007ApJ...659L.111R, 2009MNRAS.393L...1R, 2009ApJ...696..160R, 2011MNRAS.410.1027R}. \citet {2011MNRAS.417..178R} pointed out that analogously to the Rossiter--McLaughlin effect, temporally resolved spectroscopic observations of an eclipse could be used to probe the apparent temperature structure of the accretion disk and the origin of the iron K$\alpha$ emission line, allowing to constrain the black hole spin and inclination.

In this paper, we examine the possibility of stars on close orbits transiting their host AGN.
There are multiple coincidences that make it possible to detect such events:
the radius of large stars is comparable to the characteristic size of an accretion disk around a $10^6\;M_\sun$ SMBH, resulting in transits deep enough to detect.
The orbital period is a few years in the innermost regions where the stellar number density is highest,
short enough for repeated observations.
Finally, the typical transit duration for these innermost stars is an hour to a few weeks (depending on the stellar population and the observing band), which is longer than the typical cadence required to observe AGNs, but still accessible on human timescales.

In order to produce a detectable signature in the lightcurve, the transiting object has to be a main sequence O or B star, a Wolf--Rayet (WR) star, an AGB star with a surrounding dust cloud (OH/IR star), a young supermassive star, or an evolved giant. Late type main sequence stars are too small to cause a transit detectable from the ground in an AGN with black hole mass $\gtrsim 10^5\;M_\sun$, unless they are ``bloated'' by irradiation of the AGN \citep{2012ApJ...749L..31L}.

High photometric and temporal resolution observations of AGN stellar transits have the potential to map the accretion disk structure with an unprecedented accuracy. Stars are optically thick in all electromagnetic bands, and unlike broad line region clouds \citep[e.g.][]{2010A&A...517A..47M}, they have a simple spherical geometry, making it easier to interpret lightcurves and to reconstruct the image of the accretion disk along the transit chord (projected stellar trajectory). Furthermore, such transits offer unique observations of individual stars in distant galaxies. Detections of multiple transits could reveal valuable information on the number density of stellar cusp central regions.

In this paper, we calculate transit depths, durations, periods, rates, and instantaneous probabilities based on physical models of nuclear stellar clusters. We also present simulated transit lightcurves and transit depth maps in multiple frequency bands. The shape of the lightcurve depends on the observing wavelength, the mass and spin of the SMBH, the accretion rate, the inclination of the accretion disk, the projected position of the transit chord, and the orbital velocity of the transiting object. Therefore observations with sufficient photometric accuracy and time resolution allow us to constrain these parameters, and to test the employed accretion and general relativity models.

In Section~\ref {sec:agn:cusps},
we review observations of large stars closely orbiting Sgr A*, the SMBH at the center of the Galaxy (\textsection\ref{sec:agn:populations}),
and summarize theoretical models of semi-major axis distribution and mass segregation (\textsection\ref{sec:agn:density}).
We set up models of the stellar population and radial distribution (\textsection\ref{sec:agn:models}),
and determine the minimum (\textsection\ref{sec:agn:rmin}) and maximum (\textsection\ref{sec:agn:rmax}) orbital radii for transits.
We present accretion disk thermal emission models and ray-tracing simulations in Section~\ref{sec:agn:observables}, showing
transit spectra (\textsection\ref{sec:agn:spectra}),
transit depth maps (\textsection\ref{sec:agn:transitmaps})
and lightcurves (\textsection\ref{sec:agn:lightcurves}).
We calculate transit probabilities in Section~\ref{sec:agn:transits}.
In Section~\ref {sec:agn:observability}, we determine the local density of AGNs of interest (\textsection\ref{sec:agn:agndensity}),
and study the feasibility of transit detections with optical (\textsection\ref{sec:agn:optical}, \textsection\ref{sec:agn:kepler})
and X-ray instruments (\textsection\ref{sec:agn:roentgen}).
The most important simplifying assumptions, caveats, and implications are discussed in Section~\ref{sec:agn:discussion}.
Finally, we conclude our findings in Section~\ref{sec:agn:conclusions}.

\section {Nuclear stellar clusters}
\label {sec:agn:cusps}

In this section we review observations and models of stellar distribution in galactic nuclei.

Let $M_\mathrm{SMBH}$ denote the mass of the SMBH, and define $M_6 = M_\mathrm{SMBH}/(10^6\;M_\sun)$ and the gravitational radius $R_\mathrm g = GM_\mathrm{SMBH}/c^2$. Then
\begin {align}
\label {eq:gravrad}
R_\mathrm g &= 4.8\times10^{-5} \;\mathrm{mpc} \; M_6 = 2.1 \; R_\sun \; M_6\,,
\end {align}
where $1\;\mathrm{mpc}=10^{-3}\;\mathrm{pc}=206\;\mathrm{AU}$.
For non-spinning black holes, the Schwarzschild radius is $R_\mathrm S=2R_\mathrm g$, and the radius of the innermost stable circular orbit is $R_\mathrm{ISCO}=6R_\mathrm g$. For maximally spinning black holes, $R_\mathrm S = R_\mathrm g$, and $R_\mathrm{ISCO} = R_\mathrm g$ for prograde orbits in the equatorial plane.

\subsection {Stellar populations}
\label {sec:agn:populations}

Many galaxies host a dense stellar cusp in their nucleus. \citet {2008ApJ...678..116S} find that galaxies with a massive nuclear cluster are more likely to be active. Stars captured and transported inwards by the accretion disk may serve to fuel the AGN \citep{2005ApJ...619...30M}.

In the Galaxy, \citet {2005ApJ...628..246E} report the orbital parameters of six B type main sequence stars orbiting the central SMBH on highly eccentric orbits with semi-major axes less than $16 \;\mathrm{mpc} \sim 10^5\;R_\mathrm g$. These stars may be remnants of stellar binaries tidally disrupted by the SMBH, as first proposed by \citet {1988Natur.331..687H} \citep[see also][] {2003ApJ...599.1129Y}. Candidates for the binary counterparts ejected with high velocity have been identified by \citet {2009ApJ...690.1639B}.

\citet {2009ApJ...697.1741B} find more than a hundred O and WR stars further from the Galactic Center, within $1\;\mathrm{pc} = 5\times10^6\;R_\mathrm g$ projected distance. About one half of them is orbiting the central SMBH in a disc-like structure (the so-called clockwise disc), while the other half is on apparently randomly oriented orbits. These stars could have formed in a massive self-gravitating gaseous disk \citep[e.g.][]{2007MNRAS.374..515L,2009MNRAS.394..191H}, or formed further away and captured in close orbits by the \citet {1988Natur.331..687H} mechanism or by a cluster of stellar mass black holes \citep {2004ApJ...606L..21A}. Young stars could also be delivered to this region by an infalling globular cluster \citep{1975ApJ...196..407T,2001ApJ...563...34M}.

Most main sequence stars in the vicinity of a SMBH eventually evolve into red giants or supergiants, our candidates for transiting the AGN. Many such giants have been observed within 1 pc of the Galactic Center \citep[e.g.][]{2010RvMP...82.3121G, 2006ApJ...643.1011P, 2009ApJ...697.1741B, 2010ApJ...708..834B}. The fraction of stars in the giant phase within a stellar population depends strongly on its initial mass function (IMF) and formation history.

A more exotic possibility is the formation of $\sim 10^5\;M_\sun$ supermassive stars in the accretion disk as suggested by \citet {2004ApJ...608..108G}. Such a star would form outside 1000 $R_\mathrm S$, have a radius of approximately $360\;R_\sun$ in case of solar metallicity, and would radiate at its Eddington limit, being luminous enough to have a detectable optical photometric signature not only when it transits the AGN but also when it is occulted behind it.

Observations of the Galactic nucleus show that the innermost cluster of young stars (so-called S-stars) is isotropically distributed \citep{2010RvMP...82.3121G}, as predicted by theoretical arguments. Even if stars are formed on an orbit coplanar with the accretion disk, \citet {1996NewA....1..149R} provide a relaxation mechanism that could rapidly randomize the orbital orientations. The possible presence of intermediate mass black holes may help catalyze this process \citep{2009ApJ...705..361G}. And even if stars cluster in disks, this coherent behaviour averages out when observing multiple galactic cores as long as the stellar disks and the accretion disks have independent orientations. Therefore we assume an isotropic distribution of stellar orbits in the cluster for the purpose of our probability estimates.

\subsection {Density profile}
\label {sec:agn:density}

We assume circular orbits for simplicity, and denote orbital radius by $r$. A star on an eccentric orbit with the same semi-major axis would produce a transit of the same depth, with a transit probability and transit length within a factor of order unity.

\citet {1976ApJ...209..214B} showed that the equilibrium spatial number density distribution of a stellar cluster around SMBH is proportional to $r^{-\alpha}$ with $\alpha = 1.75$ if the stars in the cluster have the same mass. Analytical and numerical investigations of multiple mass populations show that the distribution for each mass bin is still likely to follow a power law. The value of $\alpha$ is predicted to be smaller (down to $\approx1.3$) for lower mass stars \citep {1977ApJ...216..883B, 2006ApJ...649...91F, 2006ApJ...645.1152H}. For massive stars representing a small mass fraction in the stellar cluster, $\alpha$ can be between 2 and $2.75$ \citep {2009ApJ...697.1861A}, or as large as 3 \citep {2009ApJ...698L..64K}. Observations of the Galaxy by \citet {2010ApJ...708..834B} show that WR/O stars from a distance of 30 mpc out to 600 mpc form an interesting anisotropic structure called the clockwise disk in the Galacit nucleus. These stars are distributed with a density exponent $\alpha=2.4\pm0.2$, while the B star population from 30 mpc to 1 pc exhibits $\alpha=2.5\pm0.2$. Note, however, that the age of main sequence O stars and WR stars is less than the two-body relaxation time $\sim0.1$--$1\;\mathrm{Gyr}$ at $r\sim100\;\mathrm{mpc}$ \citep {2009MNRAS.395.2127O}, therefore they are not expected to represent the equilibrium state. The observed mass distribution of solar mass stars in the Galactic nucleus is fit by a broken power law with $\alpha=1.2$ and $1.75$ inside and outside of $0.22\;\mathrm{pc}$, respectively \citep{2007A&A...469..125S}.

To estimate the total number of stars, we first define the radius of influence $r_\mathrm i$ as the radius of the sphere centered on the SMBH containing a total mass of $2M_\mathrm{SMBH}$ in stars and stellar remnants. In case of a singular isothermal sphere ($\alpha=2$), this equals to ${GM_\mathrm{SMBH}}/{\sigma^2}$, where $\sigma$ is the velocity dispersion of stars further than $r_\mathrm i$ \citep {2004cbhg.symp..263M}. In order to get an estimate of the number of stars, we set $r_\mathrm i = {GM_\mathrm{SMBH}}/{\sigma^2}$, independently of $\alpha$.

Using the $M$--$\sigma$ relation now allows us to express the radius of influence as a function the supermassive black hole mass only, in the form
\begin {equation}
\label {eq:radiusofinfluence}
r_\mathrm i = r_0 \; M_6^\gamma\,.
\end {equation}

Here $r_0$ and $\gamma$ depend on the coefficients of the $M$--$\sigma$ relation. In particular, $\gamma=1-2/\beta$, where $\beta$ is the slope of $\log M$--$\log \sigma$, as defined by e.g.~\citet {2002ApJ...574..740T}. For example, the best fit of \citet {2002ApJ...574..740T} (with $\beta=4.02$) results in $r_0=1.234\;\mathrm{pc}$ and $\gamma=0.50$, and the best fit of \citet {2005SSRv..116..523F} (with $\beta=4.86$) results in $r_0=0.881\;\mathrm{pc}$ and $\gamma=0.59$. For our numerical results, we adopt the best fit values of \citet {2009ApJ...698..198G} (with $\beta=4.24$): $r_0=1.075\;\mathrm{pc}$ and $\gamma=0.53$ (but use 1 pc to normalize $r_0$ in our parametric expressions).

Now let us consider the stellar population in the vicinity of the SMBH. We assume that this population contains a species of stars with mass $M_\star$, having a radius $R_\star$ large enough to produce detectable transits. We also assume that the number density of these stars follows the power law $r^{-\alpha}$. Let $b$ denote the mass fraction of these large stars within the sphere of influence relative to the total mass of all stars and stellar remnants.

Typically the inner cutoff radius $r_\mathrm{min}$ for the stellar distribution is much smaller than the radius of influence (see \textsection~\ref {sec:agn:rmin} for numerical justification). Under this assumption, the spatial number density of the stars in question as a function of orbital radius is
\begin {equation}
\label {eq:density}
n(r) = b \frac {3-\alpha}{2\pi} \frac{M_\mathrm{SMBH}}{M_\star} \frac{r^{-\alpha}}{r_i^{3-\alpha}}\,.
\end {equation}

Note that this argument has two weaknesses: first, the $M$--$\sigma$ relation has a relatively large scatter. For a given SMBH mass, the intrinsic scatter of the bulge velocity dispersion is $\sim0.3$ dex \citep{2009ApJ...698..198G}. Second, we applied results for the isothermal sphere to power law distributions with different exponents. This limits the accuracy of the transit rate estimates presented in Section~\ref{sec:agn:transits}.

\subsection {Adopted models}
\label {sec:agn:models}

\newlength{\tablecolumnspacing}
\setlength{\tablecolumnspacing}{1.9mm}

\begin{deluxetable}{
  c@{\hspace{\tablecolumnspacing}}
  c@{\hspace{\tablecolumnspacing}}
  c@{\hspace{\tablecolumnspacing}}
  c@{\hspace{\tablecolumnspacing}}
  c@{\hspace{\tablecolumnspacing}}
  c@{\hspace{\tablecolumnspacing}}
  c@{\hspace{\tablecolumnspacing}}
  c@{\hspace{\tablecolumnspacing}}
  c@{\hspace{\tablecolumnspacing}}
  c@{\hspace{\tablecolumnspacing}}
  c@{\hspace{\tablecolumnspacing}}
  c@{\hspace{\tablecolumnspacing}}
  c@{\hspace{\tablecolumnspacing}}
  c@{\hspace{\tablecolumnspacing}}
}
\tablewidth{0pc}
\tabletypesize{\scriptsize}
\tablecaption{Orbital radius and period limits for AGN transits for the adopted stellar population models \label{tab:rlimits} ($b=0.01$ for each model)}
\tablehead{ $M_\mathrm{SMBH}$ & model & $R_\star$ & $R_\star$ & $M_\star$ & $\alpha$ & $r_\mathrm {tid}$ & $r_\mathrm{coll}$ & $r_\mathrm{min}$ & $r_\mathrm{lens}$ & $r_\mathrm i$ & $r_\mathrm{max}$ & $T_\mathrm{orb}\left(r_\mathrm{min}\right)$ & $T_\mathrm{orb}\left(r_\mathrm{max}\right)$ \\
$(M_\sun)$ && $(R_\sun)$ & $(R_\mathrm g)$ & $(M_\sun)$ && (mpc) & (mpc) & (mpc) & (mpc) & (mpc) & (mpc) & (yr) & (yr) }
\startdata
$10^5$ & O stars    &  11 & 51.8  & 30  & 2.5  & 0.006 & 0.16 & 0.16 &      11 &    320 &     11 & 0.57 &      340 \\
$10^5$ & red giants & 110 & 518   & 1.5 & 1.75 & 0.16  & 0.54 & 0.54 & 22\,000 &    320 &    320 & 3.7  &  53\,000 \\
$10^6$ & O stars    &  11 & 5.18  & 30  & 2.5  & 0.013 & 0.40 & 0.40 &      11 & 1\,075 &     11 & 0.75 &      110 \\
$10^6$ & red giants & 110 & 51.8  & 1.5 & 1.75 & 0.35  & 1.3  & 1.3  & 22\,000 & 1\,075 & 1\,075 & 4.2  & 100\,000 \\
$10^7$ & O stars    &  11 & 0.518 & 30  & 2.5  & 0.028 & 1.0  & 1.0  &      11 & 3\,600 &     11 & 1.0  &       34 \\
$10^7$ & red giants & 110 & 5.18  & 1.5 & 1.75 & 0.76  & 3.0  & 3.0  & 22\,000 & 3\,600 & 3\,600 & 4.9  & 210\,000
\enddata
\end{deluxetable}

We consider two simple models for the transiting stellar populations around AGNs, summarized in Table~\ref{tab:rlimits}.

First, we assume a young stellar population, motivated by the observations of a young population with an extremely top-heavy initial mass function of mean stellar mass $\approx 30\;M_\sun$ following a density profile with $\alpha \approx 2.5$ in the central parsec of the Galaxy \citep{2010ApJ...708..834B}. Thus in our model, we assume that a fraction of the stars are O type, with stellar mass $M_\star= 30\;M_\sun$, radius $R_\star=11\;R_\sun$, and $\alpha=2.5$. This exponent is consistent with observations and theoretical predictions reviewed in Section~\ref{sec:agn:density}. We assume that these O stars constitute a mass fraction $b=0.01$ of the population. This is consistent with the estimated total mass of WR/O stars in the Galactic Center if we consider that these stars are confined in the center part of the sphere of influence. However, note that this mass fraction depends sensitively on recent star formation rate and initial mass function of stars in the neighborhood of the SMBH.

For the second model, we consider an evolved, relaxed population of stars, and optimistically assume that a $b=0.01$ mass fraction of them are giants (or main sequence stars otherwise enlarged, like ``bloated'' or surrounded by a dust cloud), which are large enough to produce detectable AGN transits. For these giants, we assume $R_\star = 110\;R_\sun$, $M_\star=1.5\;M_\sun$, and a Bahcall--Wolf equilibrium value of $\alpha=1.75$ (see Section~\ref{sec:agn:density}).

Note that depending on the star formation history, O stars and red giants might coexist in the cusp, in which case their contributions to transits add up.

\subsection {Minimum orbital radius}
\label {sec:agn:rmin}

The inner orbital radius cutoff of the stellar distribution, $r_\mathrm{min}$, is set by gravitational wave inspiral, and tidal and collisional disruption. While gravitational wave inspiral is the limiting factor for compact objects \citep{1964PhRv..136.1224P}, tidal or collisional disruption sets a tighter constraint for stars.

The tidal disruption radius is
\begin {align}
r_\mathrm t &= \left( \eta^2 \frac {M_\mathrm {SMBH}} {M_\star} \right) ^{\frac13} R_\star\nonumber\\
&= 0.013\;\mathrm{mpc} \; M_6^{\frac13} \left( \frac {M_\star} {30\;M_\sun} \right) ^{-\frac13} \left( \frac {R_\star} {11\;R_\sun} \right)\,,
\end {align}
where $\eta$ ranges from $0.8$ to $3.1$ depending on the compressibility and polytropic mass distribution index of the star \citep {1995MNRAS.275..498D}. We adopt the moderate value $\eta=2$. The tidal disruption radius is given here normalized for a massive main sequence star.

Collisional disruption might be responsible for the depletion of luminous giant stars in the inner 80 mpc of the Galactic Center \citep {1999ApJ...527..835A}. Following \citet {2007MNRAS.378..129H}, one can write that the rate at which collisions decrease stellar density is
\begin {equation}
\frac{\partial n(r,t)}{\partial t} = - \frac {n^2(r,t) v(r) \Sigma}{N_\mathrm{coll}},
\end {equation}
where $\Sigma = \pi R_\star^2$ is the collisional cross-section, and on average, $N_\mathrm{coll}\approx30$ collisions are required to disrupt a star \citep{2006ApJ...649...91F}. We define the radius limit for collisional disruption as the radius where the stellar density $e$-folds in time $t$ if the initial value is given by Equation \ref {eq:density}:
\begin {equation}
r_\mathrm{coll} = r_0 \left[ \vphantom {\left( \frac{R_\star}{11\;R_\sun}\right)^2}
1.33 \times 10^{-8} \; b (3-\alpha) M_6^{\frac32 - \gamma(3-\alpha)}\right. \times
\end {equation}
\begin {equation*}
\times \left. \left( \frac {t}{10^7\;\mathrm{yr}} \right) \left( \frac {r_0}{1\;\mathrm{pc}} \right)^{-\frac72} \left( \frac {M_\star}{30\;M_\sun} \right)^{-1} \left( \frac{R_\star}{11\;R_\sun}\right)^2 \right]^{\frac2{2\alpha+1}}.
\end {equation*}
We set the timescale to be $t=10^7\;\mathrm{yr}$: this is within an order of magnitude of both the main sequence lifetime of early type stars, and the lifetime of the giant phase for less massive stars. Note that here we only consider collisions within the large species, not with other, much smaller stars, which are less likely.

The minimum orbital radius is
\begin {equation}
r_\mathrm{min} = \max ( r_\mathrm t, r_\mathrm{coll})\,.
\end {equation}
Table~\ref{tab:rlimits} lists the minimum and maximum orbital radii for the two stellar species and three different SMBH masses. We find that in every case, collisions set a tighter constraint than tidal disruption for the potentially transiting stars.

Table~\ref{tab:rlimits} also lists the Keplerian orbital periods for stars at the minimum and maximum radii, which can be calculated as
\begin {align}
\label {eq:torb}
T_\mathrm{orb} &= 3\;\mathrm{yr} \; \left(\frac r {1\;\mathrm{mpc}} \right)^{\frac32} \; M_6^{-\frac12}.
\end {align}
We find that the closest main sequence stars have orbital periods of approximately one year, making repeated transit observations feasible. However, collisions set a larger radius limit for giants, resulting in longer orbits. This inner radius limit depends on the number fraction of giants: smaller $b$ implies less frequent collisions and thus allows closer orbits. However, a smaller value for $b$ would also mean smaller transit probabilities for a given AGN (see Section~\ref{sec:agn:transits} below).

\subsection {Maximum orbital radius}
\label {sec:agn:rmax}

For the maximum orbital radius $r_\mathrm{max}$ of stars capable of producing a transit, we have to consider two factors: gravitational microlensing due to the star, and the validity range of the presumed number density power law.

Gravitational microlensing caused by the transiting star can bend the light rays of the AGN which may significantly distort the transit lightcurve \citep{1986ApJ...304....1P}. This can happen if the transiting object is farther from the AGN than the radius $r_\mathrm{lens}$ at which the Einstein radius equals the apparent angular radius of the transiting object:
\begin {equation}
\label {eq:microlensing}
r_\mathrm{lens} = 11\;\mathrm{mpc} \left(\frac{R_\star}{11\;R_\sun}\right)^2 \left(\frac {M_\star}{30\;M_\sun}\right)^{-1}\,.
\end {equation}
Therefore we restrict our transit probability calculations to the contribution of stars within orbital radius $r_\mathrm{lens}$.

Note that this is a different configuration than a galaxy microlensing a distant quasar, which can also be used to probe the spatial structure of accretion disks around AGNs \citep[e.g.~for the case of Q2237+0305, see][and references therein]{2004ApJ...605...58K}.

The power law distribution discussed above only applies to the stellar populations within the radius of influence from the SMBH, where its gravitational field dominates. In this paper, we do not investigate the distribution of stars outside the sphere of influence, but conservatively ignore their contribution to transit probabilities.

The maximum orbital radius is thus the smaller of the microlensing radius and the radius of influence:
\begin {equation}
r_\mathrm{max} = \min (r_\mathrm{lens}, r_\mathrm i)\,.
\end {equation}
Table~\ref{tab:rlimits} shows that typically microlensing is the limiting factor among main sequence O stars, whereas the radius of influence limits transits of the red giants. The reason for this is that the $R_\star^2/M_\star$ factor that determines the microlensing limiting radius accoding to Equation (\ref {eq:microlensing}) is very different for these two species of stars.

\section {Transit observables: spectra and lightcurves}
\label {sec:agn:observables}

Next we derive the AGN spectra and the transit observables.

The AGN luminosity is bounded by the Eddington limit $L_\mathrm{Edd} = 1.3 \times 10^{44} \;\mathrm{erg}\;\mathrm s^{-1}\; M_6$ \citep {1983bhwd.book.....S}. We assume an Eddington ratio of $0.25$ as our fiducial value, following \citep {2006ApJ...648..128K,2012MNRAS.tmp...16S}. Then the AGN luminosity is
\begin {equation}
\label {eq:ledd}
L_\mathrm{AGN} = 3.6\times10^{43}\;\mathrm{erg}\;\mathrm s^{-1}\; M_6\,.
\end {equation}

\subsection {AGN spectra}
\label {sec:agn:spectra}
We adopt the general relativistic, radiatively efficient, steady-state thin accretion disk model of \citet {1973blho.conf..343N}. This model describes the flux of thermal radiation from the disk as a function of radius in Equation (5.6.14b) \citep [see][for the explicit form of $\mathscr Q$] {1974ApJ...191..499P}. We assume no radiation from within $R_\mathrm{ISCO}$. In addition to the thermal component, AGN spectra typically exhibit emission lines, excess infrared radiation from dust reprocessing UV emission, and a hard X-ray component usually assigned to inverse Compton scattering in a hot corona \citep {1993ApJ...413..507H}. We do not account for these phenomena, but choose our observing bands so that their effect is minimal: an observation window around 200 eV is low enough so that the thermal component dominates over coronal emission, but it is higher than helium Lyman absorption and detector lower energy limits. It is important to note that little is known about the geometry of the corona, and simultaneously observing a stellar transit in hard X-ray might provide feedback to the models.

We follow the accretion disk photosphere model described by \citet {2005ApJ...622L..93M} and \citet {2010ApJ...714..404T}: we assume that the dominant absorption mechanism is the bound-free process, with an opacity that depends on the frequency and temperature. We assume that the temperature and thus the absorption opacity are constant down to an optical depth of one (the ``bottom'') in the photosphere. We can calculate the total flux from the absorption and electron scattering opacities and the photosphere bottom temperature, using Equations (A13--A15) and (A17) of \citet {2010ApJ...714..404T}, but using the relativistic angular frequency given by \citet {1973blho.conf..343N} instead of Keplerian velocity. However, as the flux is known and the temperature is sought for, we have to use a simple iterative process to solve this implicit equation for the temperature. We find that usually electron scattering dominates the total opacity, but in the hottest parts of the accretion disk, absorption takes over. Note that the functional form of the specific flux differs from a blackbody spectrum, because the absorption opacity depends on frequency. For simplicity, we assume that the emerging radiation is isotropic in the frame comoving with the accretion disk.

\begin {figure}
\begin{center}
\includegraphics*[width=\threequartersfigurewidth]{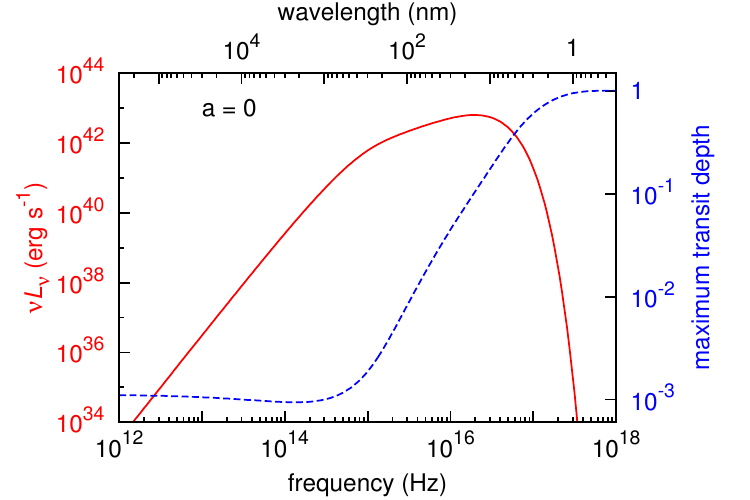}
\includegraphics*[width=\threequartersfigurewidth]{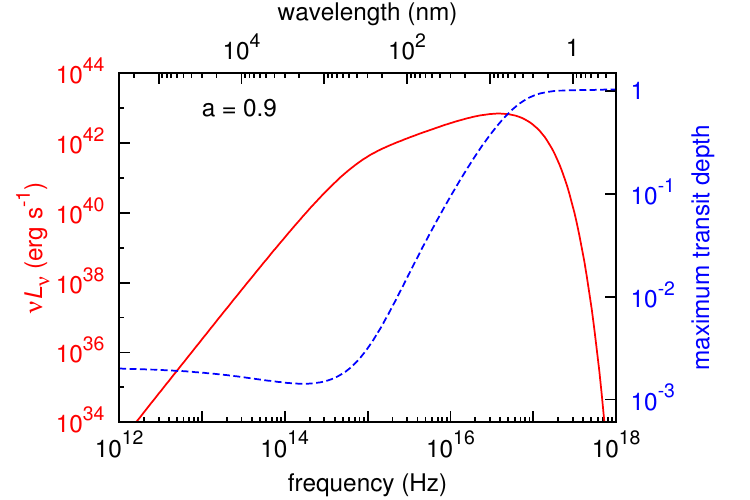}
\end{center}
\caption {Spectrum and transit depth of an accretion disk around a $10^6\;M_\sun$ black hole. Solid red curve: equivalent isotropic luminosity per logarithmic bins of frequency. Dashed blue curve: maximum possible narrow-band transit depth caused by a star with $R_\star=11\;R_\sun$ as a function of source frequency. Top (bottom) panel presents the case of a Schwarzschild (Kerr) BH with spin $a=0$ ($a=0.9$).}
\label {fig:freq}
\end {figure}

Given the specific intensity of the accretion disk as a function of radius and frequency, the observed spectrum is determined by Doppler shift, gravitational redshift, and gravitational lensing. To account for these effects, we apply the transfer function method as described by \citet {1975ApJ...202..788C} and implemented by \citet {1995CoPhC..88..109S}\footnote {available at \url{http://www.tat.physik.uni-tuebingen.de/~speith/publ/photon_transferfct_dble.f}}. We calculate the radiative efficiency as a function of spin as described by \citet {1983PhT....36j..89S} to convert luminosity to mass accretion rate, which is the input parameter of this code. We fix the inclination angle $\vartheta$ (the angle between the spin axis and the line of sight) at $60^\circ$, so that $\cos\vartheta=0.5$. The calculated spectrum of the accretion disk is shown in Figure~\ref{fig:freq} for a non-spinning black hole (with dimensionless spin $a=0$) in the top panel, and for a highly spinning black hole with a prograde disk with a conservative value $a=0.9$ in the bottom panel. The frequency is in the source rest-frame accounting for gravitational redshift, but a possible cosmological redshift for distant AGNs is not considered. The specific luminosity value displayed here is what an isotropic source would have to have in order to produce the same flux as the AGN does at this specific inclination.

Transit depth is defined as the blocked flux to out-of-transit flux ratio in a given band. Therefore the transit depth is between zero and one: zero if the transiting object does not cover any part of the accretion disk; one if the object completely blocks radiation (in which case it is called an occultation or eclipse). The transit depth varies with frequency and the location of the transiting object in projection. To be able to efficiently calculate transit depths at different positions, we implement a linear approximation to the radius--gravitational redshift grid generated by the above code to determine these values for a light ray parametrized by its projected position far from the AGN. Then we employ high order numerical approximation\footnote {code available from \url{http://www.holoborodko.com/pavel/?page_id=1879}} to integrate over the stellar disk in the projection plane. This gives the blocked specific flux, which we then divide by the total specific flux to obtain the narrow-band transit depth.

In addition to AGN spectra, Figure~\ref{fig:freq} also depicts the maximum possible narrow-band transit depth caused by an early type main sequence star with $R_\star=11\;R_\sun$ as a function of frequency. The maximum depth of a general transit can be smaller if the star does not transit the most luminous part of the projected accretion disk. At high frequencies, the most luminous region is more compact, therefore the transit is deeper. The transit depth becomes constant at infrared frequencies less than the peak frequency of a black body spectrum with temperature of the outer edge of the disk (we set $10^3\;R_\mathrm g$ in the simulations), because here the Rayleigh--Jeans approximation applies to every part of the disk. Larger stars cause deeper transits (see below).

\subsection {Transit maps}
\label {sec:agn:transitmaps}

\begin {figure}
\includegraphics*[width=\halffigurewidth]{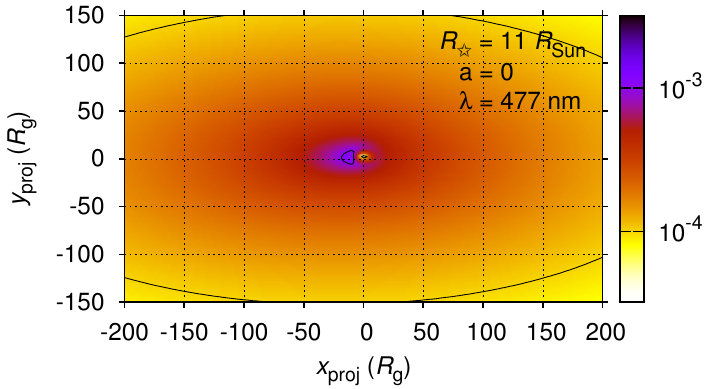}
\includegraphics*[width=\halffigurewidth]{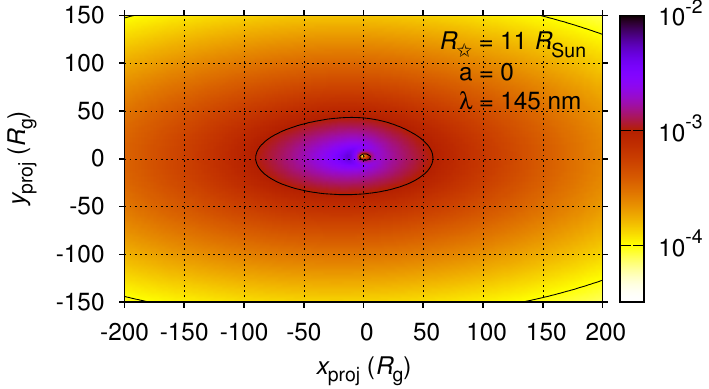}
\includegraphics*[width=\halffigurewidth]{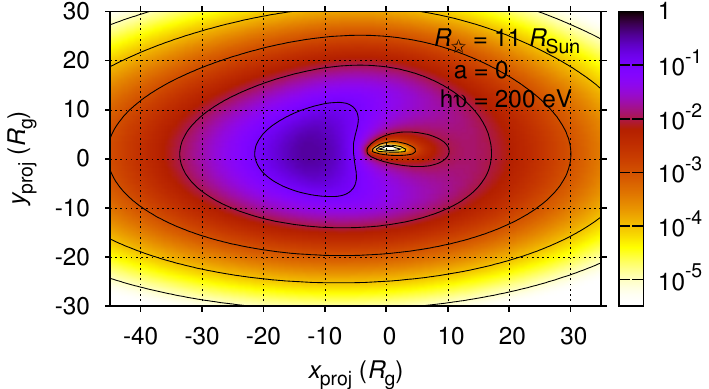}
\caption {The instantaneous narrow-band transit depth as a function of projected position of the transiting object at three different frequencies optical ({\it top panel}), EUV ({\it middle panel}), and soft X-rays ({\it bottom panel}). A transit lightcurve corresponds to the values along the stellar trajectory in the image plane shown. A non-spinning $10^6\;M_\sun$ black hole is at the origin, the observation angle is $60^{\circ}$ relative to the accretion disk, and transit depths are shown for a main sequence O-type star with $R_\star=11\;R_\sun=5.18\;R_\mathrm g$. The top half of the disk image is more severely distorted by gravitational lensing since it is farther from the observer than the black hole. The left side of the disk rotates towards the observer, thus appearing more luminous due to beaming, which results in a deeper transit. The transit is deepest if the projected position of the star crosses the most luminous regions closest to the SMBH. The black hole silhouette and the curved spacetime geometry becomes visible in the X-ray transit map.
}
\label {fig:transitmapostar}
\end {figure}

\begin {figure}
\includegraphics*[width=\halffigurewidth]{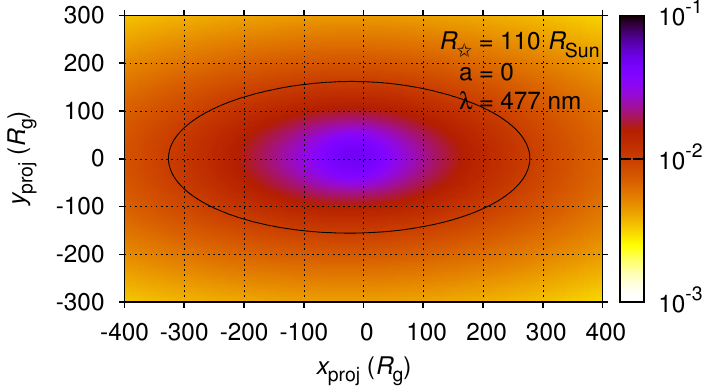}
\includegraphics*[width=\halffigurewidth]{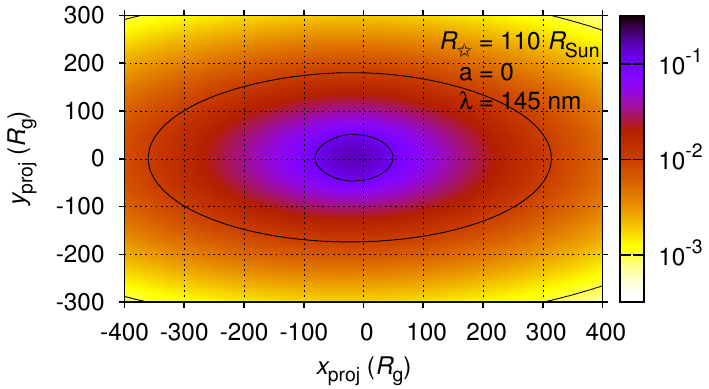}
\includegraphics*[width=\halffigurewidth]{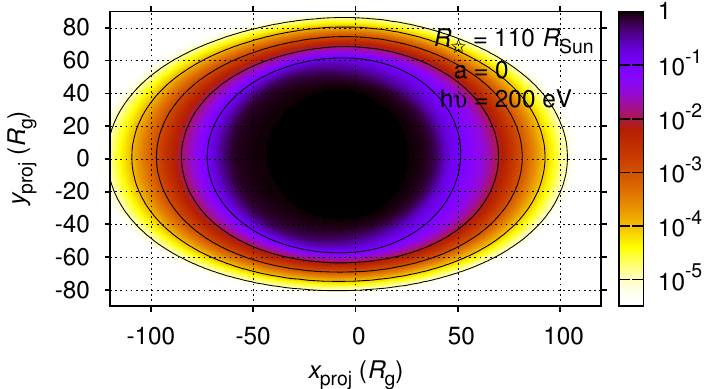}
\caption {Same as Figure~\ref {fig:transitmapostar}, but for a giant star with $R_\star=110\;R_\sun=51.8\;R_\mathrm g$. Here, the star is so large that the transit is much deeper, and the image of the accretion disk is smoothed out (e.g.~the left-right asymmetry of beaming is not apparent).}
\label {fig:transitmapgiant}
\end {figure}

\begin {figure}
\includegraphics*[width=\halffigurewidth]{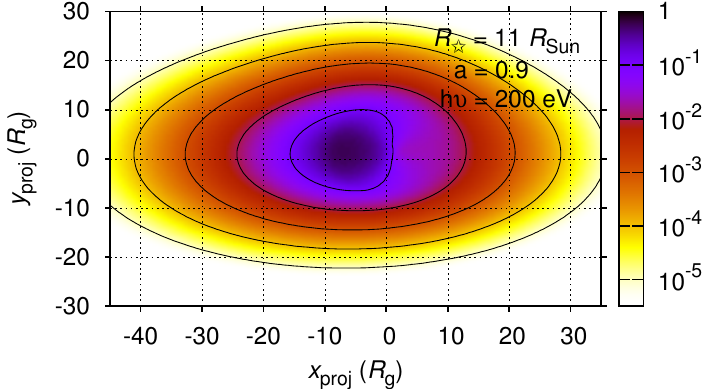}
\caption {Soft X-ray transit depth map for an AGN with a Kerr-BH spin $a=0.9$ for $R_\star=11\;R_\sun$, same as the bottom panel of Figure~\ref {fig:transitmapostar} but for a spinning SMBH. The details of the accretion disk are smoothed since the ISCO radius in this case is smaller than the stellar radius. The accretion disk image is further modified because of the Kerr geometry.}
\label {fig:transitmapspin}
\end {figure}

As stated in Section~\ref {sec:agn:introduction}, AGN transit observations can be used to map distant AGNs along the transit chord with unprecedented resolution. We now elaborate on this point. Figures~\ref{fig:transitmapostar} and \ref{fig:transitmapgiant} show the transit depth as a function the projected position of the transiting object in front of the accretion disk around a non-spinning BH for an early type main sequence star ($R_\star=11\;R_\sun$) and a giant star ($R_\star=110\;R_\sun$), respectively. The three panels in both figures show the transit depth maps for different observing frequency: optical $g$ band (top panel, $\lambda = 477 \;\mathrm{nm}$, $\nu = 6.3 \times10^{14} \;\mathrm{Hz}$), extreme ultraviolet (EUV, middle panel, $\lambda = 145 \;\mathrm{nm} = 1450$ \AA, $\nu = 2.1 \times10^{15} \;\mathrm{Hz}$), and soft X-ray (bottom panel, $h\nu=200 \;\mathrm{eV}$, $\nu = 4.8 \times10^{16} \;\mathrm{Hz}$, $\lambda = 6.2 \;\mathrm{nm}$). The spatial variations in the transit depth maps imply time-variations of the observed AGN brightness as a transiting object moves across the image. A transit lightcurve corresponds to the values along the projected stellar trajectory in Figures~\ref{fig:transitmapostar} and \ref{fig:transitmapgiant}. Conversely, observations of the transit lightcurve can be used to reveal the geometry of the accretion disk along a line in this image.

The maximum possible resolution of such a reconstructed image is set by the size of the transiting object and the spatial variations of the disk brightness. Since the disk temperature increases inwards, the emission at higher frequencies is confined to the innermost regions, implying that transit measurements at higher frequencies can give a higher resolution image (see different panels in Figures~\ref{fig:transitmapostar} and \ref{fig:transitmapgiant}). Transits of smaller objects also yield a higher resolution image. However, transits of smaller objects are less deep and hence more difficult to detect.

The left-right asymmetry in the figure is due to different Doppler shifts for regions of the disk moving towards or away from the observer. The spacetime geometry leaves an imprint on the image, the top half of the disk image is distorted by gravitational lensing close to the BH. Remarkably, the black hole silhouette (i.e., the lack of emission within the ISCO) becomes directly visible in the X-ray image of a transiting O star (see Figure~\ref{fig:transitmapostar} bottom panel).

The transit depth map is also sensitive to the spacetime geometry both directly through gravitational lensing and indirectly through the change in the ISCO radius. Figure~\ref{fig:transitmapspin} shows the soft X-ray transit map of a Novikov--Thorne accretion disk around a Kerr BH with spin $a=0.9$ (c.f.~bottom panel of Figure~\ref{fig:transitmapostar}). In this case, the ISCO is smaller than the radius of the transiting object and the BH silhouette does not appear in the image.

\subsection {Transit lightcurves}
\label {sec:agn:lightcurves}
To get a handle on the characteristic transit duration, let us consider the timescale for the center of a star to cross a disk of radius $R_\mathrm{AGN}$ centered on the most luminous part of the accretion disk image for a given frequency. We choose the value of $R_\mathrm{AGN}$ based on the transit depth map, depending on the desired transit depth. Typically $R_\mathrm{AGN}\sim5$--$1000\;R_\mathrm g$.

The transit duration is
\begin {align}
\label{eq:Deltat}
\Delta t &= 2R_\mathrm{AGN} \sqrt{\frac r{\mathrm{G} M_\mathrm{SMBH}}}\nonumber \\
&=4\;\mathrm{hours} \; M_6^{\frac12} \left( \frac {R_\mathrm{AGN}}{10\;R_\mathrm g} \right) \left(\frac r {1\;\mathrm{mpc}} \right)^{\frac12}.
\end {align}
The transit duration is of the order of hours for massive main sequence stars closest to the AGN, and weeks for the closest giants further out in the circumnuclear star cluster (see Table~\ref{tab:transit} below).

\begin {figure}
\includegraphics*[width=\halffigurewidth]{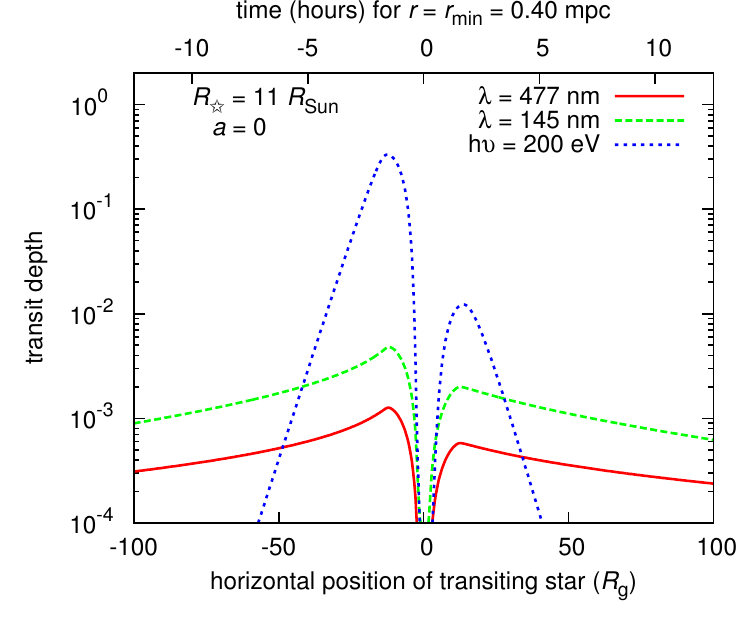}
\includegraphics*[width=\halffigurewidth]{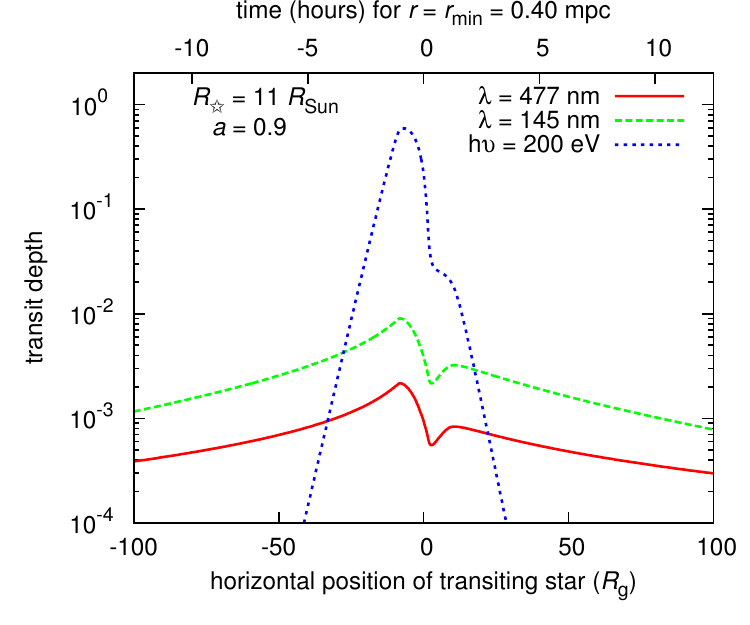}
\includegraphics*[width=\halffigurewidth]{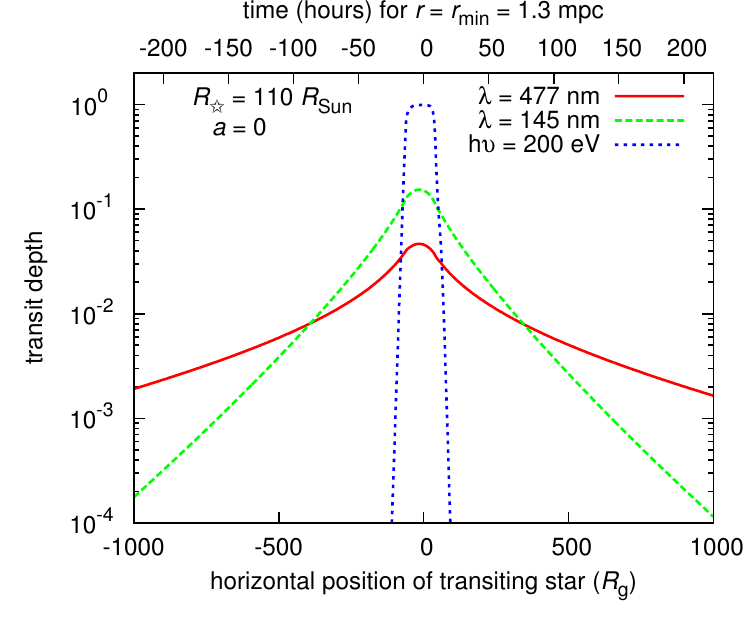}
\caption {Lightcurves of a transit: transit depth on logarithmic scale as a function of the horizontal position of the transiting star in front of the AGN in units of $R_\mathrm g$ (bottom axis), or time in case $r=5\;\mathrm{mpc}$ (top axis). Stellar radius is $R_\star = 11\;R_\sun = 5.2\;R_\mathrm g$ (top and middle panels) and $110\;R_\sun = 52\;R_\mathrm g$ (bottom panel). The SMBH spin is $a=0$ (top and bottom panels) and 0.9 (middle panel). The BH silhouette is larger than the star in the top panel only.}
\label {fig:lightcurve}
\end {figure}

Figure~\ref{fig:lightcurve} shows multiband transit lightcurves of an accretion disk due to stars of radius $R_\star = 11$ (top and middle panel) and $110\;R_\sun$ (bottom panel). The horizontal axis shows the horizontal position of the transiting star in the image plane of Figures~\ref{fig:transitmapostar}--\ref{fig:transitmapspin}. Physical distance in the projection plane is displayed in $R_\mathrm g$ units on the bottom axis, and the top axis shows time in hours for a star with minimum orbital radius as given in Table~\ref{tab:rlimits}. Note that further stars will cause a similar lightcurve, only scaled in time. The dimensionless black hole spin is zero on the top and bottom panels and $0.9$ in the middle panel. The projected stellar trajectory is horizontal with $y_\mathrm{proj}=1.5\;R_\mathrm g$, which approximately passes through the hottest part of the accretion disk. The black hole is behind the origin. The curves show the transit depth for the same three frequencies as the transit depth maps in Figures~\ref{fig:transitmapostar}--\ref{fig:transitmapgiant}. Note that these graphs are different from the usual planetary transit lightcurves showing flux, but plotting transit depth is more convenient when using a logarithmic scale. A larger value corresponds to a deeper transit, that is, a larger photometric signature. Figure~\ref{fig:lightcurve} is consistent with the expectations on the frequency dependence described above: at higher frequencies, the AGN is more compact, implying shorter and deeper transits. The sharp features near the center of the transit on the top and middle panels are due to the inner edge of the disk at the ISCO, and the left-right asymmetry is mostly due to Doppler shift of the approaching and receding parts of the disk. However, a transiting giant filters out this spatial feature due to its size, as seen on the bottom panel. The black hole silhouette is clearly visible in the top panel where the ISCO radius is larger than the transiting object. Comparing the top two panels, we infer that if the mass of the SMBH is known, observing a transit light curve allows us to put an upper limit on the spin.

We conclude that for an accretion disk around a $10^6\;M_\sun$ SMBH, $\sim0.1\%$ photometric accuracy is required in the optical, $\sim1\%$ accuracy in the extreme ultraviolet, and $\sim10\%$ accuracy in soft X-ray to detect a transit due to a 11 $R_\sun$ star. In order to detect a transit due to a 110 $R_\sun$ giant, $\sim1\%$ accuracy is sufficient in the optical, and $\sim10\%$ accuracy in extreme ultraviolet to X-ray (see Section~\ref{sec:agn:agndensity} for a comparison with intrinsic variability).

For comparison, we ran simulations for two different SMBH masses: $10^5$ and $10^7\;M_\sun$. Table~\ref {tab:transit} shows the maximum possible transit depth in each case for different sizes of transiting stars. The transit depth for a fixed stellar size decreases with the increasing spatial extent of the accretion disk around black holes with increasing mass, making the transit feature more prominent for smaller BH masses. However, AGNs with less massive SMBHs are also intrinsically fainter, and $10^5\;M_\sun$ AGNs also have a smaller local number density which makes low mass AGN transit observations more challenging (see Section~\ref {sec:agn:observability} for a discussion).

\section {Transit rates}
\label {sec:agn:transits}

In this section, we estimate the expected rates of stellar transits in AGNs. Recall that $R_\mathrm{AGN}$ denotes the radius of the circular area that the center of the star has to transit in the projection plane for a given transit depth. Let $P_\mathrm{inst}(r)$ denote the probability of a single star on a circular orbit with radius $r$ transiting this circular area at a given instance, and let $P_\mathrm{ever}(r)$ denote the probability that the orbit is aligned so that the star transits this circular area at some point during its orbit. By geometric arguments, these probabilities are
\begin {eqnarray}
P_\mathrm{inst}(r) &=& \frac {\pi R_\mathrm{AGN}^2}{4\pi r^2} = \frac14 \frac {R_\mathrm{AGN}^2}{r^2}\,, \\
P_\mathrm{ever}(r) &=& \frac {2 R_\mathrm{AGN} \times 2\pi r} {4\pi r^2} = \frac {R_\mathrm{AGN}}r\,.
\end {eqnarray}

For a single AGN,
the expected value of the number of transits at a given instance, $N_\mathrm{inst}$,
the expected value of the transit rate (number of transits observed per time for asymptotically long observations), $\dot N$,
and the expected value of the number of stars on orbits such that they ever transit, $N_\mathrm{ever}$,
can be calculated by integrating for the entire stellar population:
\begin {align}
\label{eq:ninstpar}
N_\mathrm{inst} &= \int_{r_\mathrm{min}}^{r_\mathrm {max}} 4\pi r^2 n(r) P_\mathrm{inst}(r) \mathrm dr\,, \\
\dot N          &= \int_{r_\mathrm{min}}^{r_\mathrm {max}} 4\pi r^2 n(r) \frac {P_\mathrm{ever}(r)}{T_\mathrm{orb} (r)} \mathrm dr\,, \\
\label{eq:neverpar}
N_\mathrm{ever} &= \int_{r_\mathrm{min}}^{r_\mathrm {max}} 4\pi r^2 n(r) P_\mathrm{ever}(r) \mathrm dr\,,
\end {align}
where $T_\mathrm{orb}$ is the Keplerian orbital period.

To interpret these probability indicators, we have to understand the relationship between the expected value $\dot N$ of the transit rate and the expected value $N_\mathrm{ever}$ of the number of stars that ever transit a given AGN. If we monitor a single target for which $N_\mathrm{ever}\ll1$, then the actual transit rate is zero with probability $1-N_\mathrm{ever}$ and $\dot N/N_\mathrm{ever}$ with a small probability $N_\mathrm{ever}$. (The probability of multiple ever transiting stars in a single AGN is negligible in this case.) However, when monitoring a large number $n \gg 1/{N_\mathrm{ever}}$ of identical targets, these effects average out: the observed total transit rate is $\approx n\dot N$ and there are $\approx nN_\mathrm{ever}$ stars causing transits among all targets in total.

We substitute Equations~(\ref{eq:radiusofinfluence}--\ref{eq:density}) into Equations~(\ref{eq:ninstpar}--\ref{eq:neverpar}), and use $1<\alpha<3$ and $r_\mathrm{min}\ll r_\mathrm{max}$ (justified by Table~\ref{tab:rlimits}) to obtain
\begin {align}
\label {eq:ninstnum}
N_\mathrm{inst} &= 3.8\times10^{-6} \; 1000^{\alpha-2} \, b \frac{3-\alpha}{\alpha-1} M_6^{3 - \gamma (3-\alpha)}\;\times \\
\nonumber
&\hspace{-10mm}\times\left(\frac{M_\star}{30\;M_\sun}\right)^{-1} \left(\frac{R_\mathrm{AGN}}{10\;R_\mathrm g}\right)^2\left(\frac{r_0}{1\;\mathrm{pc}}\right)^{\alpha-3}
\left( \frac {r_\mathrm{min}}{1\;\mathrm{mpc}} \right) ^{1-\alpha} \\
\dot N &= \frac1{93\;\mathrm{yr}} \; 1000^{\alpha-2} \, b \frac{3-\alpha}{\alpha-\frac12} M_6^{\frac52-\gamma(3-\alpha)}\;\times \\
\nonumber
&\hspace{-10mm}\times\left(\frac{M_\star}{30\;M_\sun}\right)^{-1} \left(\frac{R_\mathrm{AGN}}{10\;R_\mathrm g}\right)  \left(\frac{r_0}{1\;\mathrm{pc}}\right)^{\alpha-3}
\left( \frac {r_\mathrm{min}}{1\;\mathrm{mpc}} \right) ^{\frac12-\alpha} \\
N_\mathrm{ever}^{\alpha>2} &= 0.032 \times 1000^{\alpha-2} \, b \frac{3-\alpha}{\alpha-2} M_6^{2-\gamma(3-\alpha)}\;\times \\
\nonumber
&\hspace{-10mm}\times\left(\frac{M_\star}{30\;M_\sun}\right)^{-1} \left(\frac{R_\mathrm{AGN}}{10\;R_\mathrm g}\right)  \left(\frac{r_0}{1\;\mathrm{pc}}\right)^{\alpha-3}
\left( \frac {r_\mathrm{min}} {1\;\mathrm{mpc}} \right) ^{2-\alpha} \\
N_\mathrm{ever}^{\alpha=2} &= 0.032 \times b (3-\alpha) M_6^{2-\gamma(3-\alpha)}\;\times \\
\nonumber
&\hspace{-10mm}\times\left(\frac{M_\star}{30\;M_\sun}\right)^{-1} \left(\frac{R_\mathrm{AGN}}{10\;R_\mathrm g}\right)  \left(\frac{r_0}{1\;\mathrm{pc}}\right)^{-1}
\ln \frac{r_\mathrm{max}}{r_\mathrm{min}} \\
\label {eq:neverless2}
N_\mathrm{ever}^{\alpha<2} &= 0.032 \times 1000^{\alpha-2} \, b \frac{3-\alpha}{2-\alpha} M_6^{2-\gamma(3-\alpha)}\;\times \\
\nonumber
&\hspace{-10mm}\times\left(\frac{M_\star}{30\;M_\sun}\right)^{-1} \left(\frac{R_\mathrm{AGN}}{10\;R_\mathrm g}\right)  \left(\frac{r_0}{1\;\mathrm{pc}}\right)^{\alpha-3}
\left( \frac {r_\mathrm{max}} {1\;\mathrm{mpc}} \right) ^{2-\alpha}.
\end {align}
Here $N_\mathrm{ever}^{\alpha>2}$, $N_\mathrm{ever}^{\alpha=2}$, and $N_\mathrm{ever}^{\alpha<2}$ denote the value of $N_\mathrm{ever}$ for the corresponding population density exponents. Stars on close orbits transit more frequently, and they dominate $N_{\mathrm{inst}}$ and $\dot N$. However, $N_\mathrm{ever}$ is dominated by stars on close or wide orbits for $\alpha>2$ and $\alpha<2$, respectively. This is determined by the exponent of $r$ in the integrand of Equations~(\ref{eq:ninstpar}--\ref{eq:neverpar}).

\setlength{\tablecolumnspacing}{2mm}

\begin{deluxetable}{
  c@{\hspace{\tablecolumnspacing}}
  c@{\hspace{\tablecolumnspacing}}
  c@{\hspace{\tablecolumnspacing}}
  c@{\hspace{\tablecolumnspacing}}
  c@{\hspace{\tablecolumnspacing}}
  c@{\hspace{\tablecolumnspacing}}
  c@{\hspace{\tablecolumnspacing}}
  c@{\hspace{\tablecolumnspacing}}
  c@{\hspace{\tablecolumnspacing}}
  c@{\hspace{\tablecolumnspacing}}
  c@{\hspace{\tablecolumnspacing}}
  c@{\hspace{\tablecolumnspacing}}
}
\tablewidth{0pc}
\tabletypesize{\scriptsize}
\tablecaption{Transit probabilities and duration for different bands and transit depths for a single AGN \label{tab:transit}}
\tablehead{&&&& maximum \\ $M_\mathrm{SMBH}$ & $a$ & model & $\lambda$ & transit & transit & $R_\mathrm{AGN}$ & $N_\mathrm {inst}$ & $1/{\dot N}$ & $N_\mathrm{ever}$ & $\Delta t\left(r_\mathrm{min}\right)$ & $\Delta t\left(r_\mathrm{max}\right)$ \\
$(M_\sun)$ &&& (nm) & depth\tablenotemark{a} & depth\tablenotemark{b} & $(R_\mathrm g)$ && (yr) && (h) & (h) }
\startdata
$10^5$ & 0   & O star    & 477 & 0.038    & $10^{-2}$ & 180 & $3.8\times10^{-6}$ &    280 & 0.009 &     89 &     750 \\
       &     &           & 145 & 0.076    & $10^{-2}$ & 235 & $6.5\times10^{-6}$ &    210 & 0.012 &    120 &     980 \\
       &     &           & 6.2 & $\sim1$  & $\sim1$   &  45 & $2.4\times10^{-7}$ & 1\,100 & 0.002 &     22 &     190 \\
$10^5$ & 0.9 & O star    & 477 & 0.049    & $10^{-2}$ & 200 & $4.7\times10^{-6}$ &    250 & 0.010 &     99 &     830 \\
       &     &           & 145 & 0.12     & $10^{-2}$ & 240 & $6.8\times10^{-6}$ &    210 & 0.012 &    120 &    1000 \\
       &     &           & 6.2 & $\sim1$  & $\sim1$   &  40 & $1.9\times10^{-7}$ & 1\,300 & 0.002 &     20 &     170 \\
$10^5$ & 0   & red giant & 477 & 0.63     & $10^{-1}$ & 900 & $1.2\times10^{-5}$ & 1\,000 & 0.11  &    830 & 20\,000 \\
       &     &           & 145 & 0.78     & $10^{-1}$ & 750 & $8.6\times10^{-6}$ & 1\,200 & 0.091 &    690 & 17\,000 \\
       &     &           & 6.2 & 0.74     & $10^{-1}$ & 530 & $4.3\times10^{-6}$ & 1\,700 & 0.064 &    490 & 12\,000 \\
$10^5$ & 0.9 & red giant & 477 & 0.68     & $10^{-1}$ & 850 & $1.1\times10^{-5}$ & 1\,100 & 0.10  &    780 & 19\,000 \\
       &     &           & 145 & 0.88     & $10^{-1}$ & 750 & $8.6\times10^{-6}$ & 1\,200 & 0.091 &    690 & 17\,000 \\
       &     &           & 6.2 & $\sim1$  & $10^{-1}$ & 500 & $3.8\times10^{-6}$ & 1\,800 & 0.060 &    460 & 11\,000 \\
$10^6$ & 0   & O star    & 477 & 0.0012   & $10^{-3}$ &   6 & $5.6\times10^{-7}$ &    320 & 0.010 &    1.5 &     7.9 \\
       &     &           & 145 & 0.0047   & $10^{-3}$ &  50 & $3.9\times10^{-5}$ &     39 & 0.083 &     13 &      66 \\
       &     &           & 6.2 & 0.33     & $10^{-1}$ &   9 & $1.2\times10^{-6}$ &    220 & 0.015 &    2.3 &      12 \\
$10^6$ & 0.9 & O star    & 477 & 0.0021   & $10^{-3}$ &  14 & $3.0\times10^{-6}$ &    140 & 0.023 &    3.5 &      18 \\
       &     &           & 145 & 0.0089   & $10^{-3}$ &  70 & $7.6\times10^{-5}$ &     28 & 0.13  &     18 &      92 \\
       &     &           & 6.2 & 0.56     & $10^{-1}$ &   8 & $9.9\times10^{-7}$ &    240 & 0.012 &    2.0 &      11 \\
$10^6$ & 0   & red giant & 477 & 0.047    & $10^{-2}$ & 220 & $8.4\times10^{-5}$ &    170 & 0.78  &     98 &  2\,900 \\
       &     &           & 145 & 0.15     & $10^{-1}$ &  55 & $5.3\times10^{-6}$ &    700 & 0.20  &     25 &     710 \\
       &     &           & 6.2 & $\sim1$  & $\sim1$   &  40 & $2.8\times10^{-6}$ &    960 & 0.14  &     18 &     520 \\
$10^6$ & 0.9 & red giant & 477 & 0.064    & $10^{-2}$ & 220 & $8.4\times10^{-5}$ &    170 & 0.78  &     98 &  2\,900 \\
       &     &           & 145 & 0.23     & $10^{-1}$ &  75 & $9.8\times10^{-6}$ &    510 & 0.27  &     34 &     980 \\
       &     &           & 6.2 & $\sim1$  & $\sim1$   &  40 & $2.8\times10^{-6}$ &    960 & 0.14  &     18 &     520 \\
$10^7$ & 0   & O star    & 477 & 0.000027 & $10^{-5}$ &  17 & $5.8\times10^{-4}$ &    4.4 & 0.95  &    2.2 &     7.1 \\
       &     &           & 145 & 0.00016  & $10^{-4}$ &   5 & $5.1\times10^{-5}$ &     15 & 0.28  &   0.64 &     2.1 \\
       &     &           & 6.2 & 0.0096   & $10^{-3}$ &   8 & $1.3\times10^{-4}$ &      9 & 0.45  &    1.0 &     3.3 \\
$10^7$ & 0.9 & O star    & 477 & 0.000055 & $10^{-5}$ &  35 & $2.5\times10^{-3}$ &    2.2 & 2.0   &    4.5 &      15 \\
       &     &           & 145 & 0.00032  & $10^{-4}$ &  12 & $2.9\times10^{-4}$ &    6.3 & 0.67  &    1.5 &     5.0 \\
       &     &           & 6.2 & 0.025    & $10^{-3}$ &   6 & $7.3\times10^{-5}$ &     13 & 0.34  &   0.77 &     2.5 \\
$10^7$ & 0   & red giant & 477 & 0.0020   & $10^{-3}$ &  17 & $5.8\times10^{-5}$ &     96 & 1.8   &    3.7 &     130 \\
       &     &           & 145 & 0.011    & $10^{-2}$ &   3 & $1.8\times10^{-6}$ &    540 & 0.32  &   0.65 &      23 \\
       &     &           & 6.2 & 0.55     & $10^{-1}$ &  12 & $2.9\times10^{-5}$ &    140 & 1.3   &    2.6 &      91 \\
$10^7$ & 0.9 & red giant & 477 & 0.0036   & $10^{-3}$ &  35 & $2.4\times10^{-4}$ &     47 & 3.7   &    7.6 &     270 \\
       &     &           & 145 & 0.021    & $10^{-2}$ &  12 & $2.9\times10^{-5}$ &    140 & 1.3   &    2.6 &      91 \\
       &     &           & 6.2 & 0.83     & $10^{-1}$ &   8 & $1.3\times10^{-5}$ &    200 & 0.84  &    1.7 &      61
\enddata
\tablenotetext{a}{The maximum transit depth is determined by model parameters: SMBH mass and spin, circumnuclear stellar population, and observing wavelength.}
\tablenotetext{b}{This transit depth is chosen to be less than the maximum transit depth. Subsequent columns show values dependent on this parameter. The same transit depth is chosen for corresponding $a=0$ and $a=0.9$ cases to demonstrate how little probability estimates depend on spin.}
\end{deluxetable}

Let us substitute specific values to make numerical predictions. For concreteness, we study the cases of AGNs with mass $10^5$, $10^6$, and $10^7\;M_{\sun}$, and spin $a=0$ and $0.9$. Table~\ref {tab:transit} presents these examples. Each line of this table is generated by selecting a SMBH mass and spin, a stellar population model, and an observing wavelength. We determine the maximum possible transit depth by running our simulation with these input parameters. Then we choose a lower value as transit depth threshold, and use the calculated transit depth map (similar to Figures~\ref{fig:transitmapostar}--\ref{fig:transitmapgiant}) to identify the contour enclosing the area that a star has to transit to produce a lightcurve signature of at least this depth. We approximate this region by an ellipse, and take $R_\mathrm{AGN}$ to be the radius of a circular disk of the same area. These values can be substituted into Equations (\ref {eq:Deltat}) and (\ref{eq:ninstnum}--\ref{eq:neverless2}) to calculate the transit duration and to obtain probability estimates corresponding to the chosen transit depth treshold.

Table~\ref {tab:transit} shows that one cannot reasonably expect to detect transits when observing only a single or even a fistful of AGNs: typical transit rate is one every few hundred years (except for the very shallow transits for $M_\mathrm{SMBH}=10^7\;M_\sun$ which may happen more frequently than one in 10 years). However, these rate estimates are sensitive to poorly constrained parameters such as $r_\mathrm{min}$, $b$, and the number density exponent $\alpha$. Indeed, if we neglected the limits imposed by stellar collisions and set $r_\mathrm{min}$ to the tidal disruption radius, $r_\mathrm{min}$ would decrease by a factor of $\sim30$ or $\sim4$ (see Table~\ref{tab:rlimits}), and the instantaneous transit rate would increase a factor of $\sim200$ or $\sim3$, for O stars and giants, respectively. The event rates in a flattened, edge-on oriented star cluster may also be much larger \citep{2008ApJ...687..997S,2009ApJ...693.1959S}.

\section {Prospects for AGN transit observations}
\label {sec:agn:observability}

In the previous section we established that it is necessary to monitor a large number of AGNs to confidently detect transit events. This can be done either by a targeted survey, or by monitoring a large region on the sky. In this section, we calculate the number density of suitable AGNs in the local universe, then discuss the feasibility of observing AGN transits with specific instruments. Note that there are other instruments potentially able to detect a transit, and archival data of previous observations might already contain transits.

\subsection {AGN density and variability}
\label {sec:agn:agndensity}

The transit lightcurves are sensitive to the assumptions on the AGN, e.g.~on its mass and Eddington ratio. Larger mass means larger luminosity in optical to soft X-ray (but the disk is cooler, therefore less luminous in hard X-ray). Larger mass also means more extended accretion disk, that is, shallower transits. To quantify these effects, we consider three decades of magnitude ranges, centered on $M_\mathrm{SMBH}=10^5$, $10^6$, and $10^7\;M_\sun$. We integrate the lognormal fit of the local active black hole mass function determined by \citet {2007ApJ...667..131G,2009ApJ...704.1743G} to estimate the number density of AGNs in each mass range. Our results are given in Table~\ref {tab:dens}.

\begin{deluxetable}{ccc}
\tablewidth{0pc}
\tabletypesize{\scriptsize}
\tablecaption{Local active black hole density in three decades \label{tab:dens}}
\tablehead{SMBH mass range & number density\tablenotemark{a} \\
& $\left(\mathrm{Mpc}^{-3}\right)$}
\startdata
$10^{4.5}\;M_\sun < M_\mathrm{SMBH} < 10^{5.5}\;M_\sun$  & $5\times10^{-7}$ \\
$10^{5.5}\;M_\sun < M_\mathrm{SMBH} < 10^{6.5}\;M_\sun$  & $7\times10^{-6}$ \\
$10^{6.5}\;M_\sun < M_\mathrm{SMBH} < 10^{7.5}\;M_\sun$  & $1\times10^{-5}$
\enddata
\tablenotetext{a}{from the lognormal fit given by \citet {2007ApJ...667..131G,2009ApJ...704.1743G}}
\end{deluxetable}

For concreteness, we assume that all of these SMBHs are highly spinning and have a prograde accretion disk. However, Table~\ref {tab:transit} shows that probabilities depend weakly on the spin, therefore this assumption does not influence our estimates. (An exception is the case of X-ray observations, because accretion disk luminosity in this frequency depends strongly on spin, resulting in more potential targets, and thus more observable events down to a given luminosity limit.)

We assume the probabilities of the inclination $\cos\vartheta=0.5$ case for all AGNs (see Section \ref {sec:agn:discussion} for a discussion of this assumption).

A transit can only be detected in the lightcurve if the AGN variability amplitude on the timescale of interest is small enough compared to the transit depth. \citet {2011A&A...525A..37M} investigate a sample of over 9000 quasars in the SDSS sample between $z=0.2$ and 3, and find that 93\%, 97\%, 93\%, 87\%, and 37\% of them are variable in the $u$, $g$, $r$, $i$, and $z$ band, respectively. However, the variability is dominated by timescales of months to years, much larger than the AGN transit
timescale of hours. A useful measure of the time dependence of the AGN variability is the structure function (SF), which essentially measures the RMS magnitude difference as a function of time lag $\tau$ between magnitude measurements. Based on the SDSS sample, \citet{2012ApJ...753..106M} find that for $\tau\lesssim 3$ year, $\mathrm{SF} = 0.02 \;\mathrm{mag}\;(\tau/10\;\mathrm{day})^{0.44}$. These observations are consistent with the damped random walk model of AGN variability \citep{2009ApJ...698..895K}.
Substituting a timescale of 1 hour, 1 day, and 1 week into this relation yields an average AGN variability of $0.0018$, $0.007$, and $0.017$ magnitudes in the optical $g$ band, which corresponds to a variability of $0.17\%$, $0.7\%$, and $1.6\%$ over these timescales, respectively. Comparing this to the AGN transit lightcurve on Figure~\ref{fig:lightcurve} for transiting giants shows that the transit can cause a larger change in luminosity than the average intrinsic optical AGN variability, provided that the observation cadence is at most a few days. Thus we conclude that red giant transits may be detectable even in typically variable AGN. However, AGN variability may be a limitation for detecting transiting O stars in the optical bands, where the transit depth is much smaller.

AGN variability is more significant on the transit timescale in X-rays \citep{2012A&A...544A..80G,2012A&A...542A..83P}. However, in these bands, the transit may be a nearly complete eclipse (Figure~\ref{fig:lightcurve}), making them detectable regardless of variability. Indeed, transits of broad line clouds have already been detected in X-rays \citep{1998ApJ...501L..29M,2010A&A...517A..47M,2011MNRAS.410.1027R,2012ApJ...749L..31L}. Broad line clouds are expected to be much more densely distributed around AGNs then stars \citep{1997MNRAS.288.1015A,2006ApJ...636...83L}. However, their transit shapes are different from those of stars due to their cometary shape, with high column density heads followed by lower column tails \citep{2010A&A...517A..47M}. Some of these transiting clouds may represent the irradiated envelopes of circumnuclear ``bloated stars'' \citep{1980MNRAS.190..757E,1988MNRAS.233..601P} on very close orbits \citep{2012ApJ...749L..31L}.

\subsection {Ground-based optical instruments}
\label {sec:agn:optical}

Pan-STARRS \citep {2002SPIE.4836..154K,2010SPIE.7733E..12K} is an optical and NIR survey project, consisting of four units with 1.8 m primary mirror diameter and a field of view of $7\;\mathrm{deg}^2$ each. The photometric precision is $\approx1$\% in most bands per $\lessapprox40$ s exposure. The first telescope has been operating in science mode since 2010. Once all four units are online, the system will survey the night sky once every week with a $5\sigma$ detection limit down to $r\approx24$. We use the conservative magnitude limit $g\approx23$ for 1\% photometric accuracy per single exposure, and calculate for ten years of operation of all four units.

An AGN with $M_\mathrm{SMBH}\sim 10^6\;M_\sun$, spin $a=0.9$, and Eddington ratio $0.3$ has $g=23$ magnitude if observed from a luminosity distance of $\approx330\;$Mpc. (For comparison, a similar AGN with $a=0$ would have the same $g$ magnitude from $390\;$Mpc.)
Using \texttt{cosmocalc.py}\footnote{by Tom Aldcroft, \url{http://cxc.harvard.edu/contrib/cosmocalc/}} to calculate the comoving volume up to this luminosity distance, and using the AGN number density in Table~\ref{tab:dens}, we find that there are approximately $600$ AGNs in the given mass range to this distance observable from a single geographic location. Main sequence O stars do not cause deep enough transits in the optical bands for $10^6\;M_\sun$ SMBHs, therefore we focus on transiting red giants. Assuming all targets host red giants, there will be $\sim35$ transits in ten years according to the transit rate $\dot{N} = 1/(170\;\mathrm{yr})$ given in Table~\ref {tab:transit}. When such an event occurs, the AGN is on the night sky with probability $\approx1/2$. The transit lasts longer than a week, therefore there will be at least one observation during the transit. Assuming $0.75$ of the dark hours have photometric conditions at the site, we estimate that the Pan-STARRS survey can detect $\sim10$ stellar transits in $\sim 10^6\;M_\sun$ AGNs. (If all targets had $a=0$, the expected number of transits would be $\sim20$.)

A similar calculation for highly spinning $10^5\;M_\sun$ AGNs gives an approximate 80 Mpc distance limit for a $g=23$ magnitude. However, in the catalog of \citet {2007ApJ...667..131G,2009ApJ...704.1743G}, there is only one active black hole with $M_\mathrm{SMBH} < 10^{5.5}\;M_\sun$ closer than this distance, therefore a Pan-STARRS detection of a transit event in the lightcurve of such an object is unlikely in the standard all-sky survey mode. Finally, there are approximately $80\,000$ AGNs with masses $\sim 10^7\;M_\sun$ within $1.65\;\mathrm{Gpc}$ corresponding to $g\approx23$ mag. We do not expect 1\% deep transits in this case. The transit rate at a $10^{-3}$ photometric level is $1/(47\;\mathrm{yr})$, therefore the Pan-STARRS survey would observe $\sim6\,000$ such events during 10 years of operation. Note however that detecting these transits may be prohibitively difficult due to systematics of ground-based observations, and also due to AGN variability.

Future ground-based optical surveys of larger collecting area might have an even better chance to detect transits of $10^6\;M_\sun$ active SMBHs. For example, LSST is a survey telescope with an equivalent primary mirror diameter of 6.68 m, and a field of view of $9.6\;\mathrm{deg}^2$, currently in design and development phase \citep {2009arXiv0912.0201L}. It is planned to have an observing strategy similar to that of Pan-STARRS, but with 3.4 times the collecting area of the four Pan-STARRS units in total, it can survey an approximately $3.4^{3/2}\approx 6$ times larger volume, increasing the transit detection expectation accordingly. Therefore we estimate that LSST may detect $\sim100$ transits per decade for $M_\mathrm{SMBH}\sim 10^6\;M_\sun$. Also, since LSST is planned to have more frequent visits than Pan-STARRS, the intrinsic variability will have a smaller amplitude between subsequent observations.

Since many survey telescopes devote a certain fraction of their time to deeper surveys of smaller areas, it is worth estimating how this changes the expected number of transit detections. Recall that the four Pan-STARRS units will be able to scan the entire night sky approximately once a week in five filters with $\approx40$ second exposures. Now assume a single Pan-STARRS class instrument spends all available time on surveying an $n$ times smaller area in $g$ filter only, with $m$ visits per week to fight AGN variability, thus exposing $n/m$ times longer. As long as the observations are photon noise limited and cosmological effects are negligible, this strategy increases the surveyed volume, and also the expected number of transit detections, by a factor of $n^{1/2}m^{-3/2}$. For example, daily visits ($m=7$) of an $n\approx1300$ smaller area (a few fields) would mean two hour exposures per pointing, and would double the expected transit detections around $M_\mathrm{SMBH}\sim 10^6\;M_\sun$ AGNs to $\sim2$--4 per year, with increased robustness against intrinsic variability.

\subsection {Kepler observations}
\label {sec:agn:kepler}

The Kepler satellite carries out almost continuous photometric measurements in the bandpass of 420 to 900 nm. It achieves $\approx10^{-4}$ photometric accuracy per 30 minute exposure on $Kp=13$ dwarf stars \citep {2010ApJ...713L..79K}.

\citet {2011ApJ...743L..12M} report observations of four AGNs, and \citet {2012ApJ...751...52E} identify 13 more in the Kepler field. However, these quasars are at redshifts between $z=0.028$ and $0.625$, therefore it is not likely that either of them has low enough SMBH mass that transits due to O stars or red giants could be observed by Kepler.

\subsection {Space X-ray instruments}
\label {sec:agn:roentgen}

The Chandra X-ray observatory has been operating in orbit since 1999. Its relevant instrument is the High Resolution Camera (HRC) has a 31 arcmin by 31 arcmin field of view, sensitive from 0.1 keV to 10 keV \citep {2002PASP..114....1W}.

The X-ray Multi-Mirror Mission (XMM) Newton satellite is another proposal instrument, operating since 2000.
It is equipped with
three imaging cameras sensitive from 0.2 keV to 12 keV \citep {2001A&A...365L..27T,2001A&A...365L..18S},
with approximately $700\;\mathrm{arcmin}^2$ field of view each.

The Spektr-RG satellite is scheduled to launch in 2013 at earliest. Its eROSITA X-ray telescope system has approximately twice the effective area of a single instrument on XMM-Newton below 2 keV \citep {2010SPIE.7732E..23P}. This observatory is planned to carry out a survey consisting of eight full scans of the sky in four years, with a mean total exposure time of 2 ks for each region \citep {2012arXiv1209.3114M}.

The Extreme Physics Explorer \citep {2011arXiv1112.1327G} is a mission concept designed specifically to study accretion disks around SMBHs. It is proposed to have more than an order of magnitude larger effective area than current missions, thus capable of observing targets farther away.

As an example, we estimate the expected number of transits detected during a single 100 ks observing campaign with the HRC-I instrument on Chandra. We consider observations in the energy range 0.1 keV to 0.4 keV, logarithmically centered on the energy 0.2 keV for which we calculated transit probabilities. Based on Section~\ref{sec:agn:spectra}, the integrated luminosity in this range is $9\times10^{41}\;{\mathrm{erg}}/{\mathrm{s}}$, $8\times10^{42}\;{\mathrm{erg}}/{\mathrm{s}}$, and $6\times10^{43}\;{\mathrm{erg}}/{\mathrm{s}}$, and the photon index is $-0.2$, $0.4$, and $1.6$, for AGNs with $a=0.9$ and mass $10^5$, $10^6$, and $10^7\;M_\sun$, respectively. (Accretion disks around non-spinning black holes are up to a factor of four less luminous in this energy range.)

Let us divide the observation time into ten bins of 10 ks each for temporal resolution. If we want the relative photon noise to be 0.1 in each bin, we need a count rate of at least $0.01$ photons per second. According to the PIMMS count rate calculator\footnote{\url {http://cxc.harvard.edu/toolkit/pimms.jsp}}, this is the case up to a luminosity distance of $\approx290$, $890$, and $2\,300\;$Mpc for SMBH masses of $10^5$, $10^6$, and $10^7\;M_\sun$, respectively. This corresponds to a redshift $z\approx0.06$, $0.2$, and $0.4$, which is small enough to neglect in terms of flux change for a simple estimate. There are $\approx40$, $\approx12\,000$, and $\approx180\,000$ AGNs in the corresponding mass ranges, out to this luminosity distance, out of which $\approx0.0003$, $\approx0.08$, and $\approx1$ fall in the Chandra field of view on average. Therefore we can only observe a single or at most a few targets in one pointing. Since the probability of detecting a transit when observing a single target for this much time is negligible based on the transit rate value of Table~\ref {tab:transit}, such a short campaign is not suitable for stellar AGN transit discovery.

Now let us consider the planned eROSITA survey. Suppose the 2 ks exposure time is evenly distributed among eight visits of 250 s on each object. We have the same 0.1 relative photon noise per visit out to approximately 200\;Mpc, 700\;Mpc, and 2\;Gpc, for AGNs with $10^5$, $10^6$, and $10^7\;M_\sun$, respectively. On the entire sky, this means approximately 20, 6\,000, and 100\,000 AGNs. For such short observations, the expected number of transits sampled is the product of the number of AGNs, the number of visits per AGN, and the instantaneous probability of a transit, $N_\mathrm{inst}$. Based on the values in Table~\ref {tab:transit}, if we assume O stars around each target, we expect $3\times10^{-5}$ almost total eclipses for AGNs with mass $\sim10^5\;M_\sun$, and $0.05$ transits of depth $0.1$ for AGNs with $10^6\;M_\sun$. In case of SMBH mass around $10^7\;M_\sun$, transits due to O stars do not reach the depth of $0.1$, therefore they are not likely to be detected in the inherent variability. If we assume a $b=0.01$ mass fraction of giant stars in a typical AGN, we expect approximately $\sim10^{-3}$, $0.1$, and $10$ transits of depth $\sim1$, $\sim1$, and $0.1$ detected for SMBH masses $\sim10^5$, $10^6$, and $10^7\;M_\sun$, respectively.

We conclude that discovering stellar transits in X-ray with proposal instruments is unlikely, whereas it may be possible with survey instruments like eROSITA for AGNs with $10^6$--$10^7\;M_\sun$.

\section {Discussion}
\label {sec:agn:discussion}

In this section we highlight the uncertain points in our argument, not only to be able to properly interpret our predictions, but also to understand the implications of future AGN transit detections or non-detections.

We based our stellar models in galactic nuclei on observations of the center of the Galaxy in \textsection\ref {sec:agn:cusps}, assuming that this is a good representation of circumnuclear star clusters in active galaxies, and on theoretical dynamical models. We set up simplified stellar population models to estimate transit rates. Our estimates of transit probabilities depend strongly on the assumed value of the stellar distribution exponent $\alpha$: the $1000^{\alpha-2}$ terms in Equations (\ref {eq:ninstnum})--(\ref {eq:neverless2}) are due to the three order of magnitude difference in the distance scales of $r_\mathrm{min}$ and $r_0$. As a consequence, steeper stellar distributions feature larger transit probabilities, consistent with the statement that most probabilistic measures are dominated by closeby stars for moderate values of $\alpha$.

\begin {figure}
\begin{center}
\includegraphics*[width=\figurewidth]{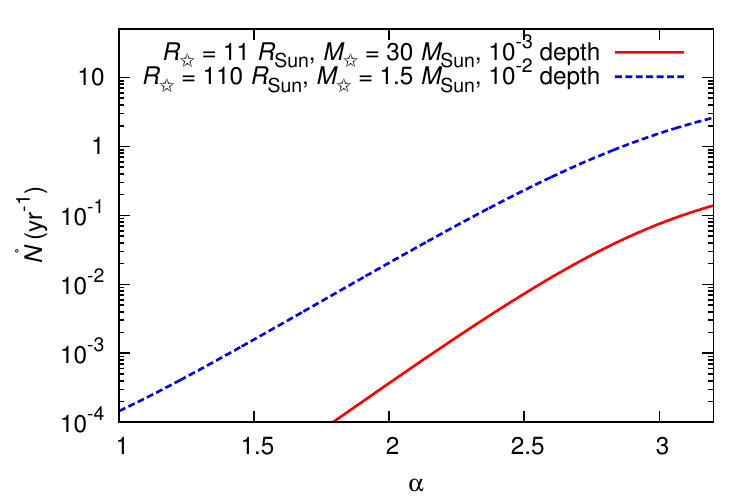}
\end{center}
\caption {Transit rate $\dot N$ for a single target AGN as a function of stellar density exponent $\alpha$, with $M_6=1$, $a=0.9$, and $b=0.01$. Solid red curve corresponds to O stars at a $g$-band transit depth threshold of $10^{-3}$, dashed blue curve corresponds to red giants or bloated stars at a $g$-band transit depth threshold of $10^{-2}$.}
\label {fig:ndotofalpha}
\end {figure}

Evaluated at $\alpha=3$, Equation (\ref {eq:density}) and thus Equations (\ref{eq:ninstnum}--\ref{eq:neverless2}) yield zero. This artifact originates in the assumption that the number of stars within $r_\mathrm{min}$ is negligible, which is not true for this value of $\alpha$. To study the case of $\alpha\approx3$, we recalculate $\dot N$ without this assumption:
\begin {align}
\dot N &= \frac b{0.093\;\mathrm{yr}} \frac{3-\alpha}{\alpha-\frac12} M_6^{\frac52}\left(\frac{M_\star}{30\;M_\sun}\right)^{-1} \times \\
\nonumber
&\times \frac{r_\mathrm{min}^{3-\alpha}}{r_\mathrm i^{3-\alpha}-r_\mathrm{min}^{3-\alpha}} \left( \frac {r_\mathrm{min}}{1\;\mathrm{mpc}} \right) ^{-\frac52}\left(\frac{R_\mathrm{AGN}}{10\;R_\mathrm g}\right).
\end {align}
Figure \ref {fig:ndotofalpha} displays this formula for $\dot N$ as a function of $\alpha$. We plot two cases: transit rate with a $g$-band threshold depth of $10^{-3}$ for O stars ($R_\mathrm{AGN}=14$), and with a $g$-band threshold depth of $10^{-2}$ for red giants or bloated stars with $R_\star=110\;R_\sun$, $M_\star=1.5\;M_\sun$ ($R_\mathrm{AGN}=220$). We fix $M_6=1$, $a=0.9$, and $b=0.01$ for both cases. The plots show us that $\dot N$ is a steeply increasing function of $\alpha$ up to and over the value 3, as expected. Comparing the plot to the values given in Table \ref {tab:transit} at $\alpha=2.5$ in the first case ($\dot N = 1/140\;\mathrm{yr}$) and at $\alpha=1.75$ in the second case ($\dot N=1/170\;\mathrm{yr}$) tells us that our approximation in Equation (\ref {eq:density}) for these values of $\alpha$ is valid.

Our method to determine the number of stars within the sphere of influence based on the $M$-$\sigma$ relation is oversimplified. The actual number of stars depends on formation, relaxation, and disruption rate and history. Furthermore, the presence of stellar remnants may dilute the stellar population within the sphere of influence for a fixed total mass, decreasing the mass fraction $b$ -- and therefore the number -- of stars that would potentially cause detectable transits.

At large orbital radii, we expect microlensing to dominate over the light obscuration due to the transiting star. We did not investigate this phenomenon in detail but conservatively ignored all transits due to stars on wide orbits where microlensing may become significant. As the transit rate is dominated by closeby stars, microlensing is not a limiting factor of transit observability after all. Also, we did not account for apsidal and Lense--Thirring precession, which may move stars in and out of transiting orbits.

AGN transits due to broad line clouds are expected to be much more frequent than the stellar transits calculated here \citep [see e.g.] [] {1998ApJ...501L..29M, 2009ApJ...695..781B, 2010A&A...517A..47M, 2007ApJ...659L.111R, 2009MNRAS.393L...1R, 2009ApJ...696..160R, 2011MNRAS.410.1027R,2011MNRAS.417..178R}. Transits of clouds may be somewhat different than stellar transits due to their cometary shape \citep{2010A&A...517A..47M}, and typically have a shorter duration due to their proximity to the SMBH and larger velocity (typically $\Delta t\sim 1\;\mathrm{hr}\;M_6$ if orbiting at $r_\mathrm{BLR}\sim 1000\;R_g$). 

Note that we have assumed that the transiting stars move on orbits much wider than the accretion disk. This assumption may be violated for stars on very close orbits ($r\lesssim 10^5 \; R_g = M_6 \; 4.8 \;\mathrm{mpc}$), where stars crossing the disk may get captured by hydrodynamic drag \citep {2001A&A...376..686K}.

We mentioned the possibility of supermassive stars possibly forming in AGN accretion disks, but do not have information on their occurrence rates. Furthermore, Wolf--Rayet stars exhibit strong stellar winds, which might form an optically thick region of radius $\sim100\;R_\sun$ \citep [Figure 5 in][and references therein] {2007ARA&A..45..177C}; OH/IR stars can form dust clouds much larger than the entire accretion disk \citep [e.g.][] {1987A&A...186..136B,2002A&A...384..585K}; and bloated stars with large irradiated envelopes \citep{1980MNRAS.190..757E,1988MNRAS.233..601P} may be present near the AGN. These objects can potentially cause much deeper transits or even eclipses.

In Section~\ref {sec:agn:spectra}, we investigated the projected thermal radiation structure of an accretion disk with specific parameters, assuming radiatively efficient accretion, modeling radiative transfer in the accretion disk photosphere, and accounting for light propagation in the Kerr metric of the SMBH. We set the Eddington ratio to $0.25$, a value based on the analyses of \citet {2006ApJ...648..128K} and \citet {2012MNRAS.tmp...16S}. This value is consistent with the findings of \citet {2007ApJ...667..131G}. However, keep in mind that magnitude limited samples are biased towards larger Eddington ratios. Also note that the Eddington ratio can in fact vary by an order of magnitude in either direction \citep[e.g.][]{2009MNRAS.397..135K}. A larger accretion rate would increase the total luminosity and the peak frequency with little effect on the size of the accretion disk. Therefore such a disk would be much brighter in X-ray than one with lower Eddington ratio, while they would exhibit transits of similar depth.

We also assumed prograde disk alignment for spinning SMBHs, but noted that the transit probabilities do not depend strongly on spin. We fixed the inclination at $\cos\vartheta=0.5$, which is the mean value for an isotropic distribution, therefore the predicted transit probabilities are typical of all inclinations. However, note that larger inclination results in a thinner image of the accretion disk, therefore deeper transits. A disk with thickness $H/R\approx 0.05$ observed from a nearly edge-on orientation results in a $\sim10$ times deeper transit, leading to a $10^3$ times larger detectable volume for photon noise limited surveys. On the other hand, there are selection effects at both inclination extrema: a coplanar torus might obscure thermal emission from edge-on AGNs, whereas jets along the rotational axis might contaminate the lightcurve of face-on AGNs and prevent transits from being detected. These are likely to confine the inclination distribution of AGNs with observable thermal radiation closer to the average values we use in our model.

Intrinsic AGN variability will pose a challenge to identifying AGN transits. Fortunately, optical AGN variability is small on the timescales of days, and thus does not rule out the possibility of transit observations for giant stars. With Pan-STARRS observations, however, the variability level at typical observation cadence is comparable to the transit depth. Transits might not be detected if the weekly observations miss the deepest part.

Simultaneous multiband observation campaigns can help distinguish variability from transit signiture, as the transit lightcurve is predicted to have a different shape in different frequency bands. Also, transit lightcurves could be contaminated by the flux reflected by clouds surrounding the AGN. Multi-wavelength campaigns following AGN for these types of events may give information about the reflecting fraction, which could allow to constrain the covering fraction of material, the location and relative sizes of the reflecting regions. Future detectors could distinguish the reflected component using polarimetry.

\section {Conclusions}
\label {sec:agn:conclusions}

In this paper, we presented simple estimates to study the prospects for detecting stellar AGN transits. We have shown that such observations would offer a novel possibility to image the accretion disk in distant AGN with unprecedented accuracy. For example, the black hole silhouette (i.e., the lack of emission within the ISCO) may be resolved using the lightcurve of an AGN transit due to an O-type main sequence star. These observations probe the accretion disk and the space-time geometry around black holes, and in particular, they are sensitive to the black hole spin. AGN stellar transit event rates offer information about the circumnuclear stellar cluster.

We predict that the Pan-STARRS survey could detect $10\sim20$ stellar transits, and LSST may detect $\sim100$ by repeated photometric observations of $\sim10^6\;M_\sun$ AGNs. We estimate that stellar transit detections in X-rays are not likely with individual campaigns on proposal instruments, but we expect a few possible detections in $\sim10^7\;M_\sun$ AGNs with X-ray surveys like eROSITA. Note that these rate estimates do not include transits by Compton thick clouds, which are observed to be common in X-rays. The transit rate corresponding to clouds could also be much more frequent in optical surveys.

However, these probability estimates are very sensitive to parameters which are based on theoretical arguments. One of these parameters is the inner radius cutoff of the nuclear stellar cluster, $r_\mathrm{min}$, which is set by stellar collisions. We have shown that if stellar collisions did not deplete the innermost regions, the less stringent limitation due to tidal disruption would imply $\sim200$ times larger transit rates for O stars. Such scenarios may be possible if the effective stellar size increases during close approach to the AGN, leading to large irradiated envelopes \citep[``bloated stars'', see][]{1980MNRAS.190..757E,1988MNRAS.233..601P}. AGN transit observations could constrain these parameters and refine circumnuclear stellar population models in distant galaxies.

\acknowledgements
The authors thank \'Akos Bogd\'an, Martin Elvis, Jeff McClintock, Barry McKernan, and Guido Risaliti for helpful discussions. BK acknowledges support from NASA through Einstein Fellowship PF9-00063 issued by the Chandra X-ray Observatory, operated by the SAO, on behalf of NASA under contract NAS8-03060.

\chapter{Stellar rotation--planetary orbit period commensurability in the \hatpeleven{} system}
\label{ch:commensurability}

\publishedarxiv{2014ApJ...788.....1}

\chapterabstract

A number of planet host stars have been observed to rotate with a period equal to an integer multiple of the orbital period of their close planet. We expand this list by analyzing \kepler{} data of \hatpeleven{} and finding a period ratio of 6:1. In particular, we present evidence for a long-lived spot on the stellar surface that is eclipsed by the planet in the same position four times, every sixth transit. We also identify minima in the out-of-transit lightcurve and confirm that their phase with respect to the stellar rotation is mostly stationary for the 48 month time frame of the observations, confirming the proposed rotation period. For comparison, we apply our methods to \keplerseventeen{} and confirm the findings of \citet{2012A&A...547A..37B} that the period ratio is not exactly 8:1 in that system. Finally, we provide a hypothesis on how interactions between a star and its planet could possibly result in an observed commensurability 
for systems where the stellar differential rotation profile happens to include a period at some latitude that is commensurable to the planetary orbit.


\section{Introduction}
\label{sec:commensurability:introduction}

Many stars have been observed to exhibit photometric variations synchronous to the orbit of their close planet. When these variations are attributed to photospheric features rotating with the stellar surface, this implies a synchronicity between stellar rotation and planetary orbit. One of the earliest robust detections of this phenomenon is by \citet{2008A&A...482..691W} in the system \tauboo{}. They report on periodic photometric variations of the host star with a period within 0.04\% of that of the planetary orbit, and they attribute this to an active region on the surface of the star. Similarly, stellar photometric variations synchronous to the planetary orbit have been detected for the planetary systems CoRoT-2 \citep{2009EM&P..105..373P,2009A&A...493..193L} and CoRoT-4 \citep{2009A&A...506..255L}. For all three stars, the rotation period inferred from spectroscopy is consistent with the period of photometric variations, indicating that the variations are due to photospheric features stationary on the stellar surface.

Another interesting example is \keplerthirteen{}. \citet{2012MNRAS.421L.122S} measure the rotational period of the star by frequency analysis of the spot-modulated lightcurve and find a 5:3 commensurability with the orbital period of the planet \keplerthirteenb{} at high significance. 

However, frequency analysis is not the only method suitable for measuring rotation rates of spots on the stellar surface. A transiting planet may eclipse spots on the surface of its host star, resulting in anomalies in the transit lightcurve. This phenomenon was observed, for example, in the systems HD 209458 \citep{2003ApJ...585L.147S}, HD 189733 \citep{2007A&A...476.1347P}, TReS-1 \citep{2009A&A...494..391R}, and CoRoT-2 \citep{2009A&A...493..193L}. Repeated transit anomaly detections due to the same spot can be used to constrain the stellar rotation period. This method was first applied by \citet{2008ApJ...683L.179S} to HD 209458.

Another application of starspot-induced transit anomalies is to constrain the spin--orbit geometry, as was first mentioned by \citet{2010ApJ...723L.223W}.  This method was developed and applied independently by \citet{2011ApJ...740...33D} and \citet{2011ApJ...743...61S} to \hatpeleven{}, by \citet{2011ApJ...733..127S} to WASP-4, and by \citet{2011ApJ...740L..10N} to CoRoT-2.

Independent measurements of the \rmeffect{} on \hatpeleven{} show that the planetary orbit normal is almost perpendicular to the projected stellar spin \citep[the projected obliquity is $\approx103^\circ$; see][]{2010ApJ...723L.223W,2011PASJ...63S.531H}. Relying only on photometric data, \citet{2011ApJ...740...33D} and \citet{2011ApJ...743...61S} independently identify two active latitudes (where spots are most prevalent) on the surface of the star, which they assume to be symmetrical around the equator, to conclude that \hatpelevenb{} is on a nearly polar orbit in accordance with the spectroscopic results, and that the stellar spin axis of \hatpeleven{} is close to being in the plane of sky.

The transit lightcurve of \keplerseventeenb{} ($P=1.49\;\mathrm{days}$) also exhibits anomalies due to spots on the surface of its host star.  In their discovery paper, \citet{2011ApJS..197...14D} analyze these anomalies to study both stellar rotation and orbital geometry. They observe that the transit anomaly pattern repeats every eighth planetary orbit, suggesting that the spots rotate once while the planet orbits eight times. They dub the phenomenon of the same spots reappearing periodically at the same phase in transit lightcurves---every eighth one in this case---the \strobo{}.  As for the orbital geometry, they found that transit anomalies in successive orbits are consistent with being caused by the same spots that rotate one eighth of a full revolution on the stellar surface with each orbit of the planet.  This implies a low projected obliquity of the planetary orbit, and also excludes frequency aliases (like the star rotating three or five times while the planet orbits eight times).

In this paper, we present evidence for a 6:1 period commensurability between the rotation of the star \hatpeleven{} and the orbit of its planet \hatpelevenb{} \citep[$P=4.89\;\mathrm{days}$;][]{2010ApJ...710.1724B}. The increasing number of systems known to exhibit such commensurability raises the question of whether this is the result of an interaction between the planet and the star.

Whenever studying stellar rotation, it is important to remember that stars with convective zones exhibit differential rotation. In this paper, the working definition of stellar rotation rate is that inferred through dominant spots on the stellar surface, either from the rotational modulation of the out-of-transit lightcurve or from transit anomalies. This way we measure the rotation rate of the stellar surface at the latitude of the spots or active regions. If spots from multiple latitudes with different rotational rates contribute significantly to the lightcurve, then we expect the inferred posterior distribution of the rotational period to have a broader profile.

Despite their usefulness in confining planetary obliquity and mapping spots, transit anomalies due to the planet eclipsing spots can also be a nuisance: they contaminate the transit lightcurve, introducing biases in the detected transit depth \citep{2009A&A...505.1277C}, time, and duration. \citet{2010A&A...520A..66B} point out that in the particular case of stellar rotation--planetary orbit commensurability, activity-induced transit timing variations can be periodic and thus can result in spurious planet detections. This further motivates the need for understanding stellar rotation--planetary orbit commensurability.

In Section \ref{sec:commensurability:acf}, we look at the periodogram and autocorrelation function of \hatpeleven{} and \keplerseventeen{} lightcurves to confine the rotational period. In Section \ref{sec:commensurability:anomalies}, we present the case of a spot on \hatpeleven{} recurring multiple times because of the \strobo{}. In Section \ref{sec:commensurability:macula}, we analyze all transit anomalies observed on \hatpeleven{} to feed the best-fit spot parameters into the rotational modulation model \macula{} \citep{2012MNRAS.427.2487K} and compare the resulting model lightcurve to observations. In Section \ref{sec:commensurability:recurrence}, we perform a statistical analysis of spot-induced anomalies in the transits of \hatpelevenb{} and \keplerseventeenb{}. In Section \ref{sec:commensurability:flipflop}, we look for the periodicity of lightcurve minima for both stars. We show evidence for two spots or spot groups at opposite longitudes on both \hatpeleven{} and \keplerseventeen{} and find that on the former, they seem to alternate in relative activity level, which is known as the ``flip-flop'' phenomenon \citep{1991LNP...380..381J}. In Section \ref{sec:commensurability:chance}, we calculate the probability of commensurate periods by chance.  In Section \ref{sec:commensurability:interaction}, we state one possible hypothesis about stellar rotation--planetary orbit resonance and discuss difficulties in proving it. Finally, we summarize our findings in Section \ref{sec:commensurability:conclusion}.

\section{Out-of-transit lightcurve}
\label{sec:commensurability:acf}

In their discovery paper, \citet{2010ApJ...710.1724B} report a strong frequency component in the HATNet lightcurve of \hatpeleven{} with a period of approximately 29.2 days. They attribute it to rotational modulation of starspots, noting that the 6.4 mmag amplitude is consistent with observations of other K dwarfs and the period is consistent with the color, activity level, and projected rotational velocity of \hatpeleven. They also note that both the secondary peaks in the autocorrelation function and the phase coherence of the lightcurve indicate that starspots (or spot groups) persist ``for at least several rotations.''

\begin{figure}
\begin{center}
\includegraphics*[width=\figurewidth]{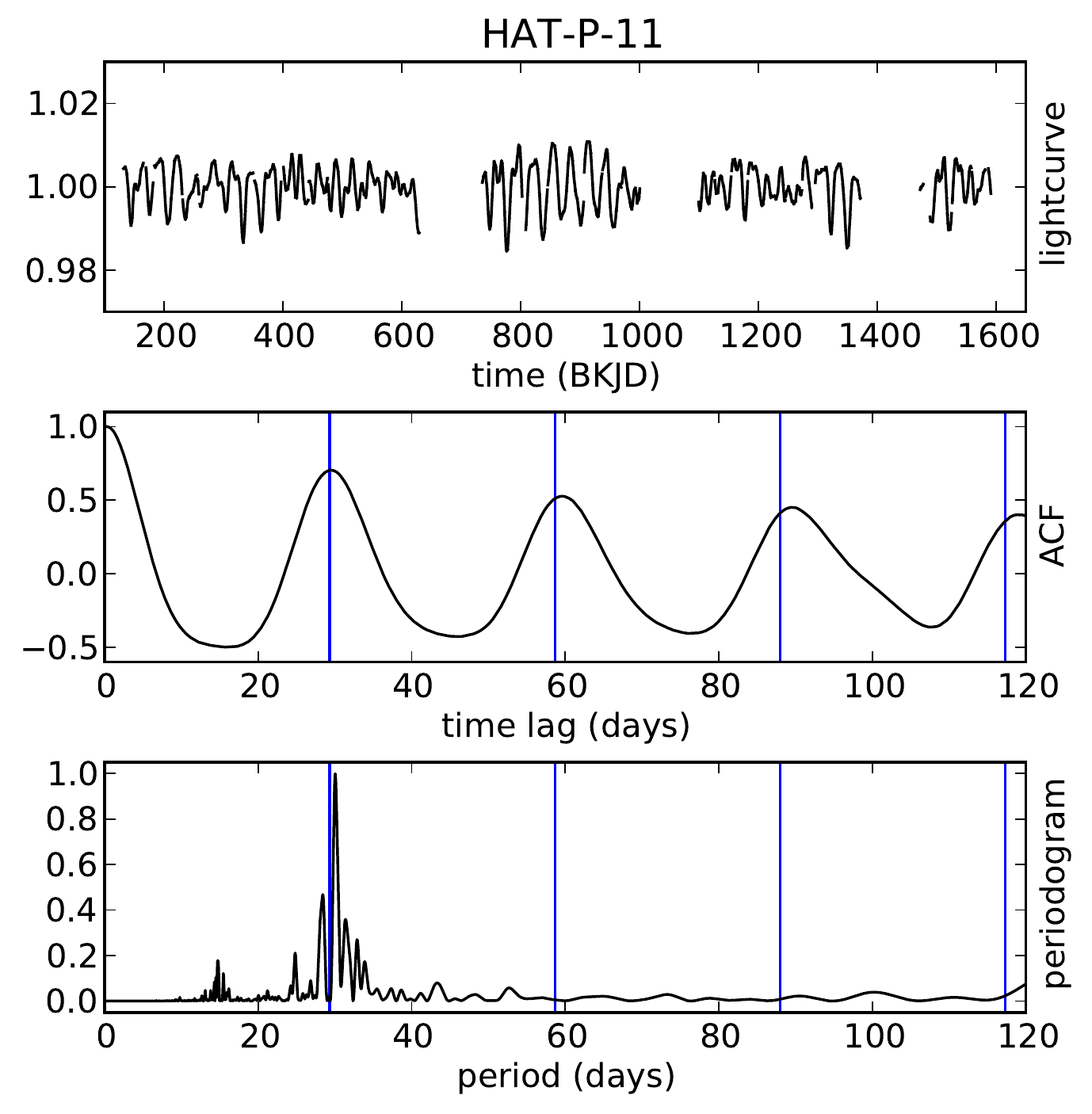}
\end{center}
\caption{Top panel: long cadence \kepler{} lightcurve of \hatpeleven, with transits of \hatpelevenb{} removed. Middle panel: autocorrelation function of the lightcurve. Bottom panel: Lomb--Scargle periodogram of the lightcurve. The blue vertical lines on the middle and bottom panels correspond to the proposed rotational period (six times the planetary orbital period) and its integer multiples.}
\label{fig:acf0}
\end{figure}

Figure \ref{fig:acf0} presents the entire \kepler{} space telescope \citep{2010Sci...327..977B} long cadence lightcurve of \hatpeleven{} (quarters 0--6, 8--10, 12--14, and 16--17, with transits of \hatpelevenb{} removed and each quarter scaled to have unit mean). Time is measured in Barycentric \kepler{} Julian Date (BKJD), which is $\textrm{BJD}_\mathrm{UTC}-2\,454\,833$. Figure \ref{fig:acf0} also displays the autocorrelation function and periodogram of the long cadence lightcurve. This analysis is similar to that performed by \citet{2010ApJ...710.1724B} but on much better quality data. We confirm their findings: we identify a peak in the autocorrelation function at a timelag of 29.32 days (with FWHM 8.05 days) and a peak in the periodogram at 30.03 days (with FWHM 0.62 days), which we identify with the rotational period of \hatpeleven{}. For comparison, six times the planetary orbital period is 29.33 days, and it is indicated along with its integer multiples on Figure \ref{fig:acf0} by blue vertical lines. We also see multiple peaks in the autocorrelation function at integer multiples of the base period, indicating that some spots must live for multiple stellar rotations.

\begin{figure}
\begin{center}
\includegraphics*[width=\figurewidth]{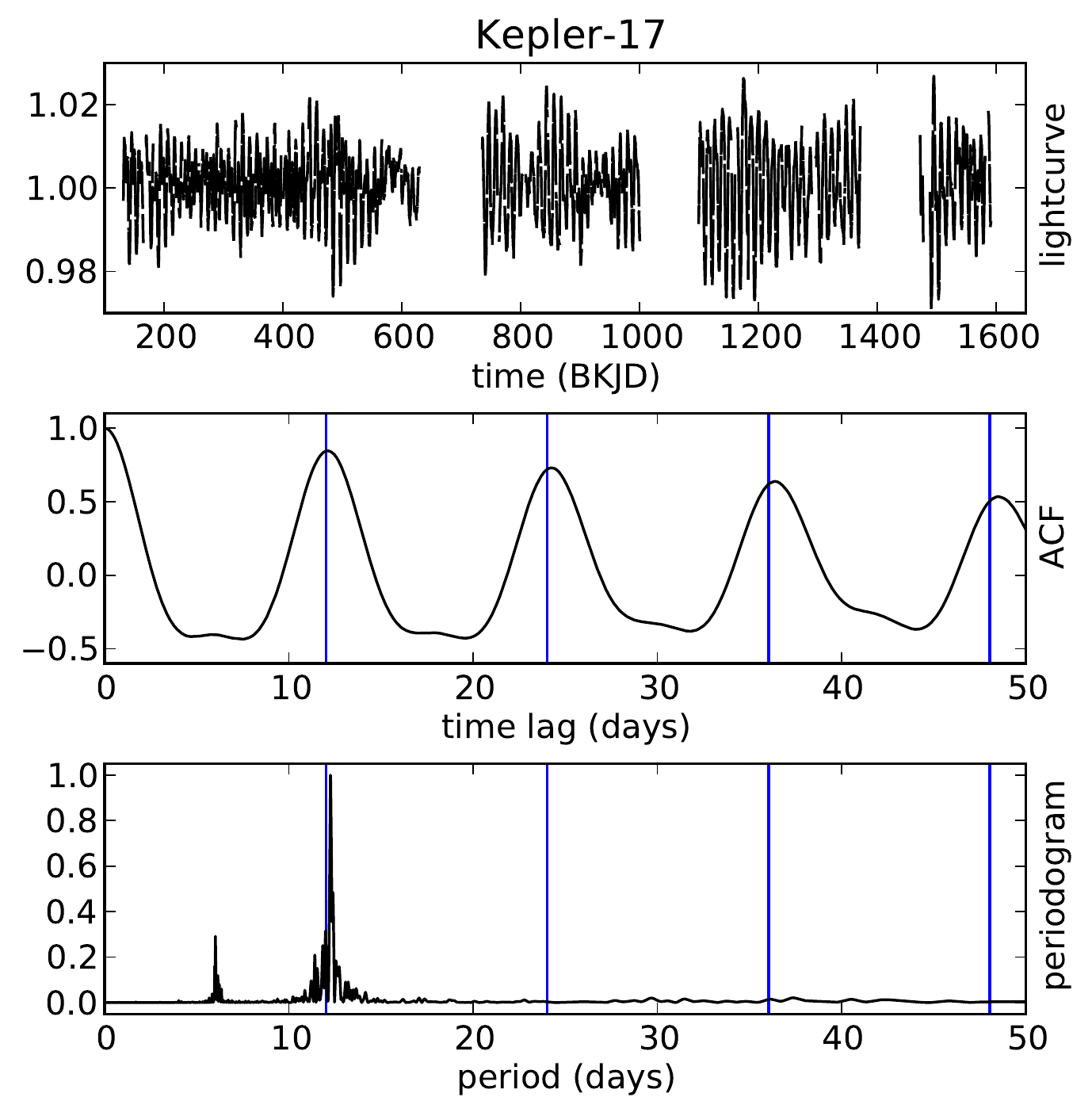}
\end{center}
\caption{Same as Figure \ref{fig:acf0}, for \keplerseventeen, with transits of \keplerseventeenb{} removed. The blue vertical lines on the middle and bottom panel correspond to the rotational period proposed by \citet{2012A&A...547A..37B} (not exactly eight times the planetary orbital period) and its integer multiples.}
\label{fig:acf1}
\end{figure}

For comparison, on Figure \ref{fig:acf1} we present the same analysis for the \kepler{} long cadence lightcurve of \keplerseventeen{} (quarters 1--6, 8--10, 12--14, and 16--17, with transits of \keplerseventeenb{} removed and each quarter scaled to have unit mean).  The blue vertical lines indicate multiples of 12.01 days, the stellar rotation period reported by \citet{2012A&A...547A..37B}, instead of eight times the planetary orbital period, which is 11.89 days. The first peak of the autocorrelation function is at 12.10 days (with FWHM 3.13 days), while the periodogram peaks at 12.25 days (with FWHM 0.11 days). It is interesting to note that \hatpeleven{} and \keplerseventeen{} are in the same \kepler{} subfield on the sky; therefore, we see gaps in both lightcurves during quarters 7, 11, and 15 due to the failure of a readout module.

The main reason for studying the autocorrelation function and the periodogram is to exclude the possibility of frequency aliases. If we interpret the half-width at half-maximum of the autocorrelation function as a direct indicator for period uncertainty \citep[as done, for example, by][]{2008A&A...488L..43A}, the resulting range is consistent with the proposed rotational periods for both stars. We refer the reader to \citet{2013MNRAS.432.1203M} for a discussion of using the autocorrelation function as a complementary method to periodograms for studying stellar rotation.

Note that \kepler{} data are dense in time, with long runs of almost continuous observations.  We confirm that the spectral window function does not have large values at periods above 30 minutes, the cadence of observations.  Therefore unlike for sparsely sampled ground based observations, frequency aliasing \citep{2010ApJ...722..937D} does not pose a problem in this analysis.

The periodograms rule out that we are dealing with an alias of the rotational rate. However, the narrow periodogram peak is located at a period slightly larger than the proposed rotational period for both stars. \citet{2013MNRAS.432.1203M} observe that spot evolution and differential rotation can cause periodogram peaks to split up into multiple narrow peaks, thus the FWHM may not correspond directly to the period uncertainty. Therefore, the periodograms are not inconsistent with the proposed rotational periods of 29.33 days for \hatpeleven{} and 12.01 days for \keplerseventeen{}.

\section{Stroboscopic effect on \hatpeleven{}}
\label{sec:commensurability:anomalies}

\citet{2010ApJ...723L.223W} were the first to note that the ratio between the stellar rotation period of \hatpeleven{} and the orbital period of \hatpelevenb{} is approximately 6:1. If it was close enough to 6:1 and there were spots that lived long enough, then one would be able to detect multiple lightcurve anomalies because of the same spot every sixth transit. However, \hatpelevenb{} has a polar orbit with respect to the stellar spin axis; therefore, if the periods were incommensurable, then the spot could not fall repeatedly under the transit chord.

\citet{2011ApJ...743...61S} pointed out that a 6:1 period ratio is \apriori{} unlikely. They were looking for recurrence of transit anomalies, but quarters 0, 1, and 2 of \kepler{} data available at the time did not provide a large enough sample for such investigations.

\begin{figure}
\begin{center}
\includegraphics*[width=\figurewidth]{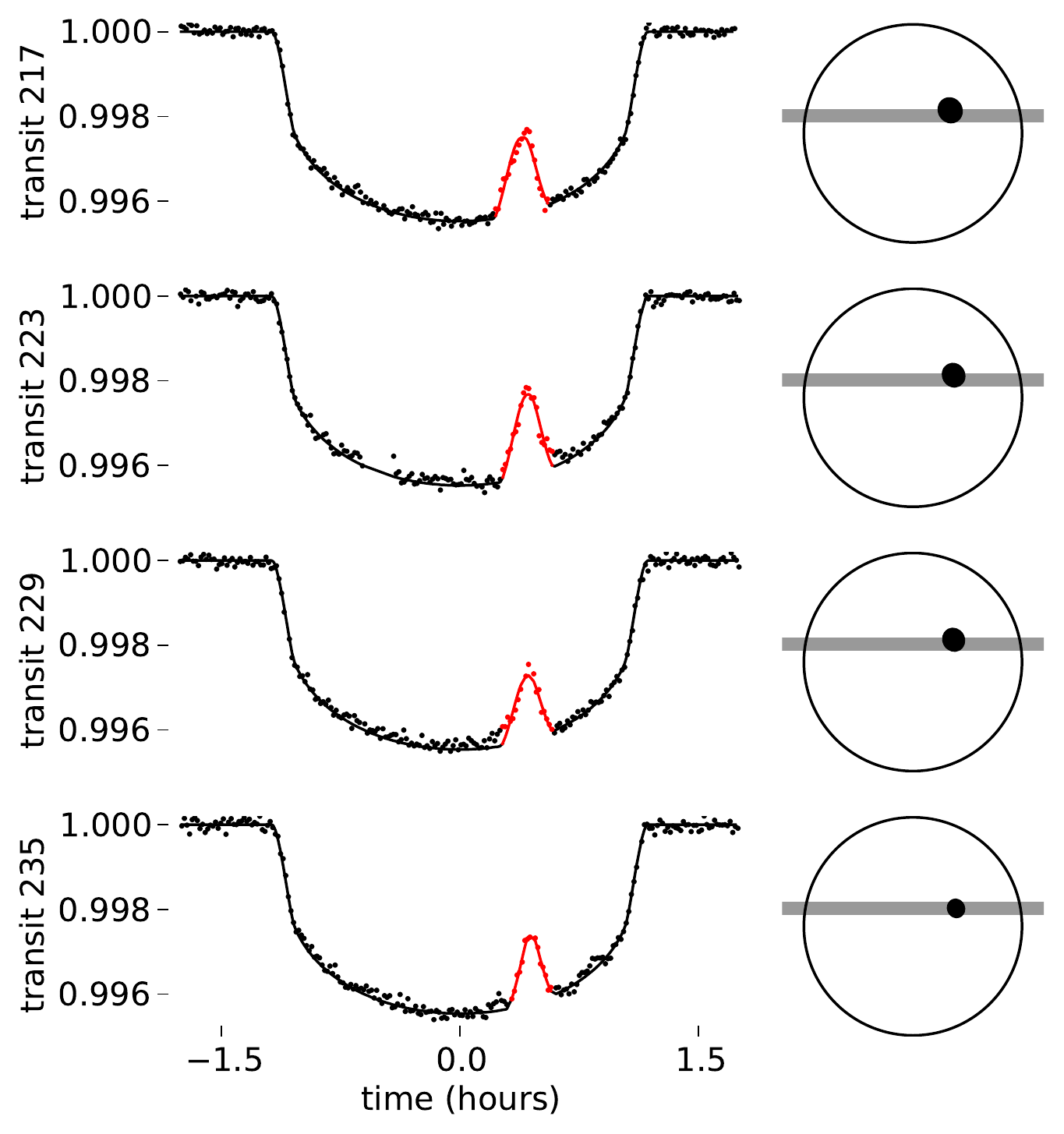}
\end {center}
\caption{Transit anomalies providing evidence for the 6:1 commensurability. The transits, from top to bottom, are separated by six planetary orbits, which is the proposed stellar rotation period. Left panels show detrended \kepler{} short cadence photometry, along with best-fit model lightcurve with single spot. Right panels show the projected stellar disk, transit chord, and best-fit spot. Note that spot seems to be stationary over this time period, which suggests a tight 6:1 commensurability.}
\label{fig:recurrence}
\end{figure}

In this section, we study a single extraordinary example of spot recurrence observed by the \kepler{} space telescope on \hatpeleven{}, presented on Figure \ref{fig:recurrence}. Lightcurves of transits 217, 223, 229, and 235 exhibit very similarly shaped spot-induced anomalies. The striking similarity between these four anomalies, spaced apart by six planetary orbits, suggests that they are caused by the same spot, which evolves little during these observations. If this is indeed the case, then we are seeing the same \strobo{} as \citet{2011ApJS..197...14D} on \keplerseventeen{}, and the similarity of transit anomalies implies that the period ratio is very close to 6:1.

\begin{figure}
\begin{center}
\includegraphics*[width=\figurewidth]{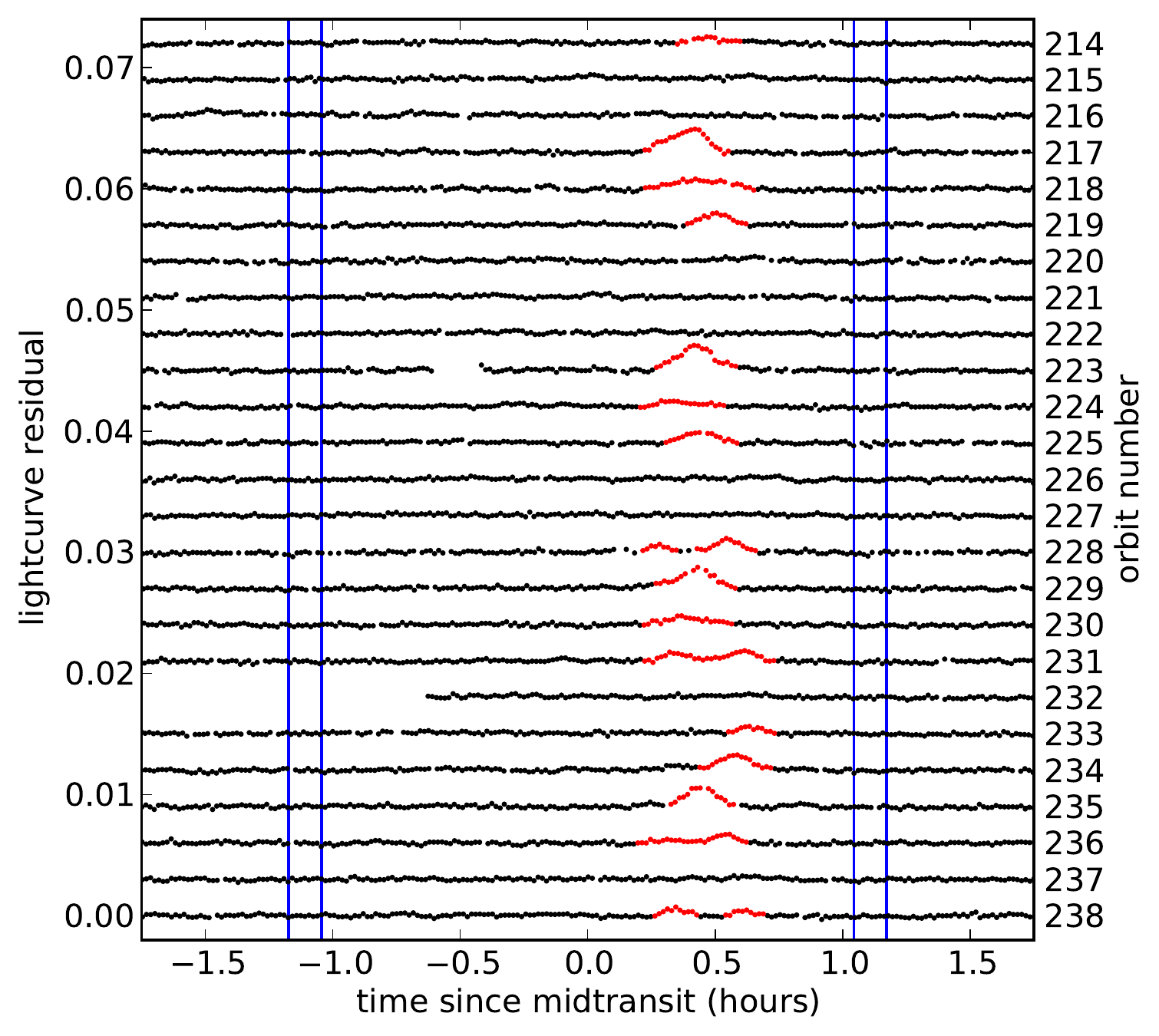}
\end {center}
\caption{\kepler{} short cadence observations minus Mandel--Agol model lightcurve for transits 214--238, as a function of time. Residuals are vertically displaced for each transit. Modeled spots (among \totalspots{} in total) are indicated with red points. The four blue vertical lines indicate first, second, third, and fourth contacts, from left to right.}
\label{fig:anomalies}
\end{figure}

However, the same transit anomaly might be caused by a continuous active band encircling the star along a constant latitude. In this case, the anomaly shape would not depend on how much the star rotates between each six transits and thus would provide no information on a possible commensurability. To exclude this possibility, we look at all transits surrounding the ones highlighted on Figure \ref{fig:recurrence}. We subtract the model transit lightcurve without spots \citep{2002ApJ...580L.171M} from the observed data and plot the residuals for each transit on Figure \ref{fig:anomalies}.

We look for anomalies in adjacent transits that are similar to the one seen on Figure \ref{fig:recurrence} in transits 217, 223, 229, and 235.  However, these adjacent transits exhibit anomalies either with much smaller amplitude (in transits 218, 224, 228, 230, and 236), or at a different orbital phase (in transit 234), or none at all (in transits 216 and 222).  Therefore we can exclude the case of a continuous dark band around the star, because such a band would cause transit anomalies of comparable amplitude at the same phase in every single transit.

Note, however, that it is not possible to determine the exact shape of the spots based on transit anomalies that only scan the star along sparse transit chords, therefore the determination of stellar rotation period hinges on the assumed shape of the spots, circular in our case.  If, on the other hand, the spots were elongated in longitude, then the shape of the transit anomaly would not be sensitive to the stellar rotation rate, therefore the \strobo{} could be observed even for incommensurable periods.

Also note that there are signs of other spots evolving on Figure \ref{fig:anomalies}, for example, between transits 225 (one small spot), 231 (now split into two), and 237 (disappeared), that are also separated by one stellar rotation each.

For our analysis, we adopt the revised transit ephemeris, planetary radius and orbital semimajor axis relative to the stellar radius, orbital inclination, and limb darkening parameters of \citet{2011ApJ...740...33D}. Their analysis accounts for eclipsed and uneclipsed spots and relies on the orbital eccentricity and argument of periastron measurement of \citet{2010ApJ...710.1724B} that uses both RV data and \hipparcos{} parallax for \hatpeleven{}.  To normalize each transit, we divide short cadence data by a linear fit to the out-of-transit observation within 0.12 days from the mid-transit time.

\section{Comparison to out-of-transit lightcurve}
\label{sec:commensurability:macula}

When analyzing the lightcurve periodicity to find the stellar rotation rate, we assumed that the lightcurve is dominated by rotational modulation of spots (as opposed to, for example, stellar pulsation). As \citet{2010ApJ...710.1724B} noted, the rotational modulation amplitude is indeed consistent with expectations based on observations of other K dwarfs. In this section, we offer an independent method to confirm this hypothesis: we first identify a number of spots via transit anomalies, then we model the rotational modulation caused by these spots and compare it to the observed lightcurve.

We adopt the analysis of \citet{2014MNRAS.442.3686B}, who manually identify \totalspots{} spots in 130 transits in the \kepler{} dataset, and run MCMC analysis to explore the spot parameters.  The model they use assumes the same quadratic limb darkening law for spots and the rest of the photosphere. It also assumes that spots are circular on the surface of the star, that is, elliptical in projection. 

The best-fitting lightcurves for transits 217, 223, 229, and 235 are shown on Figure \ref{fig:recurrence}, involving a single independent spot for each transit. The strikingly similar best-fit position of the spot as shown on the right panels further supports the hypothesis of stellar rotation--planetary orbit commensurability. We also highlight data points that are considered to be part of a spot anomaly according to the best-fit model in red on Figures \ref{fig:recurrence} and \ref{fig:anomalies}.

We feed the parameters of the spots derived from the transit anomalies into the rotational modulation model \macula{} \citep{2012MNRAS.427.2487K}. Since a long-lived stationary spot would be detected each stellar rotation (like the spot appearing in multiple transits above), we model each detected spot as if it lived for a single stellar rotation only, coming to life on the far side of the star half a rotation before we detect the transit anomaly it causes and ceasing to exist half a stellar rotation later, also on the far side. Since we see \hatpeleven{} almost equator-on \citep{2010ApJ...723L.223W,2011PASJ...63S.531H}, every spot we model gets to the far side of the star half a stellar rotation after it is eclipsed by the planet. If the same spot causes another transit anomaly one stellar rotation later, we model it as a separate spot that is created when the first one dies. This is the simplest way of treating spot evolution: properties of a long-lived spot are described by piecewise constant functions, with the jumps happening when the spot is not in sight, resulting in a continuous model lightcurve. In this treatment, we do not have to investigate whether two transit anomalies separated by an integer number of stellar rotations are due to the same spot or different spots, since we treat them as separate spots in both cases.

For the \macula{} model, we adopt the projected obliquity and inclination distribution of \citet{2011ApJ...743...61S} derived from spot crossing events that also accounts for the results of \citet{2010ApJ...723L.223W} based on the \rmeffect{}.

\begin{figure}
\begin{center}
\includegraphics*[width=\threequartersfigurewidth]{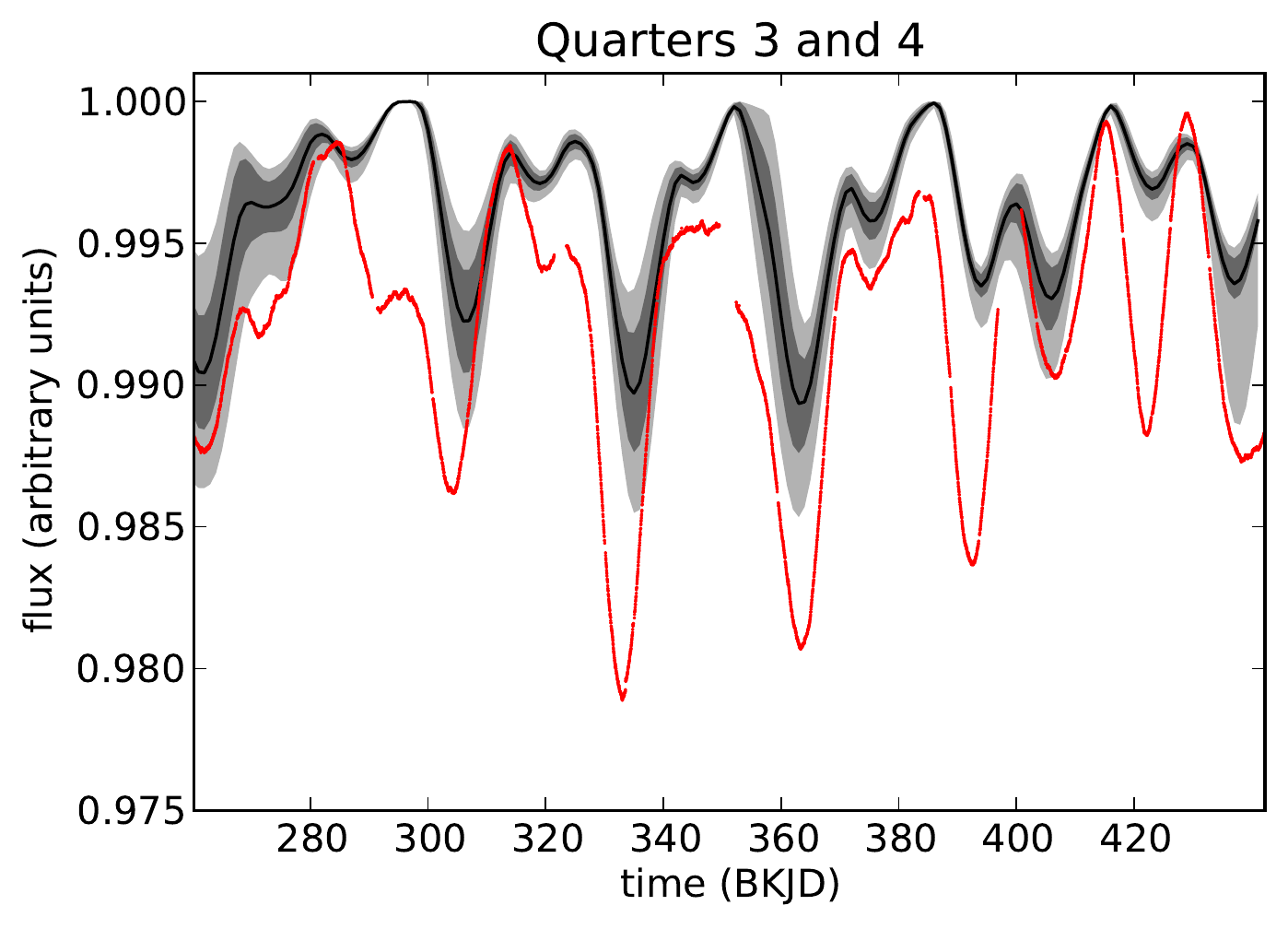}
\includegraphics*[width=\threequartersfigurewidth]{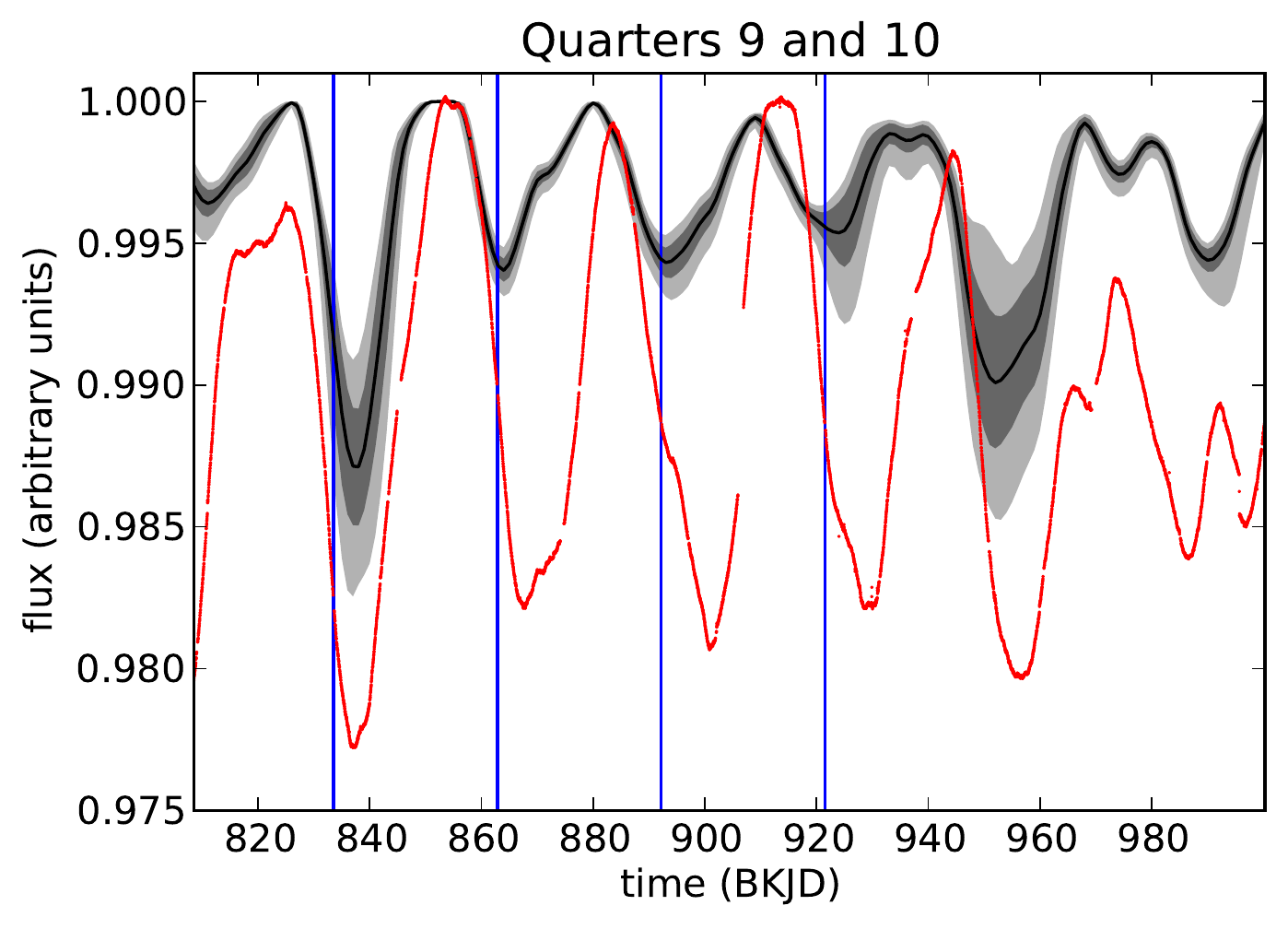}
\end {center}
\caption{Red dots: \kepler{} long cadence observations of \hatpeleven{} with hand-adjusted quarterwise scaling. Black curve and gray regions: \macula{} lightcurve model based on spots detected via transit anomalies, and its $1\sigma$ and $2\sigma$ confidence regions, accounting for the uncertainty in stellar inclination, projected obliquity, and spot parameters. This is not a fit for the out-of-transit lightcurve but rather a model generated from spot parameters based on transit anomalies. Top panel shows quarters 3 and 4, and bottom panels shows quarters 9 and 10. On the bottom panel, times of transits 217, 223, 229, and 235 are indicated by solid blue vertical lines.}
\label{fig:oot}
\end{figure}

Figure \ref{fig:oot} shows the long cadence observations in red, along with the \macula{} model lightcurve in black, for quarters 3, 4, 9, and 10. We also calculate the $1\sigma$ and $2\sigma$ confidence regions for the model lightcurve, accounting for the uncertainties of the inclination and projected obliquity as reported by \citet{2011ApJ...743...61S} and the uncertainties of the spot parameters calculated from the transit anomalies.  For the latter, we resample from the MCMC chains of \citet{2014MNRAS.442.3686B}.  The resulting confidence regions are highlighted in gray on the figure.

It is important to remember that this is not a fit for the out-of-transit lightcurve but rather a model lightcurve using spot parameters inferred from a different phenomenon as input. We see that the model is a fair match to the observations in terms of qualitative features. In particular, the deepest lightcurve minima are corretly predicted to occur after the transits drawn in blue on the bottom panel of Figure \ref{fig:oot}.  Projected obliquities of 90$^\circ$ and 270$^\circ$, both corresponding to a polar orbit, can be distinguised by the time of lightcurve minima, which would occur after or before the transit with the spot anomaly, respectively.  Our lightcurve analysis thus confirms the projected obliquity measurements based on the \rmeffect{}.

However, the observed flux variations have an amplitude approximately two to six times larger than that of our model. It is likely that there are spots on the stellar surface that are never transited by the planet; therefore, this model does not account for them. Such spots could contribute to the deeper minima in the observations, explaining the amplitude discrepancy.

For reference, the times of the four transits from Figure \ref{fig:recurrence} are also indicated on the bottom panel of Figure \ref{fig:oot} with blue vertical lines.

Dark spots simultaneously increase transit depth and decrease the total brightness of the star \citep{2009A&A...505.1277C,2011ApJ...740...33D}.  Therefore we expect a negative correlation between these two quantities.  Note, however, that the variation in out-of-transit brightness and thus in transit depth is in the order of one percent, therefore the expected change of brightness during transit is a factor of few smaller than the noise of individual short cadence photometric data points.

To investigate this correlation, we calculate the out-of-transit brightness of \hatpeleven{} at the middle of each of the 204 transits by dividing the linear fit to short cadence out-of-transit data by mean intensity across the entire quarter.  We remove all data points that are affected by transit anomalies according to the best fit model, and fit a single transit depth scaling factor to the remaining data points, using a nominal Mandel--Agol lightcurve.

We find a Pearson correlation coefficient of $r=-0.20$ between out-of-transit intensity and transit depth, which does not indicate significant correlation.  In addition to the small expected variation of transit depth, we attribute this negative result to eclipsed spots that we do not identify during transits.  Note that an uneclipsed spot increases the transit depth, whereas an eclipsed one, when not accounted for, results in a shallower transit fit:  a bias in the opposite sense.  This potentially hinders the study of the correlation between out-of-transit brightness and transit depth.

\section{Spot recurrence}
\label{sec:commensurability:recurrence}

\citet{2014MNRAS.437.1045S} apply the method of hierarchical clustering to look for recurrence of spot anomalies in \keplerthirteenb{} transits, and they find a periodicity of three orbits with a high statistical significance. This implies that after three orbits, the planet rescans the same part of the stellar surface, supporting their hypothesis that an integer number of stellar rotations (five in this case) takes place during this time.

We aim to perform a statistical analysis of the same phenomenon on \hatpeleven{} in this section, using a different approach.  To analyze similarities between transits, we devise the following method: first, we calculate the deviation of the normalized transit lightcurves from the spotless model of \citet{2002ApJ...580L.171M}. Then we run a moving boxcar average of seven data points to decrease independent noise in the data. After that, we set a threshold and flag transits with data points above it as anomalous. The next step is to pick a period and count pairs of observed transits that are spaced apart by this period. Finally, we plot the ratio of the ones among these pairs where both transits are flagged. If the planet could not eclipse the same spot in different transits, then anomalies would be independent, thus this ratio would not depend on the period.  In particular, if we flag $p$ fraction of total transits, then one randomly chosen transit is flagged with probability $p$, therefore two independent transits are simultaneously flagged with probability $p^2$ (as long as the number of transits is large).  Strong deviation of the ratio of flagged pairs of transits from $p^2$ as a function of period indicates correlated transit anomalies.

Note that this method of identifying transit anomalies is different from manually picking them for fitting in Section \ref{sec:commensurability:macula}. Using a uniform threshold has the advantage that detection does not rely on human decisions. We chose a large threshold (yielding fewer anomalies than what one can see by eye in the lightcurves) to avoid spurious detections.  It is important to note that the actual occurrence rates depend strongly on the choice of the threshold, although we find that the general features are persistent across a range of thresholds.

\begin{figure}
\begin{center}
\includegraphics*[width=\threequartersfigurewidth]{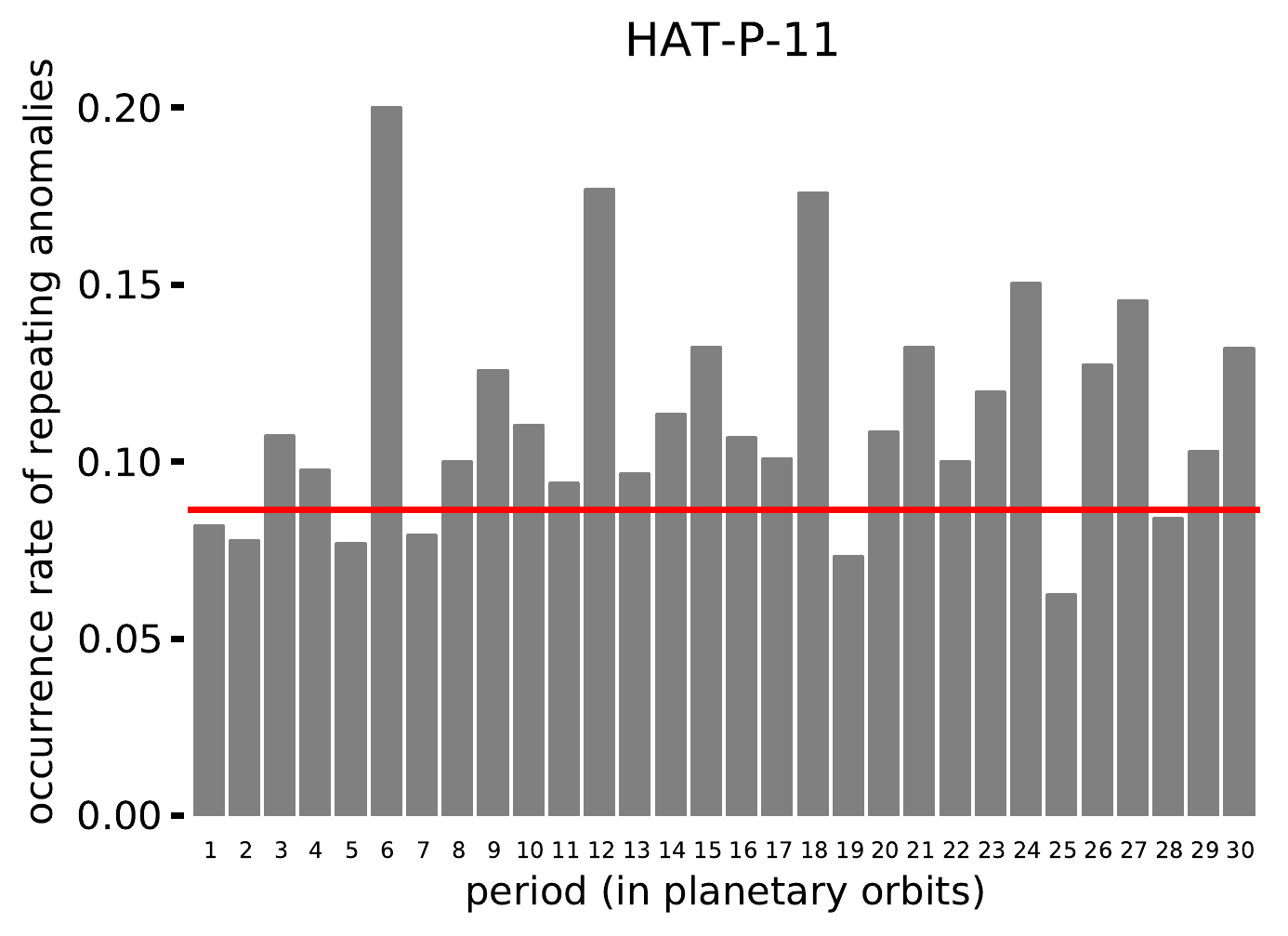}
\includegraphics*[width=\threequartersfigurewidth]{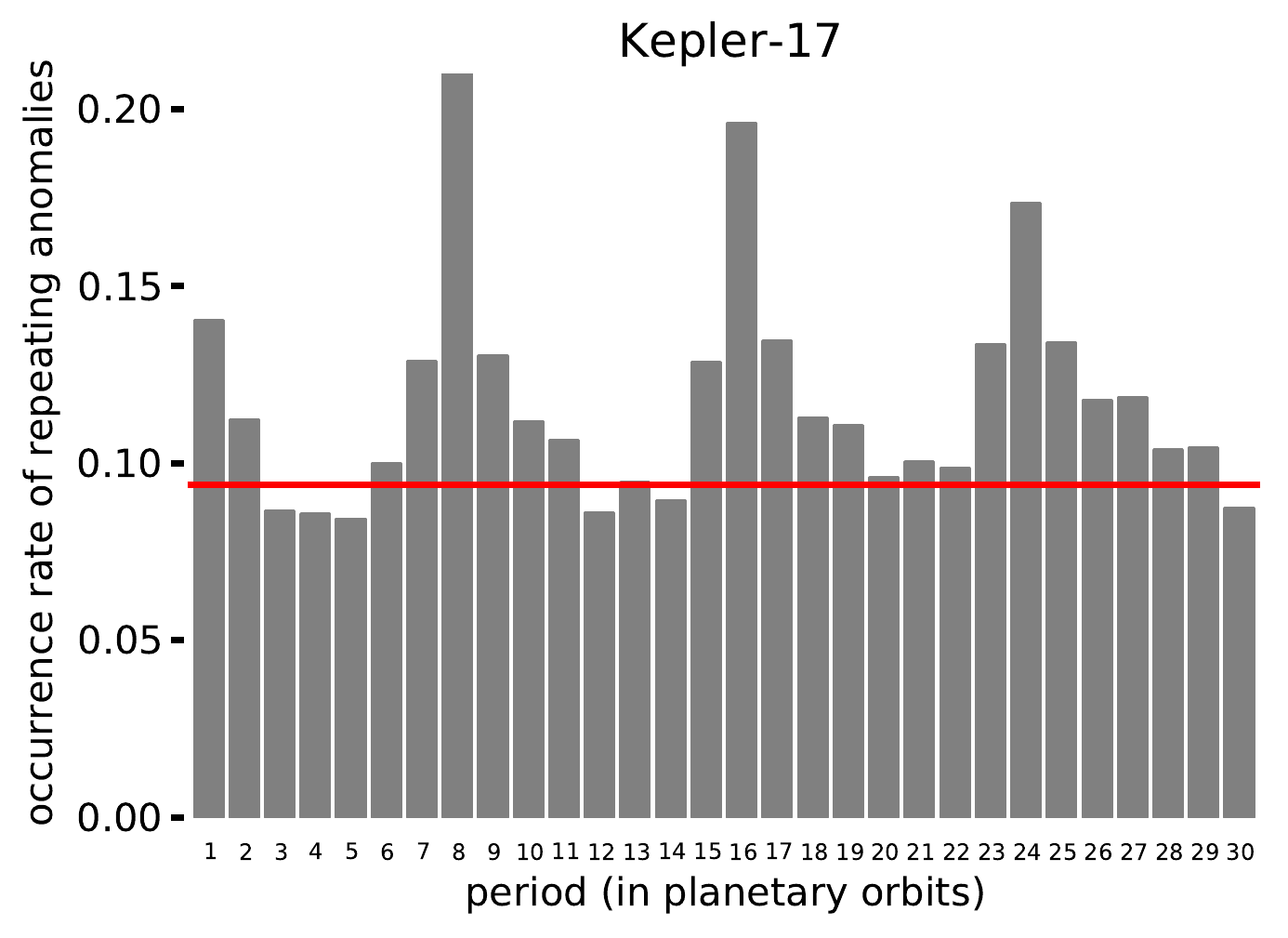}
\end{center}
\caption{Bars: occurrence fraction of transit pairs separated by a given number of orbits both exhibiting spot anomalies relative to the total number of pairs with the same spacing in the short cadence \kepler{} data, as a function of period. The ``statistical background,'' the fraction expected if such spot anomalies were independent, is represented by red horizontal lines. Top panel is for \hatpeleven{}, and bottom panel is for \keplerseventeen{}.}
\label{fig:barchart}
\end{figure}

Good quality observations exist for 204 transits of \hatpelevenb{} in the \kepler{} dataset. We pick a threshold of $1\times^{-4}$, which results in 60 flagged transits. That is, the occurrence rate of transit anomalies above this threshold is $p=0.29$.  Figure \ref{fig:barchart} presents the ratios for a number of periods on the top panel, with the statistical background of $p^2=0.09$ overplotted as a horizontal red line.  For example, there are 165 pairs of observed transits that are six orbits apart.  If transits in each pair were flagged independently, we would expect to find $165p^2=14$ pairs of transits separated by six orbits with both transits flagged.  However, there are 33 such pairs in the dataset, more than two times as many.

We perform the same analysis on 587 good quality transit lightcurves of \keplerseventeen{}. Since \keplerseventeen{} is a fainter target, we use a longer moving boxcar average, with 21 data points, to suppress photon noise. We use the same threshold as for \hatpeleven{}, resulting in 180 flagged transits. In this case, the occurrence rate is $p=0.31$, and the statistical background is $p^2=0.09$. The ratios of flagged pairs of transits as a function of period are presented on the bottom panel of Figure \ref{fig:barchart}.

We find the highest occurrence rate at periods of six and eight planetary orbits for \hatpeleven{} and \keplerseventeen{}, respectively: around 2.3 times the statistical background in both cases. We identify the next largest peaks as aliases of this frequency: at 12, 18, and 24 orbits for \hatpeleven{} and 16 and 24 for \keplerseventeen{}. These are due to long-lived spots and exhibit decreasing strength, because not all spots that live for one stellar rotation continue to live for another one.

On the basis of these observations, we can exclude period aliases: if the star rotated two, three, or four times while the planet orbits six times, we would see a strong peak at three, two, or three orbital periods, respectively, in the case of \hatpeleven{} on Figure \ref{fig:barchart}. A similar argument holds for \keplerseventeen{}. These and higher harmonics can also be readily excluded on the basis of the periodogram, the autocorrelation function, the projected rotational velocity of the star, and \apriori{} expectations of rotation rates \citep[see][for \hatpeleven{} and \keplerseventeen{}, respectively]{2010ApJ...710.1724B,2011ApJS..197...14D}.

An important difference between the interpretation of the results for the two planetary systems is due to their different geometry: \keplerseventeenb{} has an orbital axis well aligned with the projected spin axis of the star; therefore, a spot would be eclipsed by the planet again even the periods were not commensurable. On the other hand, \hatpelevenb{} is known to be on a nearly polar orbit, therefore---as \citet{2010ApJ...723L.223W} pointed out---period commensurability is required for transit anomalies to recur, otherwise the planet would scan a different part of the stellar surface, missing the spot that it had eclipsed a stellar rotation earlier.

Another consequence of the orbital alignment of \keplerseventeenb{} with its host star's rotation is the excess on the side of each peak on the bottom panel of Figure \ref{fig:barchart}: at 1, 7, 9, 15, 17, etc.~planetary orbits.  As the star rotates, each spot seems to move parallel to the transit chord, thus spots are eclipsed in multiple subsequent transits \citep[see Figure 11 of][]{2011ApJS..197...14D}.  Therefore if a spot recurs eight, sixteen, twenty-four, etc.~planetary orbits later, it is likely to also cause an anomaly in the preceding and succeeding transits.

On the other hand, transits of \hatpelevenb{} not spaced apart by an integer multiple of six orbits are expected to show spot-induced anomalies independently, because spots on the stellar surface rotate perpendicularly to the transit chord. This is indeed the case, except for secondary peaks at 3, 9, 15, 21, and 27 planetary orbits. The reason for these is the two opposite longitudes where spots seem to occur, as discussed in Section \ref{sec:commensurability:flipflop}.

Finally, we note that the orbital period of \hatpelevenb{} is 3.3 times that of \keplerseventeenb{}; therefore, periods in the upper panel represent correspondingly longer time than those in the lower panel. If we assume that spots have similar lifetime on the two stars, this explains why we see more noise for long periods for \hatpelevenb{} than for \keplerseventeenb{}.

\section{Flip-flop}
\label{sec:commensurability:flipflop}

A lightcurve rotationally modulated by a single starspot has a well-defined minimum when the spot seems to be closest to the center of the stellar disk, and a flat maximum when the spot is behind the stellar limb. For a non-evolving spot, these minima happen repeatedly with the rotational period of the star. In this section, we make use of this effect, together with the assumption that lightcurve variations are mostly due to starspots (supported by the matching order of magnitude of amplitudes shown in Section \ref{sec:commensurability:macula}), to confirm the rotational periods of \hatpeleven{} and \keplerseventeen{}.

\begin{figure}
\begin{center}
\includegraphics*[width=\threequartersfigurewidth]{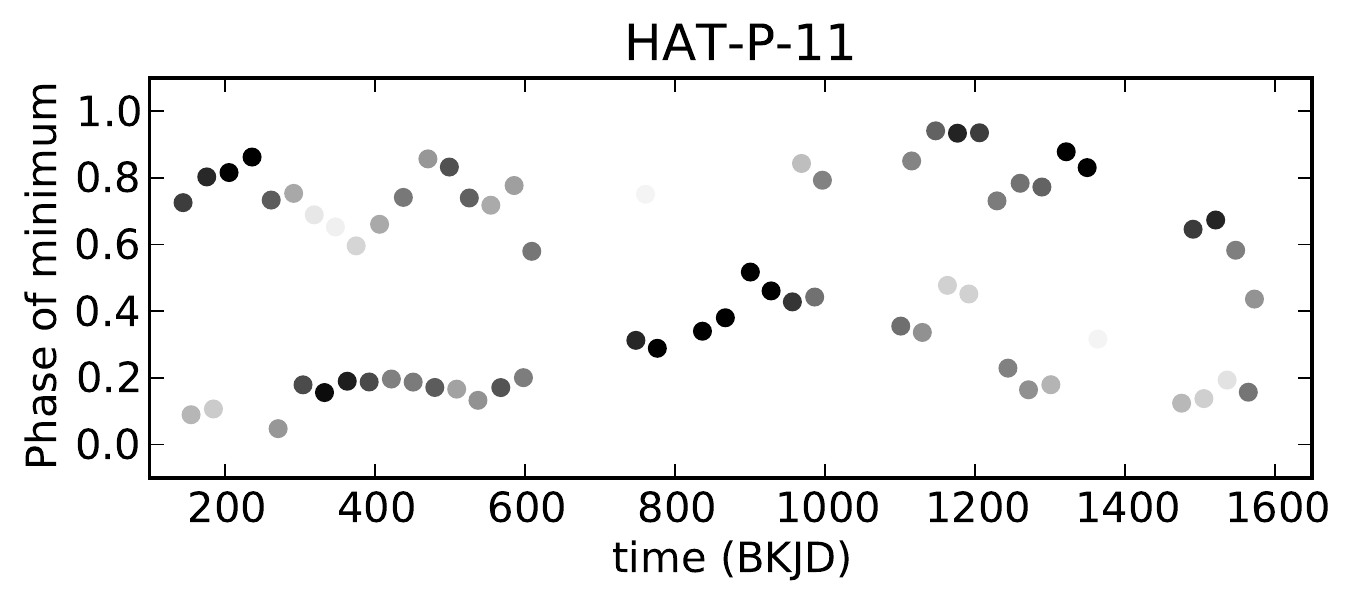}
\includegraphics*[width=\threequartersfigurewidth]{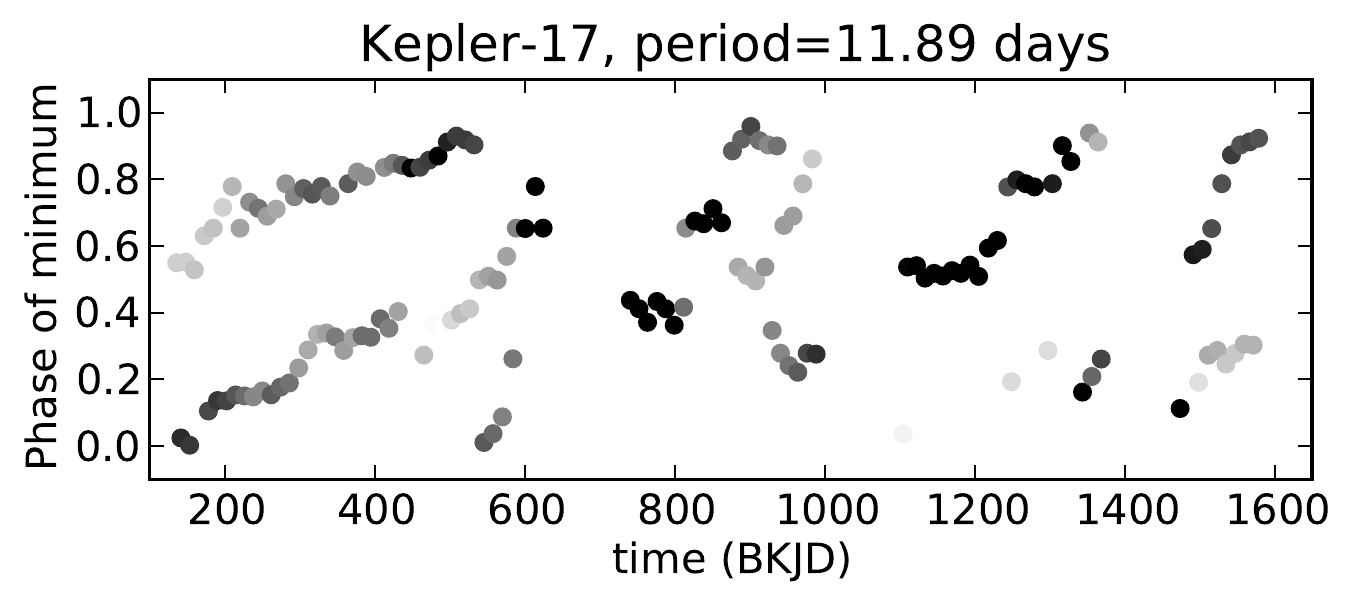}
\includegraphics*[width=\threequartersfigurewidth]{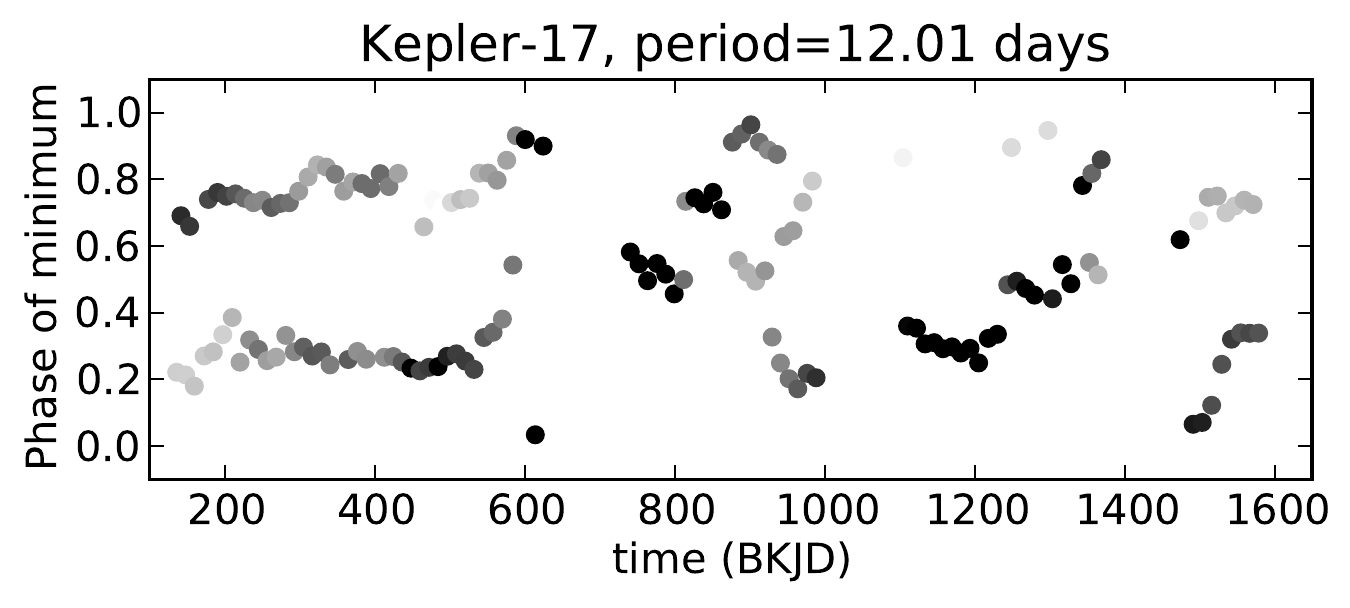}
\end{center}
\caption{Phases of lightcurve minima with respect to the proposed stellar rotation period as a function of time. The color of each filled circle corresponds to the difference between the minimum brightness and the brightness of the next smallest local minimum within 0.65 times the stellar rotation period: darker circles indicate relatively deeper minima. Top panel shows \hatpeleven{}, with the stellar rotation period being six times the planetary orbital period. The dominant phase changes from $0.7$ to $0.2$ around 300 BKJD, and it changes back around 1100 BKJD, which we interpret as flip-flop events. Middle panel shows \keplerseventeen{}, with minimum phases calculated using eight times the planetary orbital period as the stellar rotation period, as suggested by \citet{2011ApJS..197...14D}. Bottom panel shows \keplerseventeen{}, with stellar rotation period proposed by \citet{2012A&A...547A..37B}.}
\label{fig:flipflop}
\end{figure}

To this end, we identify local minima in their lightcurves. Figure \ref{fig:flipflop} shows the results for the two stars, indicating not only the time of each minimum on the horizontal axis but also their phase relative to the stellar rotation on the vertical axis with the proposed stellar rotational period. For both stars, we find two minima during most stellar rotations, indicating two large spots (or spot groups) at opposite longitudes. This structure is responsible for the spurious signal in the periodograms at half the rotation period, shown on Figures \ref{fig:acf0} and \ref{fig:acf1}. On some other stars, this phenomenon might lead to incorrect identification of rotational periods \citep[see, for example,][]{2009MNRAS.400..451C}.

We find a very stable phase in case of \hatpeleven{} with the proposed stellar rotation period of six times the planetary orbital period, further supporting the proposed 6:1 commensurability (top panel). On the other hand, in case of \keplerseventeen{}, the phases of minima exhibit a large drift if we choose to calculate them with respect to eight times the planetary orbital period as the stellar rotation period (middle panel). This phase drift indicates that the real rotational period is different from what we used to calculate the phases. Indeed, we use the period 12.01 days as suggested by \citet{2012A&A...547A..37B} to recalculate the phases, and we confirm that this yields phases of the minima without a significant drift (bottom panel). For both stars, the phase fluctuations of the minima might be due to spot migration, evolution, or new spots appearing at different longitudes.

The first discovery of phase jumps in a stellar lightcurve was reported by \citet{1991LNP...380..381J} on FK Comae Berenices and named ``flip-flop behaviour.'' \citet{2001A&A...379L..30K} attribute this phenomenon to two active regions on the star at opposite longitudes, with changing relative activity level. This results in minima in the lightcurve at two phases, with alternatingly one or the other being stronger. This phenomenon is exhibited by a large range of stars: RSCVn binaries, fast rotating G and K dwarfs, and the Sun \citep{2002AN....323..349H,2005LRSP....2....8B,2009A&ARv..17..251S}.

To determine whether a similar phenomenon takes place on the two stars we study, we quantify how much deeper each minimum is than the deepest neighboring minimum. On Figure \ref{fig:flipflop}, we represent minima that are much deeper than the ones half a stellar rotation earlier and later with black spots and ones that are not so deep with lighter gray spots. Isolated minima, that is, ones that are not preceded or succeeded with one within less than one stellar rotation, are also black.

We interpret the results for \hatpeleven{} as evidence for two flip-flop events: the dominant phase changes from $0.7$ to $0.2$ around 300 BKJD, and it changes back around 1100 BKJD. The 2 yr interval between these events is consistent with the flip-flop period on other stars \citep{2005LRSP....2....8B}. On the other hand, we are not able to interpret the results for \keplerseventeen{} as flip-flop cycles with a reasonable period.

\section{Probability of commensurability by chance}
\label{sec:commensurability:chance}

A number of systems are known to exhibit commensurability between the planetary orbit and stellar rotation, for example, \tauboo{} {\citep{2008A&A...482..691W}, CoRoT-2 {\citep{2009EM&P..105..373P,2009A&A...493..193L}, CoRoT-4 {\citep{2009A&A...506..255L}, and \keplerthirteen{} {\citep{2012MNRAS.421L.122S}.  To understand how likely this commensurability is to occur by chance, we analyze the stellar rotational period of 24\,124 active drawfs in the \kepler{} field measured by \citet{2013A&A...560A...4R, 2013yCat..35609004R}, and the orbital period of 965 confirmed \kepler{} planets\footnote{Retrieved from \url{http://archive.stsci.edu/kepler/confirmed_planets/search.php} on 2014-04-25.}.  We draw stars and planets independently from the two sets, 1\,000\,000 times, and calculate the ratio of the stellar rotational and planetary orbital periods.  The top panel of Figure \ref{fig:ratio} presents a histogram of the fractional part of such ratios (subtracting the rounded ratio from itself), 0.0 corresponding to an integer ratio.  The distribution is flat, consistent with the hypothesis that the prior rotational and orbital period distributions, when considered independently, do not inherenty carry a preference for, neither against, integer period ratios.  The underlying reason is that the period distributions are spread out, and taking the fractional part of the ratio averages out small scale variations.

\begin{figure}
\begin{center}
\includegraphics*[width=\threequartersfigurewidth]{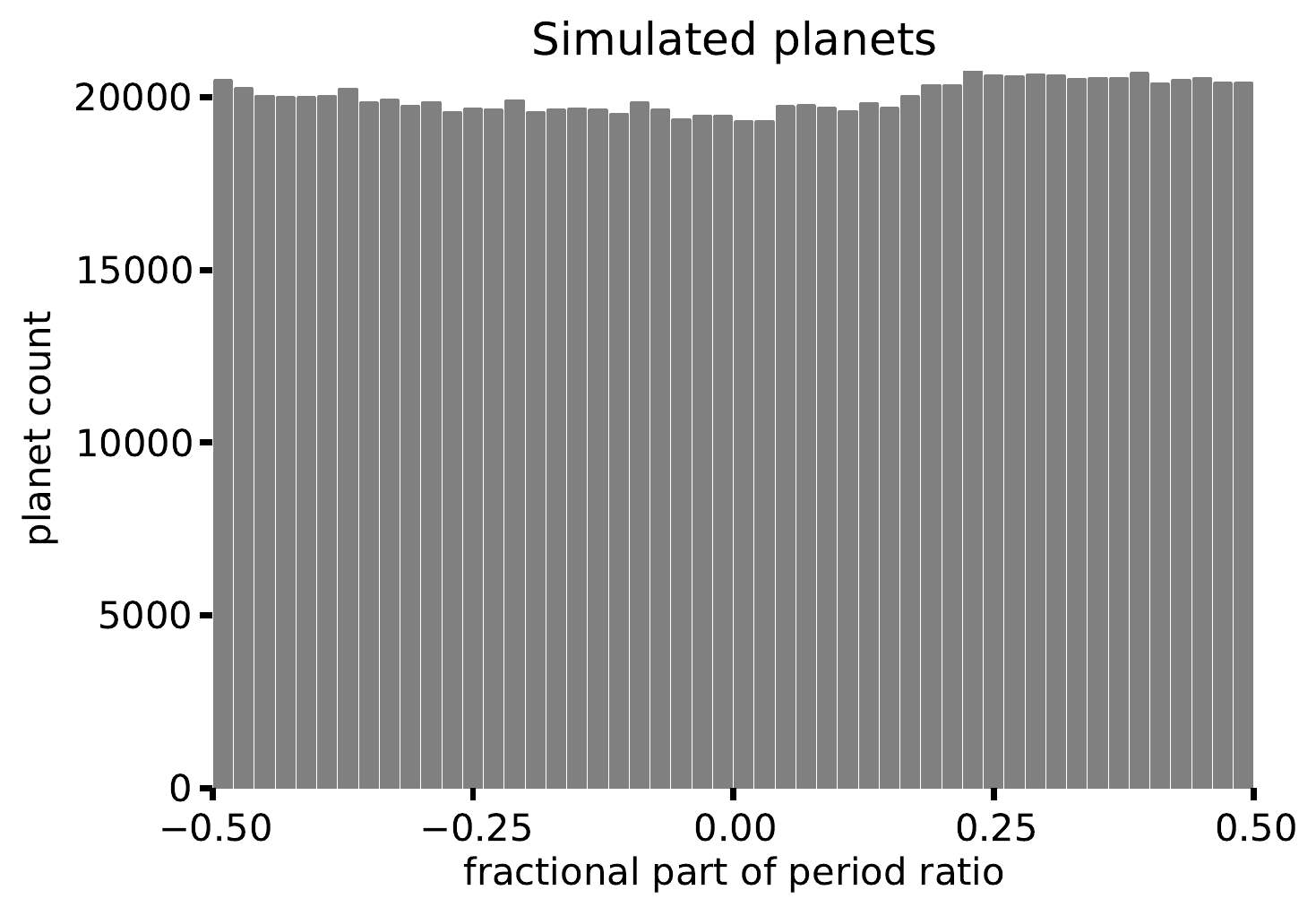}
\includegraphics*[width=\threequartersfigurewidth]{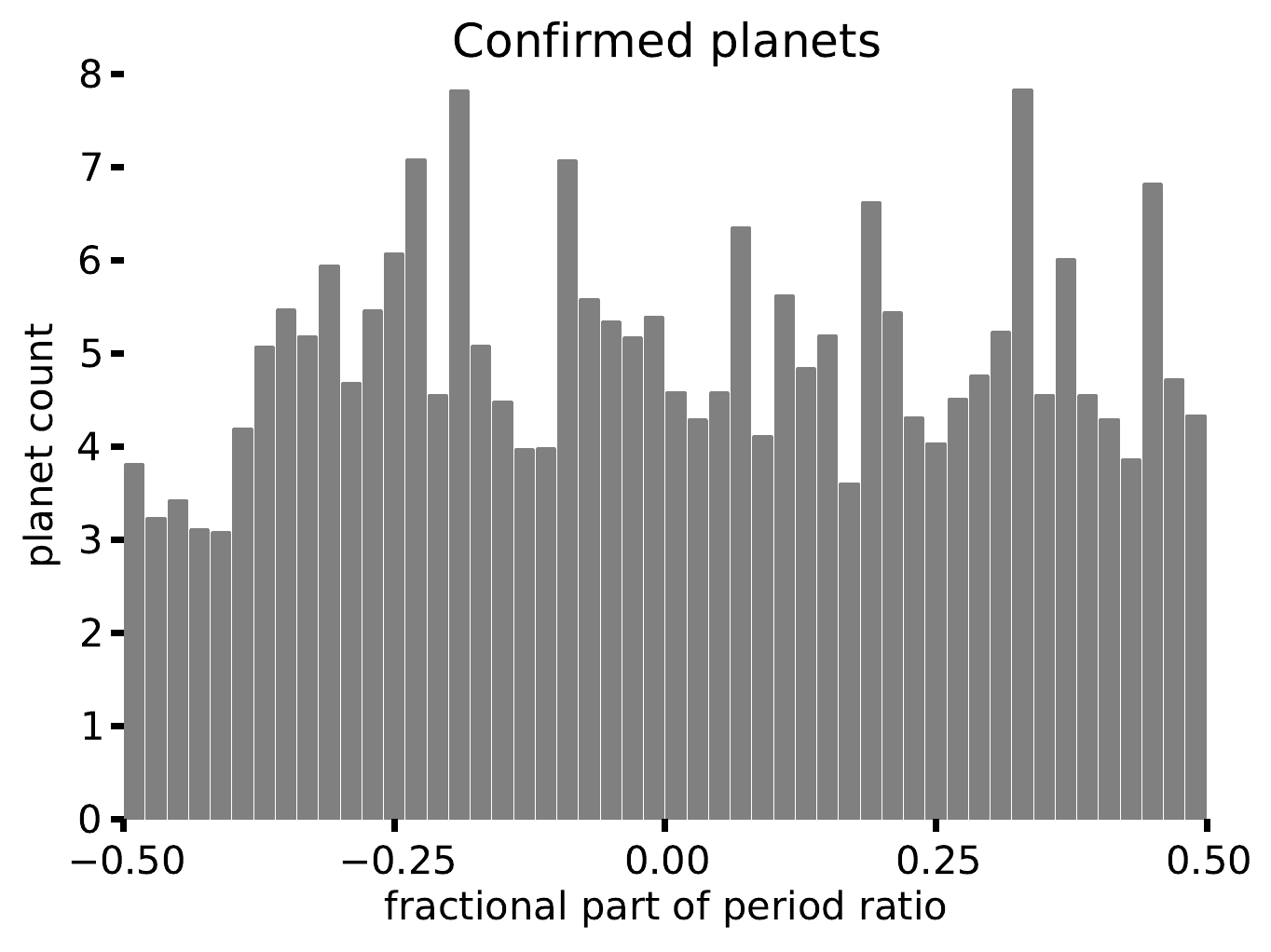}
\end{center}
\caption{Top panel: histogram of the fractional part of stellar rotation and planetary period ratios for 1\,000\,000 random and independent draws.  Bottom panel: histogram of the fractional part of stellar rotation and planetary period ratios for 251 confirmed \kepler{} planets.}
\label{fig:ratio}
\end{figure}

Among these 965 confirmed planets, there are 251 for the host star of which \citet{2013A&A...560A...4R, 2013yCat..35609004R} reports a stellar rotation period.  Their period ratio histogram is presented on the bottom panel of Figure \ref{fig:ratio}.  The main difference from the top panel is that here each stellar rotational period and planetary orbital period belong to the same physical system, instead of being drawn independently.  Note that the histogram is generated by spreading the contribution of each system across several bins according to its period ratio uncertainty, which is dominated by the stellar rotational period uncertainty reported by \citet{2013A&A...560A...4R, 2013yCat..35609004R}.

This second histogram does not exhibit a significant preference for (or against) integer period ratios either.  This would be unexpected if there was a prevalent star--planet interaction resulting in period commensurability, suggesting that the systems with such commensurability are either special or they happen by coincidence, or that the effect of such hypothetical interactions is so small that detection is not possible with this sample size.

The apparently flat period ratio fractional part prior distribution allows us to estimate the probability that commensurability happens by chance.  For example, \citet{2008A&A...482..691W} reports a period synchronicity within 0.04\% for \tauboo{}.  The probability of this happening by coincidence is 0.08\% (the uncertainty is two sided).  Similarily, the periodogram peak identified in Section \ref{sec:commensurability:acf} indicates that the period ratio for the \hatpeleven{} system is 6:1 within 0.034\%, while the autocorrelation functions tells us that the periods are commensurate within 2.4\%.  A period ratio so close to 6:1 could happen by chance with a probability of 0.068\% and 4.8\%, respectively.

Investigating the fractional part of the period ratio only reveals whether it is close to $n$:1, where $n$ is an integer.  However, the example of \keplerthirteen{} reminds us that other fractions might be of interest too.  Therefore we calculate the probability that the period ratio is within 0.034\% or within 2.4\% (corresponding to the periodogram and autocorrelation function peaks, respectively) of a rational number that can be expressed as the ratio of two integers not exceeding 6.  (These ratios are 1:6, 1:5, 1:4, 1:3, 2:5, 1:2, 3:5, 2:3, 3:4, 4:5, 5:6, 1:1, 6:5, 5:4, 4:3, 3:2, 5:3, 2:1, 5:2, 3:1, 4:1, 5:1, 6:1, in increasing order.)  We find that among 1\,000\,000 draws of independent rotational and orbital periods, 0.13\% and 8.1\% of them have a ratio that falls within the given tolerances, respectively.  For comparison, among the 251 planets in our sample, 0.0\% and 4.4\% of them are so close to any of these ratios, respectively.

\section{Star--planet interaction}
\label{sec:commensurability:interaction}

To explore the effect of a hypothetical star--planet interaction, first consider two systems without period commensurability. One is CoRoT-6, for which \citet{2011A&A...525A..14L} report that the spot covering factor gets enhanced when spots on the stellar surface cross a particular longitude with respect to the planet CoRoT-6b. \citet{2013A&A...553A..66H} find a similar behavior on the surface of LHS 6343 A, with a photospheric activity enhancement of existing spots, again at a particular position relative to its brown dwarf companion. 
Both research groups suggest that it is magnetic interactions that cause the enhancements; see, for example, the models of \citet{2006MNRAS.367L...1M}, \citet{2006A&A...460..317P}, and \citet{2008A&A...487.1163L}.

A resonance effect might exist even if magnetic (or other) interactions between a planet and its host star were too weak to transfer enough angular momentum to make the planet migrate or to change the spin of the star. The same way the companions of CoRoT-6 and LHS 6343 A might cause an enhancement synchronous to their orbit, it is conceivable that if there was a latitude on the surface of a star with a period matching that of its companion, this effect, continuously acting on the same part of the stellar surface, would result in preferential spot formation at that latitude. For example, after measuring the differential rotation of \tauboo{}, \citet{2007MNRAS.374L..42C} note that this is such a system: the planetary orbital period falls between the stellar rotation periods at the equator and the pole and therefore there is an intermediate latitude with a period matching that of the planet. By this hypothesis, a relatively weak interaction might result in photospheric activity preferentially at this latitude, which would then cause photometric variations synchronous to the planetary orbit. Such variations were later detected by \citet{2008A&A...482..691W}, who indeed suggest magnetic interactions between the star and the planet as the cause of this phenomenon.

We extend this hypothesis from matching periods to general commensurability. For example, if the differential rotation profile of \hatpeleven{} happens to be such that at some intermediate latitude, the rotational period is exactly six times the orbital period of \hatpelevenb{}, then we propose that spot formation might be enhanced at this latitude by resonance with the planet, resulting in a lightcurve reflecting this commensurability, as we have shown in this paper.

It is possible that a number of small spots might form randomly at different latitudes on the surface of \hatpeleven{}. These spots might be too small to be detected through their contribution to rotational modulation or transit anomalies. We speculate that interactions with the planet might influence the growth or merger of such small spots, preferentially creating larger ones that we detect at resonant latitudes. Even though many of these large spots might only live for relatively few stellar rotations, the buildup phase during which large spots form from smaller ones might take much longer, possibly long enough for the hypothetical resonance to have a noticeable effect in the resulting large spot distribution as a function of latitude.

That we have presented another planetary system with a tight commensurability is not enough by itself to prove that there is an interaction at force between certain stars and their close-in planets. While \hatpeleven{} is a system with some very unique features, the tight 6:1 commensurability can still be purely by coincidence. One way to confirm our hypothesis is by a statistical analysis of a large number of planet hosts. We compared the occurrence rate of detected commensurability with the rate predicted by our hypothesis using reasonable prior distributions for planetary orbital periods and stellar rotational periods.  This statistical method could be extended by accounting for differential rotation profiles, possibly by using the sample of 18\,616 stars with differential rotation parameters given by \citet{2013A&A...560A...4R, 2013yCat..35609004R}. Note, however, that this prediction is very sensitive to the assumed differential rotation profiles.  Another difficulty lies in the method of determining tight commensurability: as we have shown, neither a periodogram nor an autocorrelation function by itself is suitable for this. Part of the problem is the evolution of spots, as well as starspots occurring---although possibly in a smaller number---at other latitudes that do not have resonant rotational periods, thereby broadening the period peaks. The third method, using repeated transit anomalies like in this work or for \keplerseventeen{} by \citet{2011ApJS..197...14D}, is limited to bright host stars. Finally, identifying lightcurve minima and looking for their periodicity by itself might not be sufficient, since it is not clear \apriori{} if a star has active regions stationary on its surface or if the minima are due to spots appearing independently at random longitudes. One virtue of \hatpeleven{} is that we could apply and compare all these methods, and we are able to conclude that there are two dominant active regions, which seem to be stationary on the surface of the star for a long time.

We note that theoretically there are other ways to prove the hypothetical interaction between \hatpelevenb{} and its host star. \citet{2011ApJ...740...33D} point out that since the planet scans different latitudes of the stellar surface, it allows us to track the evolution of active latitudes with time. These observations would lead to a butterfly diagram, named after the characteristic migration pattern of active latitudes first observed on the Sun. \citet{2007ARep...51..675K} find similar behavior on some G and K dwarfs.

However, if interactions with \hatpelevenb{} induce preferential spot formation on \hatpeleven{} at fixed latitudes, this migration pattern might be suppressed. Therefore observing constant active latitudes instead of a butterfly-shaped migration pattern would be a strong indication of interactions between the planet and stellar surface activity. Unfortunately, the activity cycles for stars most similar to \hatpeleven{} in color and activity level as reported by \citet{1995ApJ...438..269B} span from 7 to 21 yr (HD 201091, 190007, and 156026), which is much longer than the timespan of \kepler{} observations. Therefore, even though we do not see strong evidence of spot migration in the \kepler{} data, we cannot yet determine whether the star exhibits Sun-like spot migration patterns on longer timescales.

It is also possible that active latitude migration is not suppressed on any star, no matter how strong the interaction with the planet is. For example, \citet{2007astro.ph..2530C} theoretically describe a mechanism that can cause the planetary interaction with the stellar magnetic fields to disappear at times (albeit for interactions with the chromosphere, not the photosphere). \citet{2008ApJ...676..628S} observationally confirm this phenomenon on both on HD 179949 and $\upsilon$ And, and dub it the \textit{on-off mechanism}. Even though there is no indication of such an event in the \kepler{} data for \hatpeleven{}, this possibility might make it impossible to confirm the effect of the planet on the bases of spot migration patterns only.

\section{Conclusion}
\label{sec:commensurability:conclusion}

The main focus of this paper is to present evidence for the 6:1 commensurability between the planetary orbit and the stellar rotation in the \hatpeleven{} system. For reference, we perform the same analysis for \keplerseventeenb{}, for which \citet{2011ApJS..197...14D} observe an 8:1 commensurability on the basis of transit anomalies. However, \citet{2012A&A...547A..37B} show that in fact, spots with a different rotational rate dominate the out-of-transit lightcurve. These results are not necessarily contradictory because of possible differential rotation: in the case of \keplerseventeen{}, the spots dominating the lightcurve might lie at a different latitude that the ones observed via anomalies in the transits of the planet with a low projected obliquity.

We calculate the autocorrelation function for the lightcurve of these two stars and present a statistical analysis of possible spot-induced transit anomaly recurrence periods, which independently exclude frequency aliases of the proposed 6:1 and 8:1 commensurabilities. In the case of \hatpeleven, the recurring transit anomalies imply a tight commensurability because of the polar orbit. We also present periodograms, and propose that the period discrepancy when looking at the FWHM of frequency peaks might be due to spot evolution causing the peaks to split.

We also present evidence for a tight 6:1 commensurability for \hatpeleven{} in the form of four observed transit anomalies presumably due to the same spot. We fit for all observed transit anomalies of \hatpeleven{} and feed the resulting spot parameters into \macula{} to show that it is plausible that rotational modulation accounts for most of the out-of-transit lightcurve variation. Furthermore, we identify minima in the lightcurve of both stars and conclude that in the case of \hatpeleven{}, there is a tight 6:1 period commensurability, whereas for \keplerseventeen{}, we confirm the period of 12.01 found by the much more sophisticated analysis of \citet{2012A&A...547A..37B}, distinct from the 8:1 commensurability. We identify two active longitudes for both stars and see an indication of two flip-flop events between these active longitudes on \hatpeleven{}.

Finally, we hypothesize that for stars with an intermediate latitude with a rotational period commensurable to the orbit of a close planet, star--planet interactions might induce spot formation preferentially at this latitude, which would show up as a resonance between the dominant period in the out-of-transit lightcurve and the planetary orbit, and also as the \strobo{} if the planet is transiting and the transit chord intersects this active latitude. However, proving this hypothesis might be difficult mostly because of the small number of bright targets and the uncertainties in differential rotation parameters.

\acknowledgements

Work by B.B.~and M.J.H.~was supported by NASA under grant NNX09AB28G from the Kepler Participating Scientist Program and grants NNX09AB33G and NNX13A124G under the Origins program. D.M.K.~is funded by the NASA Carl Sagan Fellowships. This paper includes data collected by the \kepler{} mission. Funding for the \kepler{} mission is provided by the NASA Science Mission directorate. The MCMC computations in this paper were run on the Odyssey 2.0 cluster supported by the FAS Science Division Research Computing Group at Harvard University. B.B.~is grateful for discussions with John A.~Johnson, Ruth Murray-Clay, Claire Moutou, and Joshua N.~Winn.

\chapter{A semi-analytic model for transits of spotted stars}
\label{ch:spotrod}

\publishedafter{2014MNRAS.442.3686B}

\chapterabstract

The Hubble Space Telescope (HST) and the \kepler{} space mission observed a large number of planetary transits showing anomalies due to starspot eclipses, with more such observations expected in the near future by the K2 mission and the Transiting Exoplanet Survey Satellite (TESS).  To facilitate analysis of this phenomenon, we present \texttt{spotrod}, a model for planetary transits of stars with an arbitrary limb darkening law and a number of homogeneous, circular spots on their surface.  A free, open source implementation written in \texttt{C}, ready to use in \python{}, is available for download.

We analyze \kepler{} observations of the planetary host star \hatpeleven{}, and study the size and contrast of more than two hundred starspots.  We find that the flux ratio of spots ranges at least from 0.6 to 0.9, corresponding to an effective temperature approximately 100 to 450 K lower than the stellar surface, although it is possible that some spots are darker than 0.5.  The largest detected spots have a radius less than approximately 0.2 stellar radii.


\section{Introduction}
\label{sec:spotrod:introduction}

In a transiting planetary system, spots on the face of the host star can result in deviations in the transit lightcurve from the well-known model described by \citet{2002ApJ...580L.171M}. An unocculted spot --- since it is darker than the stellar surface --- causes a blend in the opposite sense as a background star, leading to a deeper transit \citep[see, for example,][]{2009A&A...505.1277C}. On the other hand, a spot causes an anomalous rebrightening when it is eclipsed by the planet, because the planet blocks less flux than it would if the spot was not behind it.

Such spot-induced transit lightcurve anomalies were first observed by \citet{2003ApJ...585L.147S} in HST observations of HD 209458. Other systems exhibiting similar features include HD 189733 \citep{2007A&A...476.1347P} and TrES-1 \citep{2009A&A...494..391R}. \citet{2009A&A...505.1277C} found a correlation between stellar brightness and transit depth in the system CoRoT-2, which they attribute to varying levels of stellar activity, and show how this effect, when unaccounted for, causes a bias in the planet size estimate.

In the era of the \kepler{} satellite, a large number of planets transiting active stars have been discovered and observed with high temporal and photometric resolution, providing further examples of transit anomalies. Two such systems are Kepler-17 \citep{2011ApJS..197...14D} and \hatpeleven{} \citep{2010ApJ...710.1724B}. Spots revealed by transit anomalies can be used, for example, to constrain the projected obliquity and the stellar inclination \citep{2011ApJ...740...33D,2011ApJ...743...61S}. Measuring the spot contrast allows one to constrain the temperature of the spots \citep{2003ApJ...585L.147S,2009A&A...494..391R}. However, there is a degeneracy between the spot size and contrast \citep{2007A&A...476.1347P,2013MNRAS.428.3671T}, which makes high quality data necessary to infer temperatures. 

The large number of photometric observations of transit anomalies motivates the development of astrophysical models. Examples include the model by \citet{2003ApJ...585L.147S}, the one by \citet{2009A&A...504..561W}, SOAP-T by \citet{2013A&A...549A..35O}, and \prism{} by \citet{2013MNRAS.428.3671T}. These models all assume homogeneous, circular spots, with four input parameters for each spot (two for position, one for size, one for darkness). The first two models simplify the geometry by assuming that the spots are circular in projection, while the other two properly account for the elliptical projected shape given spots that are circular on the stellar surface. These models all define a large resolution two dimensional grid either on the stellar surface or in the projection plane, and numerically integrate over two coordinates to calculate the transit lightcurve.

Integration in two dimensions can be computationally expensive. \citet{2012MNRAS.427.2487K} introduced \macula{}, an analytic model for a related but different phenomenon: spots on the rotating stellar surface modulating out-of-transit lightcurves of spotted stars. Its analytic nature makes \macula{} faster than numerical models for the same phenomenon, like SOAP \citep{2012A&A...545A.109B}.

In this paper, we present \spotrod{}, a counterpart of \macula{} for transit lightcurves of spotted stars. We describe the problem as a two dimensional integral in polar coordinates in the projection plane. Using assumptions similar to those of previous models, we derive an analytic formulation for the integral with respect to the polar angle, so that numerical integration needs to be performed only with respect to the radial coordinate. This semi-analytic nature provides improved speed over previous models requiring two dimensional numerical integration.  In particular, if the resolution of the integration grid is $n$ in each dimension, then a double numerical integral takes $\mathcal O\left(n^2\right)$ time to evaluate, whereas \spotrod{} runs in $\mathcal O(n)$ time.  Typical values are 
$n\approx300$ for SOAP,
$n\approx750$ for the model of \citet{2003ApJ...585L.147S},
and the grid spacing being one hundredth of the planet diameter for \prism{}, resulting in $n\approx1000$ for a typical hot Jupiter or $n\approx2000$ for \hatpelevenb{}.  In this work we use $n=1000$.

We describe the semi-analytic model in Section \ref{sec:spotrod:geom}: we state the simplifying assumptions, describe the two dimensional integral in polar coordinates, and introduce the subroutines of \spotrod{}, the free and open source implementation available for the astronomical community.  In Section \ref{sec:spotrod:application}, we apply \spotrod{} to \kepler{} observations of \hatpeleven{}, investigate model artifacts like observational biases and correlations of fit parameters for individual spots, validate \spotrod{} on synthetic data generated by \prism{}, study the distribution of spot size and contrast on \hatpeleven{} that we believe to be physical, and look at the model residuals for validation.  Section \ref{sec:spotrod:conclusion} concludes our findings.  Technical details of the model pertaining to calculating angles and handling spots that are partially behind the limb are given in Appendix \ref{sec:spotrod:derivations}.

\section{Spot anomaly model}
\label{sec:spotrod:geom}

\subsection{Assumptions}
\label{sec:spotrod:assumptions}

Our model has two major assumptions. The first one is that the boundary of each spot is a circle on the surface of the spherical star. We define the radius $a$ of the spot to be the radius of this circle in three-dimensional space, in units of stellar radius. We assume that $0<a<1$. Note that the analytic rotational modulation model \macula{} \citep{2012MNRAS.427.2487K} takes as input parameter the half angle $\alpha$ of the cone with this circle as its directrix and the center of the star as its apex, which is related to the radius by $a = \sin\alpha$.

We define the center of the spot as the intersection point of the surface of the sphere and the axis of this cone. Note that the center of the spot does not lie in the plane of the boundary. The advantage of this definition is that the center is on the stellar surface, allowing for easier conversion between input parameters of \spotrod{} and \macula{}, and easier treatment of stellar rotation.  A further advantage of characterizing the spot location with the projection of a point on the stellar surface over the projection of the geometrical center of the spot boundary in the interior of the star is that its domain does not depend on the spot radius, which also makes it easier to define an isotropic prior for the location of the spot.

The second assumption is that each spot is homogeneous and observes the same limb darkening law as the star. This means that as viewed by the observer, the ratio of the flux from a spot and the flux from the unspotted stellar surface at the same projected distance $r$ from the center of the star does not depend on the distance $r$. We denote this dimensionless flux ratio by $f$ in accordance with \citet{2012MNRAS.427.2487K}.  Note that flux ratio is sometimes called contrast, for example, by \citet{2013MNRAS.428.3671T}.

Assuming a constant flux ratio across the stellar disk is consistent with the findings of \citet{2003SoPh..213..301W}: they study a sample of 18\,000 spots on the Sun, and observe no dependence of spot contrast on where the spot is seen.  Note that what they call contrast can be expressed as $f-1$ using our notation.  As for the homogeneity of spots, we shall see in Section \ref{sec:spotrod:penumbra} how to compose more complicated structures, like a spot with umbra and penumbra, using two homogeneous spots.


\subsection{Integration}

Let $I(r)$ denote the stellar intensity according to the limb darkening law up to an arbitrary scaling factor, where $0\leqslant r \leqslant 1$ is the projected distance from the center of the star in units of stellar radius. Let $C_r$ denote the circle of radius $r$ in the projection plane concentric with the stellar disk. Then the total out-of-transit flux of the unspotted stellar surface can be calculated as a two dimensional integral over the polar coordinates $(\vartheta, r)$, with the inner integral along $C_r$, and the outer integral with respect to the radial coordinate:
\begin{align}
\label{eq:nospot}
F_0 &= \int_0^1\int_0^{2\pi} I(r) \mathrm d\vartheta r \mathrm dr = \int_0^1 2\pi I(r) r\mathrm dr.
\end{align}
Here $\mathrm d\vartheta r \mathrm dr$ is the area of an infinitesimal element in the projection plane. The integrand does not depend on $\vartheta$, therefore the inner integral can be evaluated as the product of the integrand $I(r)$ and the length $2\pi$ of the integration interval.

\begin{figure}
\begin{center}
\includegraphics*[width=\figurewidth]{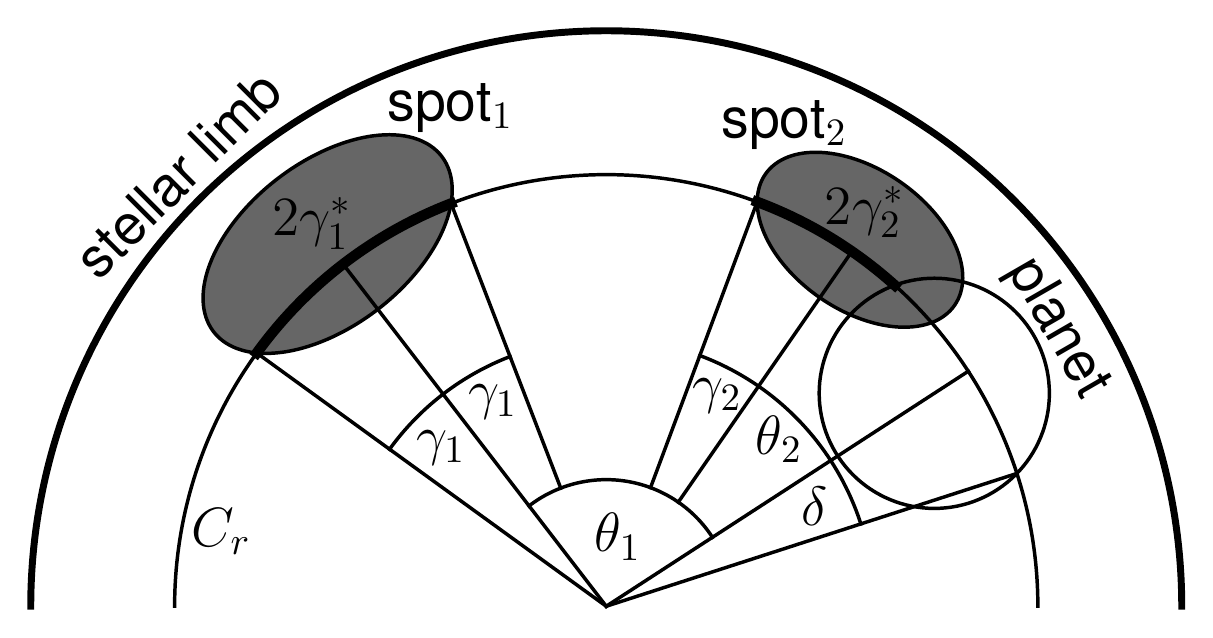}
\end{center}
\caption{Example configuration with transiting planet and two spots, showing $\gamma_i$, $\gamma^*_i$, and $\delta$ for a given value of $r$. $\theta_i$ is used to calculate $\gamma^*_i$, see Appendix \ref{sec:spotrod:calc}.}
\label{fig:observer}
\end{figure}

Now consider the case of a single spot visible on the star. Let $\gamma(r)$ be half the central angle of the arc of $C_r$ that overlaps with a spot, as shown on Figure \ref{fig:observer}. Then the total flux is
\begin{align}
\nonumber
F_1 &= \int_0^1\left(\int_0^{2\gamma(r)} fI(r)\mathrm d\vartheta + \int_{2\gamma(r)}^{2\pi}I(r)\mathrm d\vartheta\right) r\mathrm dr \\
\nonumber
&= \int_0^1\left(2\gamma(r)fI(r) + (2\pi - 2\gamma(r))I(r)\right)r\mathrm dr \\
\label{eq:factoredout}
&= \int_0^12(\pi + (f-1)\gamma(r))I(r)r\mathrm dr.
\end{align}
Here the inner integral is composed of two parts: inside the spot, on an arc of total length $2\gamma(r)$, the intensity is $fI(r)$, whereas outside the spot, on the remaining arc of length $2\pi-2\gamma(r)$, the intensity is $I(r)$. The integrands do not depend on $\vartheta$, therefore each integral reduces again to the product of the integrand and the length of the corresponding interval. In the final step, we factor out $2I(r)$, and collect the terms with $\gamma(r)$. 

If there are $s$ non-overlapping spots, with corresponding flux ratios $f_i$ and half central angle functions $\gamma_i(r)$, then each inner integral evaluates to $2\gamma_i(r)f_iI(r)$, and the unspotted stellar surface will have an arc length of $2\pi-\sum_{i=1}^s2\gamma_i(r)$, giving a total flux of
\begin{align*}
F_s &= \int_0^1\left(\sum_{i=1}^s2\gamma_i(r)f_iI(r) + \left(2\pi - \sum_{i=1}^s2\gamma_i(r)\right)I(r)\right)r\mathrm dr.
\end{align*}
Just as we factored out $2I(r)$ and collected $\gamma(r)$ in Equation (\ref{eq:factoredout}), we can do the same for each $\gamma_i(r)$ to account for the contribution of multiple spots in a single summation:
\begin{align}
\label{eq:multispot}
F_s &= \int_0^12\left(\pi + \sum_{i=1}^s(f_i-1)\gamma_i(r)\right)I(r)r\mathrm dr.
\end{align}

During transit, let $\delta(r)$ be the half central angle of the arc of $C_r$ that is obscured by the planet. Let $\gamma^*_i(r)$ be the half central angle of the arc on the same circle that overlaps with spot $i$, but is not obscured by the planet. See spot 2 on Figure \ref{fig:observer} for an example. Then the total flux can be calculated by substituting $\gamma^*_i(r)$ for $\gamma_i(r)$ in Equation (\ref{eq:multispot}), and subtracting $2\delta(r)$ from the arc length of the unspotted stellar surface:
\begin{align}
\nonumber
F_\mathrm{transit} &= \int_0^1\left(\sum_{i=1}^s2\gamma^*_i(r)f_iI(r) + \right. \\
\nonumber
&\phantom{=} \quad + \left.\left(2\pi - \sum_{i=1}^s2\gamma^*_i(r) - \delta(r)\right)I(r)\right)r\mathrm dr \\
\label{eq:transit}
&= \int_0^12\left(\pi - \delta(r) + \sum_{i=1}^s(f_i-1)\gamma^*_i(r)\right)I(r)r\mathrm dr.
\end{align}
The contribution of the planet in this formula is formally equivalent to a spot with flux ratio $f=0$. 

The final product of our proposed model is the dimensionless normalized transit lightcurve
\begin{align}
\label{eq:normalized}
F_\mathrm{normalized} &= \frac{F_\mathrm{transit}}{F_s}.
\end{align}
At this step, the arbitrary scaling factor in $I(r)$ cancels out.

Note that we define $\gamma_i(r)$ and $\gamma^*_i(r)$ to be zero in case the corresponding arcs do not exists, that is, $C_r$ does not intersect the spot in projection. Similarly, $\delta(r)$ is understood to be zero if the planet does not eclipse $C_r$. See Appendix \ref{sec:spotrod:calc} on calculating $\gamma_i$, $\gamma^*_i$, and $\delta$.

In case there are no spots on the stellar surface, even though our model still yields the correct lightcurve asympotically for large grid resolution $n$, we suggest using the fully analytic algorithm of \citet{2002ApJ...580L.171M} if speed is a consideration.

\subsection{Implementation}

The model described in this paper is implemented as a software package called \spotrod{}. It provides four functions:
\begin{itemize}
\item[] \texttt{elements} takes the planetary period, semi-major axis, $k=e\cos\varpi$, $h=e\sin\varpi$, and an array of observation times of a transiting planet with respect to the time of midtransit as input parameters, and calculates the arrays of planar orbital elements $\xi$ and $\eta$ using the formalism of \citet{2009MNRAS.396.1737P}.
\item[] \circleangle{} takes the planetary radius $R_\mathrm p$, the distance $z$ of the centers of the planet and the stellar disk in projection plane, and an array of radii $r$ as input parameters, and calculates the array $\delta(r)$.
\item[] \ellipseangle{} takes the projected spot semi-major axis $a$, the distance $z$ of the centers of the projected spot boundary and the stellar disk in projection plane, and an array of $r$ as input parameters, and calculates the array $\gamma(r)$. This function is executed internally by \integratetransit{}, so the user does not have to call it direcly.
\item[] \integratetransit{} takes $R_\mathrm p$, arrays of the projected coordinates of the planet and the spots, spot radii and flux ratios, an array of the radii $r$ and weights for numerical integration (the latter are calculated from the integration quadrature and the limb darkening law), and precalculated values of $\delta(r)$ for each value of $r$ and each observation time as input parameters. It calculates the normalized lightcurve $F_\mathrm{normalized}$, using analytic integration with respect to the polar angle, and numerical integration with respect to $r$.
\end{itemize}

The software can employ any integration quadrature, that is, numerical method that works by evaluating the integrand at given values of $r$ and summing up using given weights.  It evaluates the inner integral analytically in the form of the sum given in Equations (\ref{eq:multispot}) and (\ref{eq:transit}) on each annulus defined by the input array of $r$ values.  Then it performs numerical integration with respect to $r$ by adding up the products of these values and the corresponding weights.

Among the simplest integration quadratures are the trapezoidal rule and the midpoint rule.  We recommend an integration mesh of $n\approx1000$ values uniformly spaced between 0 and 1.  More complicated rules can also be prescribed.  For example, since the integrand of the outer integral grows rapidly with $r$ in Equations (\ref{eq:nospot}-\ref{eq:transit}), one might wish to use a nonuniform mesh that is coarser for small $r$ and finer for large $r$.

\spotrod{} can also handle arbitrary limb darkening laws, even ones without an analytic formula: it only relies on limb darkening values evaluated at the values of $r$ used in the integration rule. The limb darkening law and the integration quadrature weights are then multiplied together before they are passed to \integratetransit{}, since it is only their product that is ever used.

Repeated evaluations of \integratetransit{} at the same observation times with fixed planetary orbital parameters are required, for example, for fitting or a Monte Carlo Markov Chain (MCMC) exploring spot parameters. The code has been optimized for such use: one needs to calculate the arrays $\xi$ and $\eta$, then the projected planetary coordinates at each observation, and finally an array of $\delta$ only once at the beginning. These values do not depend on spot parameters, and recalculating the same values of $\delta$ in each iteration would be very costly, since it has to be calculated for each observation time and each $r$: hundreds of thousands of times in a typical application. Instead, we evaluate $\delta$ hundreds of thousands of times only once, before we start the fit or MCMC, and then use these precalculated values.

We can also avoid evaluating the function \ellipseangle{} hundreds of thousands of time in each iteration if we neglect the effect of stellar rotation during a single transit. In this case, for a given set of spots, $\gamma_i(r)$ does not depend on time, therefore we only need to calculate it $n$ times: once for each value of $r$.

It is, of course, also possible to model a transit in the extreme case of a star that rotates so rapidly that spots move substantially during the duration of the planetary transit. In this case, one needs to recalculate the spot positions and call the function \integratetransit{} with a time array of length one for each observation. This method, however, is slower than if we assumed that spots were stationary within the duration of a single transit.

\spotrod{} is free and open source software, released under the GNU General Public License. It is implemented in \texttt{C} in the interest of speed, and provides bindings for use in \python{}. Bindings for different programming languages should be reasonably easy to add.

\spotrod{} is publicly available for download at \url{https://github.com/bencebeky/spotrod}, including the \texttt{C} code, \python{} bindings, two example programs in \python{}, compilation instructions, and a copy of the license.

\subsection{Umbra, penumbra, and faculae}
\label{sec:spotrod:penumbra}

Note that in Equations (\ref{eq:multispot}--\ref{eq:transit}), the contribution of spots add up, regardless of whether they overlap or not. As \citet{2012MNRAS.427.2487K} points out in his Section 2.4, this feature can be used to build a composite spot with a central umbra of radius $a_\mathrm u$ and flux ratio $f_\mathrm u$ and a surrounding penumbra of radius $a_\mathrm p$ and flux ratio $f_\mathrm p$ by feeding two concentric spots with radius-flux ratio pairs $(a_\mathrm p, f_\mathrm p)$ and $(a_\mathrm u, 1 - f_\mathrm p + f_\mathrm u)$ into the model.

Spots should have flux ratio between 0 and 1, $f=0$ for a completely dark spot, and $f=1$ for one indistinguishable from the stellar surface. We note that as the model can handle any value of flux ratio $f$, faculae and plages (bright areas on the stellar photosphere and chromosphere, respectively) can also be modelled using a flux ratio value exceeding 1, as suggested, for example, by \citet{2012A&A...545A.109B, 2012MNRAS.427.2487K}.

\section{Application to \hatpeleven}
\label{sec:spotrod:application}

\hatpeleven{} is 9.6 visual magnitude K4 dwarf star in the field of the \kepler{} space telescope \citep{2010Sci...327..977B}. It has been known to host \hatpelevenb{}, a transiting hot Neptune on a 4.9 day orbit \citep{2010ApJ...710.1724B}, before the launch of the \kepler{} mission, and therefore has been observed with short (one minute) cadence from the beginning, during quarters 0--6, 9--10, 12--14, and 16--17. Missing data in quarters 7, 11, and 15 are due to the failure of a readout module in 2010 January. We perform our analysis on this dataset, using the flux values in the \texttt{SAP\_FLUX} column, and dividing each transit by a linear fit to the out-of-transit data within 0.12 days from the midtransit time to normalize the transit lightcurves.  Visual inspection shows that a linear fit is satisfactory, because the out-of-transit lightcurve modulation timescale is the rotation period of \hatpeleven, which is 29.2 days \citep{2010ApJ...710.1724B,
2014ApJ...788.....1}, much larger than the 0.24 days of the total width of our window.

For our analysis, we adopt the orbital eccentricity and argument of periastron values reported by \citet{2010ApJ...710.1724B} based on RV data and Hipparcos parallax for \hatpeleven{}. However, we use the revised transit ephemeris, planetary radius and orbital semi-major axis relative to the stellar radius, orbital inclination, and limb darkening parameters of \citet{2011ApJ...740...33D}.  Their treatment relies on the above eccentricity and argument of periastron values, but accounts for eclipsed and uneclipsed spots, that biased earlier analyses.  We number the transits according to this ephemeris, with the midtransit time of transit 0 being $T_0 = 2\,454\,605.891\,55 \pm 0.000\,13$ (barycentric dynamical time).

Rebrightening events in the transit lightcurve of \hatpeleven{} due to spots were first predicted by \citet{2010ApJ...723L.223W}, and first reported independently by \citet{2011ApJ...743...61S} and \citet{2011ApJ...740...33D}.  They are used to constrain the stellar rotational period by \citet{2014ApJ...788.....1}, who also compare a model out-of-transit lightcurve based on the MCMC chains described in this section, and find that it is consistent with the assumption that out-of-transit variation is dominated by rotation of spots.

\subsection{Analysis of individual spots}
\label{sec:spotrod:individual}

First, we present the analysis of two individual starspots in order to study correlations between spot parameters inherent to the model. Each spot is described by four parameters: $x$ and $y$ are the coordinates of the spot center in stellar radius units, as seen by the observer, in a Cartesian coordinate system whose origin is the center of the stellar disk, and where the planet is moving approximately in the positive $x$ direction during transit. (More precisely, for inclined orbits the $y$ axis is defined by projecting the line of sight on the orbital plane, then projecting that on the sky plane. For an inclined eccentric orbit, the projected velocity of the planet is not exactly parallel to the $x$ axis at mid-transit, except for special values of the argument of periastron.) The other two parameters are the spot radius $a$ and the flux ratio $f$ described in Section \ref{sec:spotrod:assumptions}.

Figure \ref{fig:transit218} presents the lightcurves of transits 74 and 218.  For both transits, we identify the deviation from the lightcurve model of \citet{2002ApJ...580L.171M} as an indication for the planet eclipsing a single spot on the surface of the star.  The lightcurve anomalies are observed about $0.4$ hours before midtransit during transit 74, and about half an hour after midtransit during transit 218.  We also plot the best fit (least sum of squared residuals) \spotrod{} lightcurves for both transits in red on this figure.

\begin{figure}
\begin{center}
\includegraphics*[width=\threequartersfigurewidth]{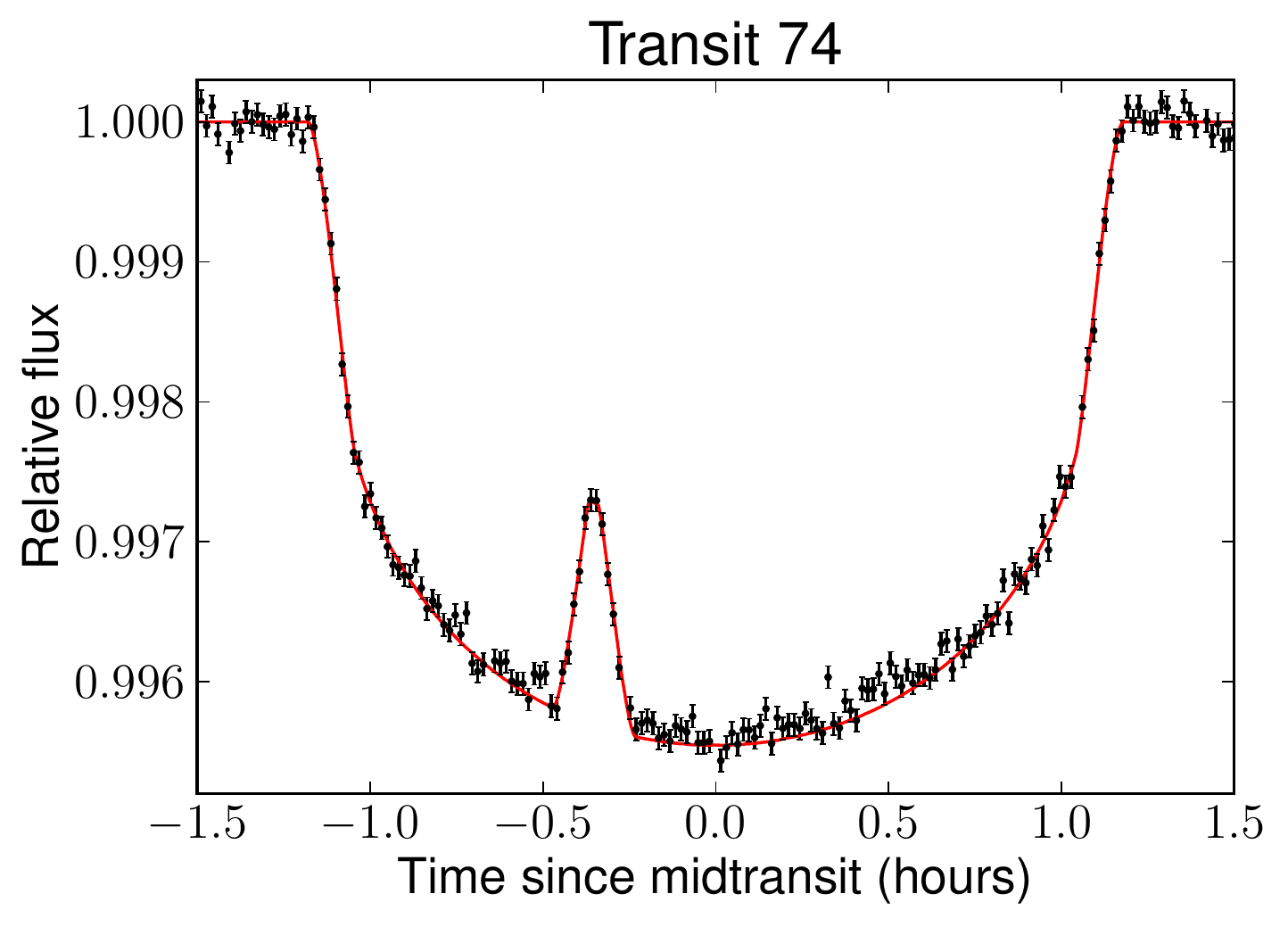}
\includegraphics*[width=\threequartersfigurewidth]{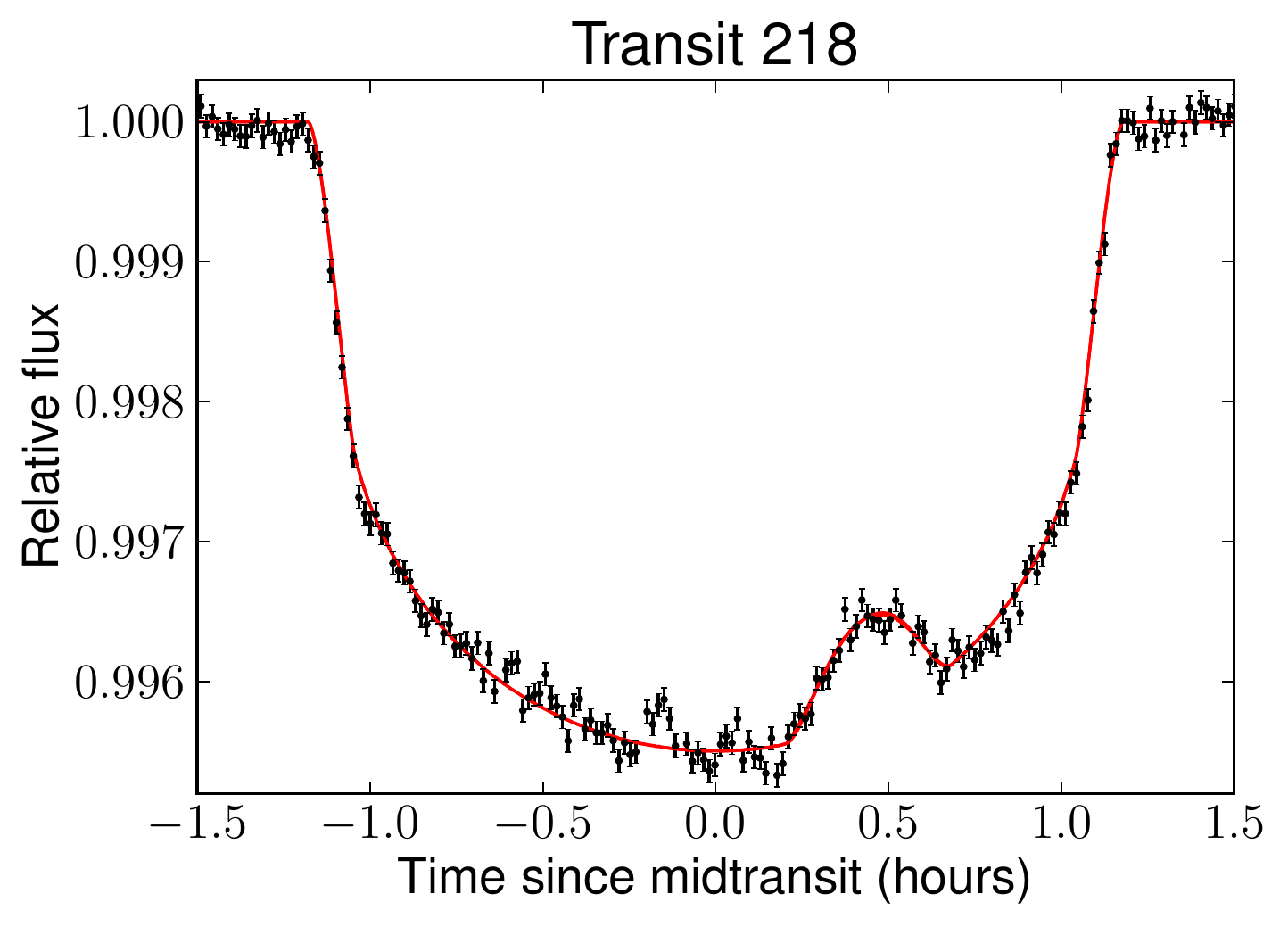}
\end{center}
\caption{\hatpeleven{} transit 74 (top panel) and transit 218 (bottom panel) lightcurves. Dots are \kepler{} short cadence observations, with errorbars given by the \texttt{SAP\_FLUX\_ERR} data column. Red curves are best fit \spotrod{} models assuming a single spot on the stellar surface for both transits.}
\label{fig:transit218}
\end{figure}

In fact, for transit 218, we find two best fit solutions: in the projection plane, the spot can either be situated above or below the transit chord.  To illustrate this bimodality, we present on Figure \ref{fig:spots} an observer's view of the star (large empty circle), the planet (black filled circle), and the best fit solutions for the single spot (gray ellipses) during transits 74 and 218.  The transit chord is also drawn, as wide as the diameter of the planet.  We do not find the same bimodality in case of transit 74, therefore only one solution is depicted.  Note that in fact we plot two model lightcurves for transit 218 on Figure \ref{fig:transit218}, corresponding to the best fits for each mode.  However, they are indistinguisable on this figure, which is closely related to our inability to infer which solution describes the spot in reality.

\begin{figure}
\begin{center}
\includegraphics*[width=\figurewidth]{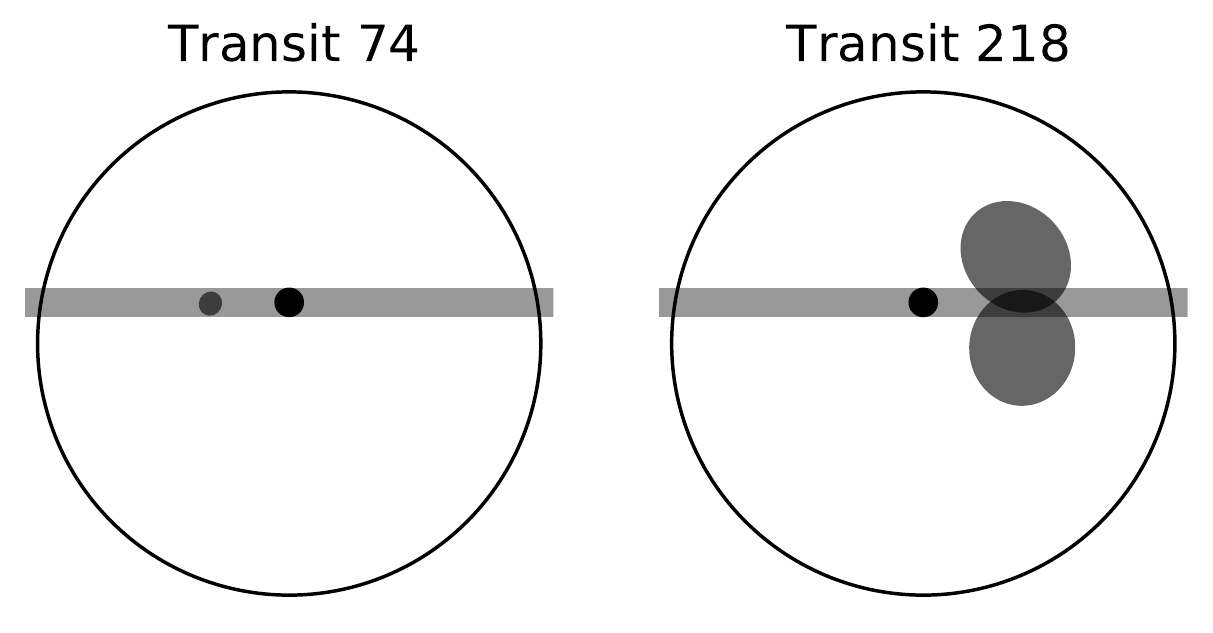}
\end{center}
\caption{Projected images of \hatpeleven{} during transits 74 (left panel) and 218 (right panel). Large circle is the star, solid black small circle is the planet \hatpelevenb{} at midtransit, gray strip is the transit chord, one or two gray ellipses are best fit solutions for the one or two modes of the spot.}
\label{fig:spots}
\end{figure}

We run parallel tempered MCMC simulations \citep{2005PCCP....7.3910E} for both transits, using the \texttt{emcee} software package \citep{2013PASP..125..306F}, at 10 different temperatures, with 100 concurrent chains at each temperature. We first run both simulations for 1000 steps for burn-in.  By inspecting the evolution of the mean and scatter of each parameter, we find that the chain already converges after about half this many steps.  Then we run the simulation for another 1000 steps to sample what we believe to be the equilibrium distribution.  The lowest, zero temperature chain provides us with the equilibrium distribution, whereas the higher temperature chains guarantee that we explore the entire parameter space, and have samples in disconnected modes in numbers proportional to the corresponding posterior probabilities.

Figure \ref{fig:mcmc} illustrates the results of the MCMC simulation for the single spot models for transits 74 and 218: joint distributions for four pairs of parameters as well as histograms for each parameter are presented.  A dashed line is drawn at the $y$ coordinate of the planet at midtransit, which is very close the the impact parameter $b$ (in fact they would be the same for a circular orbit), to help distinguish the two modes and inspect symmetry with respect to the transit chord.  We immediately confirm that there is only one mode for transit 74, and two modes for transit 218.  The reason for this is that the anomaly during transit 74 can be well described by a spot that lies under the transit chord, therefore the two modes overlap.  On the other hand, the lightcurve of transit 218 can only be well modelled if the spot is further away from the transit chord, in which case the two modes are disconnected.  We note that we experience bimodality for roughly one quarter of all spots in the entire \kepler{} dataset for \hatpeleven{}.

\begin{figure}
\begin{center}
\includegraphics*[width=\figurewidth]{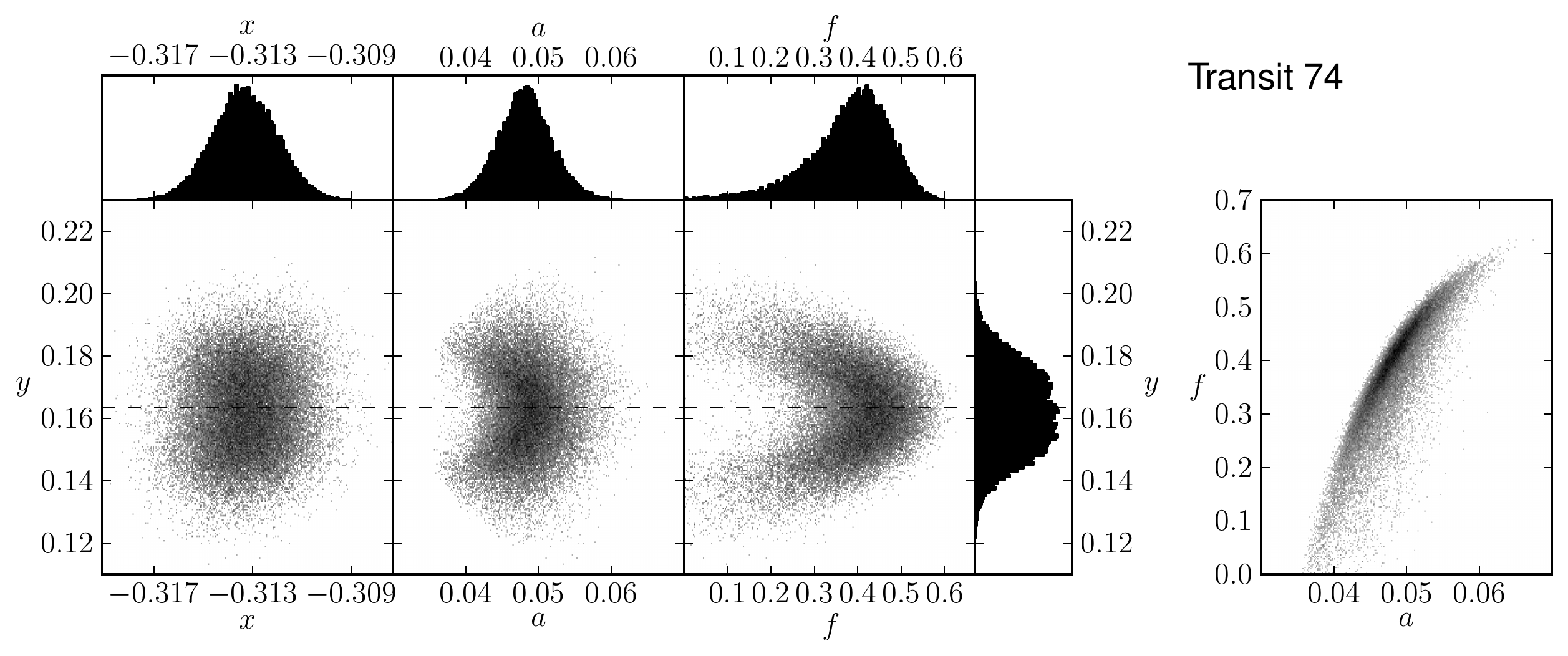}
\includegraphics*[width=\figurewidth]{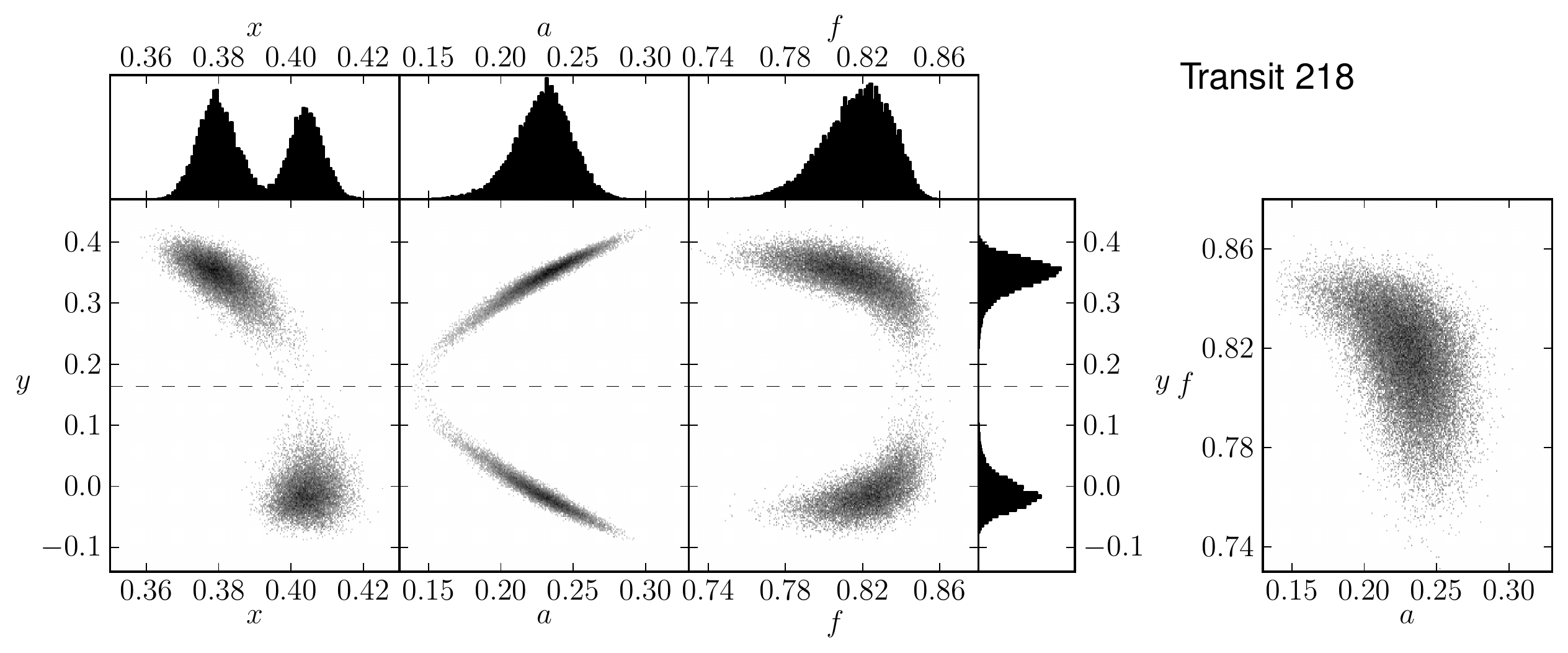}
\end{center}
\caption{Spot parameter distributions of 100\,000 MCMC samples for transits 74 (top panels) and 218 (bottom panels) of \hatpelevenb{}.  Four scatterplots show joint distributions of pairs of parameters for each transit: $x$--$y$, $a$--$y$, $f$--$y$, and $a$--$f$.  Thin dashed line indicates $y$ coordinate of planet at midtransit.  The four side panels on top and right present histograms of $x$, $a$, $f$, and $y$.}
\label{fig:mcmc}
\end{figure}

Figure \ref{fig:mcmc} also tells us about the correlations of spot parameters.  First, we note that in both cases, $x$ is better constrained than $y$, even within a single mode.  Recall that $x$ is the coordinate (almost exactly) parallel to the transit chord, therefore it directly relates to when the anomaly is observed, which is well defined by the observations.  On the other hand, we shall see that $y$ correlates with other parameters that together shape the transit anomaly, resulting in a larger uncertainty.

We also note that the two solutions of transit 218 have slightly different best fit $x$ values, even though $x$ is fairly well constrained.  The explanation for this is that since the spot is elliptical in projection, with its semi-major axis not quite parallel to the $y$ axis, therefore the two spot solutions must have different $x$ coordinates in order to intersect the transit chord at roughly the same $x$ coordinate.  This effect can also be observed on Figure \ref{fig:spots}.

The joint distribution of $a$ and $y$ for the spot in transit 218 on Figure \ref{fig:mcmc} tells us that the larger the spot is, the further it has to be from the transit chord.  This is expected from geometrical arguments, as the duration of the transit anomaly can be well constrained from the observations.  This correlation between spot parameters has been first pointed out by \citet{2003ApJ...585L.147S}.  However, this effect does not show for transit 74, probably due to the spot being close to the transit chord.

If we inspect the joint distribution of $f$ and $y$ for either transit, we notice that if the spot is brighter (has a larger flux ratio), then it is likely to be closer to the transit chord.  Such a correlation was first noted by \citet{2009A&A...504..561W}, and it remains to be explained.

Finally, we notice that there is a strong correlation between $a$ and $f$ for transit 74: the brighter the spot is, the larger it has to be.  This correlation has been reported by \citet{2007A&A...476.1347P}, \citet{2009A&A...504..561W} and \citet{2013MNRAS.428.3671T}.  This phenomenon is naïvely explained by that the rebrightening amplitude is proportional to how much the flux blocked by the planet is less than it would be for the unspotted photosphere.  If a spot is not that much darker than the typical stellar surface, a larger area is required to produce the same flux deficit.  This argument is expected to hold for spots centered on the transit chord, for which the spot radius directly determines the occulted area.  On the other hand, we do not observe the same $a$--$f$ correlation for transit 218, becase $a$ correlates strongly with $y$, resulting in a more complex effect on the occulted spot area.

\subsection{Test on synthetic lightcurves}
\label{sec:spotrod:synthetic}

In this section, we generate synthetic transit lightcurves, and apply to them the same analysis as in the last section.  We use \prism{} \citep{2013MNRAS.428.3671T} for generating lightcurves, and analyze them using \spotrod{}.  Using different models for data generation and analysis allows us to validate them agains each other, that is, make sure that parameters like spot radius and flux ratio are interpreted identically, and they produce the same result.

We take the best fit parameters of the spots in transits 74 and 218, convert the projected coordinates to equatorial coordinates as expected by \prism{}, and generate two model transits.  The difference from the \spotrod{} model with the same input parameters has mean $10^{-8}$ and standard deviation $2\cdot10^{-6}$ across all observation times, which indicates a good agreement between the two models.  For comparison, the mean photon noise is $8\cdot10^{-5}$ for the same data points.

We find that the \spotrod{} best fit has residuals with a standard deviation approximately 1.3 times that predicted by the \texttt{SAP\_FLUX\_ERR} data column of the \kepler{} dataset, therefore we add independent, normally distributed noise scaled to 1.3 times the corresponding \texttt{SAP\_FLUX\_ERR} value to each data point calculated by \prism{}.  We run an MCMC simulation on the resulting synthetic lightcurves for both transits, in a fashion identical to that explained in the previous section.  The resulting chain distribution is presented on Figure \ref{fig:synthetic}.

\begin{figure}
\begin{center}
\includegraphics*[width=\figurewidth]{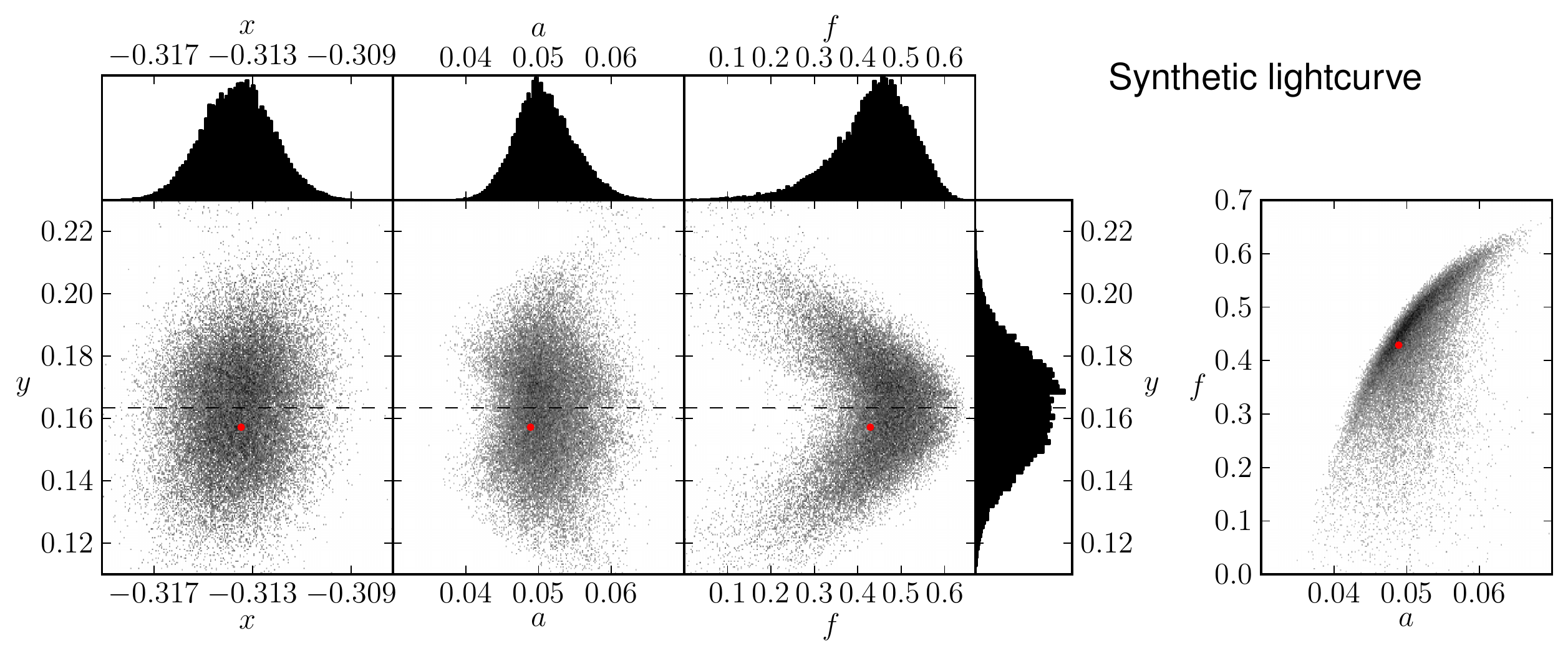}
\includegraphics*[width=\figurewidth]{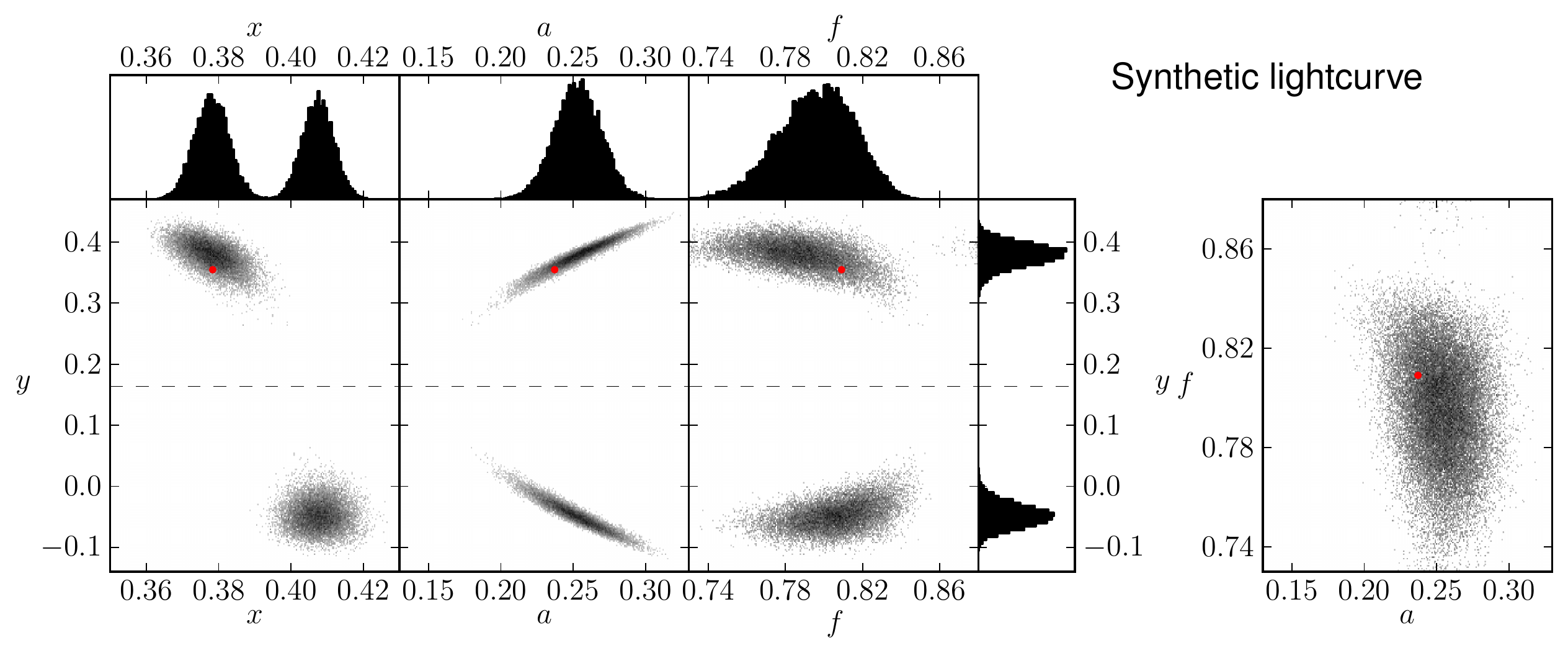}
\end{center}
\caption{Same as Figure \ref{fig:mcmc}, for synthetic transit lightcurves generated using \prism{}, with the best fit spot parameters of the spot from transit 74 (top panels) and 218 (bottom panels) of \hatpelevenb{}.  Red points indicate the best fit spot parameters.}
\label{fig:synthetic}
\end{figure}

Comparing Figures \ref{fig:mcmc} and \ref{fig:synthetic}, we find that the chains converge to roughly the same parameters, further confirming that \prism{} and \spotrod{} interpret input parameters in compatible ways.  We also find that the extent and shape of the equilibrium distributions, that is, the correlations between spot parameters, are roughly the same.  In case of transit 218, even though the synthetic lightcurve is generated based on only one of the two modes, we are unable to determine which mode it is from the MCMC analysis: the distribution is bimodal, just like it was for \kepler{} observations.

The ultimate test to decide whether bimodality and parameter correlations are inherent properties of lightcurve models, and not unique to the implementation we use, would be to run MCMC analysis using \prism{} or another model, other than \spotrod{}.  However, since all previously known models require numerical integration in two dimensions, this would be prohibitively computationally expensive.  Instead, we rely on the very small difference of the two lightcurve models when run with the same input parameters to conclude that \spotrod{} reproduces the results of \prism{}.

\subsection{Distribution of spot parameters}
\label{sec:spotrod:spotparameterdistribution}

In order to study the spot ensemble distribution, we look for anomalies in 204 transits of \hatpelevenb{} of which there are complete, high quality short cadence \kepler{} data.  First, we identify transit anomalies by visual inspection, fit spot parameters using guesses as initial values, and run MCMC simulations.  In a few cases, we find that the chain abandons our initial fit and converges to a solution of a much larger spot with flux ratio close to one, representing a much longer duration and smaller amplitude lightcurve anomaly.  We attribute this to either noise or the effect of a large number of small spots, neither of which we prefer to incorrectly treat as a single, very large spot.  We therefore decrease the number of modelled spots by one for such transits, or omit transits that exhibit such a behaviour with a single spot model, in five cases in total.  For another ten transits, at most 25\% samples of the chain are in an isolated mode representing such an unphysical spot, which samples we discard while keeping the rest of the chain.

We end up with \totalspots{} spots in 130 transits.  For each transit, we independently run the same parallel tempered MCMC simulations as described in Section \ref{sec:spotrod:individual}, using the \texttt{emcee} software package, at 10 different temperatures.  For 73 transits with one spot each, we use 100 concurrent chains, for 43 transits with two spots each, 200 concurrent chains, for 12 transits with three spots each, we employ 300 chains, and for 2 transits with four spots each, 400 chains.  We have four dimensions of parameter space for each spot: $x$, $y$, $a$, and $f$.  We run the simulations for 1000 steps that we discard.  Again, inspection of the chain shows that roughly half of this is already enough for convergence.  Then we run the chain for another 1000 steps to sample the supposedly equilibrium distribution.  This yields the dataset that we use for our analysis in this paper, and the same dataset is used by \citet{2014ApJ...788.....1}.

The next step is to quantify how much better fit these models provide than if we modelled the lightcurve without the spots.  In order to do so, we calculate the Bayesian Information Criterion (BIC), which is the sum of $\chi^2$ and an additional term penalizing extra model parameters to avoid overfitting \citep {Schwarz1978}.  We find that every single one of the \totalspots{} spots yields a BIC value at least $25.0$ lower than the model without that spot, which is a very significant improvement, justifying every spot in our analysis.  This suggests that probably more spots could be carefully included.

We present the spot radius--flux ratio distribution of all spots on Figure \ref{fig:mcmcall}, left middle panel.  Note that the distribution of most individual spots overlap, except for a few, which appear as separate clusters on the figure.  The distribution is bounded from the side of small radius and large flux ratio by an observational bias: whether we visually inspect the lightcurves or apply an algorithmic transit anomaly search, small amplitude anomalies will be lost in photon noise.  The maximum eclipsed spot area is the smaller of $R_\mathrm p^2$ and $a^2$ relative to the stellar disk, therefore the maximum transit anomaly amplitude for given $a$ and $f$ is $A_\mathrm{max} = \min\left(R_\mathrm p^2, a^2\right) (1-f)$ (not accounting for limb darkening and projection distortion).  The actual amplitude of the transit anomaly can be less, depending on $y$.  Figure \ref{fig:mcmcall} shows the $A_\mathrm{max}=0.0002$ curve in red, where $0.0002$ is an arbitrarily chosen value, which seems to act as an approximate amplitude threshold for anomalies that we detected.

\begin{figure}
\begin{center}
\includegraphics*[width=\threequartersfigurewidth]{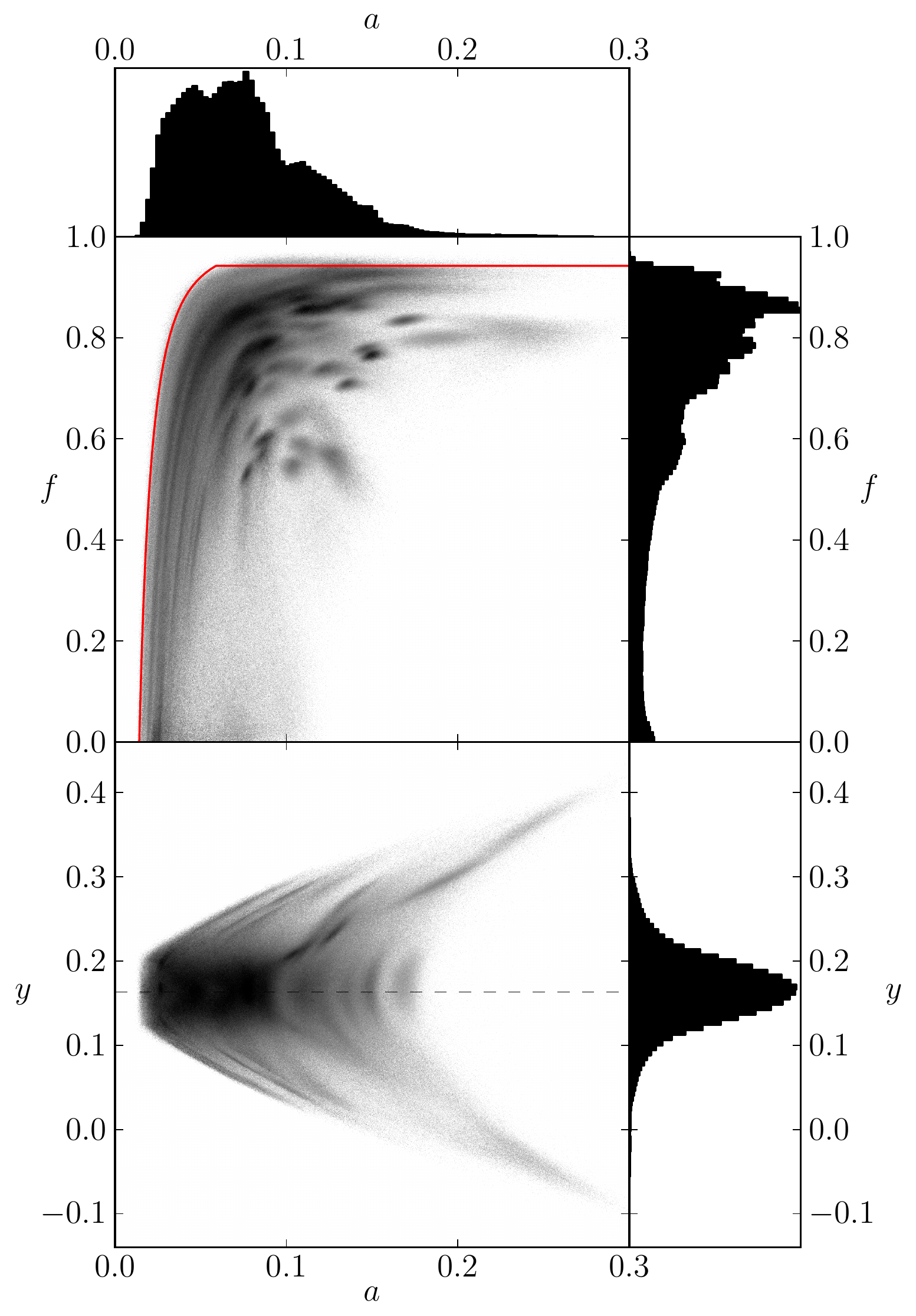}
\end{center}
\caption{Top panel: histogram of spot radius $a$.
Left middle panel: joint distribution of spot radius $a$ and flux ratio $f$.  Red curve is the transit anomaly amplitude threshold $A_\mathrm{max}=0.0002$.
Right middle panel: histogram of flux ratio $f$.
Left bottom panel: joint distribution of spot radius $a$ and spot center projected $y$ coordinate.  Thin dashed line indicates $y$ coordinate of planet at midtransit.
Right bottom panel: histogram of spot center projected $y$ coordinate.
All distributions and histograms combine the entire MCMC chain of all \totalspots{} spots.}
\label{fig:mcmcall}
\end{figure}

We note that for small spots, individual spot distributions spread along constant $\min\left(R_\mathrm p^2, a^2\right)(1-f)$ curves.  The reason is that this quanitity describes the total amount of flux missing due to the spot, which is directly related to the shape of the transit lightcurve anomaly, and therefore can be well constrained.  For small spots, the anomaly does not last long, therefore there are fewer data points than for large spots.  This makes it difficult to resolve the degeneracy between radius and flux ratio.  For large spots, however, the transit anomaly has a larger amplitude, which can directly be used to infer the flux ratio.  Note that consistently with this argument, the rightmost panels of Figure \ref{fig:mcmc} show that the flux ratio of the small spot seen in transit 74 has a much larger uncertainty than that of the large spot seen in transit 218, even though they have comparable relative uncertainties of their radii.

This effect results in small spots having a weaker constraint on flux ratio.  However, as we will see later in this section, this does not indicate the presence of small, dark spots.  Because of this large flux ratio uncertainty of small spots, we are not able to detect a significant correlation between spot radius and flux ratio for the \totalspots{} spots studied.

The bottom left panel of Figure \ref{fig:mcmcall} shows the joint distribution of spot radius and $y$ coordinate for all spots.  Because of the polar orbit of \hatpelevenb{}, the latter roughly corresponds to the longitude of the spot on the stellar surface.  In this case, we can assume that there is no physical correlation between the two parameters, therefore we have to interpret the joint distribution in terms of observational biases.  The observed radius is bounded from below, because very small spots would not cause a detectable signal in the lightcurve.  For values of $y$ further from the transit chord, the smallest detectable radius increases for geometrical reasons: the spot has to overlap with the strip that the planet scans on the stellar surface.  Finally, the spot radius is bounded from above by the physical distribution of spots, and we expect this to be independent from $y$.  However, this is not reflected on Figure \ref{fig:mcmcall}: we see that MCMC states with radius above $0.2$ prefer values of $y$ further from $b$.  We interpret this as an artifact: we suspect that the lightcurve anomaly due to irregularly shaped spots or spot groups, when mistakenly interpreted as a single spot, results in a large spot further from the transit chord.  We believe this radius is unphysical, because we never see transit anomalies that last long enough to require a spot of similar size with $y\approx b$ as an explanation.  Therefore we conclude that the upper limit of physical spot radius distribution is at most somewhat lower than $0.2$, the radius of the largest detected spot with $y\approx b$.  It is also possible that this is not a single spot either, so the actual upper radius limit might be smaller than this.

The three side panels on Figure \ref{fig:mcmcall} show histograms of spot radius, flux ratio, and $y$ coordinate, generated from the MCMC chains for all spots.  The same observational biases are reflected here: there are no very small ($a\approx0$) spots, and no very bright ($f\approx1$) spots, because these would be undetectable.  $y$ is concentrated around the impact parameter, because this is where even small spots are eclipsed by the planet.

To investigate the radius and flux ratio distribution of spots, we plot the median of these parameters in increasing order on Figure \ref{fig:afsorted}.  The top panel presents spot radius, the bottom shows flux ratio.  The horizontal axes indicate the rank of the spot in the order of median values.  In addition to the median, we shade the $1\sigma$ and $3\sigma$ intervals of the parameter distribution of each individual spot.  The advantage of this presentation over the histograms of Figure \ref{fig:mcmcall} is that we can disentangle the spread of a spot parameter for an individual spot due to uncertainties from the spread of the ensemble distribution due to spots being different.

Together with the histograms of Figure \ref{fig:mcmcall}, Figure \ref{fig:afsorted} helps us confirm the observational biases agains small size and large flux ratio.  In addition, we note that most spots are smaller than $a\approx0.15$ (three times the radius of \hatpelevenb{}), with very few spots around the size of $a\approx0.2$.  We believe that larger spots are artifacts, because they are only seen with $y$ different from $b$.

On the other hand, even though bright spots are more frequent than dark ones, our first impression is that there is a number of almost black spots, that is, spots with $f$ close to zero.  Note that a large number density in terms of a spot parameter, that is, large bin count in the histogram corresponds to a less steep curve when data points are plotted in increasing order of that parameter.  However, when closely inspecting the parameter uncertainties of individual spots, we see that the ones that seem to be consistent with being very dark show a flux ratio distribution that includes much larger flux ratios as well.  In fact, only two spots have a $3\sigma$ confidence interval that excludes flux ratios above $0.5$, even though 30 spots have median flux ratio and 36 spots have best fit (least sum of squared residuals) flux ratio lower than $0.5$.  On the other hand, more than half of all spots studied are consisent with having a flux ratio not exceeding $0.5$ at the $3\sigma$ level.  Therefore we cannot either prove or disprove the existence of spots darker than $f=0.5$.  On the other hand, many spots with flux ratios from approximately $0.6$ to $0.9$ have small flux ratio uncertainties, suggesting the existence of spots brighter than $f=0.5$.

\begin{figure}
\begin{center}
\includegraphics*[width=\figurewidth]{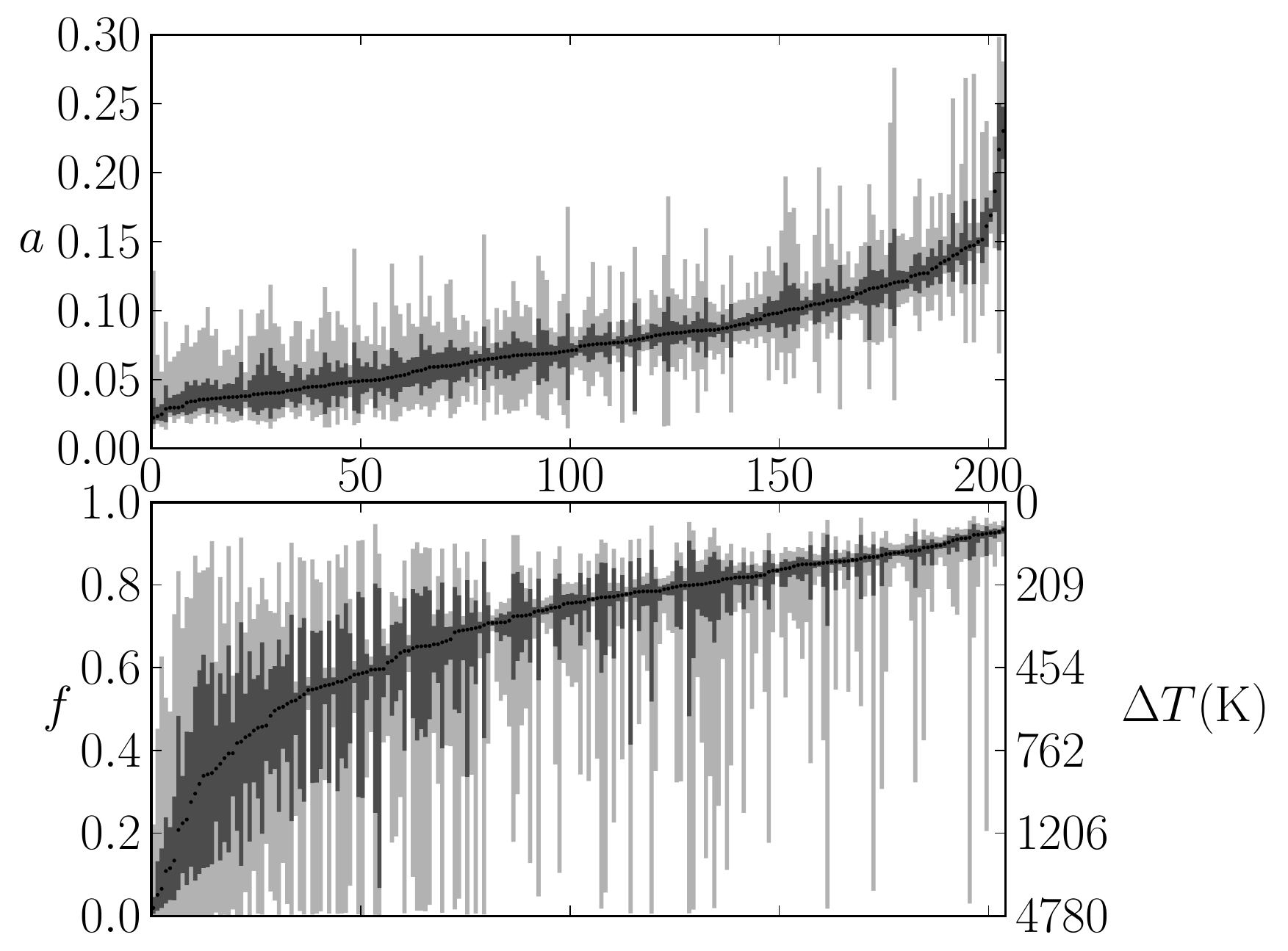}
\end{center}
\caption{Median values of spot radius $a$ (top panel) and flux ratio $f$ (bottom panel).  Horizontal axis shows rank of spot.  Each panel is ordered by the median of the corresponding parameter.  Shaded regions indicate the $1\sigma$ (dark gray) and $3\sigma$ (light gray) confidence intervals based on the MCMC simulation for each spot.  Right vertical axis of bottom panel shows inferred temperature difference of photosphere and spot assuming that both radiate as black bodies.}
\label{fig:afsorted}
\end{figure}

\subsection{Spot temperature}

Assuming that the stellar photosphere and the spots both radiate as black bodies, one can calculate the spot temperature $T_\mathrm s$ from the flux ratio $f$ and the effective photospheric temperature $T_\mathrm{eff}$. This method was first applied to infer the spot temperature in the context of spot-induced transit lightcurve anomalies by \citet{2003ApJ...585L.147S}.

We can use the same method to calculate spot temperatures on \hatpeleven{} from \spotrod{} MCMC results.  We use the photospheric effective temperature result $T_\mathrm{eff} = 4780\,\mathrm K \pm 50\,\mathrm K$ of \citet{2010ApJ...710.1724B}, and we integrate Planck's law over the \kepler{} response function to determine the flux ratio as a function of spot temperature.  The temperature difference $\Delta T = T_\mathrm{eff} - T_\mathrm s$ corresponding to certain values of flux ratio $f$ are displayed on the right vertical axis of the bottom panel on Figure \ref{fig:afsorted}.

As we noted in Section \ref{sec:spotrod:spotparameterdistribution}, we cannot infer anything about the existence of spots with flux ratio less than $0.5$.  Therefore we cannot determine whether there are spots with spot temperature below the corresponding temperature of $T_\mathrm s = 4180\pm40\;\mathrm K$, or a temperature difference of $\Delta T = 600\pm10\;\mathrm K$.

On the other hand, Figure \ref{fig:afsorted} tells us that there are spots with well constrained flux ratios ranging approximately from $f=0.6$ to $0.9$.  In terms of temperature, this means that there exist spots ranging approximately from $T_\mathrm s = 4330\;\mathrm K$ to $4680\;\mathrm K$ ($\Delta T = 450\;\mathrm K$ to $100\;\mathrm K$).  It is possible that \hatpeleven{} also has brighter spots (that are not detected due to observational bias), and darker spots.

For comparison, \citet{2003SoPh..213..301W} show that the Sun exhibits spots ranging from $f\approx0.15$ (with $a\approx0.03$) up to $f\approx0.7$ (with $a\approx0.01$), where the flux ratio is measured at 672.3 nm with 10 nm bandpass.  This shows that \hatpeleven{} is not the only star where individual spots are thought to exhibit a large range of flux ratios.

An advantage of transit anomalies over spectroscopic methods is the ability to measure temperatures of individual spots.  For example, \citet{2004AJ....128.1802O} study TiO absorption features to determine the spot temperature to be $T_\mathrm s = 3350\pm115\;\mathrm K$ on EQ Vir, a BY Dra-type flare star with $T_\mathrm{eff} = 4380\;\mathrm K$, of spectral type K5 Ve, close to K4 of \hatpeleven{}.  This corresponds to a flux ratio of $f=0.21\pm0.05$ in the \kepler{} bandpass.  We note that this value is lower than the flux ratio of the majority of spots we found on \hatpeleven{}, though we were not able to either confirm or exclude the existence of spots with such low flux ratios.  However, the result of \citet{2004AJ....128.1802O} is averaged over all spots visible on the stellar disk, and the temperature range of individual spots cannot be determined by their method.


\subsection{Transit anomaly duration and amplitude}
\label{sec:spotrod:durationamplitude}

In addition to investigating the parameters taken by \spotrod{}: $x$, $y$, $a$, and $f$, it is also interesting to study directly observable properties of spot eclipses: the duration and amplitude of the transit anomaly.  Given the results of MCMC calculations, the easiest way to extract these properties is to measure them on model lightcurves.  We therefore draw 1000 states from each chain, and generate model lightcurves with one spot each.  We calculate the amplitude of the transit anomaly as the maximum deviation from a spotless model, and the duration of the transit anomaly as the length of the time interval on which the model with one spot predicts more flux than the spotless model.  The resulting distribution is plotted on Figure \ref{fig:durationamplitude}.

The transit anomaly duration--amplitude distribution is confined from three sides.  The amplitude is bounded from below, as described in Section \ref{sec:spotrod:spotparameterdistribution}.  We draw a red line on Figure \ref{fig:durationamplitude} at $A=0.0002$, the same amplitude value as on Figure \ref{fig:mcmcall}.  Note, however, that the transit anomaly amplitude is represented directly on Figure \ref{fig:durationamplitude}, whereas on Figure \ref{fig:mcmcall}, we could only calculate the maximum possible value $A_\mathrm{max}$ for given values of $a$ and $f$.

Figure \ref{fig:durationamplitude} tells us that visual inspection as performed by the authors results in an amplitude limit that is mostly constant across anomaly durations.  Anomalies selected programmatically, however, might have a different boundary, as it is possible to identify an anomaly with a smaller amplitude in noisy data if it lasts sufficiently long.

From the side of short anomalies, the distribution is bounded by geometrical arguments.  The amplitude of the anomaly tells us how much flux is missing with respect to a spotless transit.  This places a lower limit on the geometrical extent of the spot, which then translates to the duration of the event.  To characterize this boundary, we feed 10\,000 random black spots with to \spotrod{}, and plot the envelope of their duration-amplitude distribution in black on Figure \ref{fig:durationamplitude}.  Gray spots ($f>0$) would cause anomalies with the same duration but smaller amplitude than if they were black ($f=0$), therefore we only draw black spots to determine this boundary.  We notice that the MCMC distribution extends close to this boundary, which means that the chains contain spot states that are almost entirely black.  As discussed in Section \ref{sec:spotrod:spotparameterdistribution}, however, this does not mean that the best fit for those spots is necessarily black.

\begin{figure}
\begin{center}
\includegraphics*[width=\figurewidth]{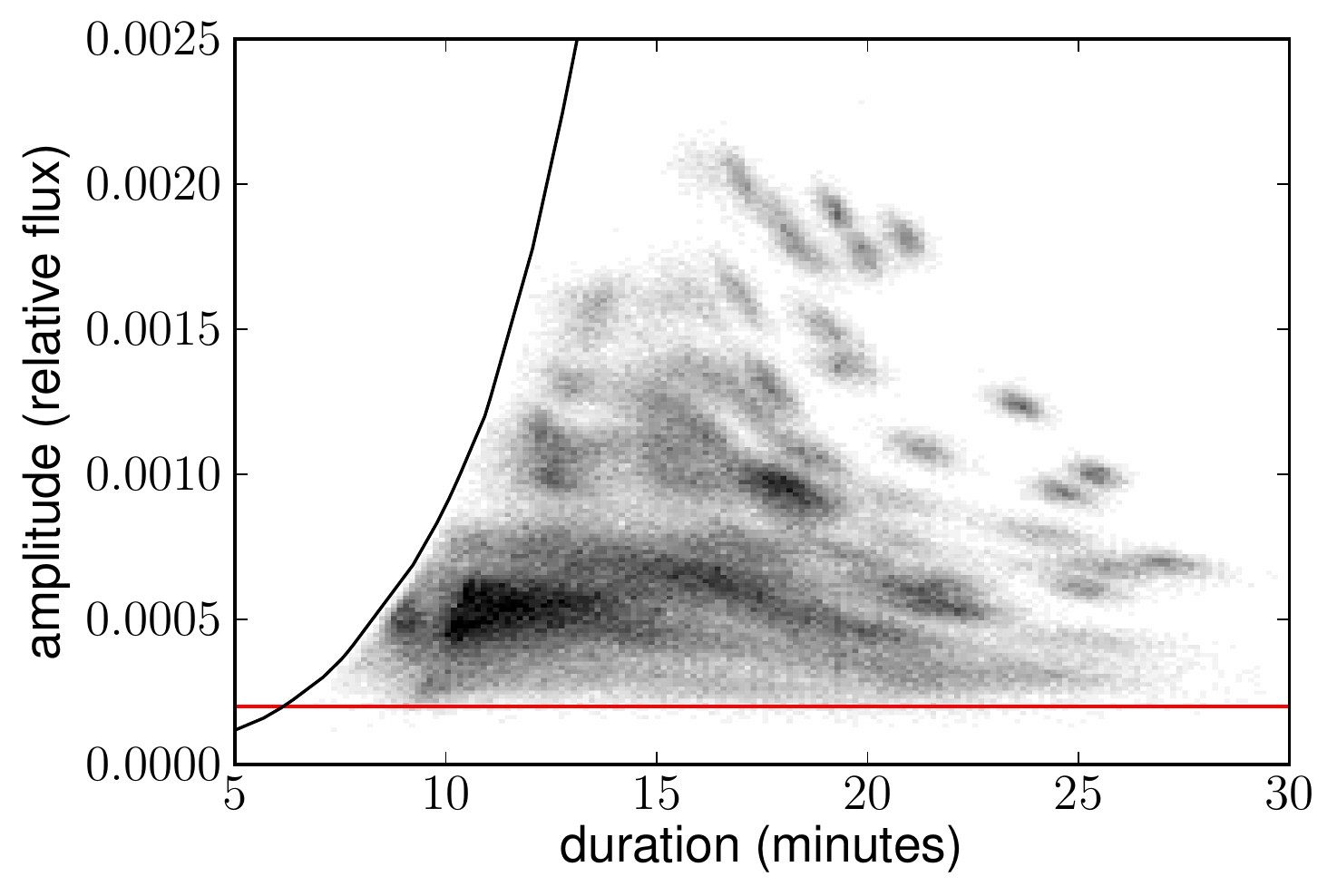}
\end{center}
\caption{Distribution of transit anomaly duration and amplitude in black dots.  Each spot is resampled 1000 times from its MCMC chain.  The $A=0.0002$ constant amplitude line is shown in red.  Black curve represents the boundary of parameter space imposed by our model.}
\label{fig:durationamplitude}
\end{figure}

From the right and above, however, the distribution does not extend to its theoretical limits: the maximum conceivable amplitude would be the transit depth (for a black spot that is larger than the planet), and the maximum conceivable transit duration would be the transit duration.  The largest amplitude we observe is roughly half of this, and the longest anomaly lasts only about one quarter of the entire transit.  We interpret these limits as indications of the actual spot parameter distribution, namely the lack of large dark spots, and the upper limit on spot size $a\lessapprox0.2$.  We also note that on Figure \ref{fig:durationamplitude}, the longest transit anomalies seem to have small amplitudes, for which we cannot offer an explanation.

While measuring directly observable quantities like transit anomaly duration and amplitude on models generated by \spotrod{} is very convenient, we note that this inevitably introduces biases.  For example, if we observe an anomaly with duration and amplitude that would place it on the left of the black curve on Figure \ref{fig:durationamplitude}, that could not be reproduced by \spotrod{}, and consequently we would measure different duration and amplitude values with our method.  However, our MCMC analysis disentangles multiple spots and measures transit anomaly durations efficiently, and still yields meaningful conclusions about the darkness of spots and the amplitude threshold for detection.

\subsection{Residuals}

Finally, we study the distribution of residuals to assess the goodness of fit, and to compare our model with \totalspots{} spots to the spotless lightcurve model.  We calculate lightcurves using \spotrod{} and the Mandel--Agol model for the two cases, respectively.  We only consider the residuals at observations that took place during transits (between first and fourth contact).  We use the \texttt{SAP\_FLUX\_ERR} column calculated by the \kepler{} Photometric Analysis module as the error estimate of the lightcurve data in the \texttt{SAP\_FLUX} column.  Figure \ref{fig:residuals} displays the normalized residual distribution histograms: in black when calculated using the spotless model, and in red when calculated using our model with spots, with a logarithmic vertical scale.  The spotless model residuals have a large excess on the positive side, which we attribute to the presence of spot-induced transit lightcurve anomalies.  On the other hand, our \spotrod{} model accounts for enough spots to make the residual distribution fairly symmetric.  We note that this might potentially be used as a detection method to identify targets that exhibit spot-induced transit anomalies.

In terms of the Mandel--Agol fit, it is interesting to note that the largest residual is $25.75\sigma$.  Comparing this to the transit depth of roughly 50 sigma, we see that the largest transit anomaly has an amplitude of half the transit depth, just as we noted in Section \ref{sec:spotrod:durationamplitude}.  (The exact transit depth expressed in units of photon noise varies due to the quarterly rotations of the \kepler{} satellite.)

The relative error of \hatpeleven{} photometry is very small ($\lessapprox 10^{-4}$).  Assuming that it is dominated by photon noise, the error distribution can be approximated with an independent normal distribution for each data point.  In this case, it is valid to calculate $\chi^2$, and from that, reduced $\chi^2$.  For the models without and with spots, we get the strikingly different values $\chi_\mathrm{spotless}^2=1.7\cdot10^5$ and $\chi_\mathrm{spots}^2=3.3\cdot10^4$, respectively.

The total number of observations taken during the 130 transits is $N=18\,135$.  Since we fix orbital parameters, transit ephemeris, and limb darkening, the spotless model has no fit parameters.  With the number of data points as the degrees of freedom, we get $\chi_\mathrm{red,spotless}^2=9.2$, which, being much larger than unity, motives a model with more free parameters.

Our spot model has four fit parameters for each spot, that is $P=4\cdot\totalspots{}=812$ fit parameters in total.  Note, however, that \citet{2010arXiv1012.3754A} prove that it is not justified to calculate the degrees of freedom as $K=N-P$ for non-linear models like \spotrod{}, and in fact no reliable method is known to calculate $K$ in general.  Therefore we give the value $\frac{\chi_\mathrm{spots}^2}{N-P} = 1.9$ for reference only, and are left with the symmetry of the histogram as the only way to quantify the goodness of the fit.

\begin{figure}
\begin{center}
\includegraphics*[width=\figurewidth]{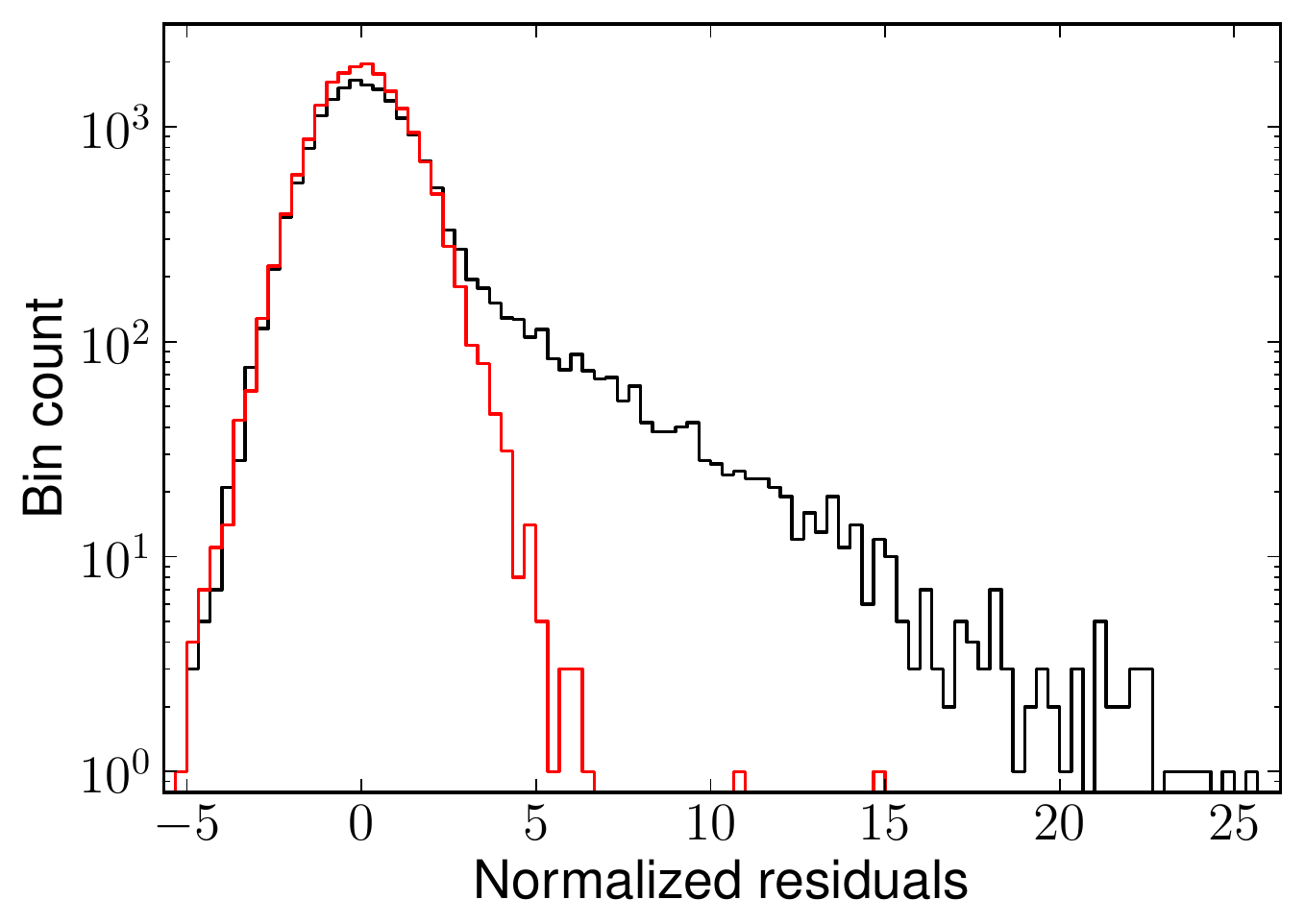}
\end{center}
\caption{Residual histograms of short cadence \kepler{} observations that were taken during transits of \hatpelevenb{}, normalized by \texttt{SAP\_FLUX\_ERR}.  Black histogram corresponds to residuals with respect to the Mandel--Agol lightcurve model, red histogram corresponds to residuals with respect to the best fit \spotrod{} model with a total of \totalspots{} spots.  Vertical axis (counts) is on a logarithmic scale.}
\label{fig:residuals}
\end{figure}

\section{Conclusion}
\label{sec:spotrod:conclusion}

In this paper, we present \spotrod{}, a transit lightcurve model accounting for both eclipsed and uneclipsed starspots.  The advantage of our model over previous methods is that in polar coordinates, we integrate analytically with respect to the polar angle, therefore time-consuming numerical integration only remains to be performed along a single dimension, the radial coordinate.  This feature makes our model fast enough not only for fitting, but also for efficient statistical investigations using, for example, MCMC.  A free and open source implementation of our model is publicly available.

The model assumes that spots follow the same limb darkening law as the stellar photosphere, consistent with observations of the Sun \citep{2003SoPh..213..301W}.  It also assumes that spots are homogeneous.  Umbra-penumbra structure can be mimicked by superimposing two concentric spots, while bright features can also be modelled using a flux ratio exceeding unity.

We apply our model to \kepler{} data of \hatpeleven{} transits.  We investigate correlations between fit parameters of individual spots, and confirm findings of previous investigations using similar models.  We also study the size and flux ratio distributions.  We establish an upper limit of $\lessapprox 0.2$ for the spot radius, and find strong indication for the presence of spot with flux ratio ranging from 0.6 to 0.9, corresponding to an effective temperature 100 to 450 K lower than that of the spotless photosphere.  We cannot prove nor disprove the existence of spots with flux ratio less than $0.5$.  We do not find a significant correlation between spot size and flux ratio.

While \hatpeleven{} is unique in its brightness and large transit anomaly amplitudes within the Kepler field, \spotrod{} can potentially be used to model \kepler{} observations of other transiting planetary hosts.  In addition, comparable quality photometric observations are expected to taken of a much larger number of stars, for example, by the K2 mission of the \kepler{} satellite \citep{2014AAS...22322801H}, and by the Transiting Exoplanet Survey Satellite \citep[TESS,][]{2010AAS...21545006R}.

While our model provides a good fit to observations of \hatpeleven{}, it is important to keep in mind that our perfectly circular and homogeneous spots are a simplified version of what the stellar surface actually looks like.  However, mapping out the projected stellar surface by, for example, two dimensional deconvolution with the planetary disk, is an underdetermined and potentially numerically unstable inversion problem.  The general advantage of model simplifications is that the small degree of freedom makes fitting robust.  This is exactly what \spotrod{} provides: a simplistic, approximate, but robust and fast way to model transit lightcurves of spotted stars.

\acknowledgements
Work by B.B.~and M.J.H.~was supported by NASA under grant NNX09AB28G from the Kepler Participating Scientist Program and grants NNX09AB33G and NNX13A124G under the Origins program. D.M.K.~is funded by the NASA Carl Sagan Fellowships. This paper includes data collected by the Kepler mission. Funding for the Kepler mission is provided by the NASA Science Mission directorate. The MCMC computations in this paper were run on the Odyssey 2.0 cluster supported by the FAS Science Division Research Computing Group at Harvard University.  B.B.~is grateful to Robert Noyes for useful comments on the manuscript, and to John A.~Johnson for suggesting the name \spotrod{}.

\setcounter{section}0
\renewcommand\thesection{\thechapter.\Alph{section}}
\section{Derivations}
\label{sec:spotrod:derivations}

\subsection{Geometry}
\label{sec:spotrod:geometry}

Let \xs{} and \ys{} denote the coordinates of the center of the spot as seen by the observer, in a Cartesian coordinate system with the center of the stellar disk as the origin, in stellar radius units. Then the angle between the plane of the spot boundary and the line of sight is 
\begin{align*}
\beta &= \arccos\sqrt{\xs^2+\ys^2}.
\end{align*}
Figure \ref{fig:sideview} shows a side view. Note that this is the same $\beta$ that \citet{2012MNRAS.427.2487K} defines in his Equation (1).

\begin{figure}
\begin{center}
\includegraphics*[width=\figurewidth]{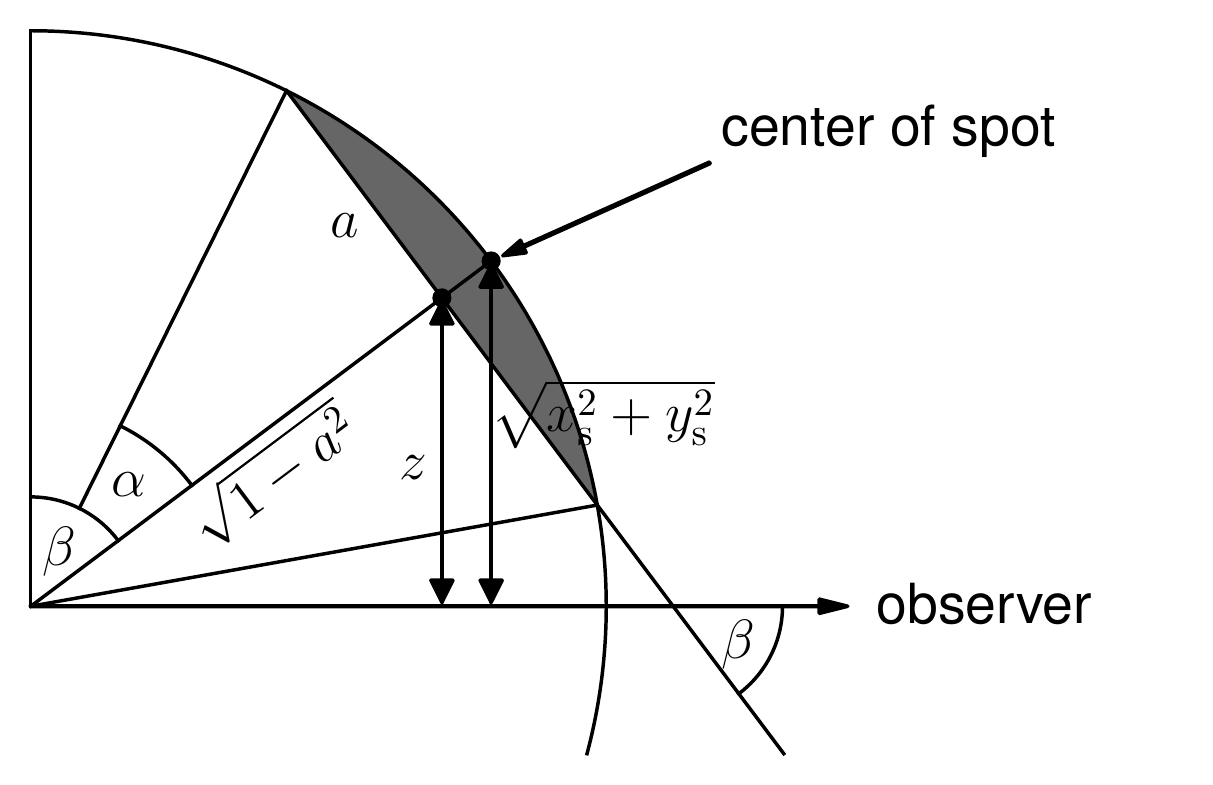}
\end{center}
\caption{A cross section in the plane containing the center of the star, the center of the spot, and the observer. The observer is to the right at an infinite distance.}
\label{fig:sideview}
\end{figure}

In projection (as seen by the observer), the spot is an ellipse with semi-major axis $a$, and semi-minor axis
\begin{align}
\label{eq:b}
b &= a\sin\beta = a\sqrt{1-(\xs^2+\ys^2)}.
\end{align}
The center of this ellipse is the projection of the intersection point of the axis of the cone and the plane in which the spot boundary lies, not the projection of the center of the spot which is on the stellar surface. The center of the ellipse lies at a distance of $\sqrt{1-a^2}$ from the center of the star, therefore in projection, the distance between the center of the stellar disk and the center of the ellipse is
\begin{align}
\label{eq:z}
z &= \sqrt{1-a^2}\cos\beta = \sqrt{1-a^2}\sqrt{\xs^2+\ys^2}.
\end{align}

Expressing $\xs^2+\ys^2$ from Equation (\ref{eq:z}) and substituting into Equation (\ref{eq:b}), we can express $b$ in terms of $a$ and $z$:
\begin{align}
\label{eq:b1}
b &= a \sqrt{1 - \frac{z^2}{1-a^2}}.
\end{align}

The input parameters of the subroutine \integratetransit{} are \xs{}, \ys{}, and $a$: it calculates $z$ from Equation (\ref{eq:z}), and passes it to \ellipseangle{}, which in turn calculates $b$ using Equation (\ref{eq:b1}).

\subsection{Calculating $\gamma$, $\gamma^*$, and $\delta$}
\label{sec:spotrod:calc}

\begin{figure}
\begin{center}
\includegraphics*[width=\figurewidth]{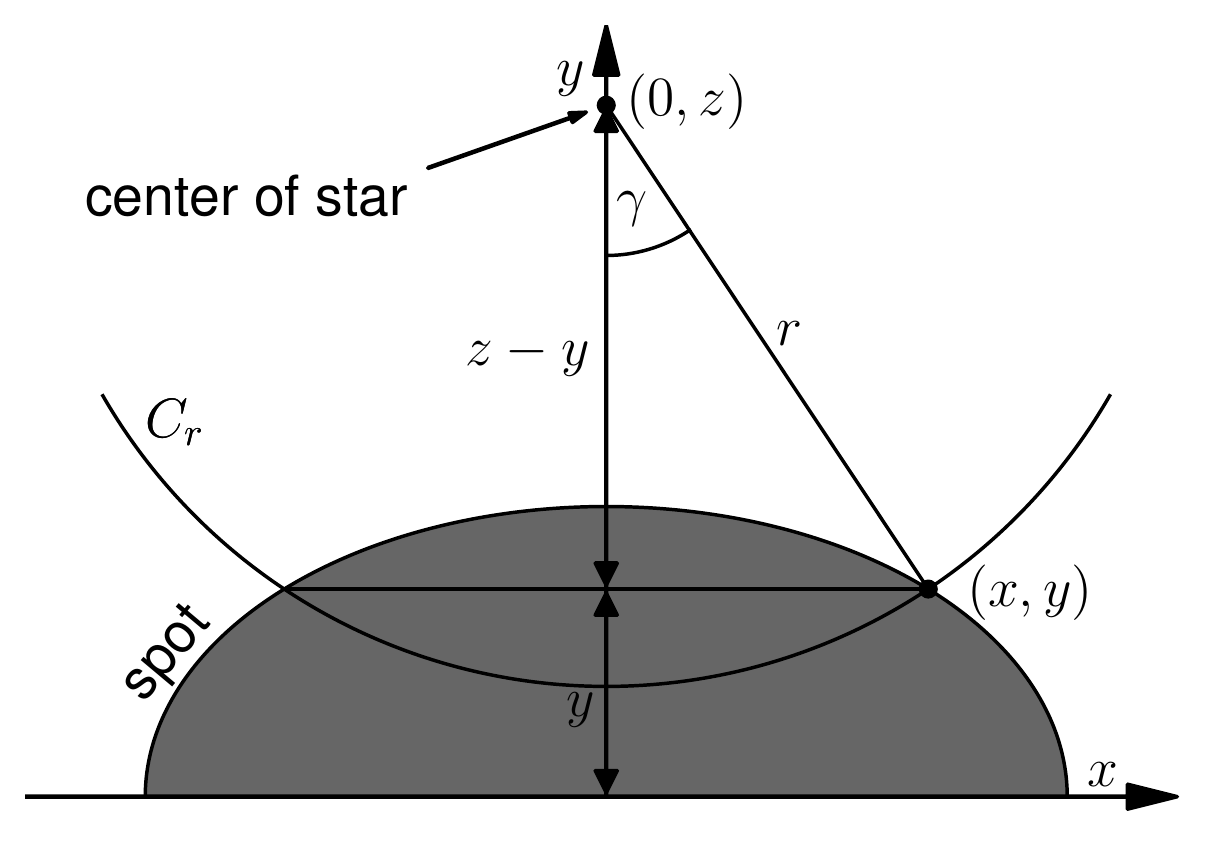}
\end{center}
\caption{The spot, the center of the star, and $C_r$, as seen from the direction of the observer, in a Cartesian coordinate system centered on the center of the spot ellipse (not the projection of the spot center).}
\label{fig:ellipse}
\end{figure}

To calculate $\gamma(r)$, consider a Cartesian coordinate system in the projection plane with the center of the ellipse as the origin, the $x$ axis parallel to the semi-major axis, the $y$ to the semi-minor axis. Let the center of the stellar disk be at $(0,z)$, and let $(x,y)$ denote an intersection point of the ellipse and $C_r$. See Figure \ref{fig:ellipse}. Then $(x,y)$ satisfies the following set of quadratic equations:
\begin{align}
\label{eq:quad1}
\frac{x^2}{a^2} + \frac{y^2}{b^2} &= 1 \\
\label{eq:quad2}
x^2 + (y-z)^2 &= r^2.
\end{align}
If there are no intersection points, then $C_r$ is either located entirely outside the ellipse, thus $\gamma=0$, or entirely outside, in which case $\gamma=\pi$. If there are one, two, three, or four intersection points, then they must be located symmetrically around the $y$ axis, because if $(x,y)$ is a solution, then so is $(-x,y)$ (and they coincide if $x=0$). After solving for $y$, we calculate $\gamma$ using
\begin{align*}
\gamma &= \arccos\frac{z-y}r.
\end{align*}
Appendix \ref{sec:spotrod:limb} discusses the case of four intersection points. One or three intersection points are singular cases between other cases, and can be treated along with the case on either side.

\begin{figure}
\begin{center}
\includegraphics*[width=\figurewidth]{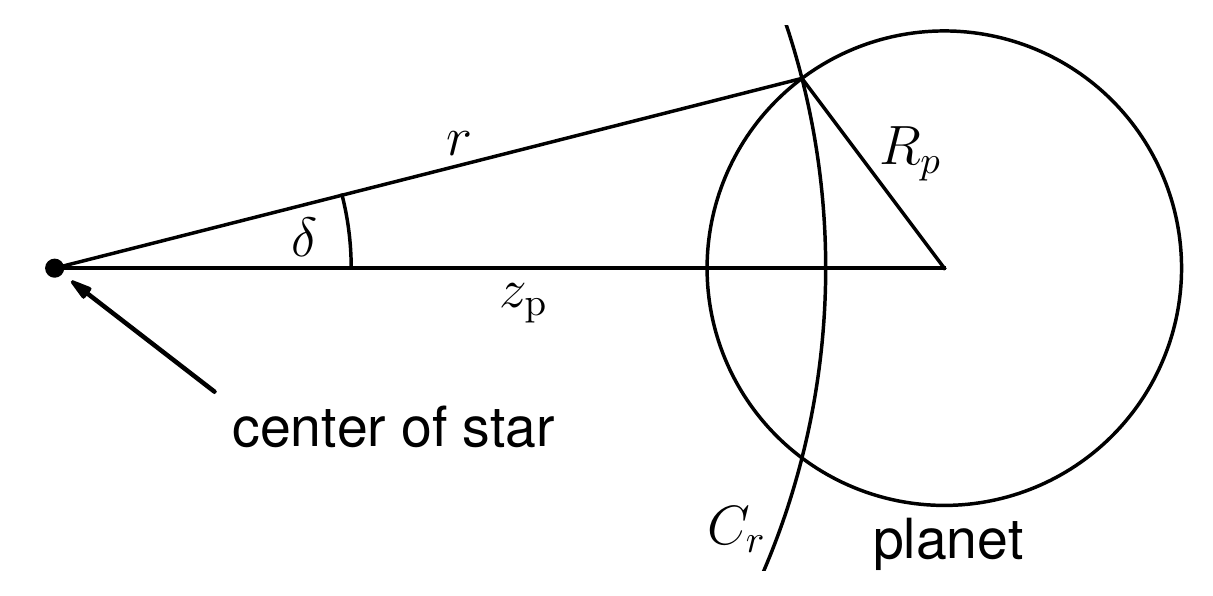}
\end{center}
\caption{The triangle in the sky plane defined by the center of the star, the center of the planet, and one intersection point of the edge of the planet with $C_r$.}
\label{fig:cosines}
\end{figure}

Let $R_\mathrm p$ denote the radius of the planet, and $z_\mathrm p$ its projected separation from the center of the stellar disk, both in stellar radius units. To calculate $\delta(r)$, we could repeat the above derivation with $a=b=R_\mathrm p$. Or we can use the law of cosines in the triangle defined by the center of the star, the center of the planet, and the intersection point of the edge of the planet with $C_r$:
\begin{align*}
R_\mathrm p^2 = r^2 + z_\mathrm p^2 - 2rz_\mathrm p\cos\delta(r),
\end{align*}
as seen on Figure \ref{fig:cosines}. Again, the cases of $C_r$ being disjoint from or entirely occulted by the planet should be tested for separately, yielding $\delta=0$ and $\delta=\pi$, respectively.

Finally, let $\theta$ denote the angle between the center of the spot and the center of the planet as seen from the center of the star, which we also calculate using the law of cosines. We always choose the angle for which $0\leqslant\theta\leqslant\pi$. Now $\gamma^*(r)$ is determined by $\gamma(r)$, $\delta(r)$, and $\theta$ according to the following cases:
\begin{align}
\label{eq:cases}
\gamma^* = \begin{cases}
\gamma & \textrm{if } \theta \geqslant \gamma + \delta \textrm{ (arcs disjoint)} \\
\gamma - \delta & \textrm{if } \gamma \geqslant \theta + \delta \textrm{ (planet arc inside spot arc)} \\
0 & \textrm{if } \delta \geqslant \gamma + \delta \textrm{ (spot arc inside planet arc)} \\
\frac{\gamma + \theta - \delta}2 & \textrm{o/w, if } \gamma + \delta + \theta \leqslant 2\pi \textrm{ (partial overlap)} \\
\pi - \delta & \textrm{o/w, if } \gamma + \delta + \theta > 2\pi \textrm{ (circular overlap)}.
\end{cases}
\end{align}
In the first case, the arcs are disjoint, therefore none of the spot arc is eclipsed by the planet. This happens to spot 1 on Figure \ref{fig:observer}. 
In the second case, the entire planet arc gets subtracted from the spot arc. The part of the spot arc that is not occulted by the planet is now composed of two arcs on either side of the planet, and we define $\gamma^*$ as half the total central angle of them. In the third case, the planet occults the entire spot arc. ``Otherwise'' for the last two cases means that the triangle inequality holds between $\gamma$, $\delta$, and $\theta$. In the fourth case, the planet and spot arcs overlap in a single arc. This happens to spot 2 on Figure \ref{fig:observer}. In the last case, they overlap in two arcs. This can happen only if at least one of the planetary disk and the spot ellipse contain the center of the stellar disk in projection. This situation is illustrated by Figure \ref{fig:circular}.

\begin{figure}
\begin{center}
\includegraphics*[width=\figurewidth]{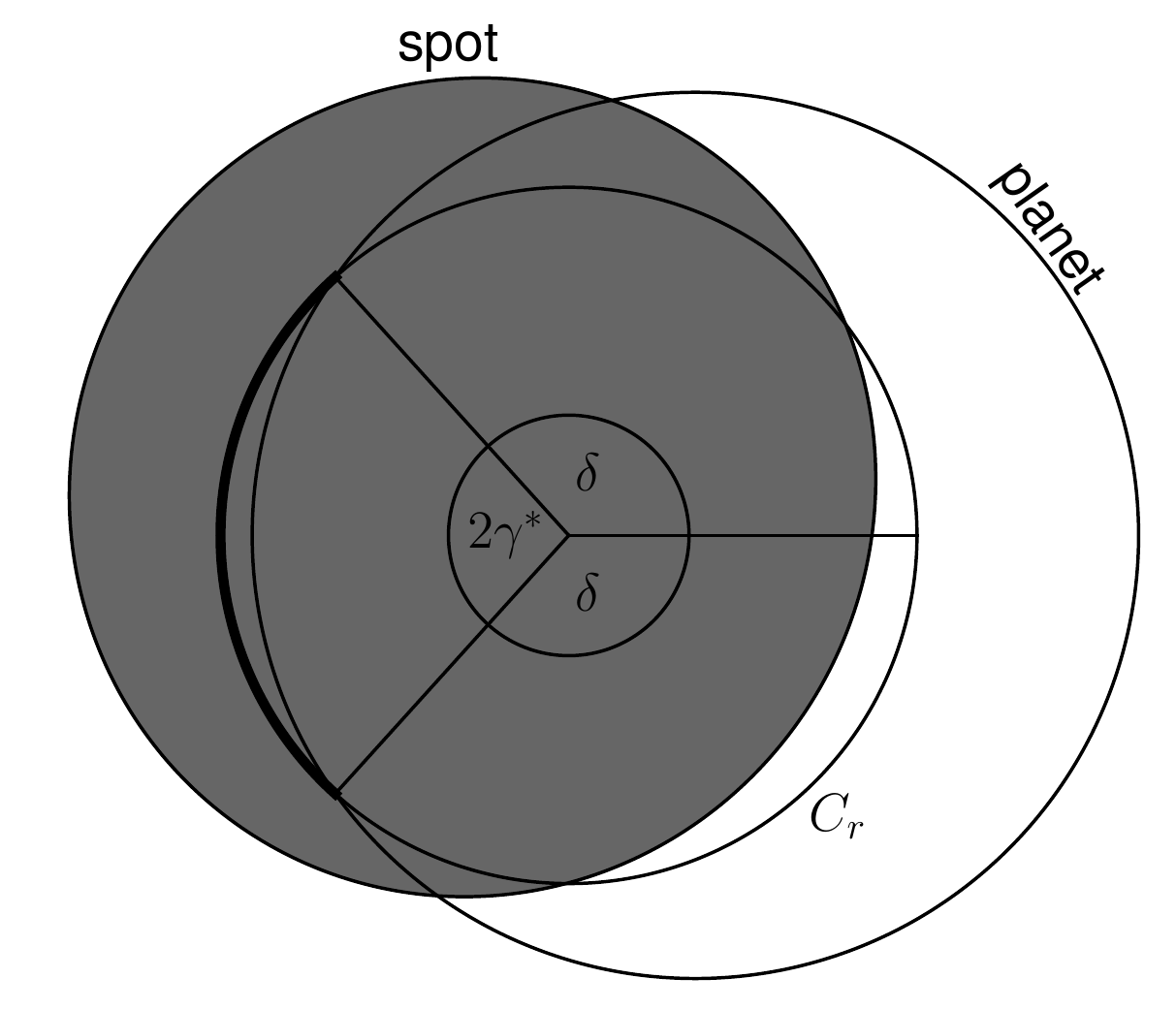}
\end{center}
\caption{Example for the spot arc and the planet arc overlapping in two arcs.}
\label{fig:circular}
\end{figure}

The function \circleangle{} calculates $\beta$ from $r$, $z_\mathrm p$, and $R_\mathrm p$, \ellipseangle{} calculates $\gamma$ from $r$, $z$, and $a$, and finally \integratetransit{} calculates $\gamma^*$ based on Equation (\ref{eq:cases}), evaluates the integrals of Equations (\ref{eq:multispot}) and (\ref{eq:transit}), and calculates $F_\mathrm{normalized}$ according to Equation (\ref{eq:normalized}).

\subsection{Spots partially behind the limb}
\label{sec:spotrod:limb}

The boundary of the spot is assumed to be a circle, therefore its projection (as seen by the observer) is an ellipse. However, we must investigate whether the boundary of what we see of the spot coincides with the projection of its boundary on the stellar surface. It is easy to see that it is indeed the case as long as no part of the spot covers another part of it. That is, as long as the entire spot is visible, we will always see it as an ellipse. (Remember the caveat that the center of this ellipse is not the projection of what we defined as the center of the spot.)

\begin{figure}
\begin{center}
\includegraphics*[width=\figurewidth]{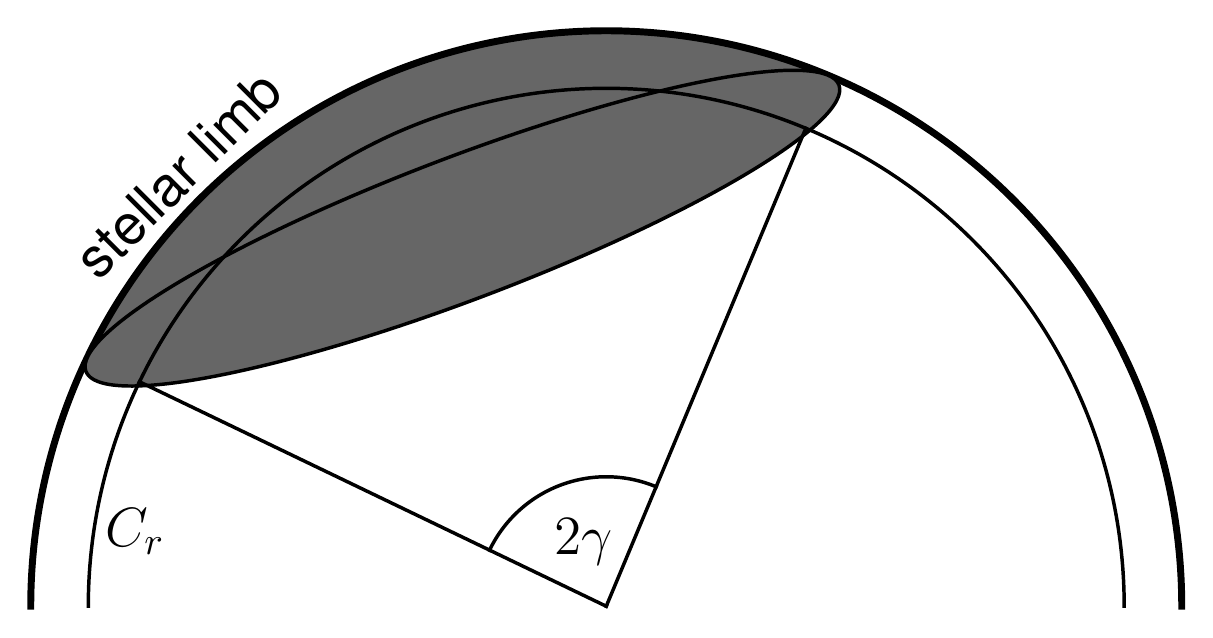}
\end{center}
\caption{Example of a spot partially hiding behind the stellar limb. Note how it is bounded partially by an arc of the ellipse that is its boundary in projection, and partially by an arc of the stellar limb.}
\label{fig:limbspot}
\end{figure}

However, when the spot partially hides behind the stellar limb, we will see it as a shape bounded by an arc of an ellipse (the projection of the spot boundary) and an arc of a circle (the edge of the stellar disk), as illustrated on Figure \ref{fig:limbspot}. We prove in the next section that in this case, the ellipse touches the stellar limb from the inside.

The two cases (the entire spot visible, or part of it is behind the limb) are delineated by a critical value of $\beta$, or equivalently, a critical value of $z$, both of which we expect to depend only on $a$. From Figure \ref{fig:sideview}, we can see that in this critical case, 
\begin{align*}
\beta_\mathrm{crit} &= \alpha \\
\zcrit &= \sqrt{1-a^2} \cos\beta_\mathrm{crit} = \sqrt{1-a^2} \cos\alpha \\
&= \sqrt{1-a^2} \cos\arcsin a = \sqrt{1-a^2} \sqrt{1-a^2} \\
&= 1-a^2,
\end{align*}
using Equation (\ref{eq:z}) to express $z$ in terms of $a$ and $\beta$.

If $z<\zcrit$, then the entire spot is visible. If $z=\zcrit$, then the ellipse touches the edge of the stellar limb at one point, as we prove in the next section. If $z>\zcrit$, then part of the spot is behind the stellar limb. Note that since we describe the spot with the parameter $z$ (implicity, through $x_s$ and $y_s$), $z=1$ corresponding to the spot center being on the stellar limb, therefore we cannot deal with spots that are partially visible, but their center is behind the limb. Such spots, however, will have a small contribution to $F_s$ and $F_\mathrm{transit}$, because only a very small part of them is visible, furthermore this small part is on the limb, which is usually darker to start with.  Furthermore, since $F_s\approx F_\mathrm{transit}$ as long as $R_\mathrm p\ll1$, omitting such a spot will have a very small effect on $F_\mathrm{normalized}$.

\subsection{Number of intersection points}
\label{sec:spotrod:number}

In this section, we prove that as long as $z<\zcrit$, the spot boundary and $C_r$ can have at most two intersection points. We also explain how to correctly model the ellipse in terms of $\gamma$ if $z>\zcrit$ and there are four intersection points. Finally, we prove that if $z=\zcrit$, then the spot boundary ellipse touches the stellar limb at one point, and if $z>\zcrit$, then at two points. 

Let us express $x^2$ from Equation (\ref{eq:quad1}) and substitute into Equation (\ref{eq:quad2}). This results in a quadratic equation for $y$:
\begin{align}
\label{eq:quadratic}
0 &= A y^2 + B y + C \\
\nonumber
A &= \frac{a^2}{b^2} - 1 \\
\nonumber
B &= 2z \\
\nonumber
C &= r^2 - a^2 - z^2 \\
\nonumber
y_\pm &= \frac{-z \pm \sqrt{z^2 - \left(\frac{a^2}{b^2}-1\right)\left(r^2-a^2-z^2\right)}}{\frac{a^2}{b^2}-1}.
\end{align}
Equation (\ref{eq:quad1}) tells us that each solution $y$ represents two intersection points if $|y|<b$, one if $|y|=b$, or none if $|y|>b$.
Now let us consider the following inequality:
\begin{align}
\label{ineq:proveme}
y_- &< -b \\
\frac{-z - \sqrt{z^2 - \left(\frac{a^2}{b^2}-1\right)\left(r^2-a^2-z^2\right)}}{\frac{a^2}{b^2}-1} &< -b.
\end{align}
This inequality is a sufficient condition for that $y_-$ does not represent real intersection points, that is, there are at most two intersection points (corresponding to $y_+$) 
Now we increase the left hand side of Inequality (\ref{ineq:proveme}). This will make it sharper, leading to a more restrictive, therefore still sufficient (but not necessary) condition. If we find at the end that this still holds whenever $z<\zcrit$, that proves our original statement: 
\begin{align}
\label{ineq:root}
\frac{-z}{\frac{a^2}{b^2}-1} &< -b.
\end{align}
Assume that $z>0$, in which case we also have $a>b$ and thus $A>0$. Then multiplying by $\frac{Ab}{az}$ does not change the direction of inequality:
\begin{align}
\nonumber
-\frac ba &< -\frac{b^2}{az} \left(\frac{a^2}{b^2}-1\right) \\
\label{ineq:samecrit}
\frac az - \frac{b^2}{az} &< \frac ba.
\end{align}
Substituting $b$ from Equation (\ref{eq:b1}), we get:
\begin{align}
\nonumber
\frac az - \frac az \left(1 - \frac{z^2}{1-a^2}\right) &< \sqrt{1 - \frac{z^2}{1-a^2}} \\
\label{ineq:gettingthere}
\frac{az}{1-a^2} &< \sqrt{1 - \frac{z^2}{1-a^2}}.
\end{align}
Both sides of Inequality (\ref{ineq:gettingthere}) are positive, therefore we can square them to get an equivalent inequality:
\begin{align}
\nonumber
\frac{a^2z^2}{\left(1-a^2\right)^2} &< 1 - \frac{z^2}{1-a^2} \\
\nonumber
a^2z^2 &< (1 - a^2)^2 - (1-a^2) z^2 \\
\nonumber
z^2 &< \left(1-a^2\right)^2 \\
\nonumber
z &< 1-a^2 \\
\label{ineq:final}
z &< \zcrit.
\end{align}
We are also allowed to multiply by $(1-a^2)^2$, which has to be positive. Finally, we arrive exactly at the critical value of $z$ that we have already established.

This derivation shows that if Inequality (\ref{ineq:final}) holds, then so does Inequality (\ref{ineq:proveme}). That is, if the spot is entirely visible, then there are at most two intersection points.

To understand the dependence of the number of intersection points on $r$, we plot $y_\pm$ as a function of $r$ on Figure \ref{fig:ypm}. So far we have proven that if $z<\zcrit$, then $y_-<-b$ for all values of $r$, which case is illustrated on the top panels. We have one intersection point if and only if $|y_+|=b$, which happens at $r=z\pm b$, when $C_r$ touches the ellipse. 

\begin{figure}
\begin{center}
\includegraphics*[width=\figurewidth]{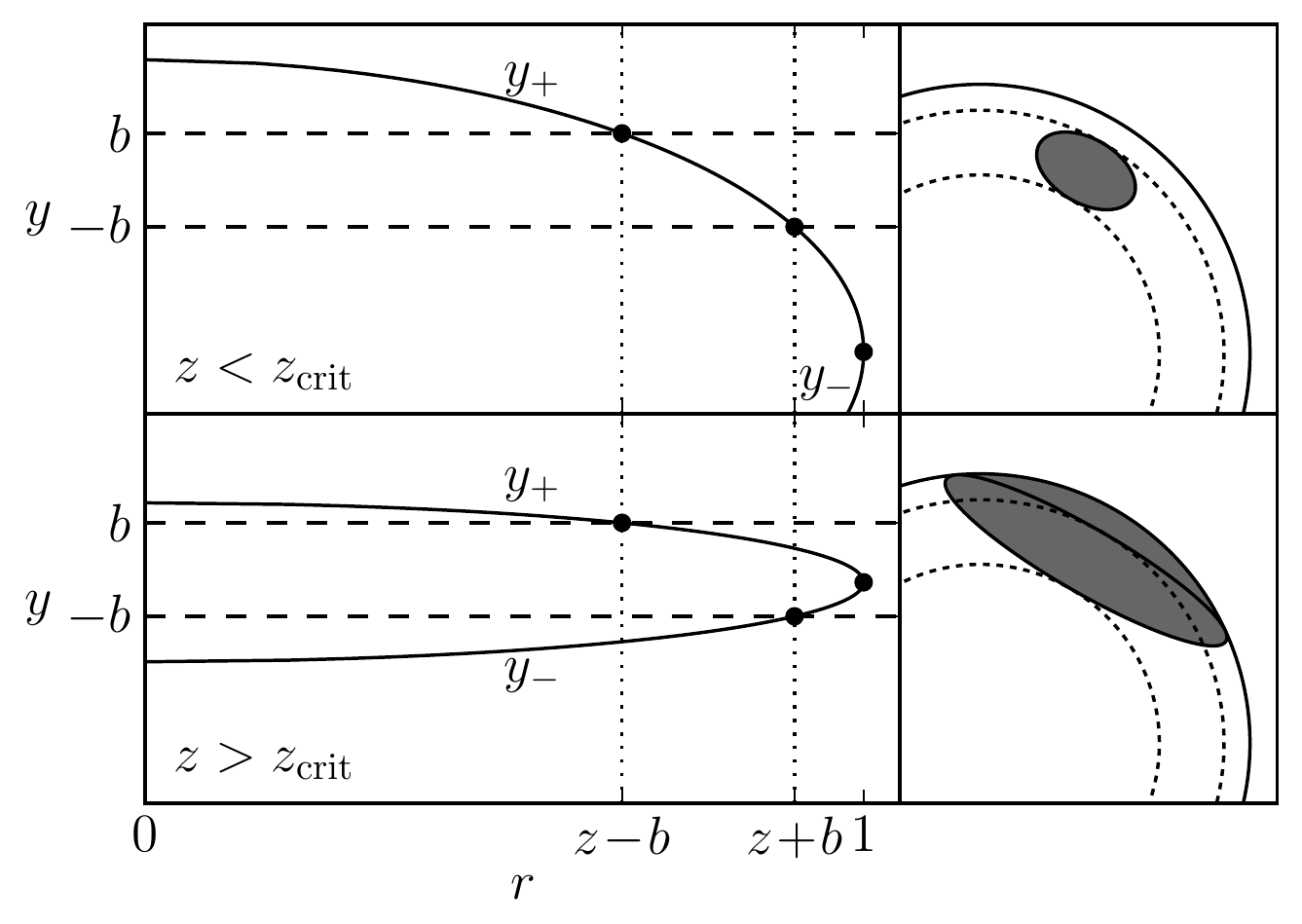}
\end{center}
\caption{Left panels: $y_\pm$ as a function of $r$. The upper branch of the parabola is $y_+$, the lower branch is $y_-$. Horizontal dashed lines are drawn at $y=\pm b$, vertical dotted lines at $r=z\pm b$ where $|y|=b$. Right panels: spots in projection, with $C_{r_1}$ and $C_{r_2}$, the two circles that touch the projection of the boundary of the spots (not the boundary of their projection), in dotted lines. Top panels: $a=0.2$, $\zcrit=0.96$, the spot is entirely visible. Bottom panels: $a=0.6$, $\zcrit=0.64$, the spot is partially behind the limb. In both cases, $b=0.12, z=0.78, r_1=0.66, r_2=0.90$.}
\label{fig:ypm}
\end{figure}

In case $z>\zcrit$, let us investigate how to properly account for the spot shape. If $r<z-b$, then $C_r$ is disjoint from the spot. At $r=z-b$, the circle $C_r$ touches the ellipse, we have $y_+=b$, yielding $x=0$ as a multiple root for the intersection point. Further increasing $r$ will result in $y_+<b$, representing two distinct real solutions for $x$. When $r$ reaches $z+b$, $C_r$ will touch the ellipse from the inside at a point corresponding to $y_-=-b$, $x=0$: this is the third point of intersection.

This can happen if and only if the radius $R$ of the osculating circle at the endpoint of the semi-minor axis is larger than $r=z+b$. As a sanity check, let us investigate what it means in terms of $z$:
\begin{align*}
R &> r \\
\frac{a^2}b &> z + b.
\end{align*}
This inequality is $\frac{az}b$ times Inequality (\ref{ineq:samecrit}) with the inequality sign in the other direction. Inequality (\ref{ineq:samecrit}), in turn, is equivalent to Inequality (\ref{ineq:final}). That is, the condition on the curvature of the ellipse is equivalent to Inequality (\ref{ineq:final}) with the inequality sign in the other direction: $z>\zcrit$. This is consistent with our previous statements.


Further increasing $r$ will yield four distinct intersection points, with $-b<y_-<y_+<b$. This scenario is also demonstrated on Figure \ref{fig:limbspot}. By continuity, the outside pair of intersection points corresponds to $y_+$, because they exist ever since $r=z-b$. The inside pair only appeared at $r=z+b$, and therefore corresponds to $y_-$, which crossed $-b$ at the same value of $r$.

Our code always calculates $\gamma$ based on $y_+$. This means that the entire arc between the outside intersection points is considered to be part of the spot, which correctly describes the shape of the spot as the observer sees it. That is, the code gives the correct result even in case $z>\zcrit$.

Finally, if $r=1$, then the inside and outside pair of intersection points coincide according to Figures \ref{fig:limbspot} and \ref{fig:ypm}: the ellipse touches the stellar limb from the inside. Another way of saying this is $y_-=y_+$, which happens exactly if the discriminant of Equation (\ref{eq:quadratic}) is zero. We now prove this statement.
\begin{align*}
B^2 - 4AC &= 0 \\
4z^2 - 4 \left(\frac{a^2}{b^2} - 1\right) \left(r^2 - a^2 - z^2\right) &= 0 \\
z^2 - \left(\frac1{1 - \frac{z^2}{1-a^2}} - 1\right) \left(1 - a^2 - z^2\right) &= 0 \\
z^2 - \left(\frac{1 - a^2}{1 - a^2 - z^2} - 1\right) \left(1 - a^2 - z^2\right) &= 0 \\
z^2 - \frac{z^2}{1 - a^2 - z^2} \left(1 - a^2 - z^2\right) &= 0.
\end{align*}
The last equation trivially holds true, which proves that if $r=1$, then $y_- = y_+$. Note that this is true regardless of the value of $z$: if $z<\zcrit$, then $y_- = y_+ < -b$ (no real solution for $x$, no intersection points); if $z=\zcrit$, then $y_- = y_+ = -b$ (touching in a single point with $x=0$, quadruple root); and if $z<\zcrit$, then $y_- = y_+ > -b$ (touching at two points, multiple roots each, like on Figures \ref{fig:limbspot} and \ref{fig:ypm}), as seen from Ineqalities (\ref{ineq:root}--\ref{ineq:final}). This concludes our proof.

\renewcommand\thesection{\thechapter.\arabic{section}}

\chapter{Outlook}

\epigraph{Yes, the universe continues to evolve.}{Neil deGrasse Tyson (1998)}
\section{Inflated hot Jupiters}

It was clear from early observations of transiting exoplanets that many
hot Jupiters have much larger radii than expected from models.
It is, by the way, not a coincidence that many of the first known transiting
exoplanets belong to this population, because of the observation bias towards
deeper transits.  The first theoretical work investigating a possible
mechanism behind the inflated radius of hot Jupiters is by
\citet{2001ApJ...548..466B}.

More than a decade later with hundreds of hot Jupiters known to date, we still
do not have a perfect understanding of inflated hot Jupiters.
\citet{2013ApJ...768...14W} classify the proposed mechanisms into three
categories: incident stellar flux-driven mechanisms, tidal mechanisms, and
delayed contraction.  See the corresponding sections of
\citet{2010RPPh...73a6901B,2014arXiv1401.4738B} for a recent review of this
puzzle.

\section{Planet discoveries and observational biases}

As astrophysical instruments and exoplanet discovery methods improved,
different parts of the exoplanet parameter space were populated by discoveries.
The first few dozen known planets were discovered by ground-based radial
velocity and transit surveys, with a strong bias for short period, and large
mass or radius.  Radial velocity observations,  the major confirmation method
for transiting planets and a discovery method by itself, are difficult or
impossible to perform on faint or fast rotating stars.

Both of these biases were reshaped with hundreds of planets discovered by the
\kepler{} satellite: continuous observations over many years reveal planets
with long orbital periods, and transit timing variations can be used to
confirm planetary systems around stars that are unsuitable for radial velocity
measurements.  Consequently, we now know a lot of tightly packed systems like
the six planets of Kepler-11 \citep{2011Natur.470...53L}, as well as some
transiting planets with very long orbital periods, like Kepler-47c with
303.2 days \citep{2012Sci...337.1511O}, although the low geometrical transit
probability hides most long period planets from transiting surveys.

The next important exoplanet discovery mission is the Transiting Exoplanet
Survey Satellite (TESS), which is expected to discover more than 1000
transiting exoplanets \citep{2010AAS...21545006R},
and will therefore reshape the parameter space of known
exoplanets yet another time.  Since most TESS targets will be observed for
only a month, the orbital period bias will be different from that of
\kepler{}.  Also, because the entire sky will be surveyed, many planets will
be found around bright stars, making follow-up observations (ground-based
radial velocity measurements, or space-based transmission spectroscopy)
easier.

\section{HST observations of HAT-P-1}

In Chapter \ref{ch:hst}, we report on Hubble Space Telescope (HST)
observations of the occultation (secondary eclipse) of the hot Jupiter
HAT-P-1b.  Since then, two other papers on HST observations of the same target
appeared in the literature: \citet{2013MNRAS.435.3481W} use the Wide Field
Camera 3 instrument to observe in the wavelength range of 1.087--1.678
$\mu$m, and find a strong water absorption feature.
\citet{2014MNRAS.437...46N} observe with the Space Telescope Imaging
Spectrograph in wavelengths ranging from 0.29 to 1.027 $\mu$m, and find strong
sodium absorption.  Combining the two sets of observations, they suggest an
overabundance of sodium, and hypothesize the presence of an absorber at high
altitudes.  They exclude the presence of clouds in the atmosphere, consistent
with our findings in Chapter \ref{ch:hst}.

\section{James Webb Space Telescope and planetary atmospheres}

The James Webb Space Telescope (JWST) is an infrared-optimized space
observatory with a 6.6 meter aperture that will be orbiting the Sun--Earth
second Lagrange point \citep{2006SSRv..123..485G}.  It is currently scheduled
for launch in 2018.  Considered to be the scientific successor to the Hubble
Space Telescope and the Spitzer Space Telescope, it will be one of NASA's
largest space projects.

With its very large aperture, JWST will be able to provide unprecedented
spectroscopic observations of planetary atmospheres during transits and
occultations.  In addition to continuing the quest for understanding
temperature inversion \citep{2010ApJ...725..261M}, it will be able to detect
various molecules, most notably CO$_2$ and O$_3$, in planetary atmospheres
\citep{2009ApJ...698..519K, 2009PASP..121..952D, 2011A&A...525A..83B}.  Its
light gathering capacity will also allow extending currently performed
spectroscopic observations to fainter targets, therefore increasing the
number of well-characterized planets.

\section{\spotrod{} investigations of systems discovered by K2 and TESS}

As stated in Chapter \ref{ch:spotrod}, a number of planets transiting active
dwarfs are expected to be discovered by the K2 mission (the repurposed
\kepler{} satellite after the failure of its second reaction wheel) and the
TESS mission.  This will provide a rich dataset for spot-induced transit
lightcurve anomaly investigations with \spotrod{}.

In this context, it is important to remember the advantages and disadvantages
of \spotrod{}.  The simplifying assumptions make the model robust and fast for
fitting, potentially allowing us to draw statistical conclusions on spot
parameters.  It is conceivable that correlations between spot temperature and
stellar spectral type, or between spot temperature and size on a single star,
will be discovered.  On the other hand, we might find out that the circular,
homogeneous treatment of spots is too limiting for intepreting the high
quality data, or that it introduces artifacts in the results.



\begin{singlespacing}
  \renewcommand{\bibname}{References}

  \bibliographystyle{apj}
  \bibliography{t}
\end{singlespacing}


\end{document}